\documentclass[a4paper,twoside,twocolumn,headsepline]{scrartcl}

\pdfoutput=1

\usepackage[blocks,affil-sl]{authblk}
\usepackage{scrpage2}
\usepackage{typearea}
\usepackage{graphicx}
\usepackage{subfigure}
\usepackage{longtable}
\usepackage{balance}

\usepackage{color}
\definecolor{light-gray}{gray}{0.7}
\def\gray{\color{light-gray}}

\usepackage{hyperref}

\hypersetup{
  pdftitle = {SuperB Progress Reports: The Detector},
  colorlinks=true,
  breaklinks=true,
  linkcolor=blue,
  anchorcolor=blue,
  citecolor=blue,
  filecolor=blue,
  menucolor=blue,
  pagecolor=blue,
  urlcolor=blue,
  pdfpagelabels
}

\input  babarsym

\def\ze#1   {\ensuremath{\zeta_{#1}}\xspace}

\def\Hz     {\ensuremath{{\rm \,Hz}}\xspace}
\def\MHz    {\ensuremath{{\rm \,MHz}}\xspace}
\def\GHz    {\ensuremath{{\rm \,GHz}}\xspace}
\def\kHz    {\ensuremath{{\rm \,kHz}}\xspace}

\def\pepii    {\pep2 \xspace}

\def\CO2  {$\mathrm{CO}_2$\xspace}

\newcommand{\wpsec}[1]
{
\section{\texorpdfstring{\boldmath#1}{#1}}
\centerline{\rule[ 0.15in]{0.9\columnwidth}{2pt}}
\bigskip
}

\newcommand{\wpsubsec}[1]
{
\subsection{\texorpdfstring{\boldmath#1}{#1}}
}

\newcommand{\wpsubsubsec}[1]
{
\subsubsection{#1}
}

\newcommand{\wpparagraph}[1]
{
\paragraph{#1}
}

\areaset[12mm]{165mm}{240mm}
\ifx \footfont \undefined
\else
\renewcommand{\footfont}{\normalfont\rmfamily\scshape\large}
\fi

\begin{document}
\pagestyle{empty}

\setcounter{page}{-5}

{\onecolumn
\hfill INFN/AE\_10/4, LAL 10-115, SLAC-R-954
\begin{center}

{\boldmath{
\phantom{\Huge{I}}
\phantom{\Huge{I}}
\phantom{\Huge{I}}
\phantom{\Huge{I}}
\phantom{\Huge{I}}

{\Huge\bf
{\vskip 48pt
Super$B$\\ Progress Reports}
}

 \ \phantom{\Huge{I}}
   \phantom{\Huge{I}}
   \phantom{\Huge{I}}
   \phantom{\Huge{I}}\\ %
{\huge
{\vskip 48pt
\gray{Physics}}
} \\
{\huge
{\vskip 12pt
\gray{Accelerator}}
} \\
{\huge\bf
{\vskip 12pt
Detector}
} \\
{\huge
{\vskip 12pt
\gray{Computing}}
} \\
\phantom{\Huge{I}}
\phantom{\Huge{I}}
{\Large\bf June 30, 2010}\\
\phantom{\Huge{I}}
\phantom{\Huge{I}}
}}
\end{center}
\bigskip
}
{\begin{center}
\large \bf Abstract
\end{center}
This report describes the present status of the detector design for Super$B$. It  is one of four separate progress reports   that, taken collectively, describe progress made on the Super$B$ Project since the publication
of the Super$B$ Conceptual Design Report in 2007 and the Proceedings of Super$B$ Workshop VI in Valencia in 2008.
}

\vfill\clearpage
\newpage
\pagenumbering{roman}
\thispagestyle{empty}
\begin{center}\mbox{E.~Grauges,}\\*  {{\bf Universitat De Barcelona, } {\it Fac. Fisica. Dept. ECM
Barcelona E-08028, Spain}}\end{center}
\begin{center}\mbox{G.~Donvito,}
\mbox{V.~Spinoso}\\*  {{\bf INFN Bari }{\it and }{\bf Universit\`{a} di Bari,} {\it Dipartimento di Fisica, I-70126 Bari, Italy}}\end{center}
\begin{center}\mbox{M.~Manghisoni,}
\mbox{V.~Re,}
\mbox{G.~Traversi}\\*  {{\bf INFN Pavia }{\it and }{\bf Universit\`{a} di Bergamo}  {\it Dipartimento di Ingegneria Industriale, I-24129 Bergamo, Italy}}\end{center}
\begin{center}\mbox{G.~Eigen,}
\mbox{D.~Fehlker,}
\mbox{L.~Helleve}\\*  {{\bf University of Bergen,} {\it Institute of Physics, N-5007 Bergen, Norway}}\end{center}
\begin{center}\mbox{A.~Carbone,}
\mbox{R.~Di Sipio,}
\mbox{A.~Gabrielli,}
\mbox{D.~Galli,}
\mbox{F.~Giorgi,}
\mbox{U.~Marconi,}
\mbox{S.~Perazzini,}
\mbox{C.~Sbarra,}
\mbox{V.~Vagnoni,}
\mbox{S.~Valentinetti,}
\mbox{M.~Villa,}
\mbox{A.~Zoccoli}\\*  {{\bf INFN Bologna }{\it and }{\bf Universit\`{a} di Bologna,} {\it Dipartimento di Fisica, I-40127 Bologna, Italy}}\end{center}
\begin{center}\mbox{C.~Cheng,}
\mbox{A.~Chivukula,}
\mbox{D.~Doll,}
\mbox{B.~Echenard,}
\mbox{D.~Hitlin,}
\mbox{P.~Ongmongkolkul,}
\mbox{F.~Porter,}
\mbox{A.~Rakitin,}
\mbox{M.~Thomas,}
\mbox{R.~Zhu}\\*  {{\bf California Institute of Technology,} {\it Pasadena, California 91125, USA}}\end{center}
\begin{center}\mbox{G.~Tatishvili}\\*  {{\bf Carleton University,} {\it Ottawa, Ontario, Canada K1S 5B6}}\end{center}
\begin{center}\mbox{R.~Andreassen,}
\mbox{C.~Fabby,}
\mbox{B.~Meadows,}
\mbox{A.~Simpson,}
\mbox{M.~Sokoloff,}
\mbox{K.~Tomko}\\*  {{\bf University of Cincinnati,} {\it Cincinnati, Ohio 45221, USA}}\end{center}
\begin{center}\mbox{A.~Fella}\\*  {{\bf INFN CNAF} {\it I-40127 Bologna, Italy}}\end{center}
\begin{center}\mbox{M.~Andreotti,}
\mbox{W.~Baldini,}
\mbox{R.~Calabrese,}
\mbox{V.~Carassiti,}
\mbox{G.~Cibinetto,}
\mbox{A.~Cotta Ramusino,}
\mbox{A.~Gianoli,}
\mbox{E.~Luppi,}
\mbox{M.~Munerato,}
\mbox{V.~Santoro,}
\mbox{L.~Tomassetti}\\*  {{\bf INFN Ferrara }{\it and }{\bf Universit\`{a} di Ferrara,} {\it Dipartimento di Fisica, I-44100 Ferrara, Italy}}\end{center}
\begin{center}\mbox{D.~Stoker}\\*  {{\bf University of California, Irvine} {\it Irvine, California 92697, USA}}\end{center}
\begin{center}\mbox{O.~Bezshyyko,}
\mbox{G.~Dolinska}\\*  {{\bf Taras Shevchenko National University of Kyiv} {\it Kyiv, 01601, Ukraine }}\end{center}
\begin{center}\mbox{N.~Arnaud,}
\mbox{C.~Beigbeder,}
\mbox{F.~Bogard,}
\mbox{D.~Breton,}
\mbox{L.~Burmistrov,}
\mbox{D.~Charlet,}
\mbox{J.~Maalmi,}
\mbox{L.~Perez Perez,}
\mbox{V.~Puill,}
\mbox{A.~Stocchi,}
\mbox{V.~Tocut,}
\mbox{S.~Wallon,}
\mbox{G.~Wormser}\\*  {{\bf Laboratoire de l'Acc\'{e}l\'{e}rateur Lin\'{e}aire,} {\it IN2P3/CNRS, Universit\'{e} Paris-Sud 11, F-91898 Orsay, France}}\end{center}
\begin{center}\mbox{D.~Brown}\\*  {{\bf Lawrence Berkeley National Laboratory, } {\it University of California, Berkeley, California 94720, USA}}\end{center}
\begin{center}\mbox{A.~Calcaterra,}
\mbox{R.~de Sangro,}
\mbox{G.~Felici,}
\mbox{G.~Finocchiaro,}
\mbox{P.~Patteri,}
\mbox{I.~Peruzzi,}
\mbox{M.~Piccolo,}
\mbox{M.~Rama}\\*  {{\bf Laboratori Nazionali di Frascati dell'INFN,} {\it I-00044 Frascati, Italy}}\end{center}
\begin{center}\mbox{S.~Fantinel,}
\mbox{G.~Maron}\\*  {{\bf Laboratori Nazionali di Legnaro dell'INFN,} {\it I-35020 Legnaro, Italy}}\end{center}
\begin{center}\mbox{E.~Ben-Haim,}
\mbox{G.~Calderini,}
\mbox{H.~Lebbolo,}
\mbox{G.~Marchiori}\\*  {{\bf Laboratoire de Physique Nucl\'{e}aire et de Hautes Energies,} {\it IN2P3/CNRS, Universit\'{e} Pierre et Marie Curie-Paris 6, F-75005 Paris, France }}\end{center}
\begin{center}\mbox{R.~Cenci,}
\mbox{A.~Jawahery,}
\mbox{D.A.~Roberts}\\*  {{\bf University of Maryland,} {\it College Park, Maryland 20742, USA}}\end{center}
\begin{center}\mbox{D.~Lindemann,}
\mbox{P.~Patel,}
\mbox{S.~Robertson,}
\mbox{D.~Swersky}\\*  {{\bf McGill University,} {\it Montr\'{e}al, Qu\'{e}bec, Canada H3A 2T8}}\end{center}
\begin{center}\mbox{P.~Biassoni,}
\mbox{M.~Citterio,}
\mbox{V.~Liberali,}
\mbox{F.~Palombo,}
\mbox{A.~Stabile,}
\mbox{S.~Stracka}\\*  {{\bf INFN Milano }{\it and }{\bf Universit\`{a} di Milano,} {\it Dipartimento di Fisica, I-20133 Milano, Italy}}\end{center}
\begin{center}\mbox{A.~Aloisio,}
\mbox{S.~Cavaliere,}
\mbox{G.~De Nardo,}
\mbox{A.~Doria,}
\mbox{R.~Giordano,}
\mbox{A.~Ordine,}
\mbox{S.~Pardi,}
\mbox{G.~Russo,}
\mbox{C.~Sciacca}\\*  {{\bf INFN Napoli }{\it and }{\bf Universit\`{a} di Napoli Federico II,} {\it Dipartimento di Scienze Fisiche, I-80126, Napoli, Italy}}\end{center}
\begin{center}\mbox{A.Y.~Barniakov,}
\mbox{M.Y.~Barniakov,}
\mbox{V.E.~Blinov,}
\mbox{V.P.~Druzhinin,}
\mbox{V.B..~Golubev,}
\mbox{S.A.~Kononov,}
\mbox{E.~Kravchenko,}
\mbox{A.P.~Onuchin,}
\mbox{S.I.~Serednyakov,}
\mbox{Y.I.~Skovpen,}
\mbox{E.P.~Solodov}\\*  {{\bf Budker Institute of Nuclear Physics,} {\it Novosibirsk 630090, Russia}}\end{center}
\begin{center}\mbox{M.~Bellato,}
\mbox{M.~Benettoni,}
\mbox{M.~Corvo,}
\mbox{A.~Crescente,}
\mbox{F.~Dal Corso,}
\mbox{C.~Fanin,}
\mbox{E.~Feltresi,}
\mbox{N.~Gagliardi,}
\mbox{M.~Morandin,}
\mbox{M.~Posocco,}
\mbox{M.~Rotondo,}
\mbox{R.~Stroili}\\*  {{\bf INFN Padova }{\it and }{\bf Universit\`{a} di Padova, } {\it Dipartimento di Fisica, I-35131 Padova, Italy}}\end{center}
\begin{center}\mbox{C.~Andreoli,}
\mbox{L.~Gaioni,}
\mbox{E.~Pozzati,}
\mbox{L.~Ratti,}
\mbox{V.~Speziali}\\*  {{\bf INFN Pavia }{\it and }{\bf Universit\`{a} di Pavia,} {\it Dipartimento di Elettronica, I-27100 Pavia, Italy}}\end{center}
\begin{center}\mbox{D.~Aisa,}
\mbox{M.~Bizzarri,}
\mbox{C.~Cecchi,}
\mbox{S.~Germani,}
\mbox{P.~Lubrano,}
\mbox{E.~Manoni,}
\mbox{A.~Papi,}
\mbox{A.~Piluso ,}
\mbox{A.~Rossi}\\*  {{\bf INFN Perugia }{\it and }{\bf Universit\`{a} di Perugia,} {\it Dipartimento di Fisica, I-06123 Perugia, Italy}}\end{center}
\begin{center}\mbox{M.~Lebeau}\\*  {{\bf INFN Perugia,} {\it I-06123 Perugia, Italy, and} \\  {\bf California Institute of Technology,} {\it Pasadena, California 91125, USA}}\end{center}
\begin{center}\mbox{C.~Avanzini,}
\mbox{G.~Batignani,}
\mbox{S.~Bettarini,}
\mbox{F.~Bosi,}
\mbox{M.~Ceccanti,}
\mbox{A.~Cervelli,}
\mbox{A.~Ciampa,}
\mbox{F.~Crescioli,}
\mbox{M.~Dell'Orso,}
\mbox{D.~Fabiani,}
\mbox{F.~Forti,}
\mbox{P.~Giannetti,}
\mbox{M.~Giorgi,}
\mbox{S.~Gregucci,}
\mbox{A.~Lusiani,}
\mbox{P.~Mammini,}
\mbox{G.~Marchiori,}
\mbox{M.~Massa,}
\mbox{E.~Mazzoni,}
\mbox{F.~Morsani,}
\mbox{N.~Neri,}
\mbox{E.~Paoloni,}
\mbox{E.~Paoloni,}
\mbox{M.~Piendibene,}
\mbox{A.~Profeti,}
\mbox{G.~Rizzo,}
\mbox{L.~Sartori,}
\mbox{J.~Walsh,}
\mbox{E.~Yurtsev}\\*  {{\bf INFN Pisa, Universit\`{a} di Pisa,} {\it Dipartimento di Fisica, and} {\bf Scuola Normale Superiore,} {\it I-56127 Pisa, Italy}}\end{center}
\begin{center}\mbox{D.M.~Asner,}
\mbox{J. E.~ Fast,}
\mbox{R.T.~ Kouzes,}\\*  {{\bf Pacific Northwest National Laboratory,} {\it Richland, Washington 99352, USA}}\end{center}
\begin{center}\mbox{A.~Bevan,}
\mbox{F.~Gannaway,}
\mbox{J.~Mistry,}
\mbox{C.~Walker}\\*  {{\bf Queen Mary,} {\it University of London, London E1 4NS, United Kingdom}}\end{center}
\begin{center}\mbox{C.A.J.~Brew,}
\mbox{R.E.~Coath,}
\mbox{J.P.~Crooks,}
\mbox{R.M.~Harper,}
\mbox{A.~Lintern,}
\mbox{A.~Nichols,}
\mbox{M.~Staniztki,}
\mbox{R.~Turchetta,}
\mbox{F.F.~Wilson}\\*  {{\bf Rutherford Appleton Laboratory,} {\it Chilton, Didcot, Oxon, OX11 0QX, United Kingdom}}\end{center}
\begin{center}\mbox{V.~Bocci,}
\mbox{G.~Chiodi,}
\mbox{R.~Faccini,}
\mbox{C.~Gargiulo,}
\mbox{D.~Pinci,}
\mbox{L.~Recchia,}
\mbox{D.~Ruggieri}\\*  {{\bf INFN Roma }{\it and }{\bf Universit\`{a} di Roma La Sapienza,} {\it Dipartimento di Fisica, I-00185 Roma, Italy}}\end{center}
\begin{center}\mbox{A.~Di Simone}\\*  {{\bf INFN Roma Tor Vergata }{\it and }{\bf Universit\`{a} di Roma Tor Vergata,} {\it Dipartimento di Fisica, I-00133 Roma, Italy}}\end{center}
\begin{center}\mbox{P.~Branchini,}
\mbox{A.~Passeri,}
\mbox{F.~Ruggieri,}
\mbox{E.~Spiriti}\\*  {{\bf INFN Roma Tre }{\it and }{\bf Universit\`{a} di Roma Tre,} {\it Dipartimento di Fisica, I-00154 Roma, Italy}}\end{center}
\begin{center}\mbox{D.~Aston,}
\mbox{M.~Convery,}
\mbox{G.~Dubois-Felsmann,}
\mbox{W.~Dunwoodie,}
\mbox{M.~Kelsey,}
\mbox{P.~Kim,}
\mbox{M.~Kocian,}
\mbox{D.~Leith,}
\mbox{S.~Luitz,}
\mbox{D.~MacFarlane,}
\mbox{B.~Ratcliff,}
\mbox{M.~Sullivan,}
\mbox{J.~Va'vra,}
\mbox{W.~Wisniewski,}
\mbox{W.~Yang}\\*  {{\bf SLAC National Accelerator Laboratory} {\it  Stanford, California 94309, USA}}\end{center}
\begin{center}\mbox{K.~Shougaev,}
\mbox{A.~Soffer}\\*  {{\bf School of Physics and Astronomy, Tel Aviv University} {\it Tel Aviv 69978, Israel}}\end{center}
\begin{center}\mbox{F.~Bianchi,}
\mbox{D.~Gamba,}
\mbox{G.~Giraudo,}
\mbox{P.~Mereu}\\*  {{\bf INFN Torino }{\it and }{\bf Universit\`{a} di Torino,} {\it Dipartimento di Fisica Sperimentale, I-10125 Torino, Italy}}\end{center}
\begin{center}\mbox{G.~Dalla Betta,}
\mbox{G.~Fontana,}
\mbox{G.~Soncini}\\*  {{\bf INFN Padova }{\it and }{\bf Universit\`a di Trento,} {\it ICT Department, I-38050 Trento, Italy}}\end{center}
\begin{center}\mbox{M.~Bomben,}
\mbox{L.~Bosisio,}
\mbox{P.~Cristaudo,}
\mbox{G.~Giacomini,}
\mbox{D.~Jugovaz,}
\mbox{L.~Lanceri,}
\mbox{I.~Rashevskaya,}
\mbox{G.~Venier,}
\mbox{L.~Vitale}\\*  {{\bf INFN Trieste}{\it and }{\bf Universit\`a di Trieste,} {\it Dipartimento di Fisica, I-34127 Trieste, Italy}}\end{center}
\begin{center}\mbox{R.~Henderson}\\*  {{\bf TRIUMF} {\it Vancouver, British Columbia, Canada V6T 2A3}}\end{center}
\begin{center}\mbox{J.-F.~Caron,}
\mbox{C.~Hearty,}
\mbox{P.~Lu,}
\mbox{R.~So}\\*  {{\bf University of British Columbia,} {\it Vancouver, British Columbia, Canada V6T 1Z1}}\end{center}
\begin{center}\mbox{P.~Taras}\\*  {{\bf Universit\'{e} de Montre\'{a}l,} {\it Physique des Particules, Montre\'{a}l, Qu\'{e}bec, Canada H3C 3J7}}\end{center}
\begin{center}\mbox{A.~Agarwal,}
\mbox{J.~Franta,}
\mbox{J.M.~Roney}\\*  {{\bf University of Victoria,} {\it Victoria, British Columbia, Canada V8W 3P6}}\end{center}

\newpage

\let\origmathversion=\mathversion
\renewcommand{\mathversion}[1]{}
\tableofcontents
\renewcommand{\mathversion}{\origmathversion}

\twocolumn
\automark[subsection]{section}
\ohead{\pagemark}
\ihead{\headmark}
\ofoot{Super{\large\it B} Detector Progress Report}

\pagestyle{scrheadings}

\pagenumbering{arabic}

\newcommand{\begsec}{\clearpage}
\newcommand{\aftsec}{\relax}

\balance

\begsec
\graphicspath{{Overview/}{Overview/}}
\wpsec{Introduction}

\wpsubsec{The Physics Motivation}
The Standard Model successfully explains the wide variety of
experimental data that has been gathered over several decades with
energies ranging from under a \gev up to several hundred \gev. At the
start of the millennium, the flavor sector was perhaps less explored
than the gauge sector, but the PEP-II and KEK-B asymmetric B Factories,
and their associated experiments \babar\ and Belle, have produced a wealth
of important flavor physics highlights during the past decade ~\cite{pdg_2008}.  The most
notable experimental objective, the establishment of the
Cabibbo-Kobayashi-Maskawa phase as consistent with experimentally
observed CP-violating asymmetries in $B$ meson decay, was cited in the
award of the 2008 Nobel Prize to Kobayashi \& Maskawa~\cite{nobel_2008}.

The B Factories have provided a set of unique, over-constrained tests of
the Unitarity Triangle. These have, in the main, been found to be
consistent with Standard Model predictions. The B factories have done
far more physics than originally envisioned; \babar\ alone has published
more than 400 papers in refereed journals to date.  Measurements of all
three angles of the Unitarity Triangle -- $\alpha$ and $\gamma$, in
addition to sin $2\beta$; the establishment of  $D^0\bar{D^0}$ mixing;
the uncovering of intriguing clues for potential New Physics in  $B
\rightarrow K^{(\star)} l^+l^-$ and $B \rightarrow K\pi$ 
decays; and unveiling an entirely unexpected new spectroscopy, are some
examples of important experimental results beyond those initially
contemplated.

With the LHC now beginning operations, the major experimental
discoveries of the next few years will probably be at the energy
frontier, where we would hope not only to complete the Standard Model by
observing the Higgs particle, but to find  signals of New Physics which
are widely expected to lie around the 1\tev energy scale. If found, the
New Physics phenomena  will need data from very sensitive heavy flavor
experiments if they are to be understood in detail. Determining the
flavor structure of the New Physics involved requires the information on rare b, c
and $\tau$ decays, and on CP violation in b and c quark decays that only
a very high luminosity asymmetric B Factory can provide~\cite{superbphysics_2008}. On the other
hand,  if such signatures of New Physics are not observed at the LHC,
then the excellent sensitivity provided at the luminosity frontier by
a next generation super B-factory provides another avenue to observing New Physics at mass scales
up to 10\tev or more through observation of rare processes involving $B$
and $D$ mesons and studies of lepton flavour violation (LFV) in $\tau$ decays.

\wpsubsec{The \superb\ Project Elements}
It is generally agreed that the physics being addressed by a next-generation B 
factory requires a data sample that is some 50---100 times larger than the
existing combined sample from \babar\ and Belle, or at least 50---75
\invab. Acquiring such an integrated luminosity in a  5 year time
frame requires that the collider run at a luminosity of at least
$10^{36} \cm^{-2}\rm{s}^{-1}$.

For a number of years, an Italian led, INFN hosted, collaboration of
scientists from Canada, Italy,  Israel, France, Norway, Spain, Poland,
UK and USA have worked together to design and  propose a high
luminosity $10^{36}$ asymmetric B Factory project, called \superb, to be
built at or near the Frascati laboratory~\cite{cdr_2007}. The project, which is managed
by a project board, includes divisions for the accelerator, the
detector, the computing, and the site \& facilities.

The accelerator portion of the project employs lessons learned  from
modern low-emittance synchrotron light sources and ILC/CLIC R\&D, and an
innovative new idea for the intersection region of the storage rings~\cite{panta_2006}, called crab waist, to reach luminosities over 50
times greater than  those obtained by earlier B factories at KEK and
SLAC. There is now an attractive, cost-effective accelerator design,
including polarized beams, which is being further refined and optimized~\cite{DSR_acc_2010}. It is designed to incorporate many PEP-II
components. This facility promises to deliver fundamental
discovery-level science at the luminosity frontier.

There is also an active international proto-collaboration working
effectively on the design of the detector. The detector team draws
heavily on its deep experience with the \babar\ detector, which has
performed in an outstanding manner both in terms of scientific
productivity and operational efficiency.  \babar\ serves as the foundation
of the design of the \superb\ detector.

To date, the \superb\ project has been very favorably reviewed by several
international committees. This international community now awaits a
decision by the Italian government on its support of the  project.

\wpsubsec{The Detector Design Progress Report}
This document describes the design and  development of the \superb\ 
detector, which is based on a major upgrade of  \babar. This is one of
several descriptive ``Design Progress  Reports (DPR)" being produced by the \superb\ project
during the first part of 2010 to motivate and summarize the development,
and present the status of each major division of the project (Physics,
Accelerator, Detector, and  Computing) so as to present a snapshot of
the entire project at an intermediate stage between the CDR, which was
written in 2007, and the TDR that is being developed during the next
year.

This ``Detector DPR'' begins with a brief overview of  the
detector design,  the challenges involved in detector operations at the
luminosity frontier,  the approach being taken to optimize the remaining
general design choices, and the R\&D program that is underway to develop
and validate the system and subsystem designs. Each of the detector
subsystems and the general detector systems are then described in more
detail, followed by a description of the integration and assembly of the
full detector. Finally, the paper concludes  with a discussion  of
detector costs and a schedule overview.

\aftsec

\begsec
\graphicspath{{Overview/}{Overview/}}
\wpsec{Overview}

The \superb\ detector concept is based on the \babar\ detector, with
those modifications required to operate at a luminosity of $10^{36}$
or more, and with a reduced center-of-mass boost.  Further
improvements needed to cope with higher beam-beam and other
beam-related backgrounds, as well as to improve detector hermeticity
and performance, are also discussed, as is the necessary R\&D required
to implement this upgrade.  Cost estimates and the schedule are
described in Section\ref{sec:Budget_and_Schedule}.

The current \babar\ detector consists of a tracking system with a five
layer double-sided silicon strip vertex tracker (SVT) and a 40 layer
drift chamber (DCH) inside a 1.5T magnetic field, a Cherenkov detector
with fused silica bar radiators (DIRC), an electromagnetic calorimeter
(EMC) consisting of 6580 CsI(Tl) crystals and an instrumented
flux-return (IFR) comprised of both limited streamer tube (LST) and
resistive plate chamber (RPC) detectors for $\KL$ detection and $\mu$
identification.

The \superb\ detector concept reuses a number of components from
\babar: the flux-return steel, the superconducting coil, the barrel of
the EMC and the fused silica bars of the DIRC. The flux-return will be
augmented with additional absorber to increase the number of
interaction lengths for muons to roughly $7\lambda$.  The DIRC camera
will be replaced by a twelve-fold modular camera using multi-channel plate (MCP) photon detectors in a
focusing configuration using fused silica optics to reduce the impact
of beam related backgrounds and improve performance. The forward EMC
will feature cerium-doped LYSO
(lutetium yttrium orthosilicate) crystals, which have a much shorter scintillation time
constant, a lower Moli\`ere radius and better radiation hardness than
the current CsI(Tl) crystals, again for reduced sensitivity to beam
backgrounds and better position resolution.
\begin{figure*}[tbh]
 \begin{center}
 \includegraphics[width=0.9\textwidth]{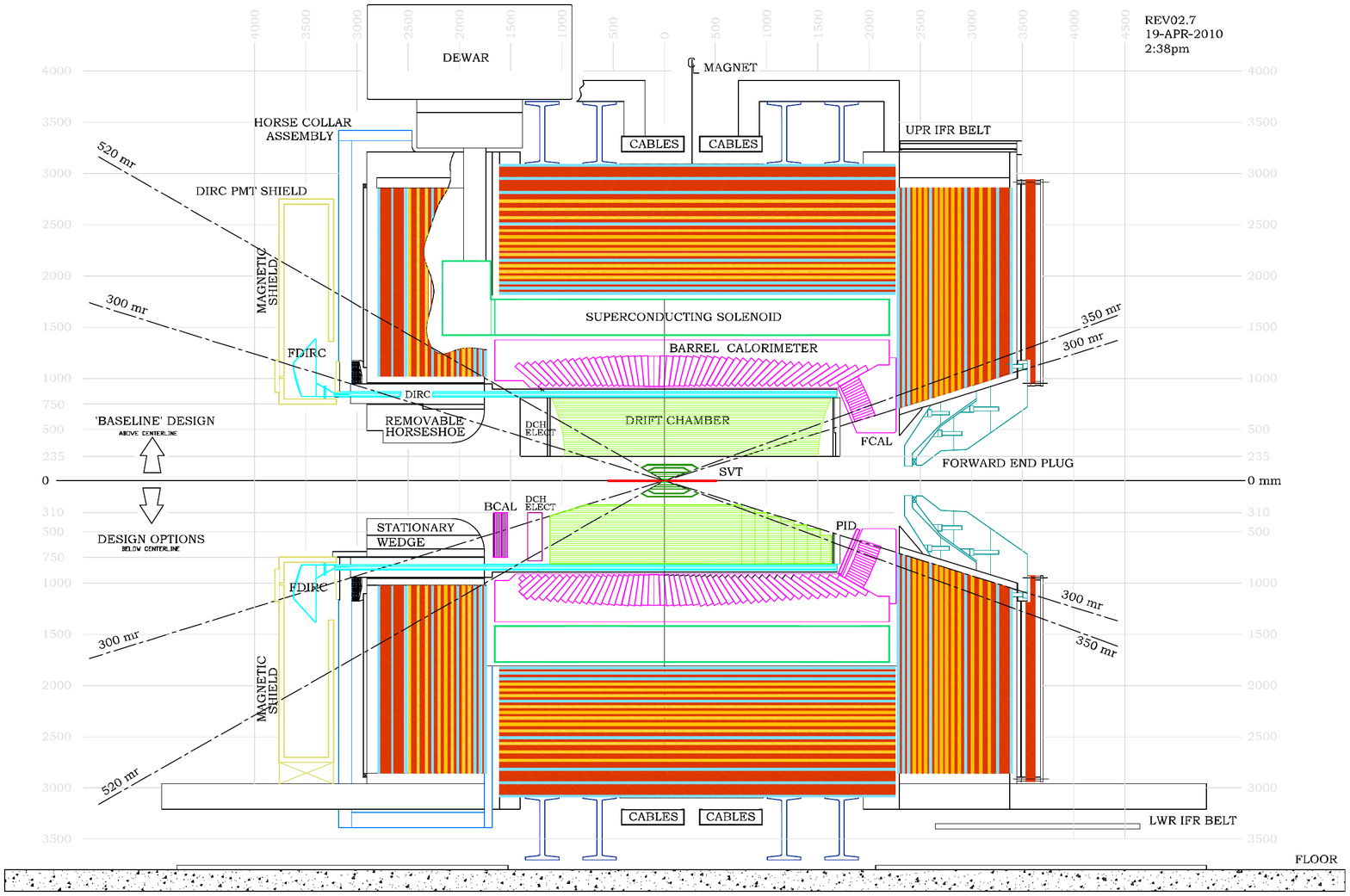}
 \vspace*{3mm}
 \caption{Concept for the \superb\ detector.  The upper half shows the
 baseline concept, and the bottom half adds a number of optional detector configurations.}
 \label{fig:det:superb}
 \end{center}
\end{figure*}

The tracking detectors for \superb\ will be new. The current SVT
cannot operate at $\mathcal{L} = 10^{36}$, and the DCH has reached the
end of its design lifetime and must be replaced. To maintain
sufficient proper-time difference ($\deltat$) resolution for time-dependent \CP violation
measurements with the \superb\ boost of $\beta\gamma=0.24$, the vertex
resolution will be improved by reducing the radius of the beam pipe,
placing the innermost layer of the SVT at a radius of roughly
$1.2\cm$.  This innermost layer of the SVT will be constructed of
either silicon striplets or Monolithic Active Pixel Sensors (MAPS)
 or other pixelated sensors, depending
on the estimated occupancy from beam-related backgrounds.  Likewise,
the design of the cell size and geometry of the DCH will be driven by occupancy
considerations. The hermeticity of the \superb\ detector, and, thus, its
performance for certain physics channels will be improved by including
a backwards ``veto-quality" EMC detector comprising a lead-scintillator
stack. The physics benefit  from the inclusion of a forward PID remains under study. 
The baseline design concept is a fast Cherenkov light based time-of-flight system.

The \superb\ detector concept is shown in Fig.~\ref{fig:det:superb}.,
The top portion of this elevation view shows the minimal set of new
detector components, with substantial reuse of elements of the current \babar\ detector; the bottom half shows a configuration with additional new components
that would  cope with higher beam backgrounds and achieve greater
hermeticity.

\wpsubsec{Physics Performance}
\label{subsec:Overview_Perf}
 The \superb\ detector design, as described in the Conceptual Design 
Report~\cite{ref:CDR},  left open a number of issues that have a large impact 
on the overall detector geometry. These include the physics impact of
of a PID device in front of the forward EMC; the need for an EMC in the 
backward region; the position of the innermost layer of the SVT; the SVT
internal geometry and the SVT-DCH 
transition radius; and the amount and distribution of absorber in the IFR. 

 These issues have been addressed by evaluating the performance of different detector configurations
 in reconstructing  charged and neutral particles as well as the overall sensitivity 
 of each configuration to a  set of benchmark decay channels. To accomplish this 
 task,  a  fast simulation code specifically developed for the \superb\ detector has 
 been used (see Section\ref{sec:computing}), combined with a complete set of analysis tools 
 inherited,  for the most part,  from the \babar\ experiment.
 Geant4-based code has been used to simulate the primary sources of  backgrounds -- including both machine-induced 
 and physics processes -- in order 
 to estimate the rates and occupancies of various sub-detectors 
 as a function of position.
The main results from these ongoing studies are summarized in this
section.

Time-dependent measurements  are an important part of the \superb\ physics
program.  In order to achieve a $\deltat$ resolution comparable
to that at  \babar, the reduced boost at \superb\ 
must be compensated by improving the vertex resolution. This requires a thin beam pipe plus SVT Layer0  that is placed as close as possible to the IP.
The main factor
 limiting the minimum radius of Layer0 is the hit rate
from $e^+e^-\to e^+e^-e^+e^-$ background events.
Two candidate detector technologies with appropriate characteristics, especially in
radiation length (\Xrad) and hit resolution, for application
in Layer0 are (1) a
hybrid pixel detector with 1.08\% \Xrad, and 14\mum hit resolution, and (2)
striplets with 0.40\% \Xrad and 8\mum hit resolution. 
Simulation studies of
$B^0\to\phi\KS$ decays have shown that with a boost  of $\beta\gamma=0.28$ the
hybrid pixels (the striplets)  reach a $\sin 2\beta_{\rm eff}$ per event
error equal to \babar\ at an inner radius of 1.5\cm (2.0\cm). With
$\beta\gamma=0.24$ the error increases by 7---8\%. Similar conclusions also
apply to $B^0\to\pi^+\pi^-$ decays. 

The \babar\ SVT five-layer design was motivated both by the need
for standalone tracking for low-$p_T$ tracks as well as the need for redundancy in case several
modules failed during operation. The default \superb\  SVT design, consisting of 
 a \babar-like SVT detector and an additional Layer0, has been
compared with two alternative configurations with a total of either five or four layers.
These simulation studies, which used the decay  $B\to D^*K$  as the benchmark channel, 
focused on the impact of the detector configuration on track
quality as well as on the reconstruction efficiency for low $p_T$ tracks. 
The studies have shown that, as expected, the low-$p_T$ tracking efficiency 
is significantly decreased for configurations with reduced numbers of SVT layers, while
the track quality is basically unaffected.  Given the importance of low momentum 
tracking efficiency for the \superb\  physics program, these results support a six-layer layout.

Studies have also shown that the best overall SVT+DCH tracking performance
is achieved if the outer radius of the SVT is kept small (14\cm as 
in \babar\ or even less) and the inner wall of the DCH is as close to the SVT as
possible. However,  as some space between the SVT and DCH is needed for 
the cryostats that contain the 
super-conducting magnets in the interaction region,
 the minimum DCH
inner radius is expected to be about 20---25\cm.

\begin{figure}[tbh]
\includegraphics[width=0.5\textwidth]{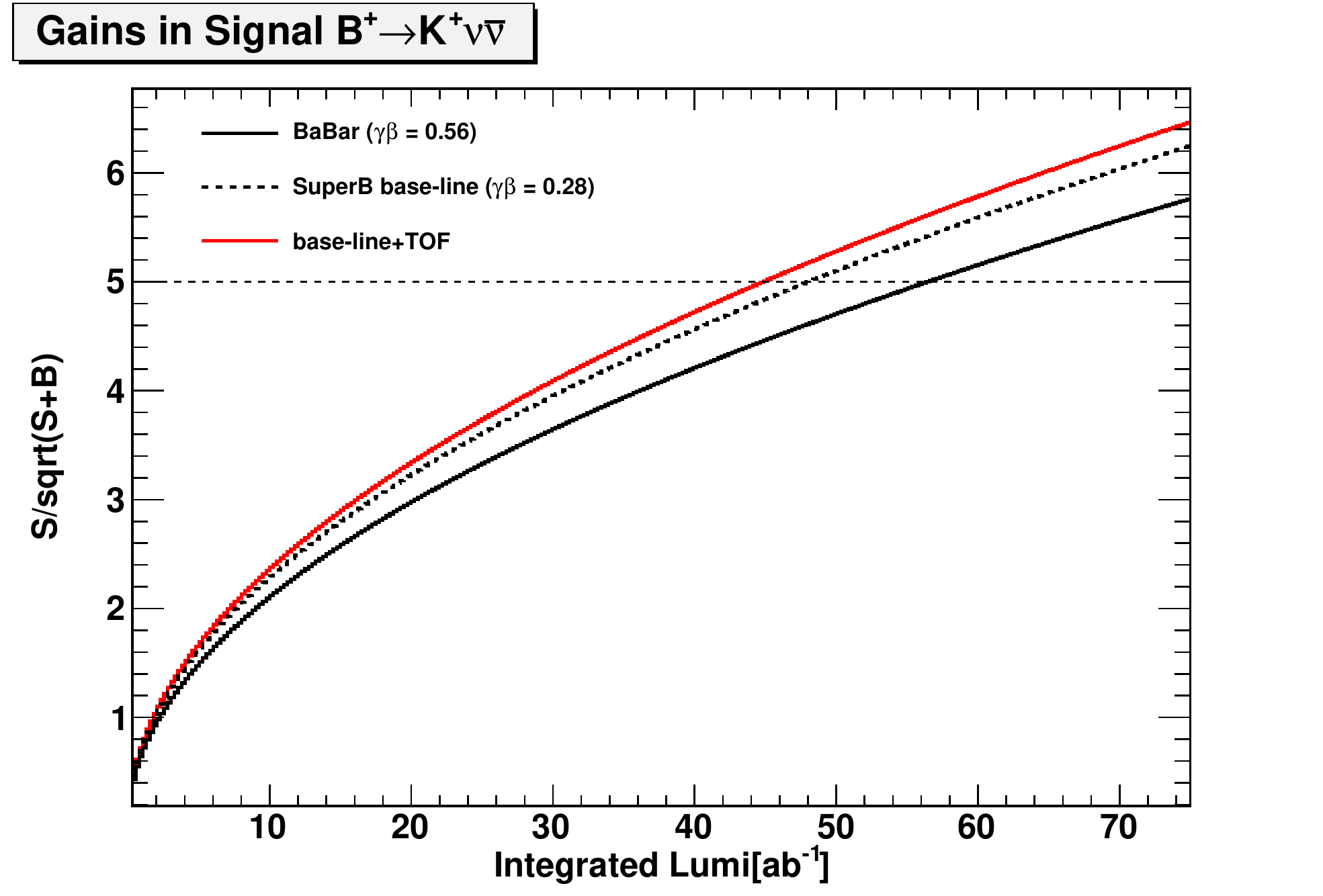}
\caption{ ${\rm S}/\sqrt{{\rm S}+{\rm B}}$ of $B\to K\nu\bar\nu$ as a function of the integrated luminosity in three different detector configurations.}
\label{fig:BtoKnunubarBF}
\end{figure}

 The impact of a forward PID device is estimated using benchmark modes such as
 $B\to K^{(*)}\nu\bar\nu$,  balancing the
 advantage of having better PID information in
 the forward region with the drawbacks arising from more
 material in front of the EMC and a slightly shorter DCH.
 Three detector configurations have been compared in these simulation 
 studies: \babar, the \superb\ 
baseline(no forward PID device), and a configuration that includes  a
 time-of-flight (TOF) detector between the DCH and the forward EMC.
 The results, presented in terms of  ${\rm S}/\sqrt{{\rm S}+{\rm B}}$,  for the decay 
 mode $B\to K\nu\bar\nu$ with the tag side
 reconstructed in the semileptonic modes, are shown in Fig.~\ref{fig:BtoKnunubarBF}.
 In summary, while the default \superb\ design leads to an improvement of
 about 7-8\% in  ${\rm S}/\sqrt{{\rm S}+{\rm B}}$, primarily due the reduced boost in \superb\, 
 the configuration with the forward TOF provides an  additional 5-6\% improved 
 sensitivity for this channel. Machine backgrounds have yet to be
 included in these simulations, but will be considered in our next updates. 
 
\begin{figure}[tbh]
\includegraphics[width=0.5\textwidth]{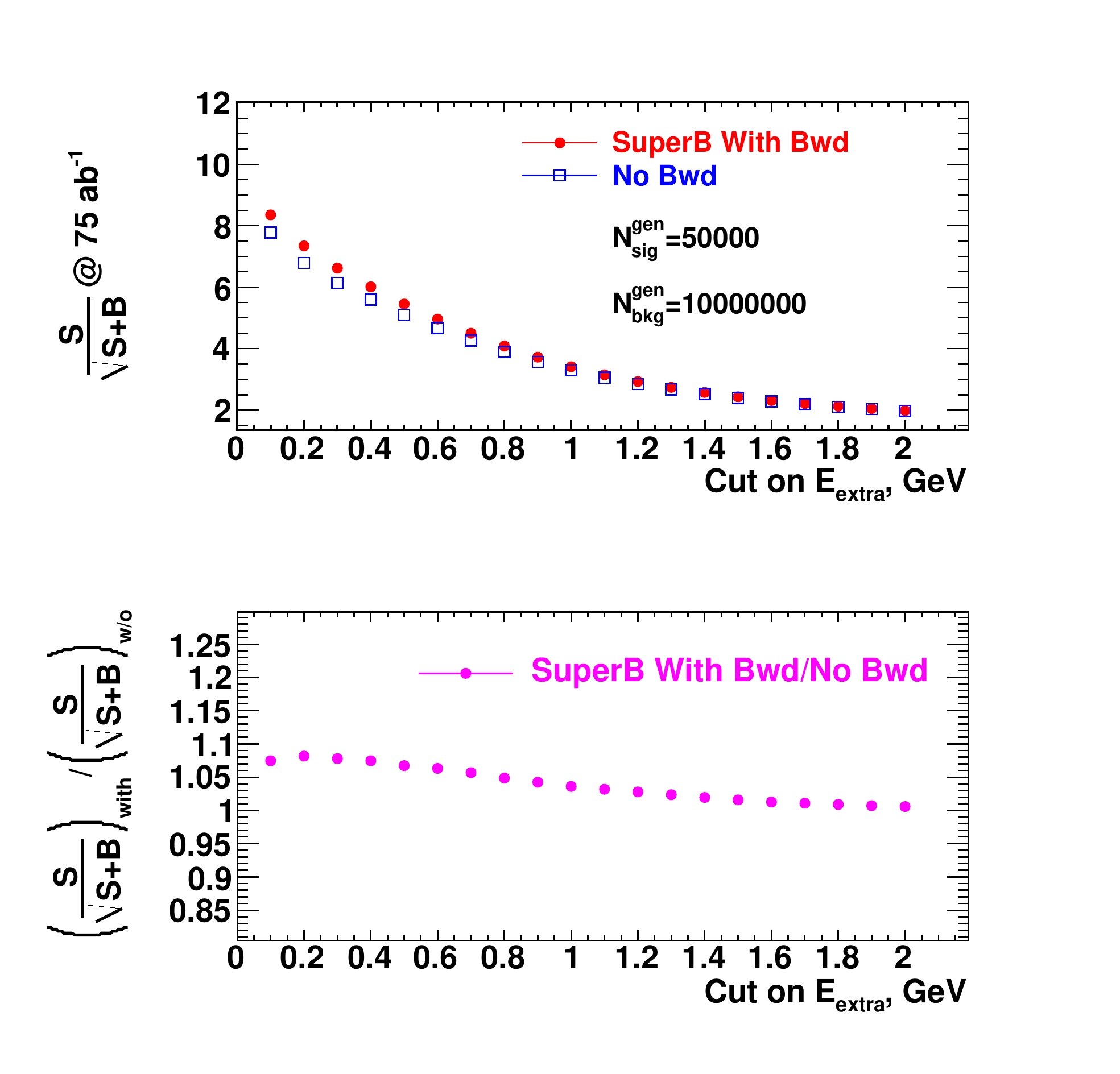}
\caption{Top: ${\rm S}/\sqrt{{\rm S}+{\rm B}}$ as a function of the cut on $E_{\rm extra}$ with (circles) and without (squares) the backward EMC. Bottom: ratio of ${\rm S}/\sqrt{{\rm S}+{\rm B}}$ for detector configurations with or without a backward EMC as a function of the $E_{\rm extra}$ cut.}
\label{fig:bwdEMC_btotaunu}
\end{figure}

The backward calorimeter under consideration is designed to be used in veto 
mode. Its impact on
physics can be estimated by studying the sensitivity of rare $B$ decays with one
or more neutrinos in the final state, which benefit from having more hermetic
detection of neutrals to reduce the background contamination.
One of the most important benchmark channels of this kind is $B\to\tau\nu$.
Preliminary studies, not including the machine backgrounds, indicate that, when 
the backward calorimeter is installed,
the statistical precision ${\rm S}/\sqrt{{\rm S}+{\rm B}}$ is enhanced by about 8\%.
The results are summarized in Fig.~\ref{fig:bwdEMC_btotaunu}. The top plot
shows how ${\rm S}/\sqrt{{\rm S}+{\rm B}}$ changes as a function of the
cut on $E_{\rm extra}$ (the total energy of charged and neutral particles that
cannot be directly associated with the reconstructed daughters of the signal
or tag $B$) with or without the backward EMC. The signal is peaked at zero. 
The bottom plot shows the ratio of ${\rm S}/\sqrt{{\rm S}+{\rm B}}$ for detector configurations with and without a backward  EMC, again as a
function of the $E_{\rm extra}$ cut.
This analysis will be repeated soon, including the main sources of machine
backgrounds, which could affect the $E_{\rm extra}$ distributions significantly.
The possibility of using the backward calorimeter as a PID time-of-flight
device is also under study.

The presence of a forward PID or backward EMC affects the maximum extension of
the DCH and therefore the tracking and $dE/dx$ performance in those
regions. The impact of a TOF PID detector is practically negligible because it
only takes a few centimeters from the DCH. On the other hand, the effect of
a forward RICH device
($\sim 20\cm$ DCH length reduction) or the backward EMC ($\sim 30\cm$) is
somewhat larger. For example,  for tracks with polar angles
 $<23^\circ$ and $>150^\circ$,  there is an increase in $\sigma_p/p$ of 
25\% and 35\%, respectively. Even in this case, however, the overall impact on the physics is generally quite
limited because only a small fraction of tracks cross the extreme forward and
backward regions.

The IFR system will be upgraded by replacing the \babar\  RPCs and LSTs with
layers of much faster extruded plastic scintillator coupled to WLS fibers
read out by APDs operated in Geiger mode. The identification of
muons and \KL is optimized with a Geant4 simulation
by tuning the amount of iron absorber and the distribution of the active
detector layers. The current baseline design has a total iron thickness of 92\cm
interspersed with eight layers of scintillator. Preliminary estimates indicate a
muon efficiency larger than 90\% for $p>1.5$\gevc at a pion 
misidentification rate of about 2\%.

\wpsubsec{Challenges on Detector Design} 
\label{subsec:Overview_challenges}

Machine background is one of the leading 
challenges of the \superb\ project: each subsystem must be 
designed so that its performance is minimally degraded 
because of the occupancy produced by background hits.
Moreover, the detectors must be protected against deterioration 
arising from radiation damage.  In effect, what is required is that each detector 
perform  as well or better than \babar\, with similar operational lifetimes,  but for two orders of magnitude higher luminosity. 

Background particles produced by beam gas scattering and by 
the synchrotron radiation near the interaction point (IP)
are expected to be
manageable since the relevant \superb\ design parameters (mainly  
the beam current) are fairly close to those in  \pepii\ .

Touschek backgrounds are expected to be larger than in \babar\
because of  the extremely low design emittances of the \superb\ beams. 
Preliminary simulation studies indicate that a system of beam collimators
upstream of the IP can reduce particle losses to tolerable
levels.

The main source of concern arises from the background particles 
produced at the IP by QED processes whose cross section is 
$\sim 200\mbarn$ corresponding
at the nominal \superb\ luminosity to a rate of $\sim 200\GHz$. 
Of particular concern is the radiative Bhabha reaction (i.e.: $\epem \to \epem \gamma$), 
where one of the incoming beam particles loses a significant fraction
of its energy by the emission of a photon. Both the photon and the radiating
beam particles emerge from the IP traveling almost collinearly with
respect to the beam-line. The magnetic elements downstream of
the IP over-steer these primary particles into the vacuum chamber walls
producing electro-magnetic showers, whose final products are the background
particles seen by the subsystems. The particles of these electromagnetic showers 
can also excite the giant nuclear resonances in the material around the beam 
line expelling neutrons from the nucleus.
Careful optimization of the mechanical apertures of the vacuum chambers and  
the optical elements is needed to keep a large stay-clear for the off-energy 
primary particles, hence reducing the background rate.

A preliminary Geant4-based Monte Carlo simulation study  of this
process at \superb\ indicates that a shield around the beamline 
will be required to keep the electrons, positrons, photons and neutrons 
away from the detector, reducing occupancies and radiation damage
to tolerable levels. 

The ``quasi-elastic Bhabha'' process has
also been considered. The cross section for producing a primary particle 
reconstructed by  the detector via this  process is $\sim 100 \nb$
corresponding to a rate of about  $100 \kHz$. It is reasonable 
to assume that this will be the driving term for the level one trigger rate.
Single beam contributions to the trigger rate are, in fact, expected to be 
of the same order as in  \babar\ , given that the nominal beam currents 
and other relevant design parameter are comparable.

\begin{figure}[tbh]
\includegraphics[width=0.5\textwidth]{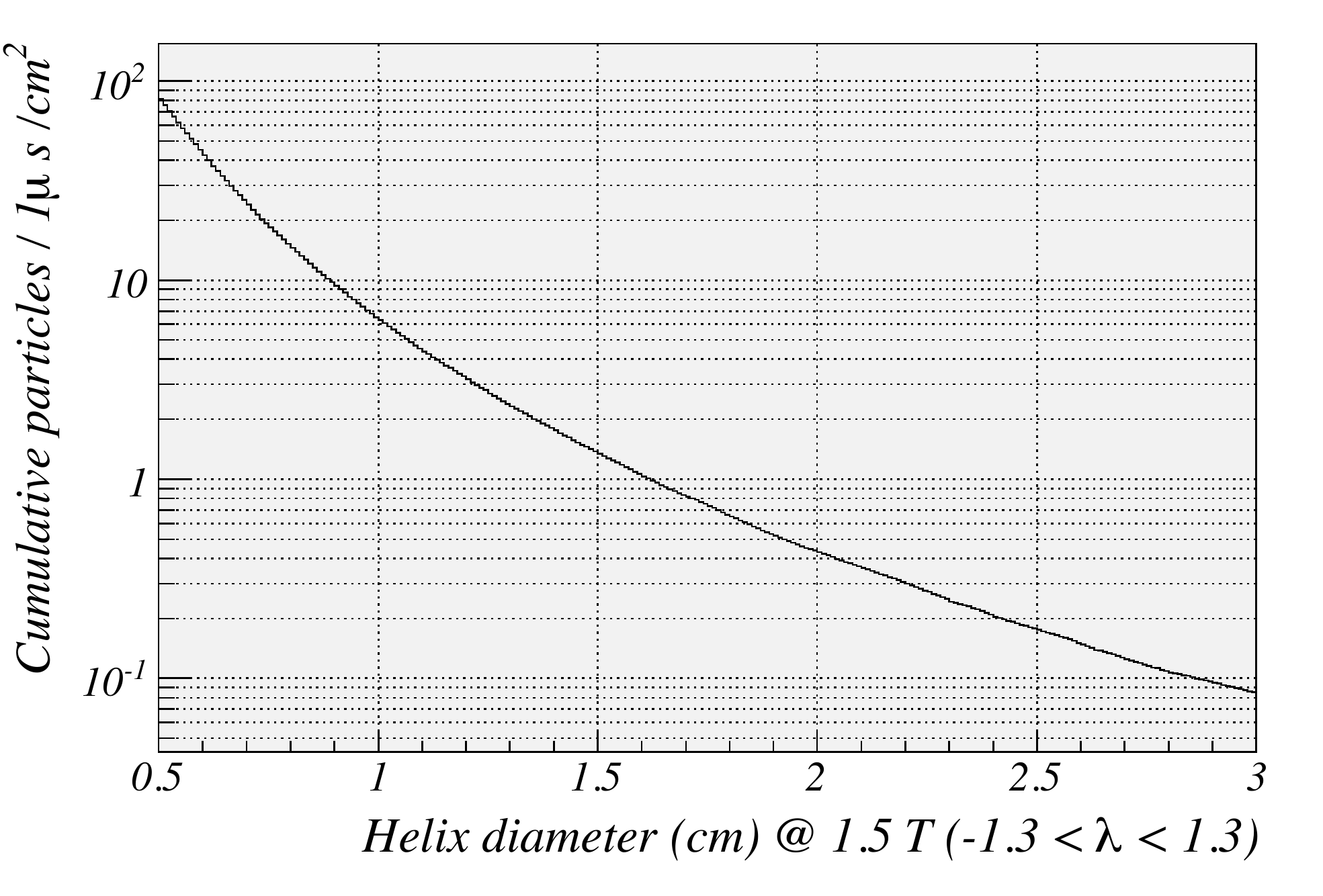}
\caption{Pairs background track rate per unit surface as a function of the SVT Layer0
radius. Multiple track hits have not been taken into account.}
\label{fig:PairsDiameter}
\end{figure}

A final  luminosity related background effect is the production of 
electron-positron pairs at the IP by the two photon process
$\epem \to \epem \epem$, whose total cross section 
evaluated at leading order with the  Monte Carlo generator 
DIAG36 \cite{Berends:1986ig} is $7.3\mbarn$  corresponding,
at nominal luminosity, to a rate of $7.3\GHz$.
The pairs produced by this process are characterized by  very soft
transverse momenta particles. The solenoidal magnetic field in the
tracking volume confines most of these background particles inside
the beam pipe. Those articles having a transverse 
momentum large enough to reach the beam pipe ( $p_T > 2.5 \mevc$)
and with a polar angle inside the Layer0 acceptance are
produced at a rate of $\sim 0.5\GHz$.
This background will be a driving factor in the design of the 
segmentation and the read-out 
architecture for SVT Layer0. The background track surface rate on
the SVT Layer0 as a function of its radius is shown in Fig.~\ref{fig:PairsDiameter}.

An effort to improve the simulation of  these background sources
 with a Geant4 based code
is underway at present. A fairly accurate model of the detector 
and beam-line elements is available to the collaboration.
Several configurations have been simulated and studied, providing
some guidelines to the detector and machine teams.
Further refinements of the interaction region and detector design will 
require development of the Geant4 background simulation tools
on the detector response side.

\wpsubsec{Open Issues} The basic geometry, structure and physics performance of the \superb\ detector is mainly predetermined by the retention of the solenoidal magnet, return steel, and the support structure from the \babar\ detector, and a number of its largest, and most expensive,  subsystems. Even though this fixes both the basic geometry, and much of the physics performance, it does not really  constrain the expected performance of the \superb\  detector in any important respect. \babar\ was already an optimized B-factory detector for physics, and any improvements in performance that could come from changing the overall layout or rebuilding the large subsystems would be modest overall. The primary challenge for \superb\  is to retain physics performance similar to \babar\ in the higher background environment described in Section\ref{subsec:Overview_challenges},  while operating at much higher ($\sim \times 50$) data taking rates.

Within these constraints, optimization of the geometrical layout and new detector elements for the most important physics channels remains of substantial interest. The primary tools for sorting through the options are: (1) simulation, performed under the auspices of a ``Detector Geometry Working Group" (DGWG), that studies basic tracking, PID, and neutrals performance of different detector configurations, including their impact on each other, and studies the physics reach of a number of benchmark channels; and (2) detector R\&D, including prototyping, developing new subsystem technologies, and understanding the costs, and robustness of systems, as well as their impacts on each other. The first item, discussed in Section\ref{subsec:Overview_Perf}, clearly provides guidance to the second, as discussed in Section\ref{subsec:Overview_RandD}
 and the subsystem chapters which follow, and vice versa.  

At the level of the overall detector, the  immediate task is to define the sub-detector envelopes. Optimization can and will continue for some time yet within each sub-detector system. The studies performed to date leave us with the default detector proposal, with only a few open options remaining at the level of the detector geometry envelopes and technology choices. These open issues are: (1) whether there is a forward PID detector, and, if so, at what z location does the DCH end and the EMC begin; and (2) whether there is a backward EMC.  These open issues are expected to be resolved by the Technical Board within the next few months following further studies by the 
DGWG, in collaboration with the relevant system groups. 

\wpsubsec{Detector R\&D}
\label{subsec:Overview_RandD}

The \superb\ detector concept rests, for the most part,  on well validated basic 
detector technology.  Nonetheless, each of the sub-detectors has may challenges due to
the high rates and demanding performance requirements with R\&D initiatives
ongoing in all detector systems to improve 
the specific performance, and optimize the overall
detector design.  These are described in more detail in each subsystem section.

The SVT innermost layer has to provide good space
resolution while coping with high background.  Although silicon
striplets are a viable option at moderate background levels, a pixel
system would certainly be more robust against background. However, keeping the
material in a pixel system low enough not to deteriorate the vertexing
performance is challenging, and there is considerable activity to
develop thin hybrid pixels or, even better, monolithic active pixels.
These devices may be part of a planned upgrade path and installed as a
second generation Layer0. Efforts are directed towards the development
of sensors, high rate readout electronics, cooling systems and
mechanical support with low material content.

In the DCH,  many parameters must be optimized for \superb\ running,
such as the gas mixture and the cell layout.  Precision measurements
of fundamental gas parameters are ongoing, as well as studies with
small cell chamber prototypes and simulation of the properties of
different gas mixtures and cell layouts. An improvement of the
performance of the DCH could be obtained by using the innovative ``Cluster Counting" method, in
which single clusters of charge are resolved time-wise and counted,
improving the resolution on the track specific ionization and the
space accuracy.  This technique requires significant R\&D to be proven
feasible in the experiment.

Though the Barrel PID system takes over major components from \babar,  the 
new camera and readout concept is a 
significant departure from the \babar\ system, requiring 
extensive R\&D. The challenges include the performance of pixelated
PMTs for DIRC, the design of the fused silica optical system, 
the coupling of the fused silica optics to the existing bar boxes, the mechanical design
of the camera, and the choice of electronics. Many of the individual components of the new camera are now under active
investigations by members of the PID group, and studies are underway with a single bar prototype located in
a cosmic ray telescope at SLAC. A full scale (1/12 azimuth ) prototype incorporating the complete optical design is planned
for cosmic ray tests during the next two years.

Endcap PID concepts are less developed, and whether they match the physics 
requirements and achieve the expected detector performance
remains to be demonstrated. Present R\&D is centered on developing a good conceptual understanding of
different proposed concepts, on simulating how their performance affects
the physics performance of the detector, and on conceptual R\&D for components of specific devices to validate 
concepts and highlight the technical and cost issues.

The EMC barrel is a well understood device at the lower luminosity of \babar. Though there will be some 
technical issues associated with refurbishing, the main R\&D needed at present is to understand the effects of pile-up 
in simulation, so as to be able to design the appropriate front-end shaping time for the readout.

The forward and backward EMCs are both new devices, using cutting edge technology. Both will require
one or more full beam tests, hopefully at the same time, within the next year or two.  Prototypes for these tests
are being designed and constructed.

Systematic studies of IFR system components have been performed in a variety of bench and cosmic ray tests, leading to the present proposed design. This design will be beam tested in a full scale prototype currently being prepared for a Fermilab beam. This device will demonstrate the muon identification capabilities as a function of different iron configurations, and will also be able  to study detector
efficiency and spatial resolution. 

At present,  the Electronics, DAQ, and Trigger (ETD), have been designed for the base luminosity of $1 \times 10^{36} \cm^{-2}\sec^{-1}$, with adequate headroom.
Further R\&D is needed to understand the requirements at a luminosity up to 4 times greater, and to insure that
there is a smooth upgrade path when the present design becomes inadequate.  On a broad scale, as discussed in the system chapter, 
each of the many components of ETD  have numerous technical challenges
that will require substantial R\&D as the design advances.

\aftsec

\begsec
\graphicspath{{SVT/}{SVT/}}
\wpsec{Silicon Vertex Tracker}

\wpsubsec{Detector Concept}

\wpsubsubsec{SVT and Layer0}

\begin{figure*}[t]
\begin{center}
\includegraphics[width=\textwidth]{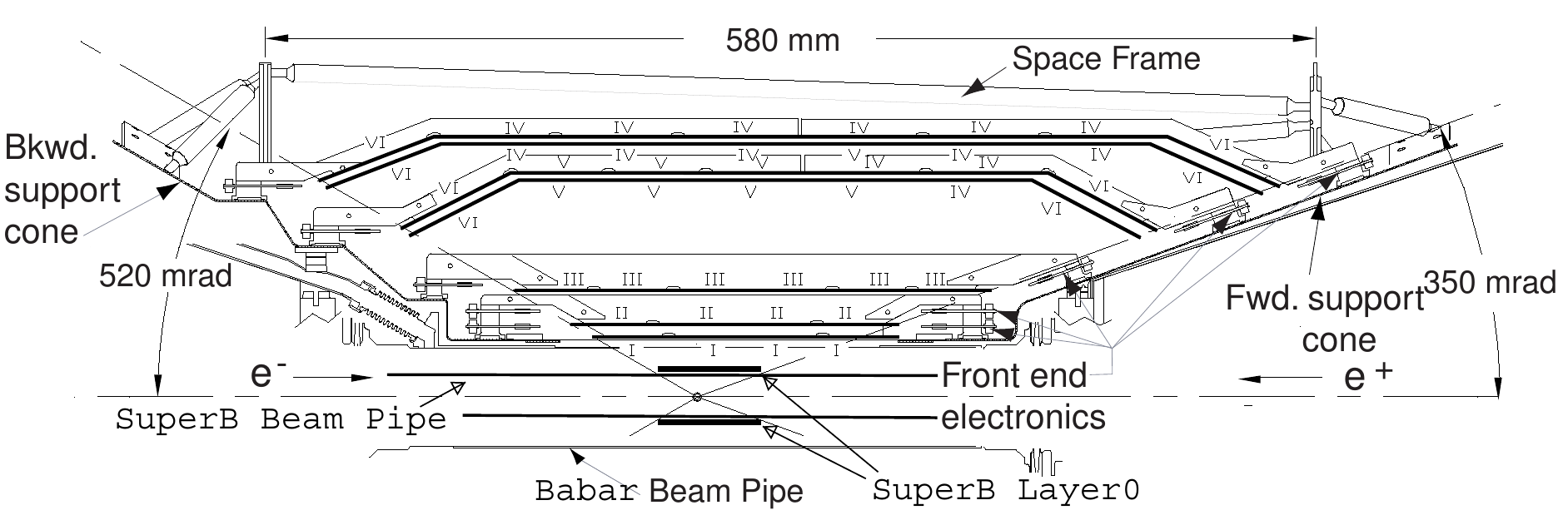}
\end{center}
\caption{Longitudinal section of the SVT}
\label{fig:svt}
\end{figure*}

The Silicon Vertex Tracker, as in \babar,  together with the drift chamber (DCH) and the solenoidal
magnet provide track and vertex reconstruction  capability for the \superb\ detector. 
Precise vertex information, primarily extracted from precise position measurements near the IP
 by the SVT, is crucial to the measurement of time-dependent CP asymmetries in $B^0$ decays, 
which remains a key element 
of the \superb\ physics program. In addition, charged particles with transverse momenta lower than
100\mevc  will not reach the central tracking chamber, so for these particles the SVT must
provide the complete tracking information.

These goals have been reached in the \babar\ detector with a five-layer  silicon strip detector,
shown schematically in Fig.~\ref{fig:svt}.
The \babar\ SVT provided excellent performance for the whole life of the
experiment, thanks to a robust design that took into account the physics requirements as
well as enough safety margin, to cope with the machine background, and redundancy
considerations. The \superb\ SVT design is based on the \babar\ vertex detector layout with
the addition of an innermost layer closer to the IP (Layer0).  
The Layer0 close-in position measurements lead to an improved vertex resolution, which is expected
 to largely compensate for the reduced boost at the \superb, thus retaining the  
$\deltat$ resolution for  B decays achieved in \babar.
 Physics studies and
background conditions, as explained in detail in the next two sections, set stringent
requirements on the Layer0 design: radius of about 1.5\cm; high granularity ($50\times 50 \muma$
pitch); low material budget (about 1\% \Xrad); and adequate radiation resistance.

For the Technical Design Report preparation, several options are under study for the Layer0
technology, with different levels of maturity, expected performance and safety margin
against background conditions. These include striplets modules based on high resistivity sensors with
short strips, and hybrid pixels and other thin pixel sensors based on CMOS Monolithic Active
Pixel Sensor (MAPS).

The current baseline configuration of the SVT Layer0 is based on the
striplets technology, which has been shown to provide the better physics performance, as
detailed in the next section. However,  options based on pixel sensors, which are more robust in high
background conditions, are still being developed with specific R\&D programs in order to meet
the Layer0 requirements, which include low pitch and material budget, high readout speed and
radiation hardness. If successful, this will allow the replacement of the Layer0 striplets modules in
a ``second phase" of the experiment. For this purpose the \superb\  interaction region and
the SVT mechanics will be designed to ensure rapid access to the detector for fast
replacement of Layer0.

The external SVT layers (1-5), with a radius
between 3 and 15\cm, will be built with the same technology used for the \babar\ SVT (double
sided silicon strip sensor), which is adequate for the machine background conditions
expected in the \superb\ accelerator scheme (\ie with low beam
currents).

The SVT angular acceptance, constrained by the interaction region design, will
be 300\mrad in both the forward and backward directions, corresponding to a solid angle
coverage of 95\% in the center-of-mass frame.

\wpsubsubsec{Performance Studies}

The ultra-low emittance beams of the  \superb\ design opens up the possibility
of using a small radius beam pipe (1\cm) in the detector acceptance, allowing to have
the innermost layer of the SVT  very close to the IP.  
The small radius of the pipe increases the heating from image charges 
and hence a water cooling channel is foreseen for the beam pipe to extract this power. 
The total amount of radial material of the beryllium
pipe, which includes a few \mum of gold foil, and the water
cooling channel, is estimated to be less than 0.5\% \Xrad.
For the proposed \superb\ boost, $\beta\gamma = 0.28$ for 
7\gev $e^-$ beam against a 4\gev  $e^+$ beam, the average
 $B$ vertex separation along the $z$ coordinate,
$\langle\Delta z\rangle \simeq \beta\gamma c \tau_{B} =  125 \mum$,
is  around half of that in \babar, where $\beta\gamma = 0.55$.
In order to maintain a suitable resolution on $\deltat$ for
time-dependent analyses, 
it is necessary to improve the
vertex resolution (by about a factor 2) with respect to that 
achieved in \babar: typically $50-80\mum$ in $z$ for exclusively
reconstructed modes and $100-150\mum$ for inclusively
reconstructed modes (typical resolutions for the 
tagging side in CPV  measurements).
The six-layer SVT solution for \superb, with the Layer0 sitting much closer
to the IP than that in \babar,
would significantly improve track parameter determination,
matching the more demanding requirements on the
vertex resolution, while maintaining the stand-alone tracking capabilities
 for low momentum particles.

The choice among the various options under consideration for the Layer0
has to take into account the physics requirements
for the vertex resolution, depending on the pitch and the
total amount of material of the modules. In addition, to assure
optimal performance for track reconstruction, the sensor
occupancy has to be maintained under  a few
percent level, imposing further constraints on the sensor
segmentation and on the front-end electronics. Radiation
hardness is also an important factor, although it is
expected not to be particularly demanding compared to
 the LHC detector specifications.

\par
The simulation program FastSim~\cite{ref:FastSim} has been used to 
study track and vertex reconstruction performance of various SVT configurations,
providing estimates of  the $B$ decay vertex resolution as well as $\deltat$ resolution 
for time-dependent CPV measurements. 
 We have considered
 several benchmark  channels, including
$B \to \pi^+\pi^-,~\phi\KS$ and also decay modes
 where the impact of the Layer0 on the decay vertex determination
is expected to be less important, such as $B \to ~\KS \KS,~\KS \pi^0$.
For each mode we have studied the resolution on  $\deltat$ and
 the per-event error on the quantity of physics interest, namely $\sin(2\beta_{eff})$.

The main conclusion is that the baseline \superb\ SVT design -- the six-layer design -- leads 
to an improved  $\deltat$ resolution over that achieved in \babar,  allowing for a  comparable (or 
 even better) per-event error on $\sin(2\beta_{eff})$, for the $B$ decay modes considered in
 this study. This conclusion is valid for all candidate technologies that have been considered for 
 Layer0,  and for reasonable values of the Layer0 radius and amount of radial material.
\begin{figure}
\begin{center}
\includegraphics[clip=true,trim=0cm 15cm 0cm 0cm, width=0.6\textwidth]{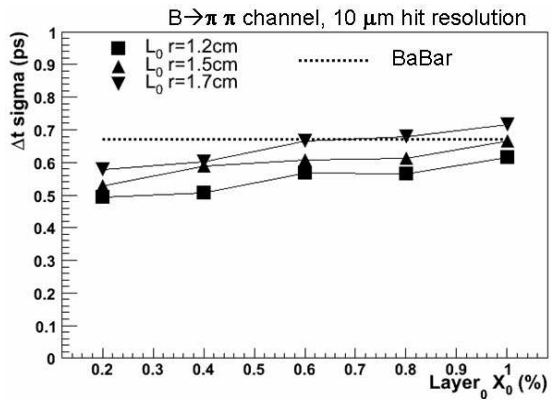}
\caption{Resolution on the proper time difference of the two B mesons
($\beta \gamma =0.28$),
for different Layer0 radii,
as a function of Layer0 thickness (in \Xrad \%).}
\label{fig:deltatresol}
\end{center}
\end{figure}
As an example, in Fig.~\ref{fig:deltatresol} is reported the resolution on $\deltat$
for different Layer0 radii as a function of the Layer0 thickness (in \Xrad \%) compared
 to the \babar\ reference value.
The dashed line represents the \babar\ reference value
using the nominal value of the boost in PEP-II , $\beta\gamma=0.55$.

We have also studied the impact of a possible Layer0 inefficiency
on $\sin(2\beta)$ sensitivity. The source of inefficiency could be related to several causes,
for example a much higher background rate than expected, causing dead time in the readout of the detector.
\begin{figure}
\begin{center}
\includegraphics[clip=true,trim=0cm 0cm 1cm 2cm, width=0.5\textwidth]{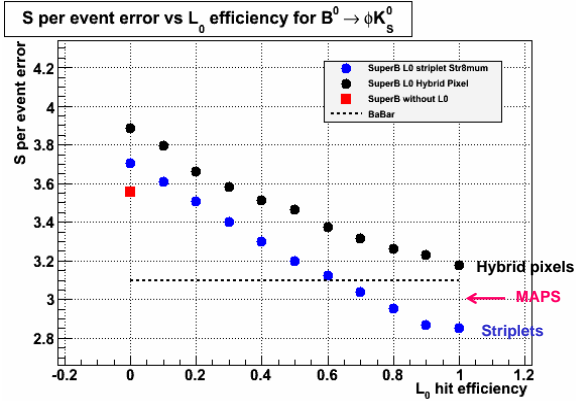}
\caption{sin2$\beta_{eff}$ per event error as a function
of the Layer0 efficiency for the dif-
ferent options (i.e. different material
budget).}
\label{fig:s_per_event_error}
\end{center}
\end{figure}
In Fig.~\ref{fig:s_per_event_error} is reported the $\sin(2\beta_{eff})$ per-event error for
 the $B\to\phi\KS$ decay mode as a function of the Layer0 hit efficiency for different Layer0 technology
 options. The Layer0 radius in the study is about 1.6\cm.
The results show that the striplet solution provides better performance, both with respect
 to \babar, even for the case of some (small)  hit inefficiency, and other Layer0 solutions.
The main advantage of the striplet solution is the smaller material budget (about 0.5\% \Xrad)
 compared to the Hybrid pixel (about 1\% \Xrad) and the MAPS solutions (about 0.7\% \Xrad).
 For particles of momenta up to a few \gevc, the multiple scattering effect is the dominant source
 of uncertainty in the determination of their trajectory, thus a low material budget detector 
 provides a clear advantage.  A striplet-based Layer0 solution would also have a better 
 intrinsic hit resolution about 8\mum) with respect to the MAPS 
 and the Hybrid Pixel (about 14\mum with a digital readout) solutions.
 For those reasons a Layer0 based on striplets has been chosen as the baseline solution for \superb,
 able to cope with the machine background according to the present estimates.

\wpsubsubsec{Background Conditions}

Background considerations influence several aspects of the 
SVT design: readout segmentation; electronics shaping time;
data transmission rate; and radiation hardness (particularly severe for Layer0).
The different sources of background have been simulated with
a detailed Geant4-based detector model and beamline description to estimate their impact
on the experiment \cite{SuperB:CDR}.
The background hits expected in the external layers of the SVT
(radius $> 3\cm$)  are mainly due to processes that scale with beam currents,
 similar to  background seen in the present \babar\ SVT.
The background at the Layer0 radius is primarily due to luminosity terms,
in particular the $e^+e^-\to e^+e^-e^+e^-$ pair
production, with radiative Bhabha events an order of magnitude smaller.
 Despite the huge cross section of the pair production process,
the rate of tracks originating from this process hitting the Layer0 sensors
 is strongly suppressed by the 1.5 Tesla magnetic field of the \superb\ detector.
Particles produced with low transverse momenta
loop in the
detector magnetic field, and only a small fraction reaches the SVT
layers, with a strong radial dependence.

According to these studies the track rate at the Layer0 at a radius of 1.5\cm is at the level of
about 5\MHz/\cma, mainly due to electrons in the \mev energy range.
The equivalent fluence corresponds to about $3.5 \times 10^{12} \rm{n}/\cma$/yr, corresponding
to  a dose rate of about 3M\rad/yr.  
A safety factor of five on top of these numbers has been considered in the design of the SVT.

\wpsubsec{Layer0 Options Under Study}
In this section we summarize the current status of the studies of the various Layer0 options, aimed at the eventual
preparation of the \superb\  Technical Design Report.

\wpsubsubsec{Striplets}

\begin{figure}
\begin{center}
\includegraphics[width=\columnwidth]{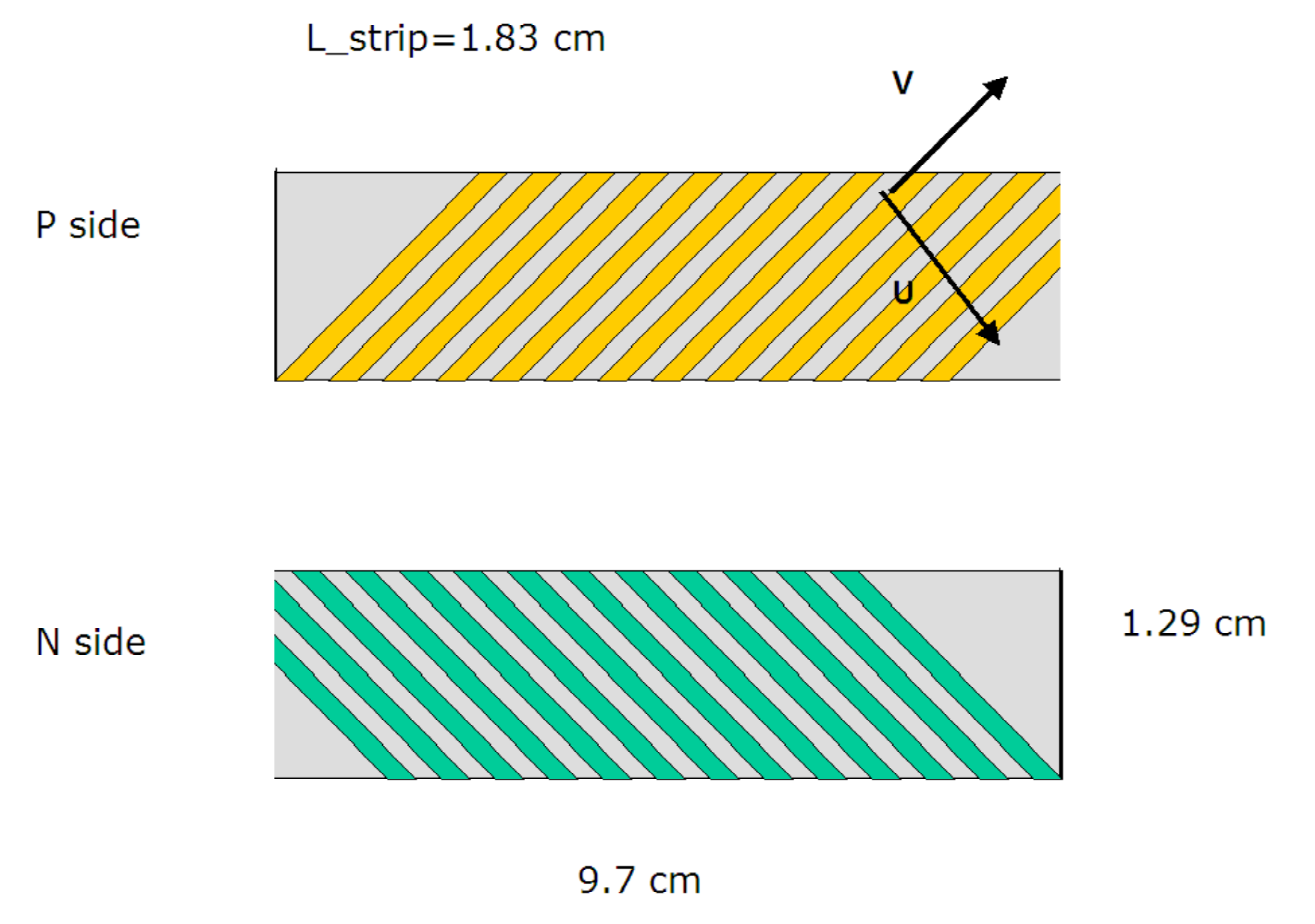}
\caption{Schematic view of the two sides of the striplets detector.}
\label{fig:svt:striplets}
\end{center}
\end{figure}

Double-sided silicon strip detectors (DSSD), $200 \mum$ thick, with
$50 \mum$ readout pitch represent a proven and robust technology, meeting 
the requirements on the SVT Layer0 design, as described in the CDR~\cite{SuperB:CDR}.
In this design, short strips will be placed at an angle 
of $\pm45^\circ$ to the detector edge on the two sides of the sensor, 
as shown in Fig~\ref{fig:svt:striplets}. 

The strips will be connected to the readout electronics through a 
a multilayer flexible circuit glued to the sensor. A standard technology
with copper traces is already available, although an aluminum
microcable technology is being explored to reduce the impact on
material of the interconnections. 

\begin{figure}[hbt]
\begin{center}
\includegraphics[width=\columnwidth]{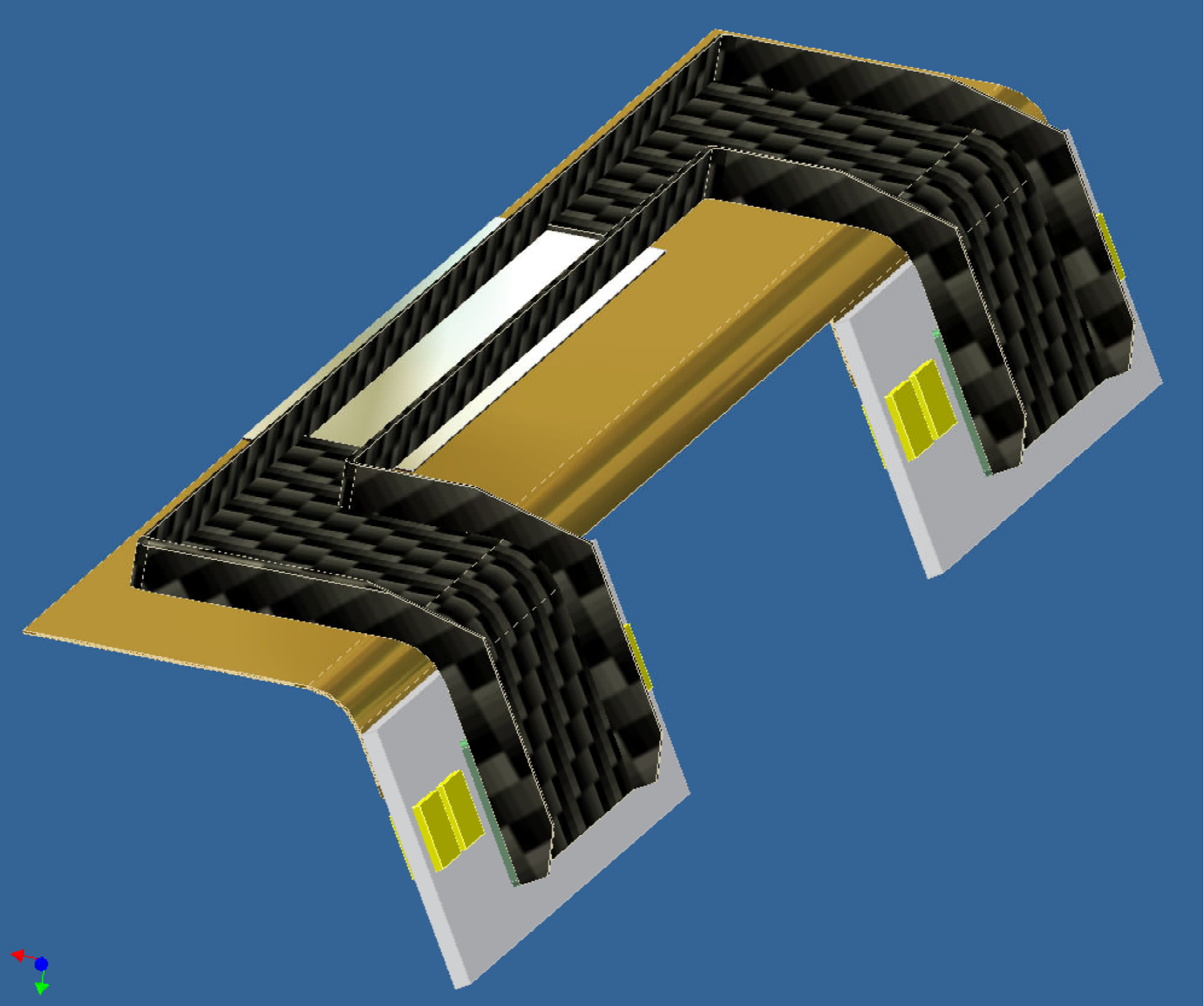}
\caption{Mechanical structure of a striplets Layer0 module.}
\label{fig:svt:striplets_module}
\end{center}
\end{figure}

The data-driven, high-bandwidth FSSR2 readout chip~\cite{svt:re:2006},
is a good match to the Layer0 striplet design and is also suitable for the readout of
the strip sensors in the outer layers. 
It has 128 analog channels providing a sparsified
digital output with address, timestamp and pulse height information for all
hits. The selectable shaper peaking time can be programmed down to 65\ns.
The chip has been realized in a $0.25\mum$ CMOS technology for high radiation
tolerance.
The readout architecture has been designed to operate with a 132\ns clock
that will define the timestamp granularity and the readout window.
A faster readout clock (70\MHz) is used in the chip, with a token pass logic,
to scan for the presence of hits in the digital section, and to transmit them off-chip,
using a selectable number of output data lines.
With six output lines, the chip can achieve an output data rate of 840\mbitsps.
With a 1.83\cm strip length the expected occupancy in the 132\ns time window 
is about 12\%, considering a hit rate of 100\MHz/\cma, including the cluster multiplicity 
and  a factor 5 safety margin on the simulated background track rate. 
The FSSR2 readout efficiency has never been measured with this occupancy.
First results from ongoing Verilog simulations indicate the efficiency
is 90\% or more. As shown in Fig.~\ref{fig:s_per_event_error} the physics 
impact of such an inefficiency is modest. Nonetheless it may be possible to redesign
the digital readout of the FSSR2 to increase the readout efficiency at high 
occupancy. A total equivalent noise charge of 600$\,e^-$ rms is expected, including
the effects of the strip and flex circuit capacitance, as well as the metal series
resistance. The signal to noise for a 200\mum detector is about 26, providing a good
noise margin. 
It is also foreseen to conduct a market survey to evaluate whether
different readout chips, possibly with a triggered readout architecture, may
provide better performance. 

Because of the unfavorable aspect
ratio of the sensors, the readout electronics needs to be rotated
and placed along the beam axis, outside of the sensitive volume
of the detector, held by a carbon fiber mechanical structure, as 
shown in Fig.~\ref{fig:svt:striplets_module}. The 8 modules forming
Layer0 will be mounted on flanges containing the cooling circuits.
For the baseline design with striplets, the Layer0 material
budget will be about $0.46 \% \Xrad$ for perpendicular tracks, assuming a
silicon sensor thickness of $200\mum$, a light module
support structure ($\sim 100\mum$ Silicon equivalent),
similar to that used for the \babar\ SVT modules, and the multilayer flex
contribution (3 flex layers/module, $\sim 45\mum$ Silicon equivalent/layer).
A reduction in the material budget to about $0.35\% \Xrad$ is possible
if kapton/aluminum microcable technology can be employed with a trace
pitch of about $50\mum$.

\wpsubsubsec{Hybrid Pixels}
Hybrid pixels technology represents a mature and viable solution but
still requires some R\&D to meet Layer0 requirements 
(reduction in the front-end pitch and in the total material
budget, with respect to hybrid pixel systems
developed for LHC experiments)
A front-end chip for hybrid pixel sensors with $50 \times 50\muma$ pitch and
a fast readout is under development. The adopted readout architecture
has been
previously developed by the SLIM5 Collaboration~\cite{slim5} for CMOS
Deep NWell MAPS~\cite{rizzo:ieee08,gabrielli:nim2007},
the data-push architecture features data sparsification on pixel
and timestamp information for the hits.
This readout has been recently optimized for the target
Layer0 rate of 100\MHz/\cma with promising results: VHDL simulation
of a full
size matrix (1.3\cma) gives hit efficiency above 98\% operating the
matrix with a 60\MHz readout clock.
A first prototype chip with 4k pixels has been submitted in September 2009
with the ST Microelectronics 130\nm process and is currently under test. The front-end chip, connected by
bump-bonding to an high resistivity pixel sensor matrix, will be then
characterized with beams in Autumn 2010.

\wpsubsubsec{MAPS}

\begin{figure}
\begin{center}
\includegraphics[clip=true,trim=1cm 7cm 0cm 2cm, width=0.55\textwidth]{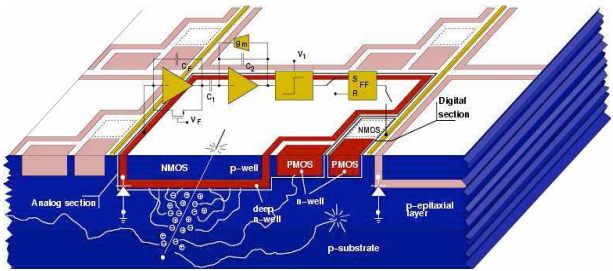}
\caption{The DNW MAPS concept.}
\end{center}
\label{fig:maps_dnw}
\end{figure}

CMOS MAPS are a newer and more challenging technology.
Their main advantage with respect to hybrid pixels is that they could be
very thin, having the sensor and the readout incorporated in a single CMOS
layer, only a few tens of microns thick.
As the readout speed is another relevant aspect for application in the
\superb\ Layer0, we proposed a new design approach to
CMOS MAPS~\cite{rizzo:ieee08} which for the first time
made it possible to build a thin pixel matrix featuring a sparsified
readout with timestamp information for the hits~\cite{gabrielli:nim2007}.
In this new design the
deep N-well (DNW) of a triple well commercial CMOS process
is used as charge collecting electrode and is extended to cover a large
fraction of the elementary cell (Fig.~\ref{fig:maps_dnw}).
Use of a large area collecting electrode
allows the designer to include also PMOS transistors in the
front-end, therefore taking full advantage of the properties
of a complementary MOS technology for the design of high
performance analog and digital blocks. However, in order to
avoid a significant degradation in charge collection effciency,
the area covered by PMOS devices and their N-wells, acting
as parasitic collection centers, has to be small with respect to
the DNW sensor area. Note, that the use of a charge preamplifier
as the input stage of the channel makes the charge sensitivity
independent of the detector capacitance.
The full signal processing chain implemented at the pixel level
(charge preamplifier, shaper, discriminator and latch) is partly
realized in the p-well physically overlapped with the area of the sensitive
element, allowing the development of a complex in-pixel logic
with similar functionalities to hybrid pixels.

Several prototype chips (the ``APSEL'' series) have been realized with the
STMicroelectronics, 130\nm triple well technology and demonstrated that the
 proposed approach is very promising for the realization of a thin pixel detector.
The APSEL4D chip, a 4k pixel matrix with $50 \times 50 \muma$ pitch,  with a new DNW cell and the
sparsified readout has been characterized during the SLIM5 testbeam
showing encouraging results~\cite{villa:elba09}. Hit
efficiency of 92\% has been measured, a value compatible with the
present sensor layout that is designed with a  fill factor
(\ie the ratio of the electrode to the total n-well area)
of about 90\%. Margins to improve 
the detection efficiency with a different sensor layout are 
being currently investigated~\cite{apsel5t:paoloni_vienna2010}

Several issues still need to be solved to demonstrate the ability to
build a working detector with this technology, which  required further R\&D.
Among others, the scalability to larger matrix size and the radiation hardness
of the technology are under evaluation for the TDR preparation.

\wpsubsubsec{Pixel Module Integration}

To minimize the detrimental effect of multiple scattering on track parameter
resolution, the reduction of
the material is crucial for all the components of the pixel module in
the active area.

The pixel module support structure needs to include a cooling
system to extract the power dissipated by the front-end electronics,
about 2W/\cma, present in the active area.
The proposed module support will be realized with a light carbon fiber support
with integrated microchannels for the coolant fluid (total material budget
for support and cooling below 0.3\%\Xrad).
Measurements on first support prototypes realized with this cooling technique
indicate that a cooling system based on microchannels can be a viable
solution to the thermal and structural problem of
Layer0~\cite{bosi:elba09}.

The pixel module will also need a light multilayer bus (Al/kapton based
with total material budget of about 0.2\%\Xrad),
with power/signal inputs and high trace density for high data speed
 to connect the front-end chips in the active area to the
HDI hybrid, in the periphery of the module.
With the data push architecture presently under study and the high background rate, the expected data with 
a 160\MHz clock need to be transferred on this bus. With triggered readout architecture (also under 
investigation) the complexity of the pixel bus, and material associated, will be reduced.

\begin{figure}
\begin{center}
\includegraphics[clip=true,trim=0cm 0cm 6cm 23cm, width=0.75\textwidth]{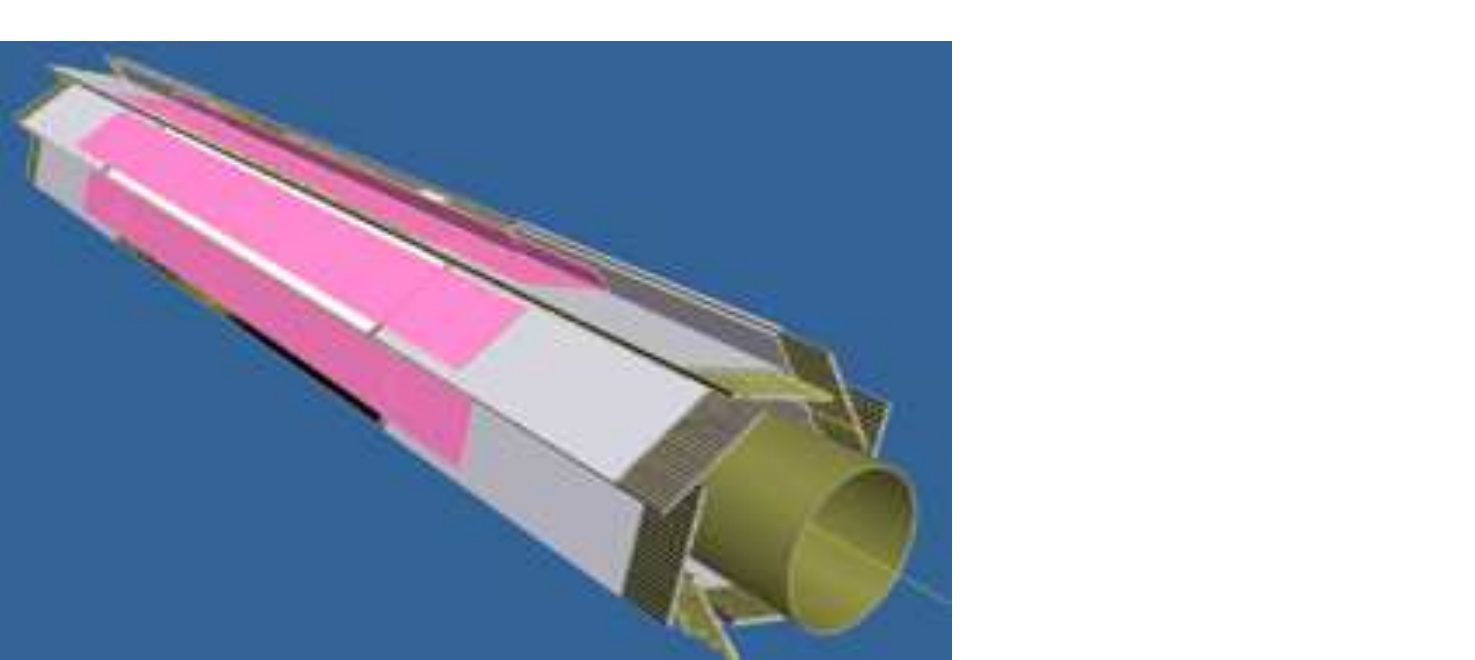}
\caption{Schematic drawing of the full Layer0 made of 8 pixel modules mounted
around the beam pipe with a pinwheel arrangement.}
\end{center}
\label{fig:layer0_pinwheel}
\end{figure}

Considering the various pixel module components (sensor and front-end with 0.4\%\Xrad,
support with cooling, and multilayer bus with decoupling capacitors) the total material in the
active area for the Layer0 module design based on hybrid pixel
is about 1\%\Xrad.
For a pixel module design based on CMOS MAPS, where  the contribution of the sensor and the integrated readout electronics become almost negligible, 0.05\%\Xrad, the total material budget is about 0.65\%\Xrad.
A schematic drawing of the full Layer0 made of 8 pixel modules mounted
around the beam pipe with a pinwheel arrangement is shown
in Fig.~\ref{fig:layer0_pinwheel}.

Due to the high background rate at the Layer0 location, radiation-hard
fast links between the pixel module and the DAQ system located outside the
detector should be adopted.
For all Layer0 options (that currently share a similar data push architecture) the untriggered data rate is 16\gbitsps per readout section, assuming a background hit rate of 100\MHz/\cma. Triggered data rate is reduced 
to about 1\gbitps per readout section.

The HDI. positioned at the end of the module, outside the active area, will be designed to host several IC
components: some glue logic, buffers, fast serializers, drivers.
The components should be radiation hard for the application at the
Layer0 location (several Mrad/yr).

The baseline option for the link between the Layer0 modules and the DAQ
boards is currently based on a mixed solution.
A fast copper link  is foreseen between the HDI and an
intermediate transition board, positioned in an area of moderate radiation
levels  (several tens of krad/yr). On this transition card the logic with LV1 buffers will store the data until the reception of the LV1 trigger signal  and only triggered data will be sent to the DAQ boards with an optical link  of 1\gbitps.
The various pixel module interfaces will be characterized in a test set-up for the TDR preparation.

\wpsubsec{A MAPS-based All-pixel SVT Using a Deep P-well Process}

\begin{figure*}
\begin{center}
\subfigure[CMOS MAPS without a deep P-well implant]{\includegraphics[width=0.45\textwidth]{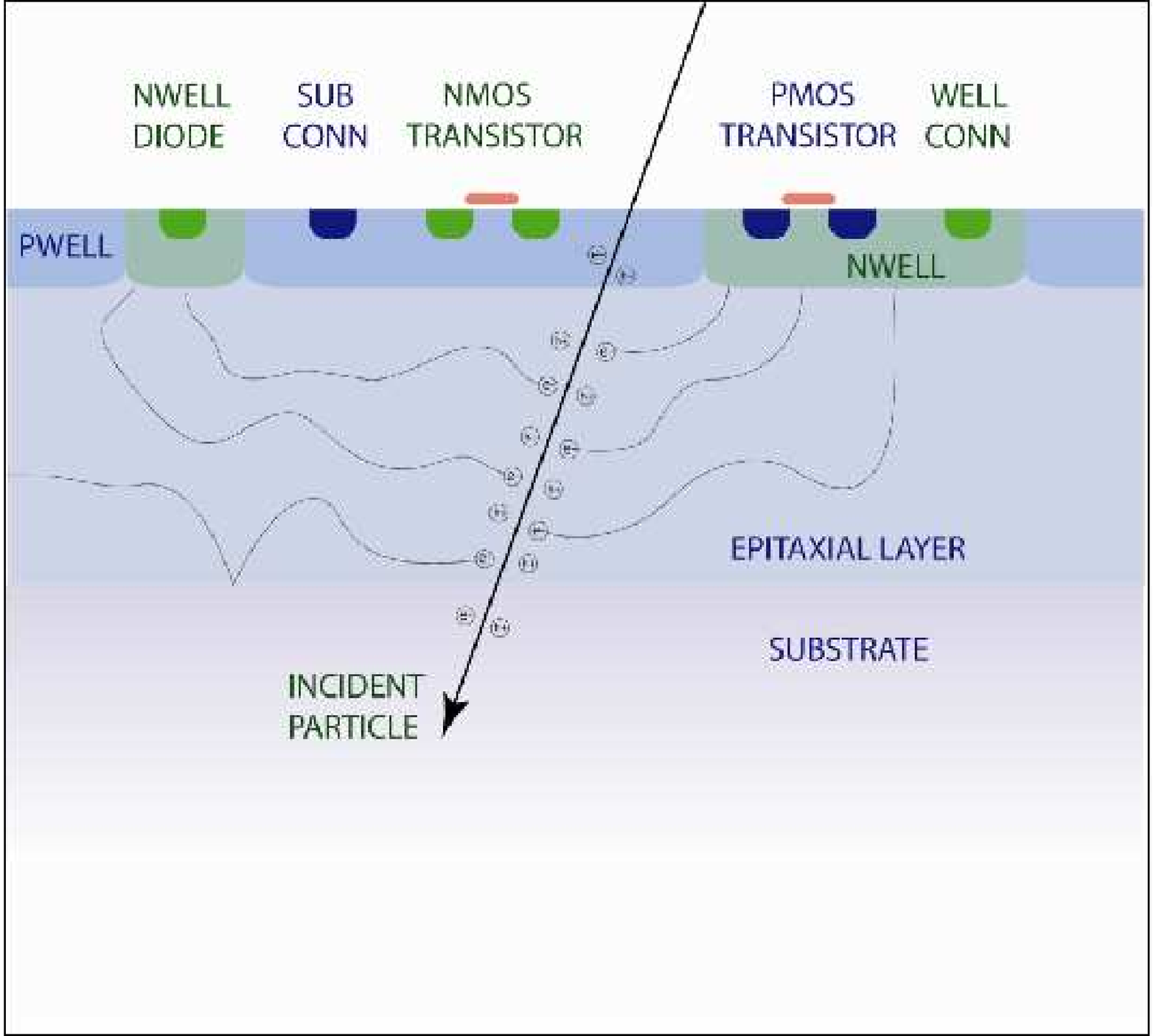}}
\subfigure[CMOS MAPS with a deep P-well implant]{\includegraphics[width=0.45\textwidth]{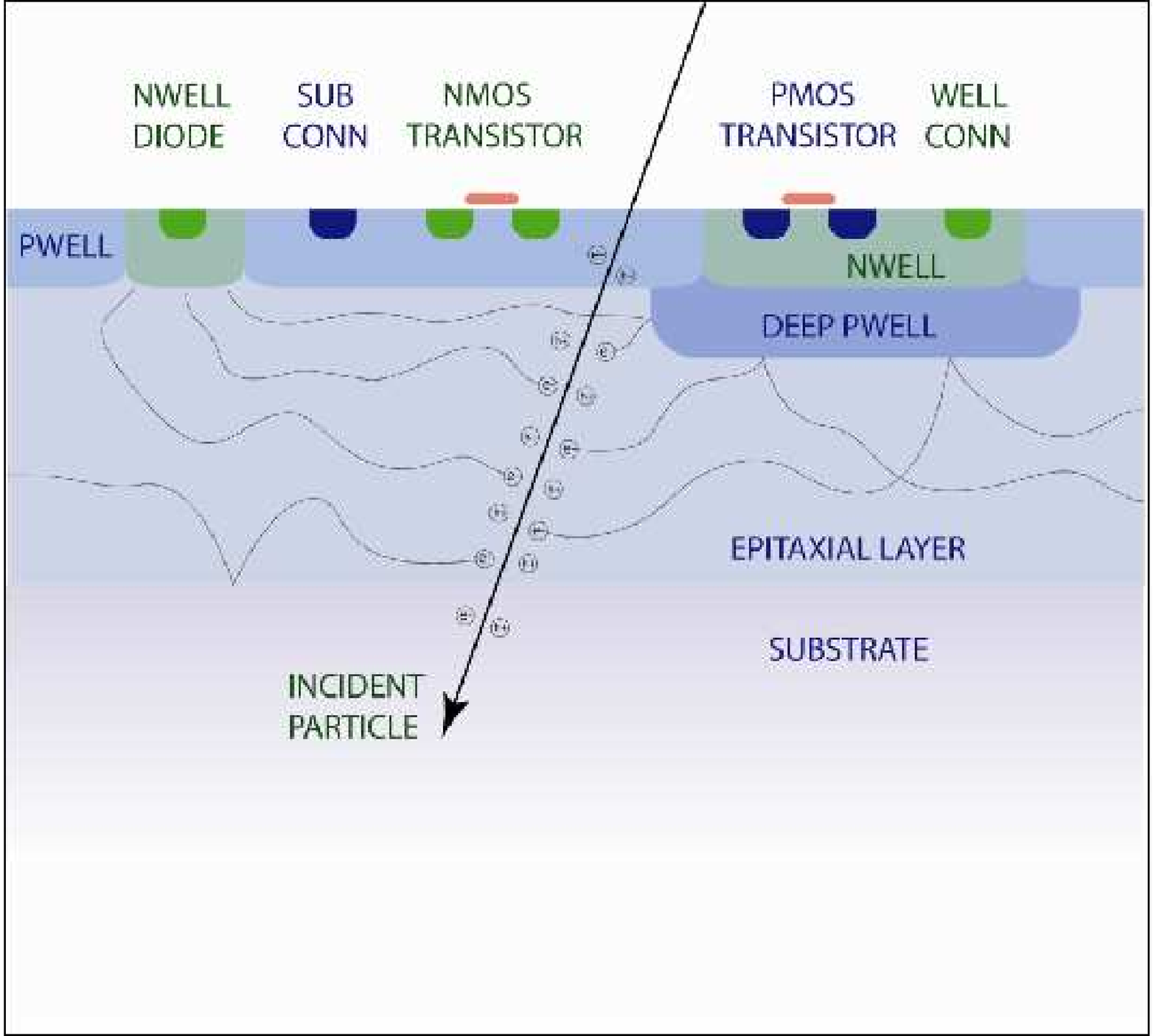}}
\end{center}
\caption{\label{fig:cmos-dpw} A CMOS MAPS without a deep P-well
implant (left) and with a deep P-well implant (right).}
\end{figure*}

Another alternative under evaluation is to have a all-pixel SVT using MAPS pixels with a pixel
size of $50 \times 50 \muma$. This approach uses the 180\nm INMAPS process which incorporates
a deep P-well. A perceived limitation of standard MAPS is not having full CMOS capability as the
additional N-wells from the PMOS transistors parasitically collect charge, thus
reducing the charge collected by the readout diode. Avoiding the usage of PMOS
transistors however does limit the capability of the readout circuitry
significantly. A limited use of PMOS is allowed with the DNW MAPS design (APSEL chips), 
which anyway accounts for
a small degradation in the collection efficiency.
Therefore, a special deep P-well layer was developed to overcome
the problems mentioned above. The deep P-well protects charge generated in the
epitaxial layer from being collected by parasitic N-wells for the PMOS.  This
then ensures that all charge is being collected by the readout diode and
maximizes charge collection efficiency. This is illustrated in
Fig.~\ref{fig:cmos-dpw}.
This enhancement allows the use of full CMOS circuitry in a MAPS and
opens completely new possibilities for in-pixel processing. The TPAC chip \cite{Ballin:2008db} for
CALICE-UK \cite{Watson:2008zzd,Crooks:2008zz}  has been designed using the
INMAPS process. The basic TPAC pixel has a size of $50 \times 50 \muma$ and comprises a
preamplifier, a shaper and a comparator \cite{Ballin:2008db}.  The
pixel only stores hit information in a Hit Flag. The pixel is running
without a clock and the timing information is provided by the logic
querying the Hit Flag. For the \superb\ application the pixel design was slightly modified.
Instead of just a comparator, a peak-hold latch was added to store the analog
information as well. The chip is organized in columns with a common ADC at the
end of each column. The ADC is realized as a Wilkinson ADC using a 5\MHz clock
rate. The simulated power consumption for each individual pixel is $12\mu$W.
The column logic constantly queries the pixels, but only digitizes the information for the pixels
with a ``Hit Flag". This allows one to save both space and reduce
the power usage and since the speed of the chip is limited by the ADC
also increases the readout speed. Both the address of the pixel being hit
and its ADC output are stored in a FIFO at the end of the column.
To further increase the readout speed, the ADC uses a pipelined
architecture with 4 analog input lines to increase throughput of the
ADC. One of the main bottlenecks is getting the data off the chip. It
is envisaged to use the Level~1 trigger information to reject most of
the events and to reduce the data rate on-chip before moving
it off-chip. This will significantly reduce the data rate and
therefore also the amount of power and services required .

For the outer layers, the requirements are much more relaxed in terms
of occupancy so, in order to reduce the power, it is planned to
multiplex the ADCs to let them handle more than one column in the
sensor. This is possible because of the much smaller hit rate in the outer
layers and the resulting relaxed timing requirements.

An advantage of the MAPS is the elimination of a lot of readout electronics,
because everything is already integrated in the sensor, which simplifies the
assembly significantly. Also since we are using an industry CMOS process, there
is a significant price advantage compared to standard HEP-style silicon and 
additional savings due to the elimination of a dedicated readout ASIC.

In order to evaluate the physics potential of  MAPS based all-pixel vertex detector,
 we are currently evaluating the
performance of the \superb\ detector with different geometries of the SVT , ranging from
the \superb\ baseline (Layer0 + 5  layers based on strip detectors), through to a 4 or 6 layer all-pixel
 detector with a realistic material budget for the support structure for all layers.

\wpsubsec{R\&D Activities}  
The technology for the Layer0 baseline striplet design is well-established. However,
 the front-end chip to be used, due 
to the expected high background occupancy, requires some deeper investigation.
Performance of the FSSR2 chip, proposed for the readout of the striplets and the outer layer
strip sensors, are being evaluated as a function of the occupancy with Verilog simulation.
Measurements are also possible in a testbench in preparation with real striplets modules
 readout with the FSSR2 chips.
The design of the digital readout of the chip will be investigate to improve its efficiency.  
The modification of the analog part of the chip for the readout of the 
long module of the external layers are currently under study.
The multilayer flexible circuit, to connect the striplets sensor to the front-end electronics, may
benefit from some R\&D to reduce the material budget: either
to reduce the minimum pitch on the Upilex circuit, or adopt
kapton/aluminum microcables and Tape Automated Bonding soldering techniques
with a 50\mum pitch.

Although silicon striplets are a viable option at moderate background levels, a pixel
system would certainly be more robust against
background. Keeping the material in a pixel system low enough not to deteriorate the vertexing
performance is challenging, and there is considerable activity to develop thin hybrid pixels
or, even better, monolithic active pixels. These devices may be part of a planned upgrade path
and installed as a second generation Layer0.

A key issue for the readout of the pixels in Layer0 is the development of 
a fast readout architecture to cope with a hit rate of the order of 100\MHz/\cma.
A first front-end chip for hybrid pixel sensor with $50 \times50\muma$ pitch and
a fast readout, data driven with timestamp for the hits, has been realized and is currently under test.
A further development of the architecture is being  pursued to evolve toward a triggered readout architecture, 
helpful to reduce the complexity of the pixel module and possibly to reduce its material budget.
  
The CMOS MAPS technology is very promising for an alternative design of
Layer0, but extensive R\&D is still needed to meet all the requirements.
Key aspects to be addressed are: sensor efficiency and its radiation tolerance; power consumption; and, as in the hybrid pixel, the readout speed of the architecture implemented.  

After the realization of the APSEL chips with the ST 130\nm DNW process, with very encouraging results, the Italian collaborators involved in the CMOS MAPS R\&D are now evaluating the possibility to improve MAPS performance with the use 
of modern vertical integration technologies~\cite{yarema:vertex07}.
A first step in this direction has been the realization of a two-tier DNW MAPS
by face to face bonding of two 130\mum CMOS wafer in the Chartered/Tezzaron
process.  Having the sensor and the analog part of the pixel cell in one tier and the digital part in the second tier can significantly improve the efficiency  of the CMOS sensor and allow a more complex in-pixel logic.
The first submission of vertically integrated DNW MAPS, now in
fabrication, includes a 3D version of a $8 \times 32$ MAPS matrix with the same
sparsified readout implemented in the APSEL chips.
A new submission is foreseen in Autumn 2010 with a new generation of the 3D MAPS implementing a faster readout architecture under development, which is still data push but could be quite easily evolve toward a triggered architecture.

The development of a thin mechanical support structure with integrated cooling for the pixel module is continuing with promising results. Prototypes with light carbon fiber microchannels for the coolant fluid (total material down to 0.15\%
\Xrad) have been produced and tested and are able to evacuate specific power up to 1.5W/\cma maintaining the pixel module temperature within the requirements. These supports could be used for either hybrid pixel or MAPS sensors.

\aftsec

\begsec
\graphicspath{{DCH/}{DCH/}}
\wpsec{Drift Chamber}
\label{sec:DCHmain}

\def \dch {DCH\xspace}
\def \Dch {Drift chamber\xspace}
\def \DCH {Drift Chamber\xspace}
\def \cc {\textit{cluster counting}\xspace}
\def \Cc {\textit{Cluster counting}\xspace}
\def \CC {\textit{Cluster Counting}\xspace}
\let\dEdx=\dedx

The \superb\ \DCH (\dch) provides measurements of the charged particle momentum 
and of the ionization energy loss used for particle
identification. This is the primary device in \superb\ to measure
velocities of particles having momenta below approximately 700\mevc.
It  is based on the \babar\ design, with 40 layers of
centimetre-sized cells strung approximately parallel to the beamline\,\cite{bib:BaBarNIM}. A subset of layers are strung at a small
stereo angle in order to provide measurements along z, the beam
axis.

The \dch is required to provide momentum measurements with the same
precision as the \babar\ \dch (approximately 0.4\% for tracks with a
transverse momentum of 1\,\gevc), and like \babar\ uses a helium-based
gas mixture in order to minimize measurement degradation from multiple
scattering.  The challenge is to achieve comparable or better
performance than \babar\ in a high luminosity environment. Both
physics and background rates will be significantly higher than in
\babar\ and as a consequence the system is required to accommodate the
100-fold increase in trigger rate and luminosity-related backgrounds
primarily composed of radiative Bhabhas and electron-pair backgrounds
from two-photon processes. However, the beam current related
backgrounds will only be modestly higher than in \babar. The nature
and spatial distributions of these backgrounds dictate the overall
geometry of the \dch.

The ionization loss measurement is required to be at least as
sensitive to particle discrimination as \babar\ which has a \dEdx
resolution of 7.5\%\,\cite{bib:BaBarNIM}. 
In \babar, conventional \dEdx drift chamber
methods were used in which the total charge deposited on each sense
wire was averaged after removing the highest 20\% of the measurements
as a means of controlling Landau fluctuations.  In addition to this
conventional approach, the \superb\ \dch group is exploring a \cc
option\,\cite{bib:ClusterCounting} which, in principle, can improve the \dEdx
resolution by approximately a factor of two.  This technique involves counting
individual clusters of electrons released in the gas ionization
process. In so doing, we remove the sensitivity of the specific energy
loss measurement to fluctuations in the amplification gain and in the
number of electrons produced in each cluster, fluctuations which
significantly limit the intrinsic
resolution of conventional \dEdx measurements.  As no experiment has
employed \cc, this is very much a detector research and development
project but one which potentially yields significant physics payoff at
\superb.

\wpsubsec{Backgrounds}
\label{subsubsec:DCH_Backgrounds}
The dominant source of background in the \superb DCH is expected to be
radiative Bhabha scattering. Photons radiated collinearly to the
initial \en or \ep direction can bring the beams off-orbit and
ultimately produce showers on the machine optic elements. This process
can happen meters away from the interaction point and the hits are in
general uniformly distributed over the whole \dch volume.  Large-angle
$\epem\to\epem(\gamma)$ scattering, on the other hand,
has the well-known $1/\vartheta^4$
cross section dependence; simulation studies are currently underway to evaluate
the need to design tapered endcaps (either conical or a with stepped
shape) at small radii to keep under control the occupancy in the very
forward region of the detector.  The actual occupancy and its
geometrical distribution in the detector depend on the details of the
machine elements, on the amount and placement of shields, on the \dch
geometry, and on the time needed to collect the signal in the
detector.  Preliminary results obtained with Geant4 simulations
indicate that in a 1\mus time window at nominal luminosity
($10^{36}\cm^{-2}\sec^{-1}$) the occupancy averaged over the whole \dch
volume is $3.5\,\%$, and slightly larger (about $5\,\%$) in the inner
layers. Intense work is presently underway to validate these results
and study their dependence on relevant parameters.

\wpsubsec{Drift Chamber Geometry}
\label{subsec:DCH_Geometry}
The \superb\ \dch will have a cylindrical geometry. The inner
radius and length of the chamber are
being re-optimized through detailed simulation studies with respect to
\babar\ since: \\
\textit{a)} in \superb\ there will be no support tube connecting the machine elements 
between the SVT and the DCH;\\
\textit{b)} the possibility is being considered to add a PID device
between the \dch and the forward calorimeter, and a calorimeter in the
backward direction.

\begin{figure}[!hbtp]
\includegraphics[width=0.50\textwidth]{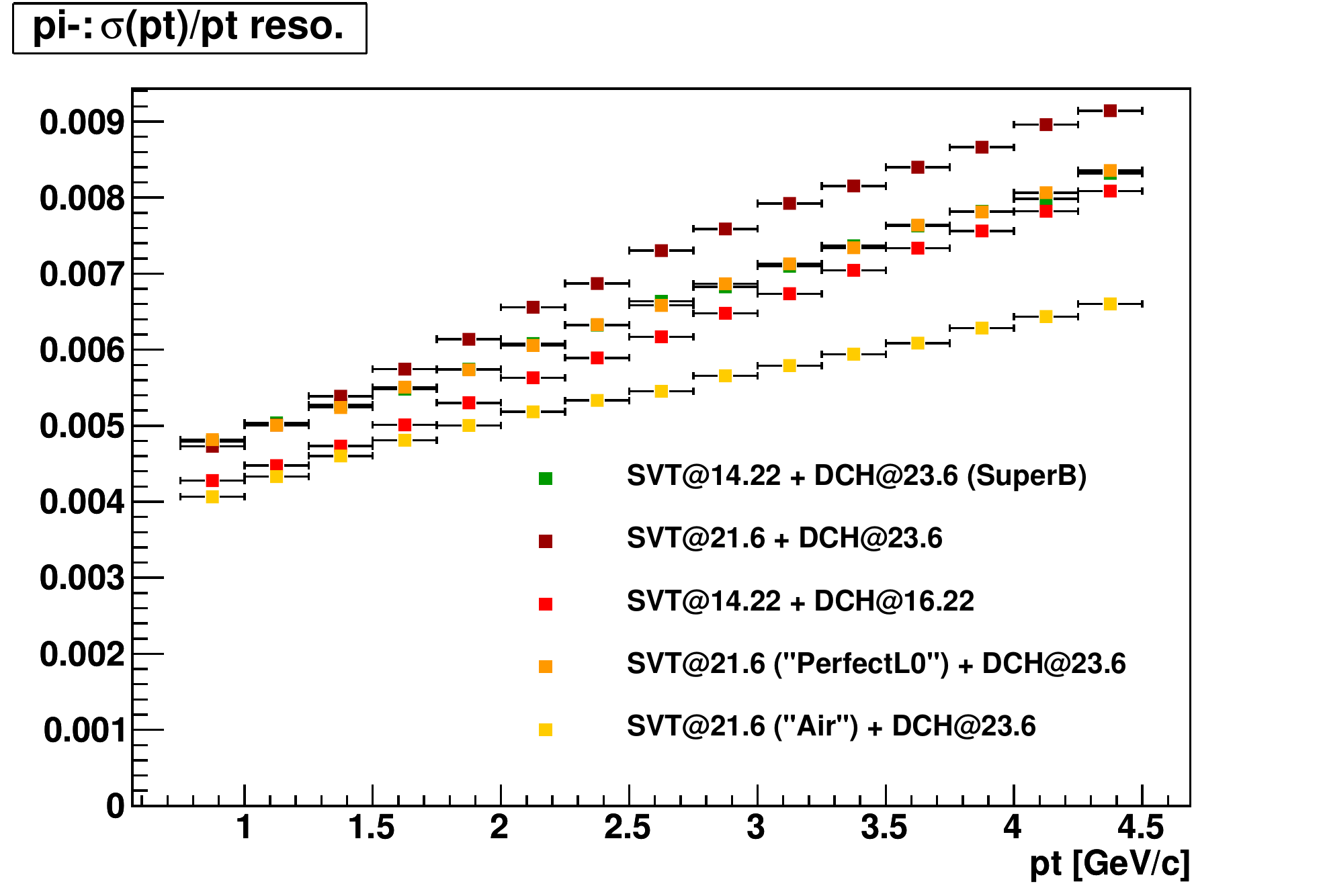}
\caption{Track momentum resolution for different values of the drift
  chamber inner radius.}
\label{fig:DCH_radius_study}
\end{figure}

Simulation studies performed on several signal samples with both high
(\eg $\,\B\to\pipi$), and medium-low  (\eg $\,\B\to\Dstar K$) momentum
tracks indicate that: \\
\textit{a)} due to the increased lever arm, momentum resolution
improves as the minimum \dch radius $R_{min}$ decreases, see
Fig.\,\ref{fig:DCH_radius_study}; $R_{min}$ is actually
limited by mechanical integration constraints with the
cryostats and the radiation shields. \\
\textit{b)} The momentum and especially the \dedx
resolution for tracks going in the forward or backward directions are
clearly affected by the change in number of measuring samples when the
chamber length is varied by $10-30\cm$. However the fraction of such
tracks is so small that the overall effect is negligible.

\begin{figure*}[!hbtp]
\begin{center}
\subfigure[Spherical endplates design.]{\includegraphics[width=0.45\textwidth]{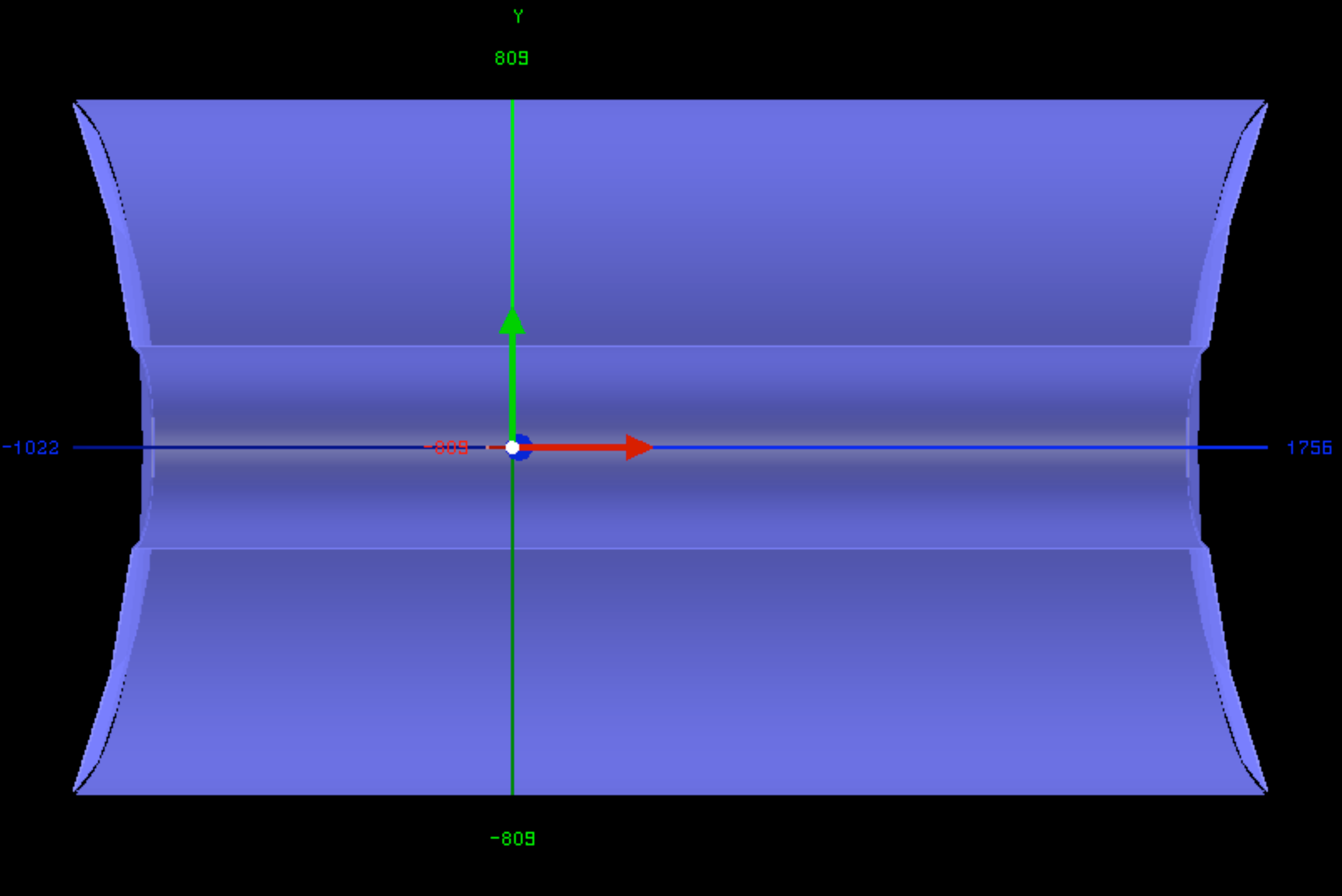} }
\subfigure[Stepped endplates design.]{\includegraphics[width=0.443\textwidth]{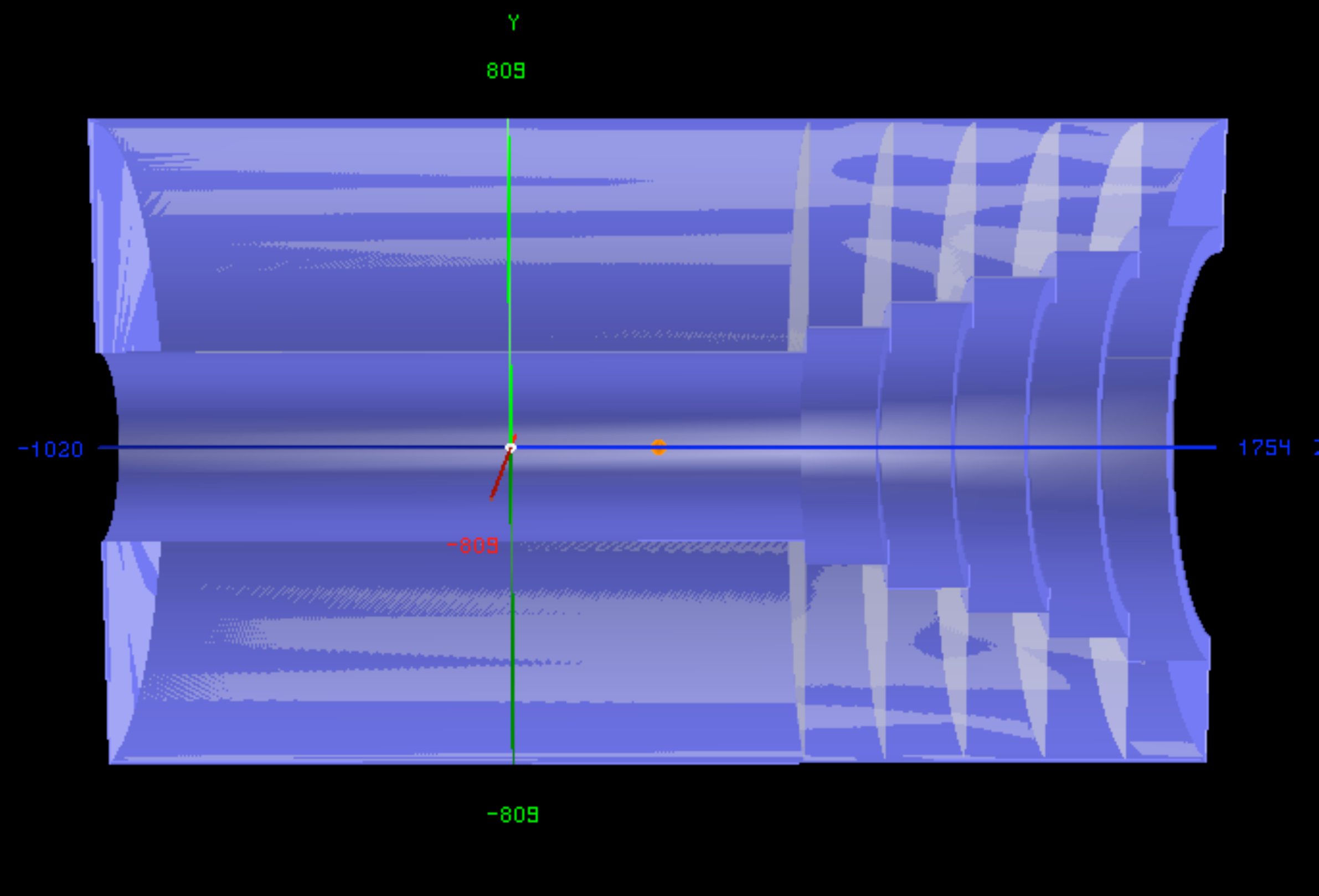}}
\end{center}
\caption{Two possible \superb\ DCH layouts.}
\label{fig:DCH_designs}
\end{figure*}

The \dch outer radius is constrained to 809\mm by the DIRC quartz
bars.  As discussed before, the DCH inner radius will be as small as
possible: since  a final design of the final focus cooling system
is not available yet, in Fig.~\ref{fig:DCH_designs} the the nominal
\babar\ DCH inner radius of 236\mm has been used. Similarly, a nominal
chamber length of 2764\mm at the outer endplate radius is used in
Fig.~\ref{fig:DCH_designs}: as mentioned above, this dimension has not
been fixed yet, since it depends on the presence and the details of
forward PID and backward EMC systems, still being discussed.  Finally,
as the rest of the detector, the drift chamber is shifted by the
nominal \babar\ offset (367\mm) with respect to the interaction point.

\wpsubsec{Mechanical Structure}
\label{subsec:DCH_MechStructure}
The drift chamber mechanical structure must sustain the wire load --
about 3 tons for 10\,000 cells -- with small deformations, while at
the same time minimizing the material for the surrounding detectors.
Carbon Fiber-resin composites have high elastic modulus and low
density, thus offering performances superior to Aluminum-alloys based
structures. Endplates with curved geometry can further reduce material
thickness with respect to flat endplates for a given deformation under
load.  For example, the KLOE drift chamber$\,$\cite{bib:KLOEDC}
features $8\,\mm$ thick Carbon Fiber spherical endplates of $4\,\m$
diameter.  Preliminary design of Carbon Fiber endplates for \SuperB\
indicate that adequate stiffness ($\le1\mm$ maximum deformation) can
be obtained with $5\mm$ thick spherical endplates, corresponding to
$0.02\Xrad$, to be compared with $0.13\Xrad$ for the \babar\ DCH aluminum
endplates.

Figure\,\ref{fig:DCH_designs} shows two possible endcap layouts,
respectively with spherical (a) or stepped (b) endplates.
A convex spherical endplate is also considered, which would provide  a
better match to the geometry of the forward PID and calorimeter systems,
and would reduce the impact of the endplate material on the performance
of these detectors, at the cost of greater sensitivity to the large-angle Bhabha background.

\wpsubsec{Gas Mixture}
\label{subsec:DCH_GasMixture}
The gas mixture for \superb should satisfy the requirements which
already concurred to the definition of the \babar\ \dch gas mixture
(80\%He-20\%iC$_4$H$_{10}$), \ie\ low density, small diffusion
coefficient and Lorentz angle, low sensitivity to photons with
$E\sim10\kev$.  To match the more stringent requirements on occupancy
rates of \superb, it could be useful to select a gas mixture with a
larger drift velocity in order to reduce ion collection times and so
the probability of hits overlapping from unrelated events.  The \cc
option would instead call for a gas with low drift velocity and
primary ionization.
As detailed in Section\ref{subsec:DCH_RandD}, R\&D work is ongoing to
optimize the gas mixture for the \superb\ environment.

\wpsubsec{Cell Design and Layout}
\label{subsec:DCH_CellLayout}
The baseline design for the drift chamber employs small rectangular
cells arranged in  concentric layers about the axis
of the chamber which is approximately aligned with the beam direction.
The precise cell dimensions and number of layers are still to be determined,
but it is expected that their side 
is between 10 and 20\mm  and that there are
approximately 40 layers as in \babar. 
The cells are grouped radially into superlayers with the
inner and outer superlayers parallel to the chamber axis (axial).
In \babar\ the chamber also had stereo layers in which
the  superlayers are oriented at a small
``stereo'' angle relative to the axis in order to
provide the z-coordinates of the track hits. The details of the stereo layer
layout in \superb\ is still to be determined on the basis of 
the cell occupancy associated with machine backgrounds.

Each cell has one 20\mum diameter gold coated sense wire surrounded
by a rectangular grid of eight field wires.  The sense wires will
be tensioned with a value consistent
with electrostatic stability and with the yield strength of the wire.
The baseline calls for a gas gain of approximately $5\times 10^4$
which requires a voltage of approximately $+2$~kV to be applied to the
sense wires with the field wires held at ground.

The field wires are aluminum with a diameter which will be
chosen to keep the electric field on the wire surface below 20~kV/\cm
as a means of suppressing the Malter effect\,\cite{bib:Malter}. These wires will be tensioned
in order to provide a gravitational sag that matches
that of the sense wires.

\begin{figure*}[hbtp]
\begin{center}
  \subfigure[\texorpdfstring{$80\%$He-$20\%i$C$_4$H$_{10}$ gas
    mixture.}{80\%He-20\%iC4H10 gas mixture.}]
  {\includegraphics[width=0.45\textwidth]{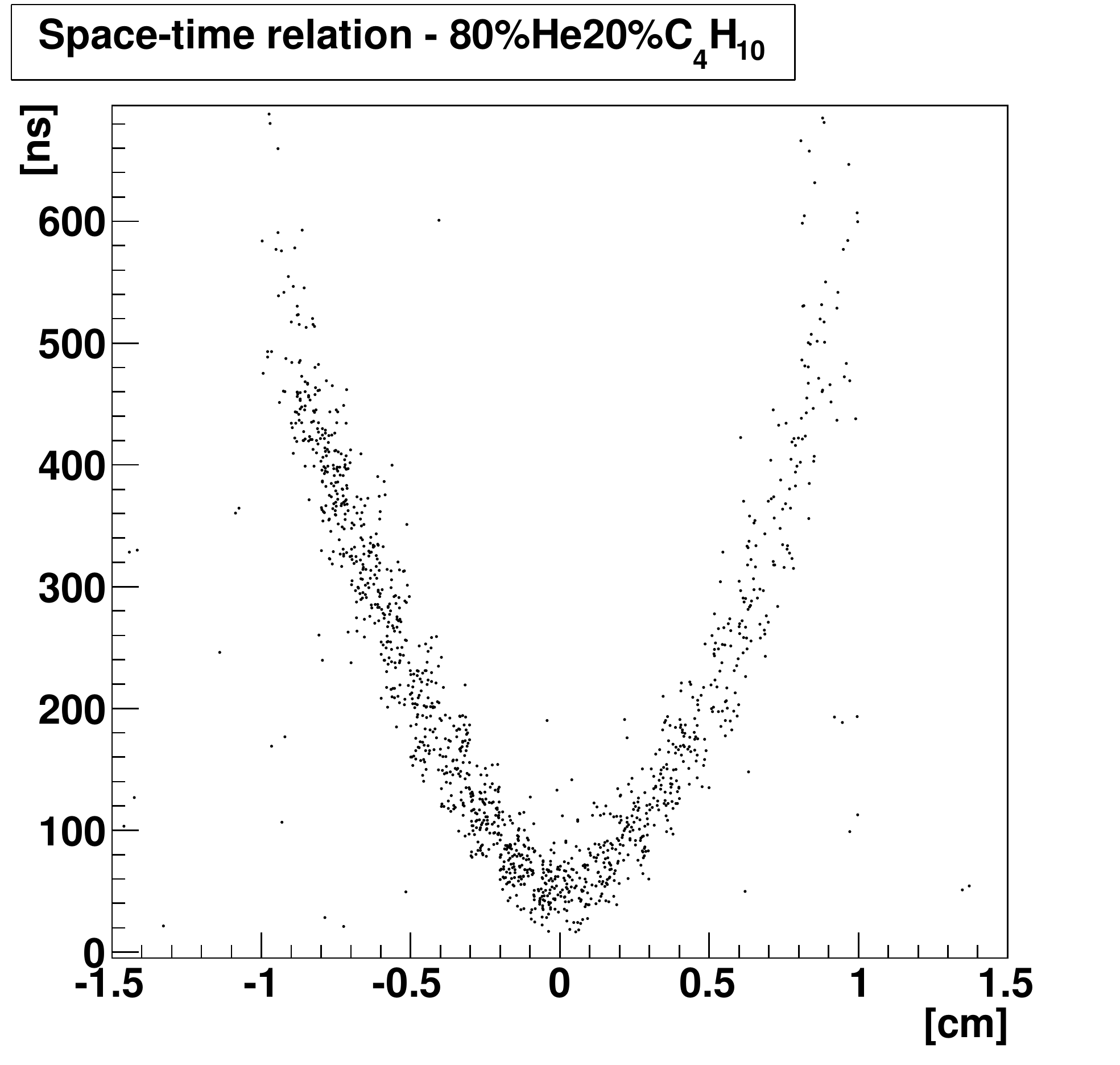}}
  \subfigure[\texorpdfstring{$52\%$He-$48\%$CH$_4$ gas
    mixture.}{52\%He-48\%CH4 gas mixture}]
  {\includegraphics[width=0.45\textwidth]{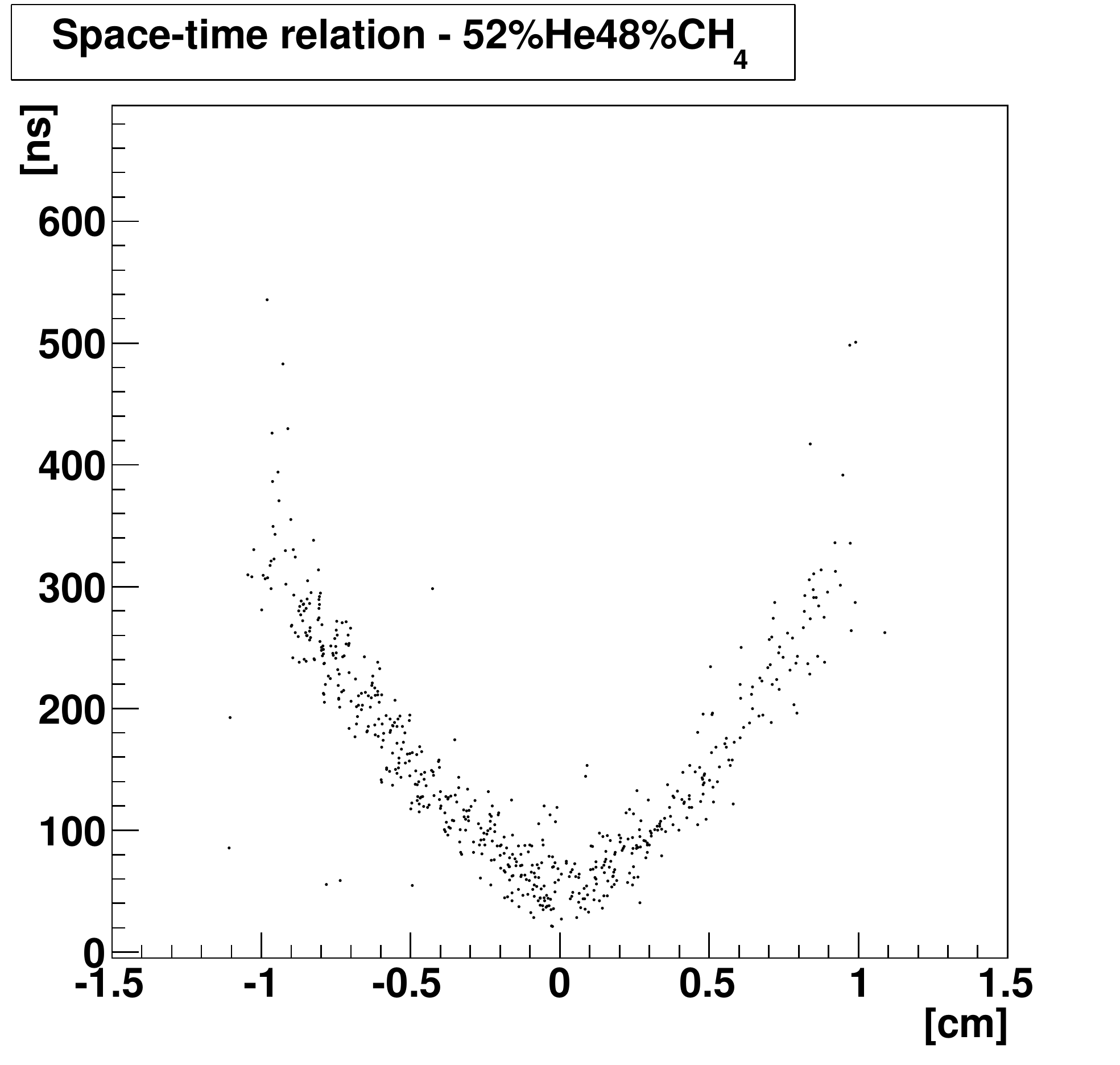}}
\end{center}
\caption{ Examples of measured space-time relation in different
  He-based gas mixtures.}
\label{fig:DC_str}
\end{figure*}

At a radius inside the innermost superlayer the chamber has an
additional layer of axially strung guard wires which serve to
electrostatically contain very low momentum electrons produced from
background particles showering in the DCH inner cylinder and SVT.
A similarly motivated layer will be considered
at the outermost radius to contain machine background related
backsplash from detector material just beyond the outer superlayer.

\wpsubsec{R\&D Work}
\label{subsec:DCH_RandD}
Various R\&D programs are underway towards the definition of an
optimal \dch for \superb, in particular: make precision measurements
of fundamental parameters (drift velocity, diffusion coefficient,
Lorentz angle) of potentially useful gas mixtures; 
study the properties of different gas
mixtures and cell layouts with small \dch prototypes and simulations; 
and verify the potential and feasibility of the \cc option.

A precision tracker made of $3\,\cm$ diameter Aluminum tubes operating
in limited streamer mode with a single tube spatial resolution of
around $100\,\mu$m has been set up.
A small prototype with a cell structure resembling the one used in the
\babar\ DCH has  also been built and commissioned. The tracker and
prototype chamber have been collecting cosmic ray data since October 2009.
Tracks can be extrapolated in the DCH prototype with a precision of
$80\,\mu$m or better.  Different gas mixtures have been tried in the
prototype: starting with the original \babar\ mixture
(80\%He-20\%iC$_4$H$_{10}$) used as a calibration point, both different
quencher proportions and different quenchers (e.g methane instead of isobutane) 
have been tested in order to assess the viability of lighter and
possibly faster operating gas.  Fig.\,\ref{fig:DC_str}a shows the
space-time correlation for one prototype cell: as mentioned before,
the cell structure is such as to mimic the overall structure of the
\babar\ DCH. Spatial resolution
is consistent with what has been obtained with the original \babar\
DCH.  A space to time relation is depicted in Fig.\,\ref{fig:DC_str}b
with a 52\%He-48\%CH$_4$ gas mixture.  This gas is roughly a factor
two faster and 50\% lighter than the original \babar\ mix: preliminary
analysis shows space resolution performances comparable to the
original mix; however detailed studies of the Lorentz angle have to be
carried out in order to consider this mixture as a viable alternative.

To improve performances of the gas tracker a possible road could be
the use of the \CC method. If the individual ionization cluster can be
detected with high efficiency, it could in principle be possible to
measure the track specific ionization by counting the clusters
themselves, providing a two-fold improvement in the resolution
compared to the traditional truncated mean method.
Having many independent time measurements in a single cell, the spatial
accuracy could also in principle be improved substantially.
Since the efficient detection of single ionization clusters requires fast risetimes
(preamplifier bandwidths of the order of 1GHz) and also sampling the signal
with rates of $\sim$2\,Gsa/$\!\sec$, these promises of exceptional
energy and spatial resolution must however fit with the available data
transfer bandwidth. A dedicate R\&D effort is required to identify
a gas mixture with well-separated clusters and high detection efficiency.
The preamplifier noise is also an issue.

Comparisons of the traditional methods to extract spatial position and
energy losses and the \cc method are being setup at the moment of
writing the present report.

\aftsec

\begsec
\graphicspath{{PID/}{PID/}}
\wpsec{Particle Identification}

\wpsubsec{Detector Concept}

The DIRC (Detector of Internally Reflected Cherenkov
light)~\cite{blair_92} is an example of innovative detector technology
that has been crucial to the performance of the \babar\ 
science program. Excellent flavor tagging will continue to be
essential for the program of physics anticipated at \superb, and the
gold standard of particle identification in this energy region is that provided by internally reflecting ring-imaging
devices (the DIRC class of ring imaging detectors). The challenge for
\superb\ is to retain (or even improve) the outstanding performance
attained by the \babar\ DIRC~\cite{adam_2007}, while also gaining an
essential factor of 100 in background rejection to deal with the much
higher luminosity.

A new Cherenkov ring imaging detector is being planned for the
\superb\ barrel, called the Focusing DIRC, or FDIRC. It will use the
existing \babar\ bar boxes and mechanical support structure. This structure
will be attached to a new photon ``camera'', which will be
optically coupled to the bar box window. The new camera design
combines a small modular focusing structure that images the photons
onto a focal plane instrumented with very fast, highly pixelated,
photon detectors (PMTs). These elements should combine to attain the
desired performance levels while being at least 100 times less
sensitive to backgrounds than the \babar\ DIRC.

Several options are also under consideration for a possible PID detector in
the forward direction. The design variables being considered should include:
(a) modest cost, (b) small mass in front of the LYSO calorimeter, 
and (c) good PID coverage at low momenta by removing the
 \dedx\ ambiguity in $\pi/K$ separation near 1\gevc. Presently, we are
considering the following technologies: (a) ``DIRC-like''
\tof (TOF)~\cite{jjv_toflike}, (b) pixelated TOF~\cite{jjv_nima2009} and (c)
Aerogel RICH~\cite{farich}. The aim is to design the best possible
\superb\ detector by optimizing physics, performance and cost,
while being constrained to the existing \babar\ geometry.

\wpsubsubsec{Charged Particle Identification at \superb}

The charged particle identification at \superb\ relies on the same
framework as the \babar\ experiment. Electrons and muons are
identified by the EMC and the IFR respectively, aided by \dedx\ 
measurements in the inner trackers (SVT and DCH).  Separation 
for low-momentum hadrons is primarily provided by \dedx. 
At higher momenta (above 0.7\gevc for pions
and kaons, above 1.3\gevc for protons), a dedicated system, the
FDIRC~-- inspired by the successful \babar\ DIRC~-- will perform the
$\pi/K$ separation. 
This new detector, described in
Section~\ref{subsection:FDIRC}, is expected to perform well over the
entire momentum range for $B$-physics. But its geometrical coverage is
limited to the barrel region. As discussed above,  there is an 
ongoing effort to determine the physics impact of a forward
PID system, together with an active R\&D effort on possible
detector technologies.

\wpsubsubsec{\babar\ DIRC}

\begin{figure}
\includegraphics[width=0.5\textwidth]{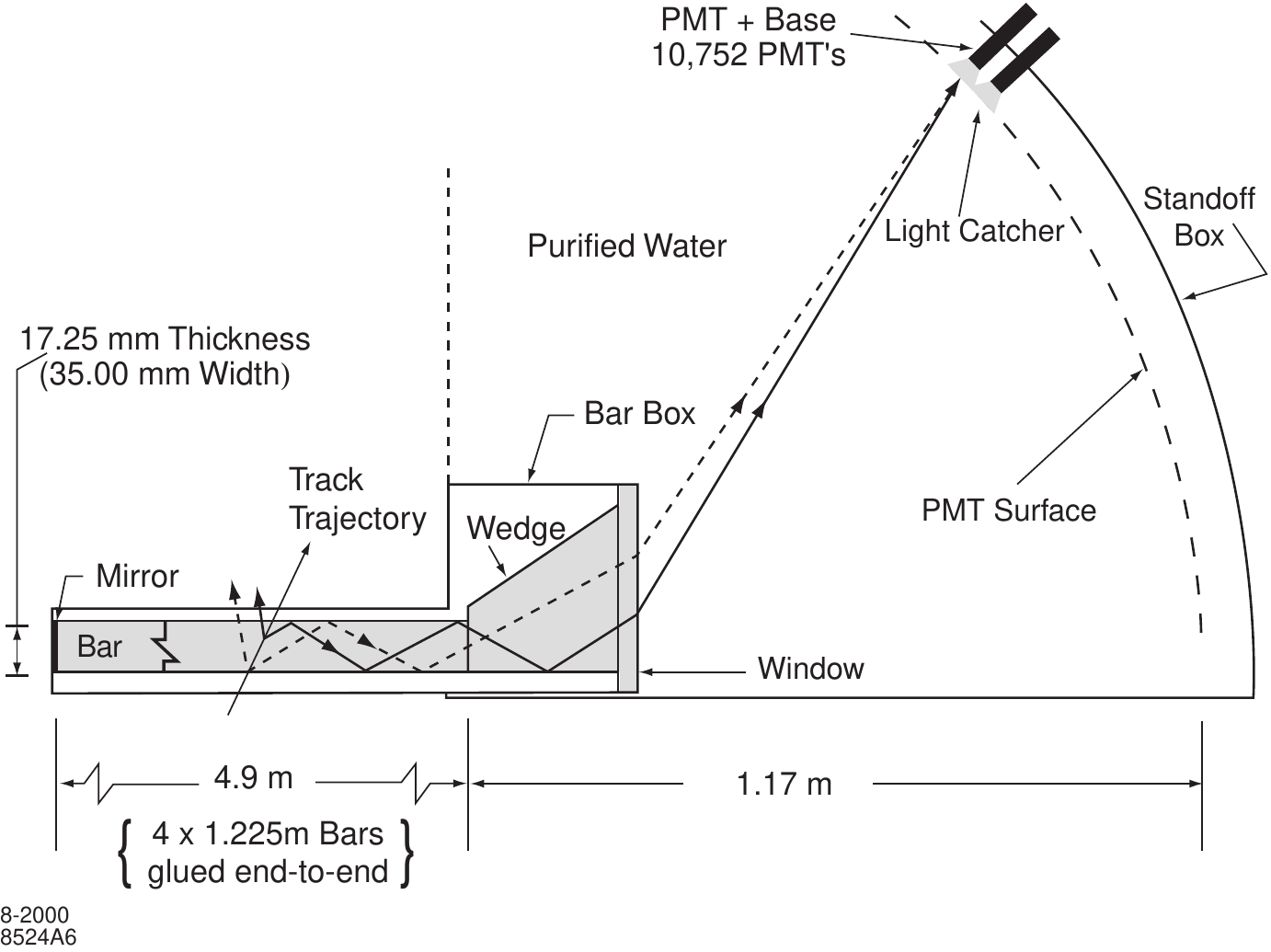}
\caption{Schematic of the \babar\ DIRC.}
\label{fig:DIRC}
\end{figure}

The \babar\ DIRC~-- see Fig.~\ref{fig:DIRC}~-- is a novel ring-imaging Cherenkov detector. The
Cherenkov light angular information, produced in ultra-pure synthetic fused silica bars, is preserved while propagating along the bar via
internal reflections to the camera (the SOB) where an image is produced and detected.

The entire DIRC has 144 quartz bars, each 4.9\m long, which are set
along the beam line and cover the whole azimuthal range. 
Thanks to an internal reflection coefficient of $\sim 0.9997$ and orthogonal bar faces,
Cherenkov photons
are transported to the back end of the bars with the magnitude of
their angles conserved and only a modest loss of photons. They exit into a  pinhole camera consisting of a large volume of purified water
(a medium chosen because it is inexpensive, transparent, and easy to clean, with average index of refraction and relative 
chromatic dispersion sufficiently close to those of the fused silica). The photon detector PMTs are located at the rear of the SOB, about
1.2\m away from the quartz bar exit window. 

The reconstruction of the Cherenkov angle uses information from the
tracking system together with the positions of the PMT hits in the
DIRC.  In addition, information on the time of arrival of hits is
used in rejecting background hits, and resolving ambiguities. 

The \babar\ DIRC performed reliably and efficiently over the whole
\babar\ data taking period (1999---2007). Its physics performance remained
consistent throughout the run period, although some upgrades, such as the addition of
shielding and replacement of electronics, were necessary to cope
with machine conditions. Its main
performance parameters are the following:
\begin{itemize}
\item measured time resolution of about 1.7\ns, close to the PMT transit time spread of 1.5\ns;
\item single photon Cherenkov angle resolution of 9.6\mrad for dimuon events;
\item Cherenkov angle resolution per track of 2.5\mrad in dimuon events;
\item $K-\pi$ separation above 2.5~`$\sigma$' from the pion Cherenkov threshold up to 4.2\gevc.
\end{itemize}

\wpsubsec{Barrel PID at \superb}
\label{subsection:FDIRC}

\wpsubsubsec{Performance Optimization}

As discussed above, the PID system in \superb\ must cope
with much higher luminosity-related background rates than in \babar~-- current estimates are 
on the order of 100 times higher. The basic strategy is to make the camera much smaller and faster.  A new photon camera imaging concept, based on
focusing optics, is therefore envisioned.  The focusing blocks (FBLOCK), responsible
for imaging the Cherenkov photons onto the PMT cathode surfaces, would be
machined from radiation-hard pieces
of fused silica. The major design constraints for the new camera are
the following: (a) it must be consistent with the existing \babar\
bar box design, as these elements will be reused in \superb; (b) it must
coexist with the \babar\ mechanical support and magnetic field constraints; (c) it
requires very fine photon detector pixelation and fast photon detectors.

Imaging is provided by a mirror structure focusing onto an image
plane containing highly pixelated photomultiplier tubes. The reduced
volume of the new camera and the use of fused silica for coupling
to the bar boxes (in place of water as 
it was in \babar\ SOB),  is expected to 
reduce the sensitivity to background by about one order of magnitude
compared to \babar\ DIRC. The very fast timing of the new PMTs is expected
to provide many additional advantages: (a) an improvement of the
Cherenkov resolution; (b) a measure of the chromatic dispersion term
in the radiator~\cite{benitez_2006,jjv_2007,benitez_2008}; (c) 
separation of ambiguous solutions in the folded optical system; and
(d), another order of magnitude improvement in background rejection.

\begin{figure*}
\begin{center}
\subfigure[FDIRC optical design (dimensions in \cm).]{\includegraphics[width=0.45\textwidth]{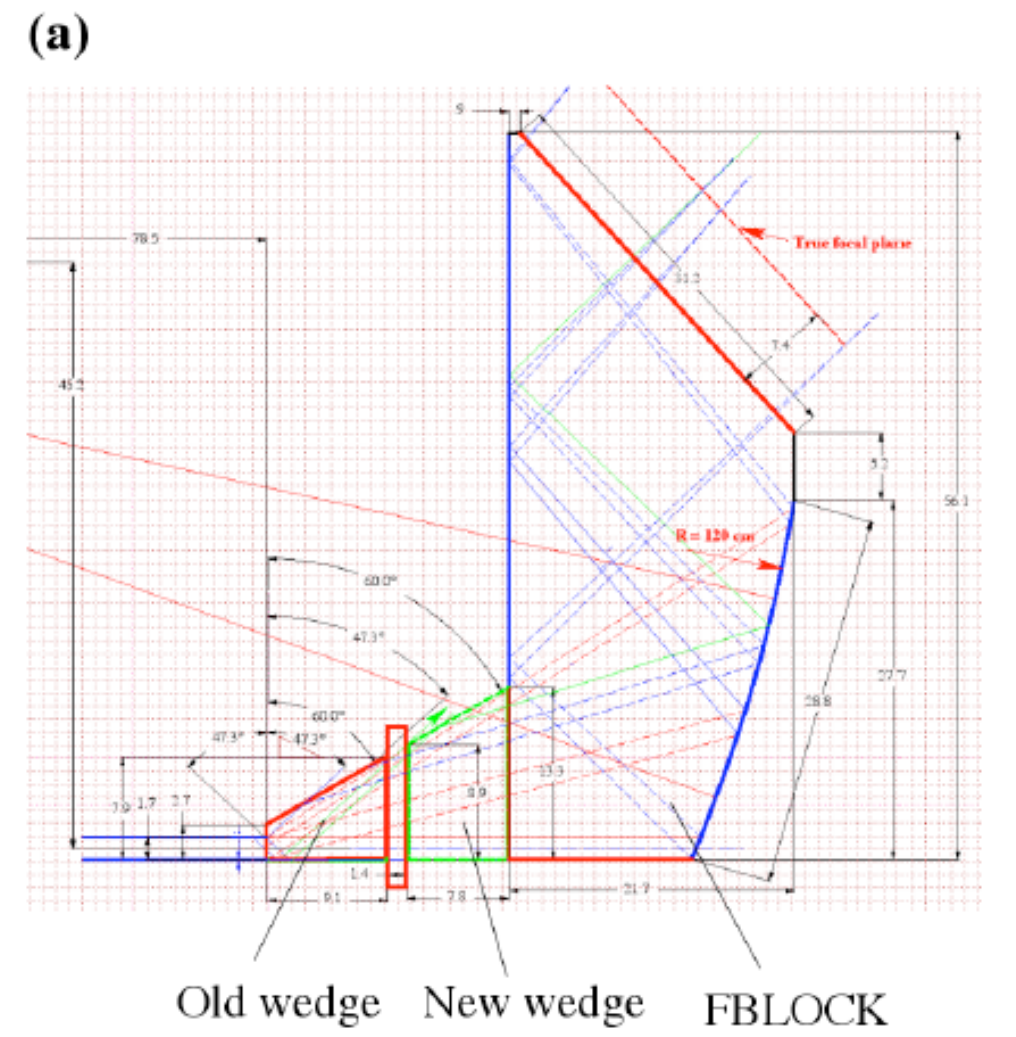} }
\subfigure[Its equivalent in the Geant4 MC model.]{\includegraphics[width=0.35\textwidth]{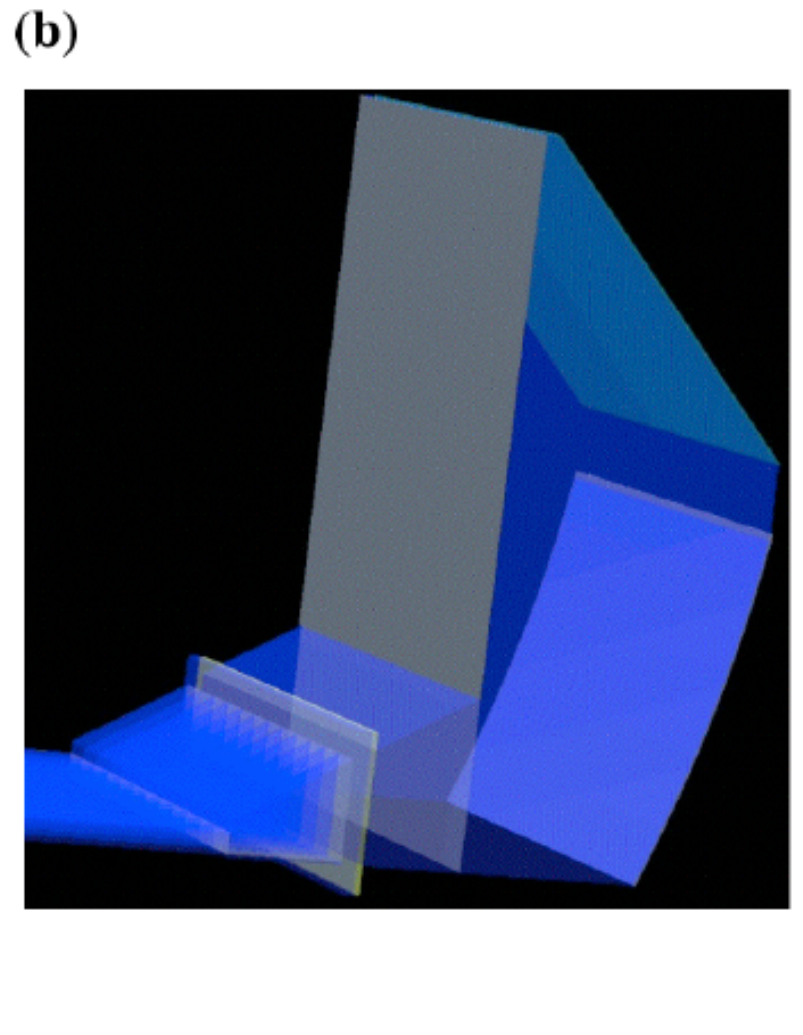}}
\end{center}
\caption{Barrel FDIRC Design.}
\label{fig:FDIRC_design}
\end{figure*}

Figure~\ref{fig:FDIRC_design} shows the new FDIRC camera design (see
Ref.~\cite{jjv_2008_2009} for more detail). It consists of two parts:
(a) a focusing block (FBLOCK) with cylindrical
and  flat mirror surfaces, and (b) a new wedge. 
The wedge at the end of the bar rotates rays with large transverse angles (in the focusing plane) before they emerge into the focusing structure. The old wedge is too short so that an additional wedge element must be added to insure that all rays strike the cylindrical
mirror.  The cylindrical mirror is rotated appropriately to make sure that all
rays reflect onto the FBLOCK flat mirror, preventing reflections
back into the bar box itself; the flat mirror then reflects rays onto
the detector focal plane with an incidence angle of almost $90^\circ$,
thus avoiding reflections. The focal plane is located in a slightly
under-focused position to reduce the FBLOCK size and therefore its
weight.  Precise focusing is unnecessary,  as the finite pixel size
would not take advantage of it. The total weight of the solid fused
silica FBLOCK is about 80kg. This significant weight requires good
mechanical support.

There are several important advantages gained in moving from the \babar\
pinhole focused design with water coupling to a focused optical design made of solid fused
silica: (a) the design is modular; (b) sensitivity to background,
especially to neutrons, is significantly reduced; (c) the pinhole-size
component of the angular resolution in the focusing plane can be
removed,  and timing can be used to measure the chromatic dispersion, thus
improving performance; (d) the total number of photomultipliers is reduced by about one half compared to a
non-focusing design with equivalent performance; (e) there is no risk
of water leaks into the \superb\ detector, and no time-consuming
maintenance of a water system, as was required to operate \babar\ safely.

Each new camera will be attached to its \babar\ bar box with an optical RTV glue, which will be injected in a liquid form between the bar box window and the new camera and cure in place. As Fig.~\ref{fig:FDIRC_design} shows, the cylindrical mirror focuses in the radial (y) direction, while pinhole focusing is used in the direction out of the plane of the schematic (the x-direction). Photons that enter the FBLOCK at large x-angles reflect from the parallel sides, leading to an additional ambiguity. However, the folded design makes the optical piece small, and places the photon detectors in an accessible location, improving both the mechanics and the background sensitivity. Since the optical mapping is 1 to 1 in the y-direction, this ``folding" reflection does not create an additional ambiguity. Since a given photon bounces inside the FBLOCK only 2---4 times, the requirements on surface quality and polishing for the optical pieces are much less stringent than that required for the DIRC bar box radiator bars. This significantly 
reduces the cost of  optical fabrication.

Each DIRC wedge inside an existing bar box has a 6\mrad angle at the bottom. This was done intentionally in \babar\ to provide simple step-wise ``focusing" of rays leaving the bar towards negative y to reduce the effect of  bar thickness. However, in the new optical system, having this angle on the inner wedge somewhat worsens the  design FDIRC optics resolution. There are two choices: (a) either leave it as it is, or (b) glue a micro-wedge at the bottom of the old wedge, inside the bar box, to correct for this angle. Though (b) is possible in principle, it is far from trivial, as the bar box must be opened.

The performance of the new FDIRC is simulated with a Geant4 based
program~\cite{doug}. Preliminary results for the expected
Cherenkov angle resolution are shown in
Table~\ref{table:FDIRC_resolution} for different layouts~\cite{doug}.
Design \#1, which has emerged as the 
 preferred one  (a $3\mm \times 12\mm$ pixel size
with the micro-wedge glued in) gives a resolution of $\sim 8.1\mrad$ per photon
for 4\gevc pions at $90^\circ$ dip angle. This can be compared
with \babar\ DIRC's measured resolution of $\sim 9.6\mrad$  per photon for di-muon
events. If we decide not to glue in the micro-wedge (design \#2), the
resolution will increase to 8.8\mrad per photon \ie, we lose about 0.7~mrad per photon.
Going to a coarser pixelization of $6\mm \times 12\mm$ will worsen the
Cherenkov angle resolution by $\sim 1\mrad$ per photon (see designs \#3 \&
\#4). On the other hand, correcting for chromatic dispersion 
using timing information on each
photon~\cite{benitez_2006,jjv_2008_2009,field_2004} may improve the FDIRC resolution by an additional 0.5---1\mrad per photon.

\begin{table}
\begin{center}
\begin{tabular}{|c|c|c|}
\hline \parbox[t][1.5cm][t]{1.4cm}{FDIRC Design} & Option & \parbox[t][1.5cm][t]{1.7cm}{$\theta_C$ \\ resolution [mrad]} \\
\hline 1 & \parbox[t][1.6cm][t]{3cm}{$3\mm \times 12\mm$ pixels with a micro-wedge} & 8.1 \\
\hline 2 & \parbox[t][1.6cm][t]{3cm}{$3\mm \times 12\mm$ pixels and no micro-wedge} & 8.8 \\
\hline 3 & \parbox[t][1.6cm][t]{3cm}{$6\mm \times 12\mm$ pixels with a micro-wedge} & 9.0 \\
\hline 4 & \parbox[t][1.6cm][t]{3cm}{$6\mm \times 12\mm$ pixels and no micro-wedge} & 9.6 \\
\hline
\end{tabular}	
\caption{FDIRC performance simulation by Geant4 MC.}
\label{table:FDIRC_resolution}
\end{center}
\end{table}

\wpsubsubsec{Design and R\&D Status}

\begin{figure*}
\begin{center}
  \subfigure[Single photoelectron response of H-9500 MaPMT with $3\mm
  \times 12\mm$
  pixels.]{\includegraphics[width=0.4\textwidth]{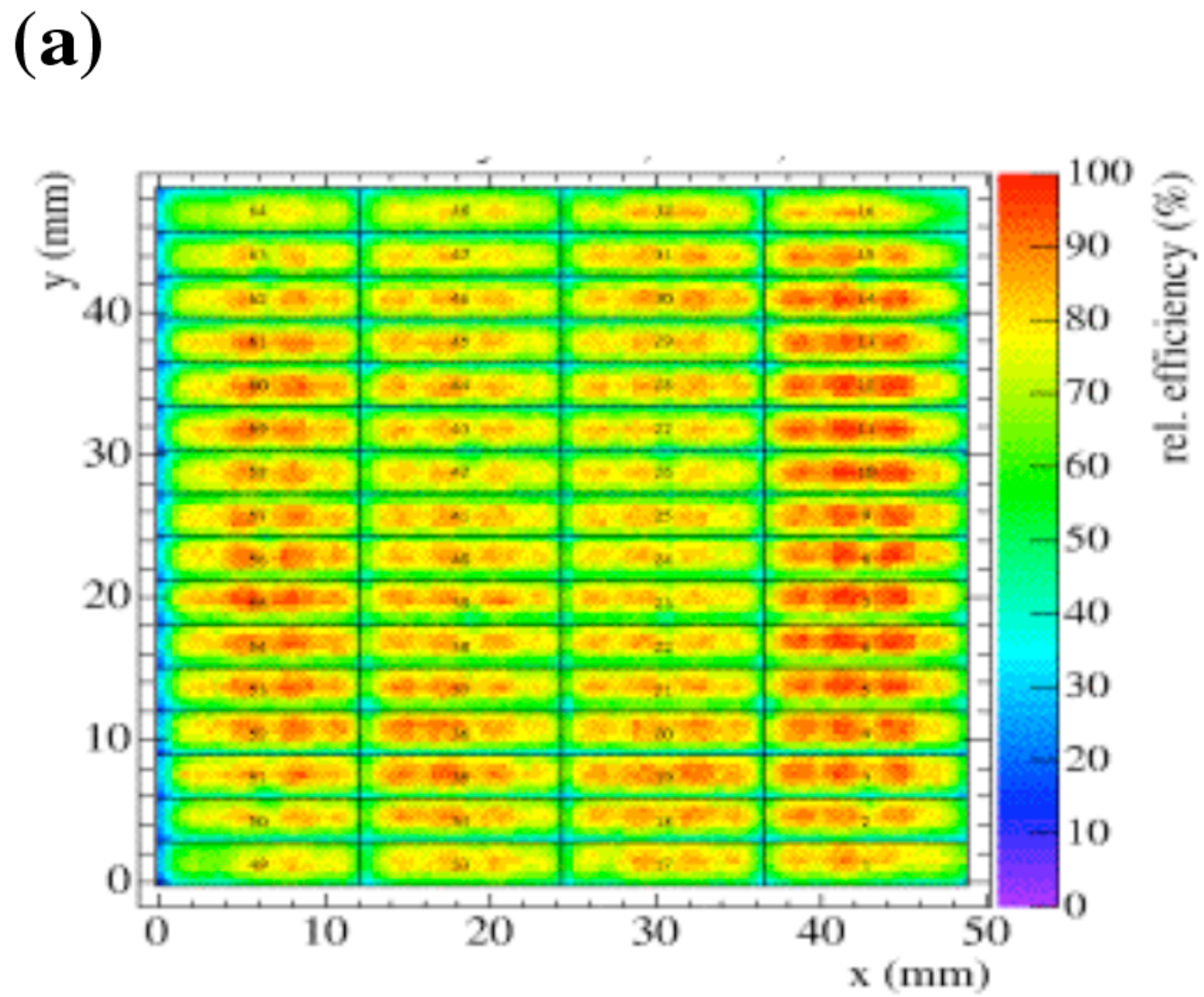}}
  \subfigure[Similar scan of H-8500 MaPMT with pixels: $6\mm \times
  6\mm$.]{\includegraphics[width=0.4\textwidth]{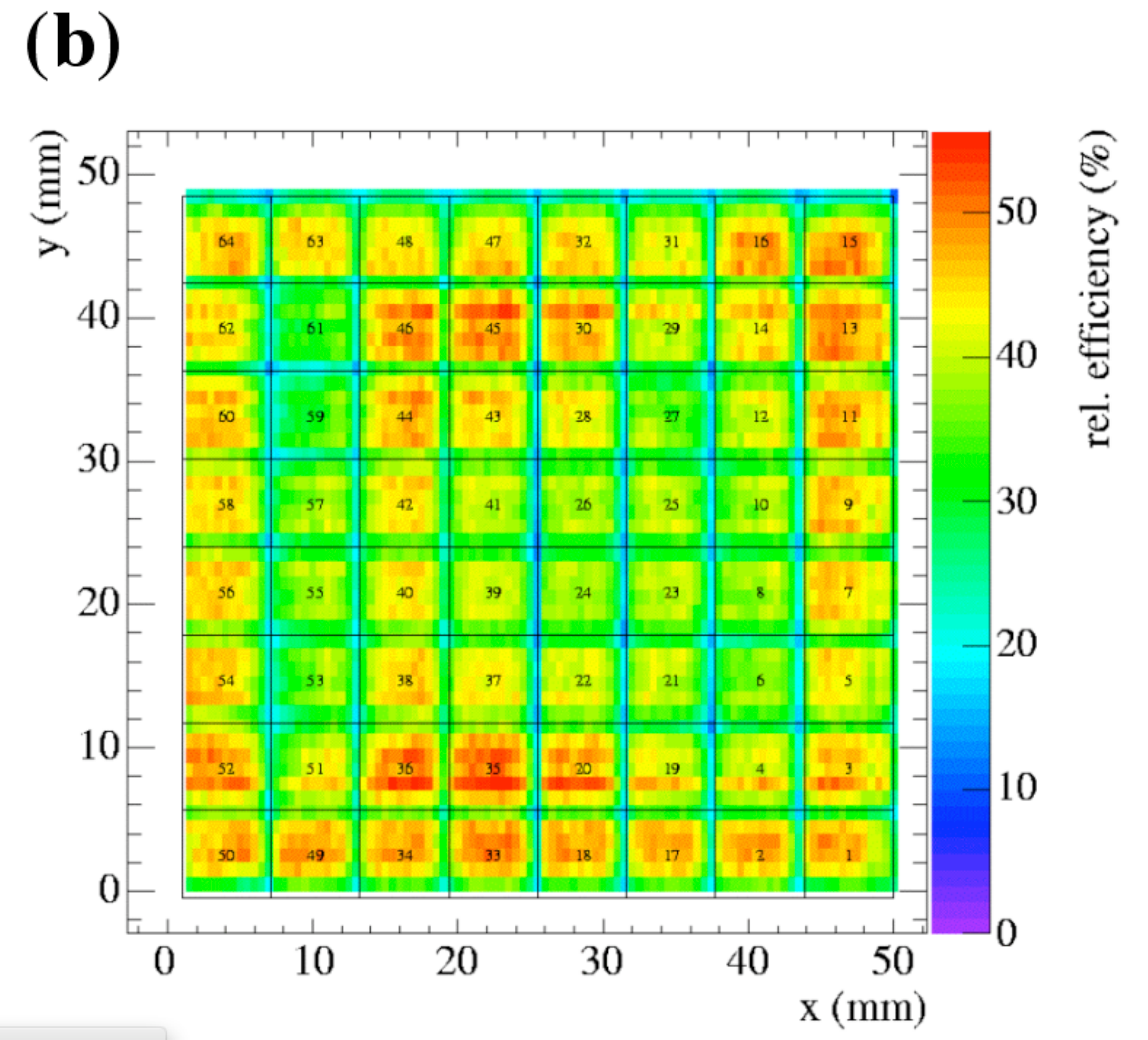}}
\end{center}
\caption{Single photoelectron response of MaPMTs.}
\label{fig:MaPMTs}
\end{figure*}

Multianode Photomultiplier Tubes (MaPMT) made by Hamamatsu are the leading choice
as photon detectors. They are highly pixelated
and about 10 times faster than the \babar\ DIRC PMTs. Their performance has
been tested and proven in high rate environments such as the HERA-B experiment.
Two PMT pixelation options are under consideration.  A pixel size of $3\mm
\times 12\mm$ can be achieved by shorting pads of the Hamamatsu 256-pixel
H-9500 MaPMT, resulting in 64 readout channels per MaPMT.
Figure~\ref{fig:MaPMTs}(a)~\cite{field_2004} shows the single
photoelectron response of this tube with such pixelization, normalized
to the Photonis Quantacon PMT. Each camera will have $\sim 48$~H-9500
MaPMT detectors, which corresponds to a total of $\sim 576$ for the
entire \superb\ FDIRC, or $\sim 36\,864$~pixels in the entire system.
Another option~-- see Fig.~\ref{fig:MaPMTs}(b)~-- is a pixel size of
$6\mm \times 12\mm$, which is achieved by shorting pads of the
Hamamatsu 64-pixel H-8500 MaPMT, resulting in 64/2 = 32 readout
channels per MaPMT, \ie\ half the total pixel count compared to the
H-9500 choice.  

Measurements with a prototype~-- a single bar
FDIRC set up  at SLAC ~\cite{benitez_2006,jjv_2007,benitez_2008}~-- confirm that the best Cherenkov angle resolution is
achieved with a pixel size of 3\mm in the vertical direction and 12\mm in
the horizontal direction, in agreement with the Monte Carlo. 
This configuration, combined with a good
single photon timing resolution, is expected to provide superior
Cherenkov angle resolution using the full three-dimensional imaging
available with the DIRC technique. 
Although the smaller pixels of the H-9500 MaPMT would  lead to better performance, a potential advantage of the H-8500 MaPMT solution could be a higher quantum efficiency (QE).  Moreover, given its wider use in the medical community,  the manufacturer ( Hamamatsu) is likely to focus efforts on this tube, leading to more reliable tubes at lower costs. For example, Hamamatsu can deliver H-8500 tubes reliably with QE $\sim 24\%$, which cannot be promised for the H-9500 tube at this point.
Furthermore, the fabrication of H-9500 tubes is likely to extend over several years~-- up to 3.5 years, according
to Hamamatsu itself. The final choice between the two MaPMTs will be
made after further R\&D.

Several options are being considered  for the FDIRC electronics. One
option is to couple a leading edge discriminator with a 100\ps/count TDC,
together with an ADC to provide the pulse height
corrections that are needed to improve timing resolution~--
aiming at a level of 150---200\ps per
single photon. An alternative choice is to use waveform digitizing
electronics,  based either on the Waveform catcher
concept~\cite{breton_2009} or the BLAB chip design~\cite{varner_2005}.
The choice between these options will be made during the R\&D period.

\begin{figure*}
\begin{center}
\subfigure[Mechanical enclosure and support of the FBLOCK with the new wedge.]{\includegraphics[width=0.4\textwidth]{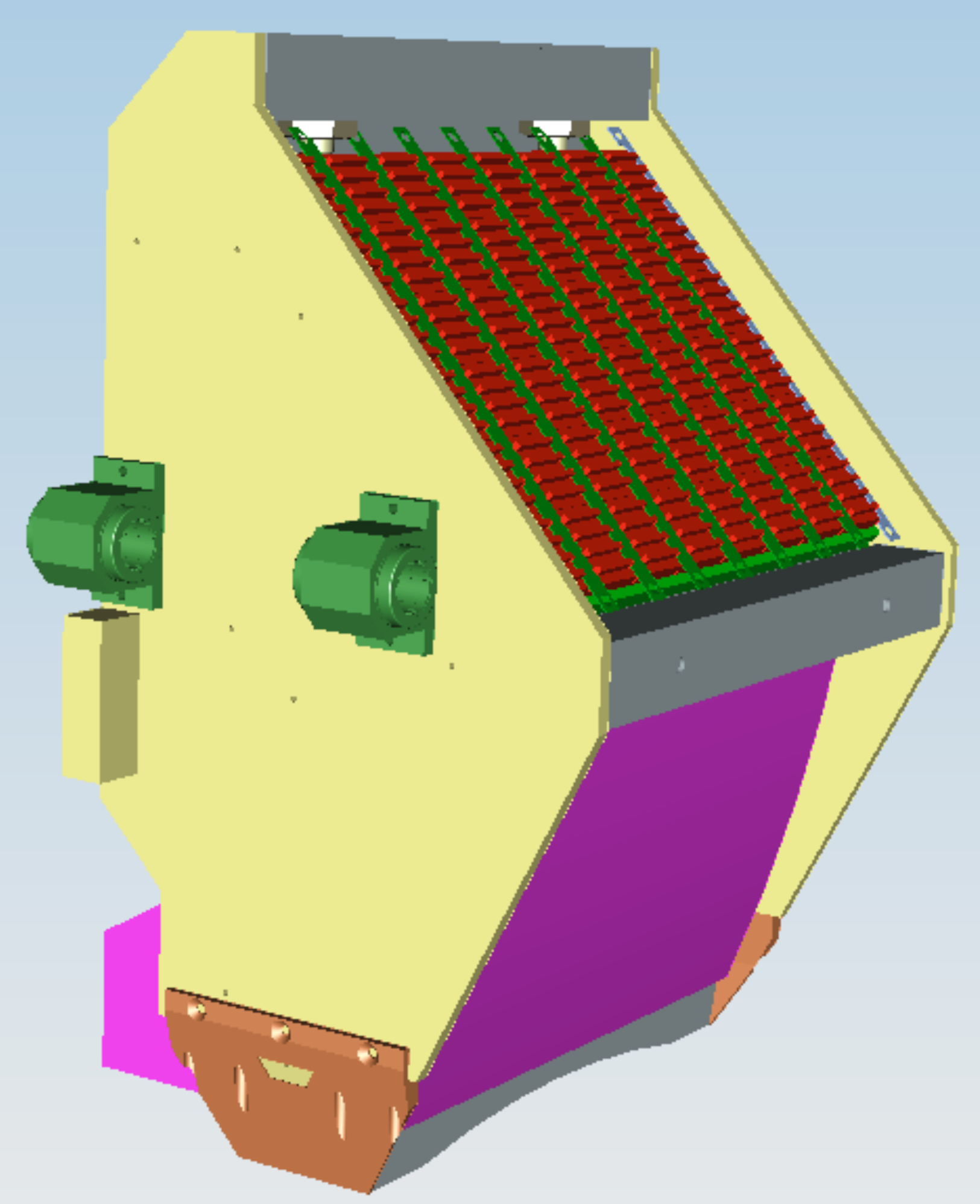}}
\subfigure[Overall mechanical support design with the new magnetic shield door.]{\includegraphics[width=0.4\textwidth]{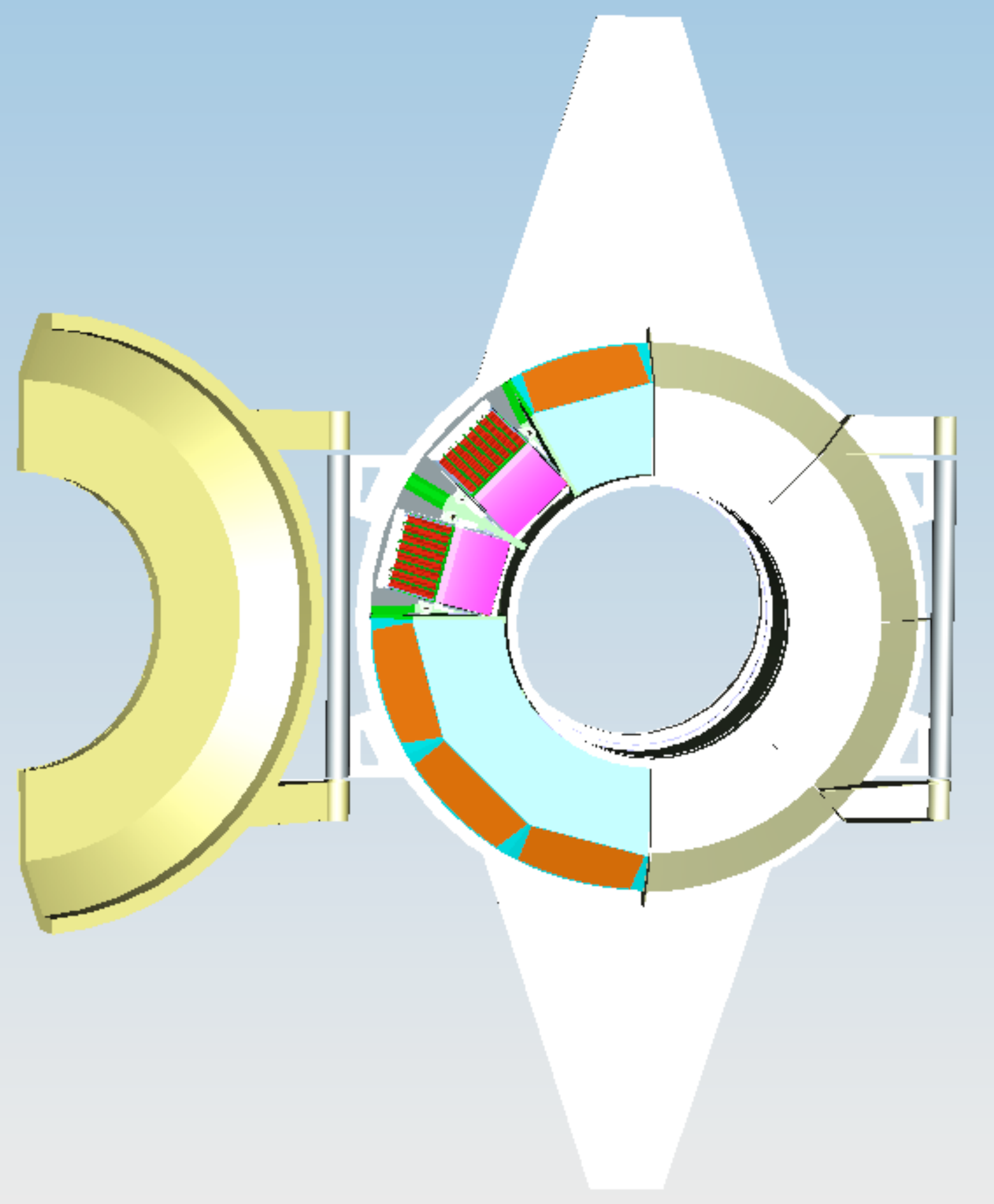}}
\end{center}
\caption{Possible mechanical design for the FDIRC.}
\label{fig:FDIRC_mechanics}
\end{figure*}

Figure~\ref{fig:FDIRC_mechanics} shows a possible design for the
mechanical support. Each bar box is a separate module with  its own FBLOCK support, light
seal, and individual access for maintenance. Each FBLOCK, weighing
almost 100kg, is supported on rods with ball bearings to provide precise control
as it is mated to the bar box. The optical coupling between
the FBLOCK and the bar box is done with an RTV coupling. Similarly,
detectors are coupled to FBLOCK with an RTV cookie. There is a common
magnetic shield mounted on hinges to allow easy access to the 
detector.

Tests of a number of these electronic scenarios continue in the SLAC cosmic ray
telescope (CRT)~\cite{jjv_2010} with the FDIRC single bar prototype.
We plan to set up a full size DIRC bar box equipped with the
new focusing optics to run  in the cosmic ray telescope in 2010-11.
In parallel, we intend to revive a scanning setup to test
photodetectors with the new electronics. Test bench setups are also
planned at LAL-Orsay and the University of Maryland.

Finally, a summary budget projecting the costs of the barrel FDIRC can be found in Section~\ref{sec:Budget_and_Schedule}.

\wpsubsec{Forward PID at \superb}

\wpsubsubsec{Motivation for a Forward PID Detector}

Though the barrel FDIRC detector combined with \dedx\ from the DCH provides good
$\pi-K$ separation up to about 4\gevc,  
hadron identification in the forward and backward regions in \superb\ is limited unless  dedicated PID devices are added there. 
Any such device needs to cover the ``cross-over" $\pi/K$ ambiguity region for \dedx\ near 1\gevc, and should also provide high momentum $\pi/K$ separation at higher momenta where the \dedx\ separation is rather poor (less than 2~`$\sigma$'). Cluster counting in the DCH, if incorporated in \superb, could provide adequate PID at the high momentum, but, of course, the cross-over ambiguity would remain.

Improved PID
performance over the entire detector solid angle increases the event reconstruction efficiency in various
exclusive
$B$-channels and helps to reduce specific backgrounds. In addition, the reconstruction of
hadronic and semi-leptonic $B$ channels~-- a key ingredient of
recoil physics analyses~-- would be improved. For some of
these channels, the reconstruction efficiencies and the purities 
improve significantly~-- the higher the number of charged particles in the
reconstructed final states, the faster the gain. Dedicated Monte-Carlo
studies aiming at quantifying these improvements are ongoing within
the \superb\ 
(DGWG).

The momenta of backward-going tracks in \superb\ is quite low on average. The EMC group is proposing a
backward veto calorimeter which may be fast enough to provide significant $\pi-K$ separation using TOF which might provide an inexpensive approach to PID in this region; R\&D continues.

The forward region covers a larger fraction of the \superb\
geometrical acceptance than the backward  because of the boost,
although it is still less than 10\% of the total. Another consequence
of the beam energy asymmetry is that the particles crossing this
region have higher average momenta. With the
help of the DGWG yhe
\superb\ PID group is investigating the option of adding forward PID coverage in detail. The status of this ongoing R\&D effort is reported
in  the following Section~\ref{subsection:forwardPidR&D}.

\wpsubsubsec{Forward PID Requirements}
 
The physics performance goals for a forward PID detector are to cover the $\pi/K$ ambiguity near 1\gevc, and, if possible, to extend the region with good $\pi/K$ separation up to 3\gevc or even above.  Space is quite limited in the forward 
area so any such \superb\ detector should be compact. A reasonable goal is a thickness of  $\sim 10\cm$. A thicker device requires either a shorter DCH, a forward shift of the EMC,  or both, which, in turn, lead to (typically modest) performance degradation for these devices. Moreover, the radiation length (\Xrad) of this new
device should be kept as low as possible, in order to avoid degrading
the reconstruction of electromagnetic showers in the EMC endcap, and the mass should be located as close as possible to the EMC.
Finally, the cost of such a detector must be small with respect to the
cost of the barrel PID, somehow in proportionality to their relative solid angle
coverage.

\wpsubsubsec{Status of the Forward PID R\&D Effort}
\label{subsection:forwardPidR&D}

Three designs for forward PID detectors are currently being investigated: a
``DIRC-like'' \tof\ device, a ``pixelated'' \tof\ detector and a Focusing
Aerogel RICH, the ``FARICH''.

\wpparagraph{``DIRC-like'' \tof\ detector concept}

In this scheme~\cite{jjv_toflike}, charged tracks cross a thin layer
of quartz in which Cherenkov photons are emitted along the particle
trajectories, at the Cherenkov polar angle. These photons are then transported through internal
reflections to one side of the quartz volume where they are detected
by PMTs located outside of the \superb\ acceptance. Unlike the DIRC,  no attempt is made to measure the Cherenkov angle directly. Instead, PID separation
is provided by TOF: at a given momentum, kaons fly
more slowly than pions,  as they are heavier. This method is challenging for several reasons, including the limited number of photons detected, and possible pattern recognition issues in the expected high background environment. Moreover,  the whole detector chain (the hardware and the reconstruction
software) must be very precisely calibrated: for instance, 3\gevc kaon and pion
are only separated by about 90\ps after 2\m~-- roughly the
expected particle flight distance in the current \superb\ layout.
On the other hand, such a detector is potentially attractive as it should fit without problems into the
available space between the DCH and EMC. But the \Xrad of this detector is the smallest and most uniform 
of the proposed layouts, and it requires a modest number of readout channels.

\begin{figure}
\includegraphics[width=0.5\textwidth]{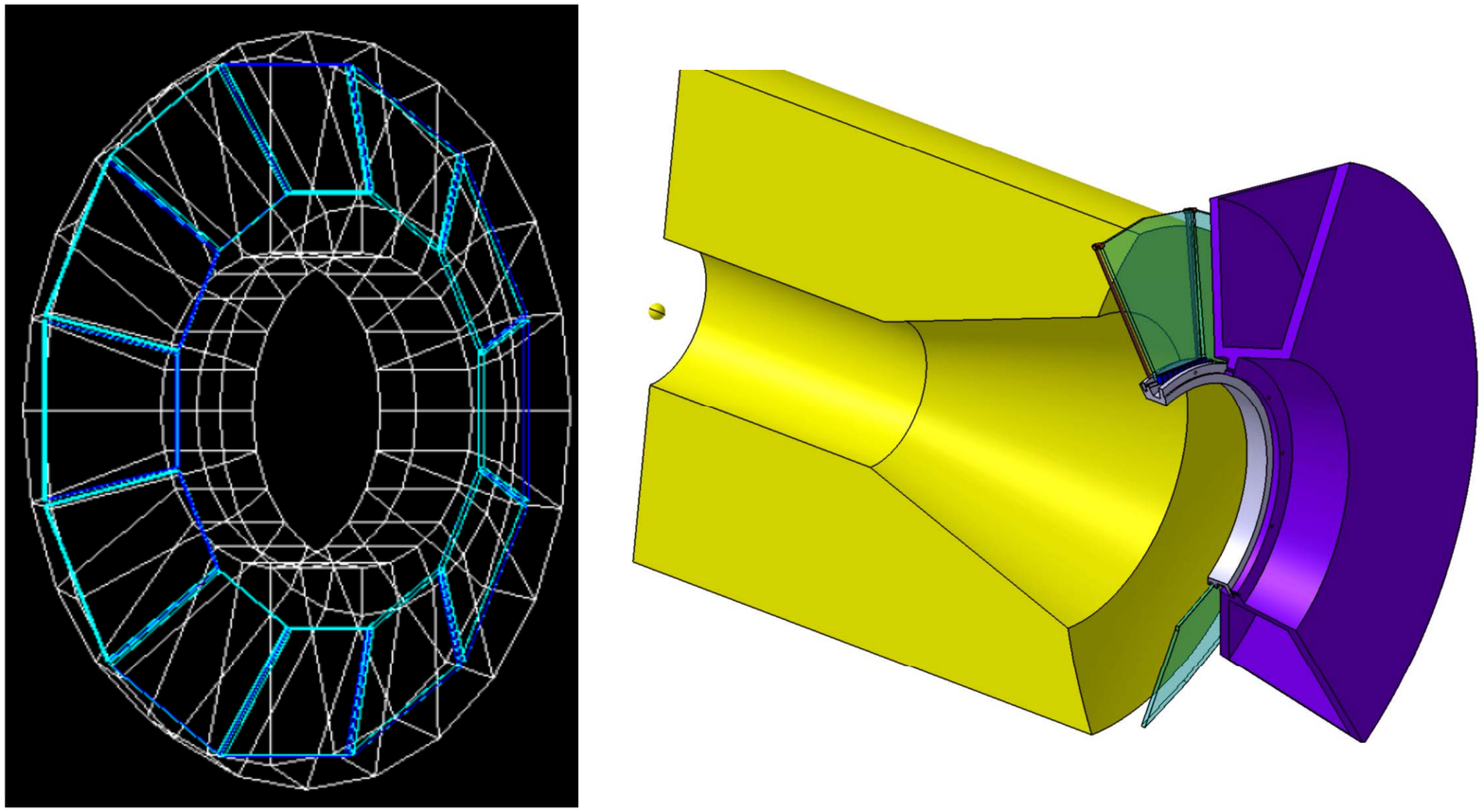}
\caption{Left: the ``DIRC-like'' TOF design, as currently implemented in
  Geant4 simulations. Right: a possible design for the mechanical
  integration of this detector (in green) in \superb. The yellow
  (magenta) volumes represent the envelope of the DCH (forward EMC).}
\label{fig:fTOF}
\end{figure}

Figure~\ref{fig:fTOF} shows the current layout of the ``DIRC-like'' \tof\
(TOF) detector, as implemented in Geant4-based simulations. Twelve
tiles (1---2\cm thick) of fused silica provide good azimuthal coverage
of the forward side of the \superb\ detector. The photons are
transported inside the fused silica volume until the inner part of the
tile where they are detected by MCP-PMTs. Simulations are in progress
to understand and optimize the detector response to signal Cherenkov
photons. 

In addition, R\&D programs are currently ongoing at the University of  Hawaii and LAL-Orsay in order to design waveform-digitizing electronics which would be able to fulfill the timing accuracy requirements of this detector (a few tens of ps at most) while being affordable and robust. Finally, as this apparatus would have to fit into a very limited space, detailed mechanical integration studies have started at LAL, in connection with the inner (DCH) and outer (EMC) subsystems.

\wpparagraph{``Pixelated'' \tof\ detector concept}

In this design,  Cherenkov light is also produced in a
quartz radiator~\cite{jjv_nima2009}. However, in this case the radiator is made
of quartz cubes, which couple directly onto matching pixelated photodetectors that cover the entire detector surface.
No photon tracking is required. This layout makes the reconstruction much easier (a given track would only produce light in a particular pixel whose
location would be predicted by the tracking algorithms); it is
insensitive to chromatic time broadening; and it is less sensitive to
background as it runs at low gain and is insensitive to single photoelectron background. 
On the other hand, the radiation length \Xrad is larger, as
the photodetectors and the electronics are located in front of the
EMC calorimeter. In addition, PMTs with excellent timing resolution
(such as MCP-PMTs) that are able to operate in a 16kG magnetic field are very expensive.
The PID performance depends on the timing resolution obtained. It should be possible, but challenging, to reach resolutions of 30\ps or better, leading to $\sim 3\sigma$ separation for 3\gevc pions and kaons. 
As a much lower cost alternative, we are looking into pixelated TOF devices that would use photon detectors
based on G-APD arrays or mesh PMTs, coupled to radiators such as LYSO,
quartz or a fast plastic scintillator, with a more modest resolution goal of $\sim 100\ps$. This would be sufficient to provide $\pi/K$ separation
near 1\gevc (where \dedx\ is useless) and help below 700\mevc. However, higher momentum PID would be  only that provided by \dedx.

\wpparagraph{FARICH concept}

The FARICH detector~\cite{farich} uses a 3-layer aerogel radiator with
focusing effect for high momentum separation and a water radiator to cover the low momentum region. The Cherenkov light is detected
by a wall of pixelated MCP-PMTs. MC simulations predict $\pi/K$
separation at the 3~`$\sigma$'~level or better up to 5\gevc,  with
$\mu/\pi$ separation up to 1\gevc. The amount of material is 
similar to the ``pixelated'' \tof\ design while the number of
channels is 4 times larger. The FARICH has the best high momentum PID performance of
all detectors proposed for the forward direction~-- it is even ``too good" at
high momenta. Its main drawbacks are thickness, mass, cost, and absence 
of beam test results.

\aftsec

\begsec
\graphicspath{{EMC/}{EMC/}}
\wpsec{Electromagnetic Calorimeter}

The \superb\ electromagnetic calorimeter (EMC) provides energy
and direction measurement of photons and electrons, and is an
important component in the identification of electrons versus
other charged particles. 
The system contains three components, shown in Fig.~\ref{fig:det:superb}:
the barrel calorimeter, reused from \babar;  
the forward endcap calorimeter, replacing the \babar\ forward endcap; 
and the backward endcap calorimeter, a new device improving the 
backward solid angle coverage.
Table~\ref{tab:SolidAngle} details the solid angle coverage of each
calorimeter. The total solid angle covered for a massless particle in the center-of-mass (CM)
is 94.1\% of 4$\pi$.

\begin{table*}[t!]
\caption{Solid angle coverage of the electromagnetic calorimeters. Values are obtained assuming
the barrel calorimeter is in the same location with respect to the collision
point as for \babar. The CM numbers are for massless particles and nominal 4 on 7\gev beam energies.
The barrel \superb\ row includes one additional ring of crystals over \babar.}
\begin{center}
\begin{tabular}{lccccc}
Calorimeter & \multicolumn{2}{c}{$\cos\theta$ (lab)} & \multicolumn{2}{c}{$\cos\theta$ (CM)} & $\Omega$ (CM)(\%)\\
            & minimum & maximum     & minimum & maximum & \\
Backward & -0.974 & -0.869&-0.985&-0.922&3.1\\
Barrel (\babar) &-0.786&0.893&-0.870&0.824&84.7\\
Barrel (\superb) &-0.805&0.893&-0.882&0.824&85.2\\
Forward &0.896&0.965&0.829&0.941&5.6
\end{tabular}
\end{center}
\label{tab:SolidAngle}
\end{table*}

In addition to the \babar\ simulation for the barrel
calorimeter, simulation packages for the new forward and backward endcaps have 
been developed, both in the form of a full simulation using the Geant4
toolkit and in the form of a fast simulation package for parametric
studies. These packages are used in the optimization of the calorimeter
and to study the physics impact of different options.

\wpsubsec{Barrel Calorimeter}

The barrel calorimeter for \superb\ is the existing \babar\ CsI(Tl)
crystal calorimeter.\cite{bib:BaBarNIM1} Estimated rates and
radiation levels indicate that this system will continue
to survive and function in the \superb\ environment.
It covers $2\pi$ in azimuth
and polar angles from 26.8\degrees to 141.8\degrees in the lab.
There are 48 rings of crystals in polar angle, with 120 crystals in each azimuthal ring, for
a total of 5,760 crystals. The crystal length ranges from $16\Xrad$ to $17.5\Xrad$.
They are read out by two redundant PIN diodes connected to a multi-range amplifier. 
A source calibration system allows calibrating the calorimeter with 6.13\mev photons from
the $^{16}N$ decay chain. 
The \babar\ barrel calorimeter will be largely unchanged
for \superb; we indicate planned changes below.

Adding one more ring of crystals at the backward
end of the barrel is under consideration. These crystals would be obtained from the current \babar\ forward calorimeter, that will not be reused in \superb. Space is already available for the
additional crystals in the existing mechanical structure, although some
modification would be required to accommodate the additional readout.

\begin{figure}[hbt]
\begin{center}
\includegraphics[width=0.9\columnwidth]{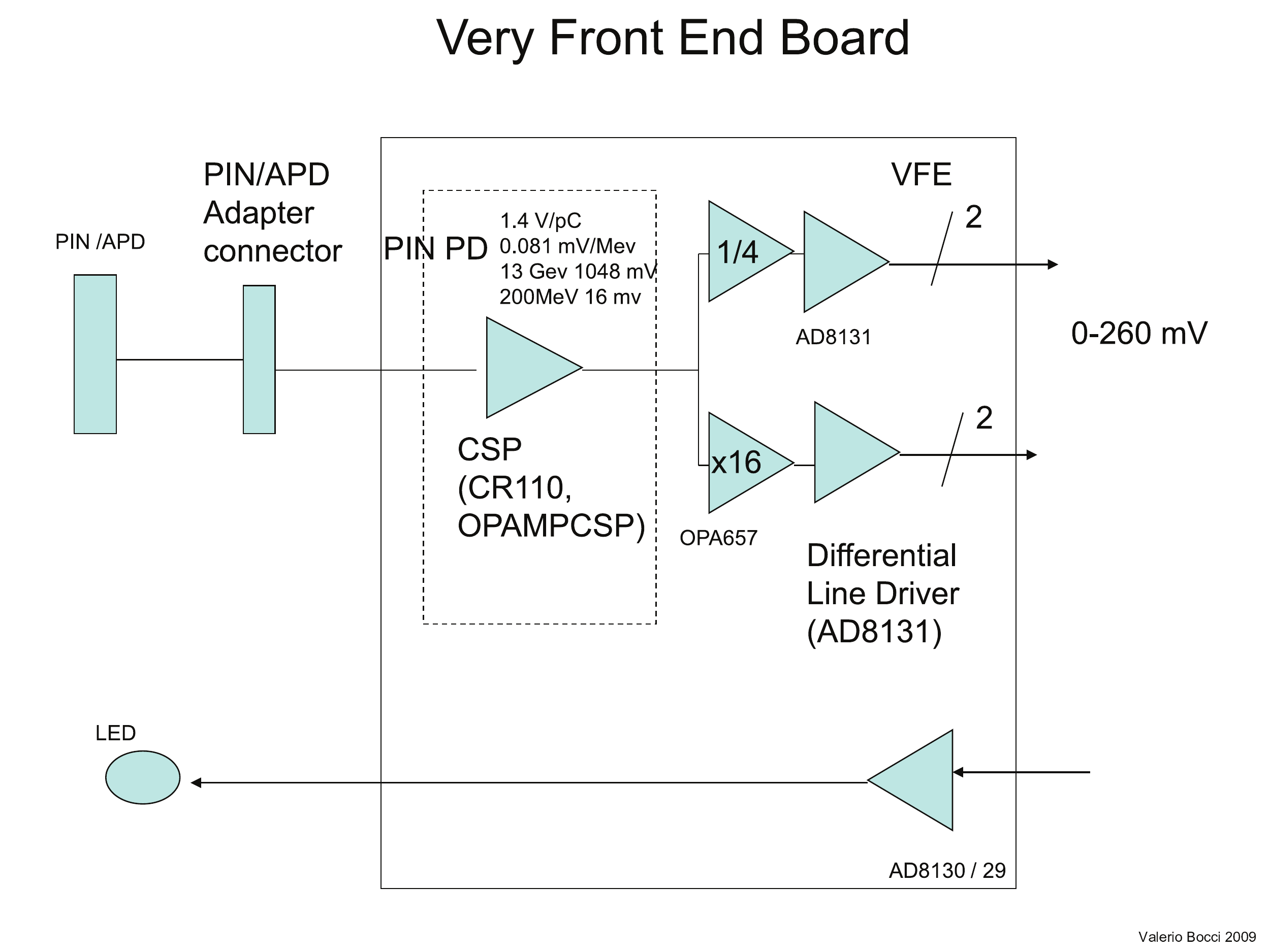}
\end{center}
\caption{Block diagram for the Very Front End board, for the barrel and
forward endcap signal readout.}
\label{fig:VFE}
\end{figure}

The existing barrel PIN diode readout is kept at \superb.
In order to accommodate the higher event rate, the shaping
time is decreased. The existing ``CARE'' chip~\cite{bib:CARE} covers the
required dynamic range by providing four
different gains to be digitized in a ten-bit ADC. However,
this system is old, and the failure rate of the analog-to-digital
boards (ADBs) is unacceptably high. Thus, a new ADB has been
designed, along with a new analog board, the Very Front End (VFE) board, shown
in Fig.~\ref{fig:VFE}.
The new design incorporates a dual-gain amplifier, followed by a
twelve-bit ADC. In order to provide good least-count resolution on the 6\mev\ 
calibration source, an additional calibration range is provided on the ADB.
The existing PIN diodes, with their redundancy, are expected to
continue to perform satisfactorily. They are epoxied to the crystals and 
changing them would be a difficult operation.

\wpsubsec{Forward Endcap Calorimeter}

The forward electromagnetic calorimeter for \superb\ is a new
device replacing the \babar\ CsI(Tl) forward calorimeter, with coverage
starting at the end of the barrel and extending down
to 270\mrad ($\cos\theta=0.965$) in the laboratory. 
Because of the increased background levels, a faster and more radiation hard
material, such as LYSO or pure CsI, is required in the forward calorimeter.
The baseline design is based on LYSO (Lutetium Yttrium Orthosilicate, with Cerium
doping) crystals. The advantages of LYSO include
a much shorter scintillation time constant (LYSO: 40\ns, CsI(Tl): 680\ns and 3.34\mus), a smaller Moli\`ere
radius (LYSO: 2.1\cm, CsI: 3.6\cm), and greater resistance to radiation damage. One radiation
length is 1.14\cm in LYSO and 1.86\cm in CsI. An alternative choice is pure
CsI~\cite{bib:BelleEMC}. However, the light output is much smaller, making
LYSO preferable.

\begin{figure}[hbt]
\begin{center}
\includegraphics[width=3in]{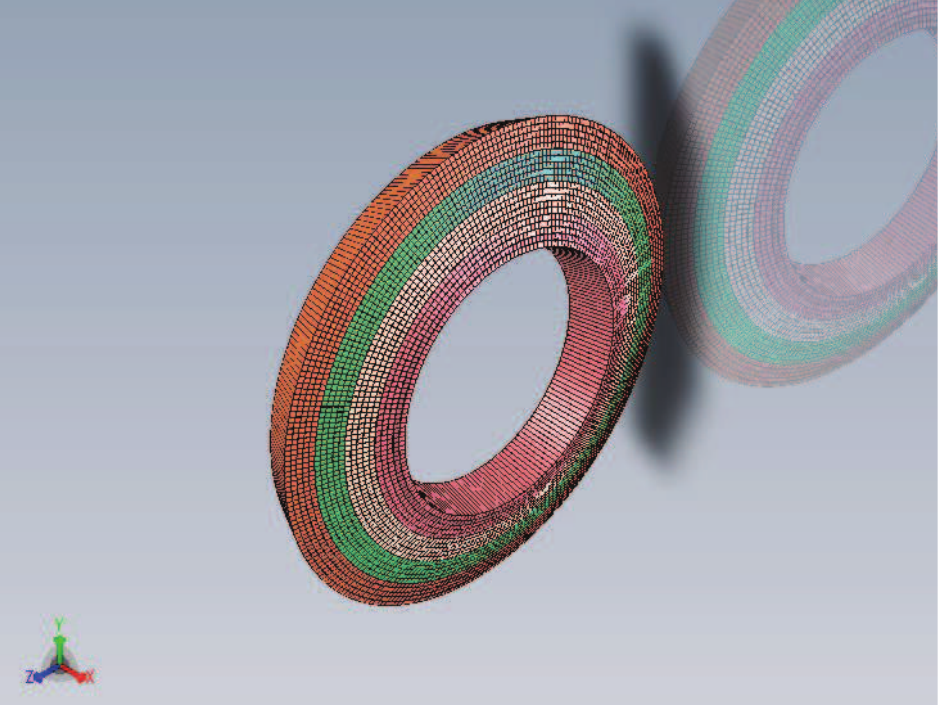}
\end{center}
\caption{Arrangement of the LYSO crystals in groups of rings.}
\label{fig:fwdEMCxtals}
\end{figure}

There are 20 rings of crystals, arranged in four groups of 5 layers each.
The crystals maintain the almost projective geometry of the barrel.
Each group of five layers is arranged in modules five crystals wide.
The preferred endcap structure is a continuous ring. However,
the numbers of modules in each group of layers are multiples of 6, allowing the
detector to be split in two halves, should that be necessary from
installation considerations. The grouping of crystals is summarized in Table~\ref{tab:LYSOlayout}
and illustrated in Fig.~\ref{fig:fwdEMCxtals}.

\begin{table}[h]
\caption{Layout of the forward endcap calorimeter.}
\begin{center}
\begin{tabular}{ccc}
Group&Modules& Crystals\\
1&36&900\\
2&42&1050\\
3&48&1200\\
4&54&1050\\
Total && 4500
\end{tabular}
\end{center}
\label{tab:LYSOlayout}
\end{table}

Each crystal is up to $2.5\times 2.5\cma$ at the back end, with a
projective taper to the front. The maximum
transverse dimensions are dictated by the Moli\`ere radius and by the desire to obtain two crystals
from a boule. The length of each crystal is approximately 20\cm, or 17.5\Xrad.

\wpsubsubsec{Mechanical Structure}
The support for the crystals is an alveolar structure (\ie, a sort of egg-crate structure, with a cell for each crystal) constructed of
either carbon fiber or glass fiber and bounded by two
conical structures at the radial extremes. 
To minimize the dead material between the endcap and the barrel, 
the outer cone is made of carbon fiber with a thickness between 6 and 10\mm.
The inner cone is instead made of 20 to 30\mm-thick aluminum.

With the inclusion of the source calibration system, described below, and the front
cooling system, the total front wall thickness may reach 20---30\mm. A good solution
that minimizes material in front of the calorimeter is to embed the
pipes  into the foam core of a sandwich panel completed by two
skins of 2---3\mm carbon fiber. A lighter alternative under investigation is 
to use depressions in pressed aluminum sheets forming the two
skins of the front wall to form the calibration and
cooling circuits.
The support at the back, providing the load-bearing
support for the forward calorimeter, is constructed in stainless steel as either
an open frame or closed plate.

\wpsubsubsec{Readout System}
Two possible readouts are under study: PIN diodes as used in the barrel and
APDs (Avalanche Photodiodes). As for the barrel, redundancy is achieved with
2 APDs or PIN diodes per crystal. APDs, with a low-noise gain of order 50,
offer the possibility of measuring signals
from sub-\mev radioactive sources. This would obviate the need for a step
with photomultipliers during the uniformity measurement process during calorimeter
construction. A concern in the \superb\ environment is the nuclear counter effect 
from background neutrons. APDs also have an advantage over PIN diodes here. Nevertheless,
it may be desirable to use the redundant photodetectors with a comparator arrangement
to eliminate spurious large signals due to this background. This is under
investigation. 
The disadvantage of APDs is the gain dependence on temperature, which can be of order $2\%/^\circ C$ (e.g.~\cite{bib:CMSTDR}). This
requires tight control of the readout temperature.  The same electronics as
for the barrel is used, with an adjustment to the VFE board gain with the
APD choice.

\wpsubsubsec{Calibration and Beam Test}
The source calibration system is a new version of the 6.13\mev calibration
system already used in \babar.
This system uses a neutron generator to produce activated
$^{16}N$ from fluorine in Fluorinert~\cite{bib:Fluorinert} coolant. The activated coolant is circulated near the front
of the crystals in the detector, where the $^{16}N$ decays with a 7\sec half-life. The 6.13\mev photons are produced in the decay chain $^{16}N\to ^{16}O^* + \beta,\ ^{16}O^*\to ^{16}O + \gamma$.

Two beam tests are planned to study the LYSO performance and the readout
options. The first beam test is at Frascati's Beam Test Facility, covering
the 50---500\mev\ energy range. The second beam test is at CERN, to cover the
\gev\ energy range.
In addition, a prototype alveolar
support structure is being constructed for the beam test.

\begin{figure*}[pt]
\begin{center}
\includegraphics[width=3.3in]{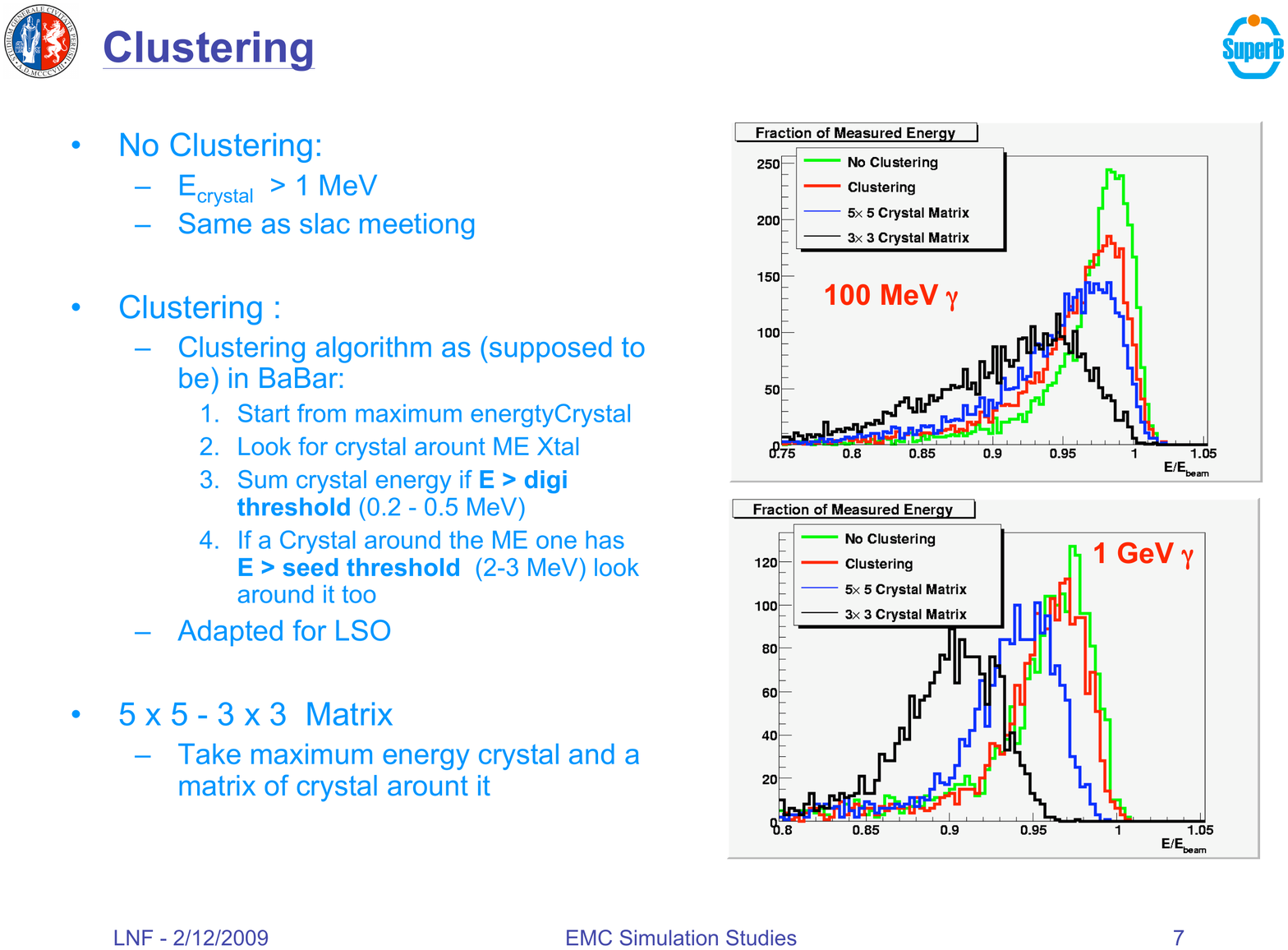}\includegraphics[width=3.3in]{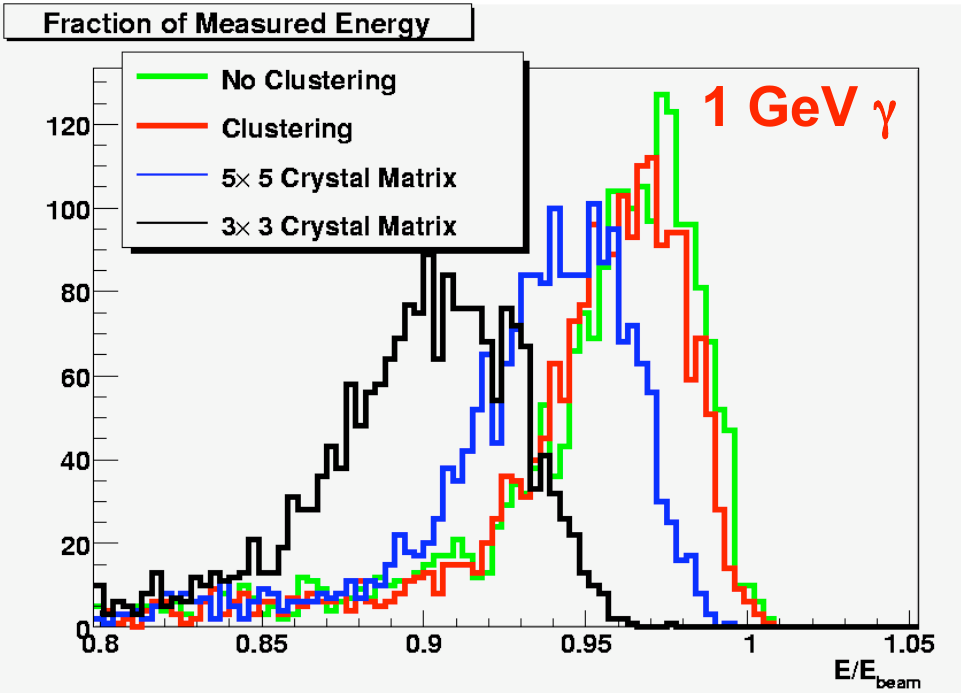}
\end{center}
\caption{Effect on the measured energy distribution for
various reconstruction algorithms. The ``No clustering'' distribution
results from simply adding all crystal energies greater than 1\mev.
The ``Clustering'' distribution results from the algorithm used in
\babar. The curves labeled $5\times 5$ crystal matrix and $3\times 3$ crystal matrix are simple sums of energy deposits in 25 or 9 crystals, respectively, centered on the crystal with the most energy. Left: 100\mev photons; Right: 1\gev photons.
}
\label{fig:FwdCluster}
\end{figure*}

\wpsubsubsec{Performance Studies}
Simulation studies are underway to optimize the detector
configuration. It is important to use a
realistic clustering algorithm in these studies, since in 
actual events multiple particles can overlap, requiring clever pattern
recognition. Fig.~\ref{fig:FwdCluster} shows how the measured energy
distribution changes for different reconstruction algorithms.

\begin{figure}[ph!]
\begin{center}
\includegraphics[width=3.3in]{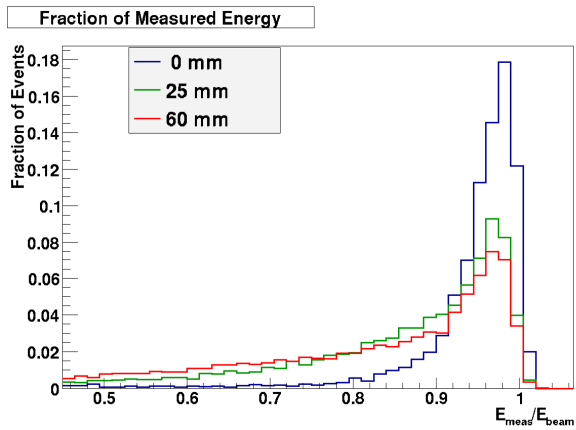}
\end{center}
\caption{Ratio of the measured/beam energy in the forward calorimeter for 100\mev\ 
photons and two different thickness of quartz, as well as no quartz, in front
of the calorimeter.}
\label{fig:EmeasEbeam}
\end{figure}

Particular attention has been devoted to the study of the effect of material in front of the forward
calorimeter, for instance due to a proposed forward PID device.
Material in front of the calorimeter enhances the low-energy tail of the measurement, although peak
width measures, such as the FWHM, are almost unaffected, as shown in Fig.~\ref{fig:EmeasEbeam}. 
for the cases of 25 and 60\mm of quartz in front of the calorimeter. 

\begin{figure}[pb!]
\begin{center}
\includegraphics[width=3.3in]{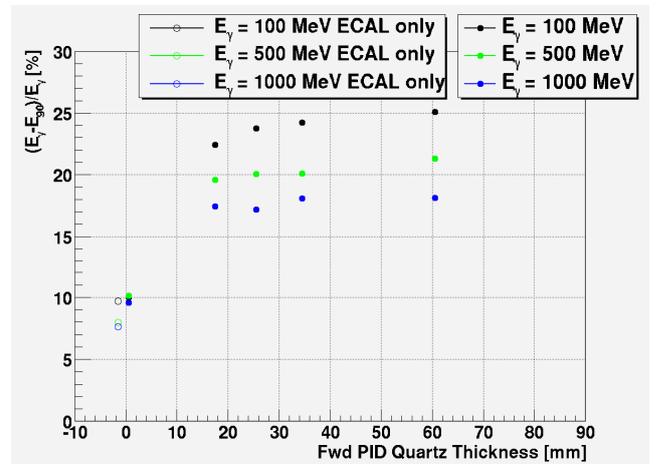}
\end{center}
\caption{The effect of quartz material in front of the forward calorimeter, as a function of thickness and photon energy. The ordinate is $f_{90}$, explained in the text, expressed as per cent.}
\label{fig:FwdMatRes}
\end{figure}

\vskip10mm

\begin{figure*}[t]
\begin{center}
\includegraphics[width=3.in]{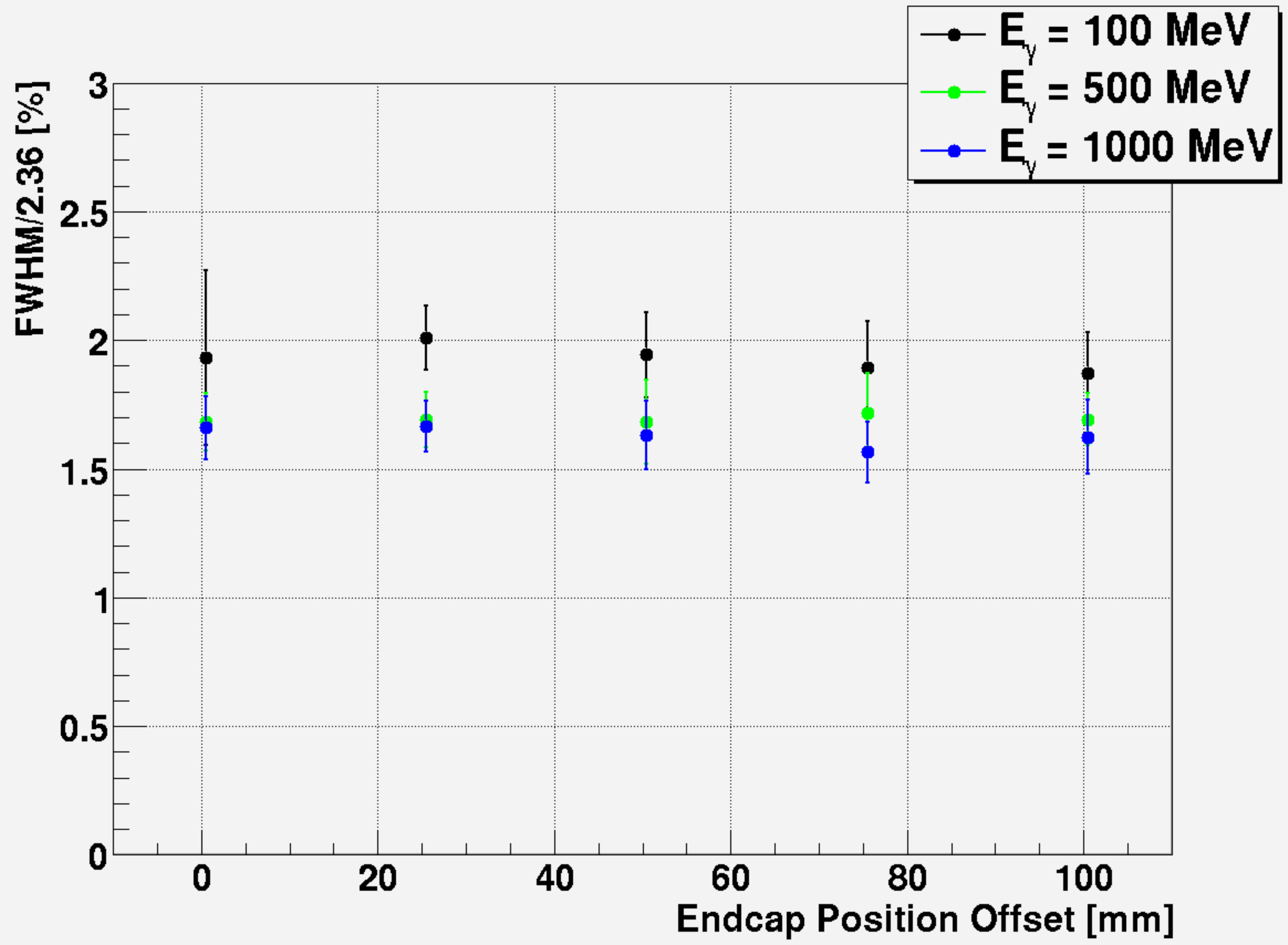}\hskip1cm\includegraphics[width=3.in]{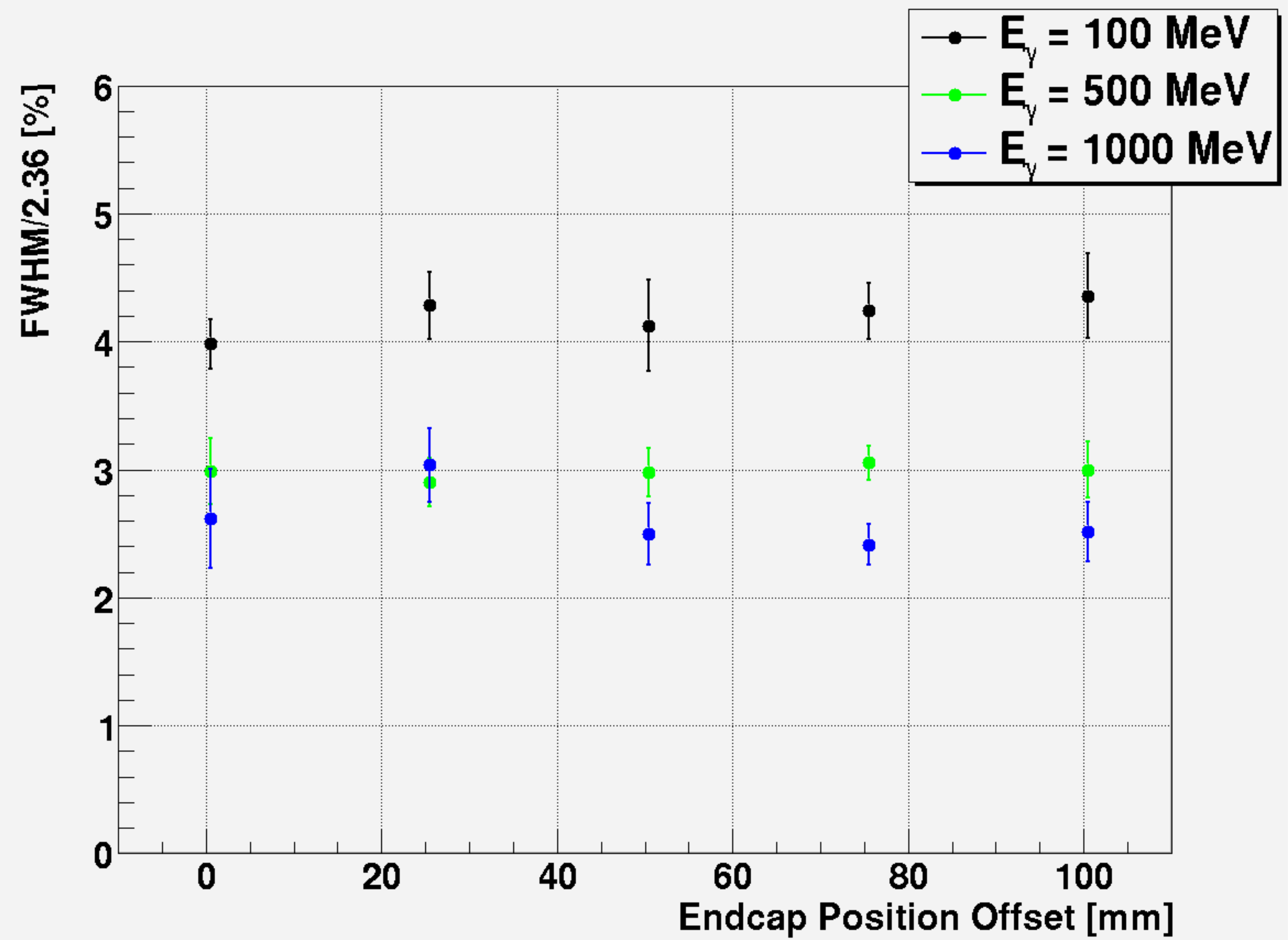}
\end{center}
\caption{Effect on resolution of z-position of forward calorimeter.
Left: Resolution as a function of position for showers away from the
edges of the forward calorimeter.
Right: Resolution as a function of position for showers in the
transition region between the barrel and forward calorimeters. Note the different scales.
}
\label{fig:FWzStudy}
\end{figure*}

\newpage

A more meaningful measure that we may use is:
$$
 f_{90}\equiv {E_{\rm true} - E_{90}\over E_{\rm true}},
$$
where $E_{\rm true}$ is the energy of the generated photon and $E_{90}$ gives
the 90\% quantile of the measured energy distribution, 
\ie, 90\% of measurements of the photon energy are above this value.
Fig.~\ref{fig:FwdMatRes} shows the effect on
the $f_{90}$ measure of resolution as a function of the quartz thickness.

Ideally, the transition between the barrel and forward calorimeters should be smooth, in
order to contain the electromagnetic showers and to keep pattern recognition simple.
Some possibilities for particle identification however require the forward calorimeter to
be moved back from the IP relative to the smooth transition point. The effect of this
on photon energy resolution has been studied, see Fig.~\ref{fig:FWzStudy}. The resolution
degrades in the barrel-endcap transition region as expected, but there is substantially
no dependence on the z-position. 

\wpsubsec{Backward Endcap Calorimeter}

The backward electromagnetic calorimeter for \superb\ is a new device with the principal
intent of improving hermeticity at modest cost. Excellent energy
resolution is not a requirement, since there is significant material
from the drift chamber in front of it. Thus a high
quality crystal calorimeter is not planned for the backward region. 
The proposed device is based on a multi-layer lead-scintillator stack with
longitudinal segmentation providing capability for $\pi/e$ separation.

The backward calorimeter is located starting at z$=-1320\mm$, allowing room for
the drift chamber front end electronics.  The inner radius is 310\mm, and the outer
radius 750\mm. The total thickness is 12\Xrad. It is constructed from a sandwich of
2.8\mm Pb alternating with 3\mm plastic scintillator (\eg, BC-404 or BC-408). The scintillator light
is collected for readout in wavelength-shifting fibers (\eg, 1\mm Y11).

\begin{figure}[thb]
\begin{center}
\includegraphics[width=3in]{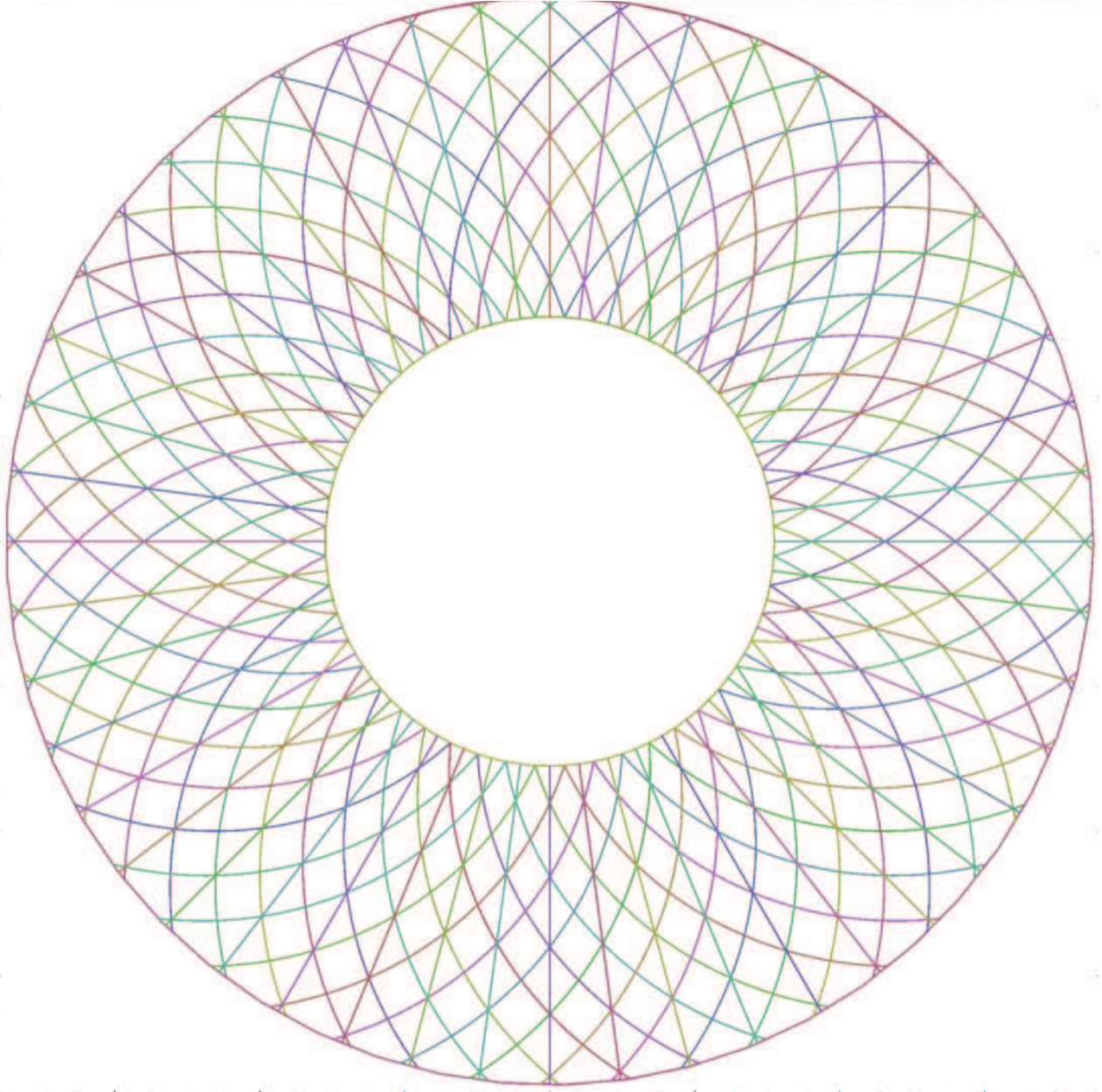}
\end{center}
\caption{The backward EMC, showing the scintillator strip
geometry for pattern recognition.}
\label{fig:backEMCgeom}
\end{figure}

To provide for transverse spatial
shower measurement, each layer of scintillator is segmented into strips. The segmentation
alternates among three different patterns for different layers:
\begin{itemize}
\item Right-handed logarithmic spiral;
\item Left-handed logarithmic spiral; and
\item Radial wedge.
\end{itemize}
This set of patterns is repeated eight times to make a total
of 24 layers.
With this arrangement, the fibers all emerge at the outer radius of the
detector. There are 48 strips per layer, for a total of 1152 strips.
The strip geometry is illustrated in Fig.~\ref{fig:backEMCgeom}.

It is desirable to maintain mechanical integrity by constructing the
scintillator layers with several strips from a single piece of scintillator,
and not completely severing them. Isolation is achieved by cutting grooves
at the strip boundaries. The optimization of this with respect
to cross-talk and mechanical properties is under investigation.

The readout fibers are embedded in grooves cut into the
scintillator. Each fiber is read out at the outer radius with a $1\times 1\mma$ 
multi-pixel photon counter (MPPC, or SiPM, for ``silicon photomultiplier'')~\cite{bib:MPPC}. A mirror
is glued to each fiber at the inner radius to maximize light collection.
The SPIROC (SiPM Integrated Read-Out Chip) integrated circuit~\cite{bib:SPIROC}
developed for the ILC is used to digitize the MPPC signals,
providing both TDC (100\ps) and ADC (12 bit) capability. Each chip contains 36 channels.

A concern with the MPPCs is radiation hardness. Degradation in performance
is observed in studies performed for the \superb\ IFR, beginning at
integrated doses of order $10^8$ 1-\mev-equivalent neutrons/\cma~\cite{bib:Faccini}.
This needs to be studied further, and possibly mitigated with shielding.

Simulation studies are being performed to investigate the performance gain  achieved by
the addition of the backward calorimeter. The $B\to\tau\nu_\tau$ decay presents an important physics channel where hermeticity
is a significant consideration. The measurement of the branching fraction has been studied in simulations to evaluate the effect
of the backward calorimeter. Events in which one $B$ decays to $D^0\pi$, with $D^0\to K^-\pi^+$, are used to tag the events,
and several of the highest branching fraction one-prong $\tau$ decays are used.

Besides the selection of the tagging $B$ decay, and one additional track for the $\tau$, the key selection criterion is on
$E_{\rm extra}$, the energy sum of all remaining clusters in the
EMC. This quantity is used to discriminate against
backgrounds by requiring events to have low values; a reasonable
criterion is to accept events with $E_{\rm extra} < 400\mev$.

\begin{figure*}[t]
\begin{center}
\includegraphics[width=3.3in]{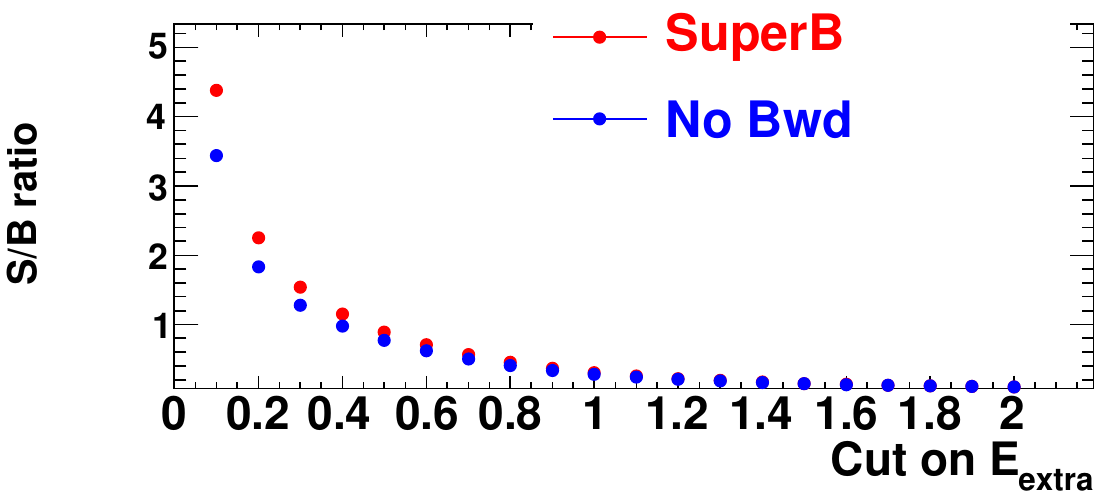}\includegraphics[width=3.3in]{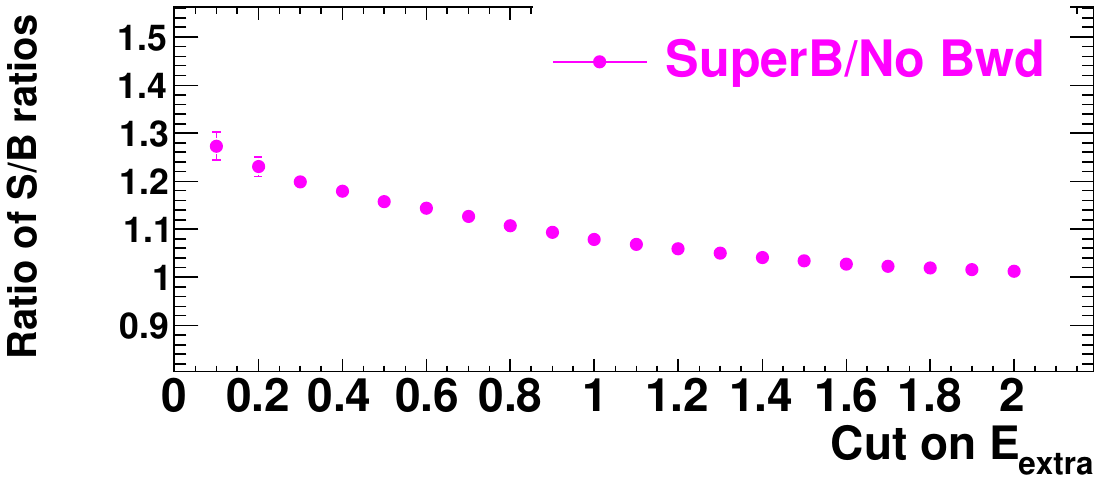}
\end{center}
\caption{Left: Signal-to-background ratio with and without
a backward calorimeter, as a function of the $E_{\rm extra}$ selection.
Right: Ratio of the S/B ratio with
a backward calorimeter to the S/B ratio without a
backward calorimeter, as a function of the $E_{\rm extra}$ selection.
}
\label{fig:BwdSB}
\end{figure*}

In this study we find that the signal-to-background ratio
is improved by approximately 20\% if the backward calorimeter
is present (Fig.~\ref{fig:BwdSB}). The corresponding improvement
in precision (${\rm S}/\sqrt{{\rm S}+{\rm B}}$) for 75\invab is approximately 8\% (Fig.~\ref{fig:bwdEMC_btotaunu}).
We note that only one tag mode has so far been investigated, and this study
is ongoing with work on additional modes to obtain results for a
more complete sample analysis. Also, the effect of background events
superimposed on the physics event has not been fully studied.

\begin{figure}[htb!]
\begin{center}
\includegraphics[width=3.3in]{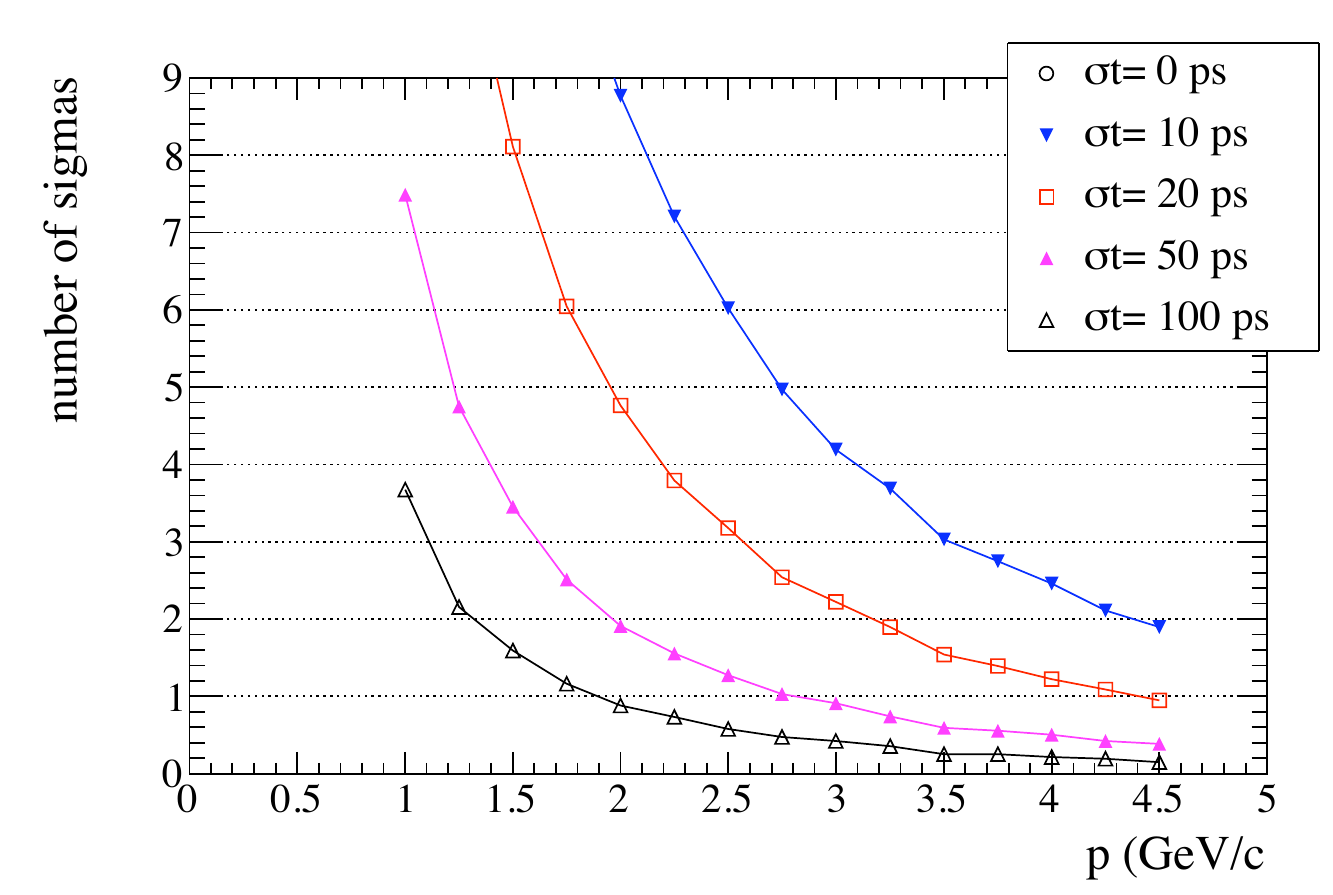}
\end{center}
\caption{Kaon-pion separation versus measured momentum for different
timing resolutions in the backward EMC region. The finite separation
for perfect timing resolution is because measured momentum is used.}
\label{fig:BwdPID}
\end{figure}

The possibility of using the backward endcap for
particle identification as a time-of-flight measuring device is also under investigation.
Figure~\ref{fig:BwdPID} shows, for example, for 100\ps timing
resolution, a separation of more than three standard deviations ($\sigma$) can
be achieved for momenta up to 1\gevc and approximately 1.5$\sigma$ up
to 1.5\gevc.

\wpsubsec{R\&D}

\wpsubsubsec{Barrel Calorimeter}

The main R\&D question for the barrel concerns the
shaping time. Simulation work is underway to
investigate  pile-up effects from backgrounds. In addition,
electronics and software issues connected with the
possibility of adding one more ring of CsI
crystals at the back end are still to be addressed.

\wpsubsubsec{Forward Calorimeter}

Since the forward calorimeter is a new device,  two beam tests are planned to test the performance of
an LYSO crystal array as well as the
solutions for the
electronics and mechanical designs. The beam tests will also investigate the
use of PIN diodes and APDs as readout options as well as 
the effect of material in front of the crystals in the beam test.
Simulation work is ongoing to predict performance and backgrounds.
Possible modifications to the electronics design to deal with
neutron nuclear-counter signals in the photodetector will be investigated.
There is an ongoing R\&D effort with vendors to produce crystals with
good light output and uniformity at an acceptable cost. 
The crystal support and integration of the calibration and cooling
circuits with the mechanical structure is under investigation in
consultation with vendors.

\wpsubsubsec{Backward Calorimeter}

A beam test of the backward calorimeter is also planned, probably concurrent
with the forward calorimeter beam test at CERN. The mechanical support and
segmentation of the plastic scintillator is being investigated for a
solution that achieves simplicity and acceptable cross-talk. The use of
multi-pixel photon counters is being studied, including the
radiation damage issue. The timing resolution for a possible time-of-flight
measurement is an interesting question. Further simulation studies are being made to
characterize the performance impact of the backward calorimeter.

\aftsec

\begsec
\graphicspath{{IFR/}{IFR/}}
\wpsec{Instrumented Flux Return}

  The Instrumented Flux Return (IFR) is designed primarily to identify muons, and, in conjunction with the electromagnetic calorimeter, to identify neutral hadrons, such as \KL. This section describes the performance requirements and a baseline design for the IFR. The iron yoke of the detector magnet provides the large amount of material needed to absorb hadrons. The yoke, as in the \babar\ detector, is segmented in depth, with large area particle detectors inserted in the gaps between segments, allowing the depth of penetration to be measured.

  In the \superb\ environment, the critical regions for backgrounds are the small polar angle sections of the endcaps and the edges of the barrel internal layers, where we estimate that in the hottest regions the rate is a few hundred \Hz/\cma. These rates are too high for gaseous detectors. While the \babar\ experience with both RPCs and LSTs has been, in the end, positive, detectors with high rate capability are required in the high background regions of \superb. A scintillator-based system provides much higher rate capability than gaseous detectors, and therefore the baseline technology choice for the \superb\ detector is extruded plastic scintillator using wavelength shifting (WLS) fibers read out with avalanche photodiode pixels operated in Geiger mode. The following subsections describe in detail all the components.

  The IFR system must have high efficiency for selecting penetrating particles such as muons, while at the same time rejecting charged hadrons (mostly pions and kaons). Such a system is critical in separating signal events in $b \rightarrow s \ell^+\ell^-$  and $b \rightarrow d  \ell^+\ell^-$  processes from background events originating from random combinations of the much more copious hadrons. Positive identification of muons with high efficiency is also important in rare  $B$ decays such as $B  \rightarrow \tau \nu_{\tau} (\gamma)$, $ B \rightarrow \mu \nu_{\mu} (\gamma)$  and  $B_d(B_s) \rightarrow \mu^+ \mu^-$ and in the search for lepton flavour-violating processes such as   $\tau \rightarrow \mu \gamma$. Background suppression in reconstruction of final states with missing energy carried by neutrinos (as in $ B \rightarrow \mu \nu_{\mu} (\gamma)$) can benefit from vetoing the presence of energy carried by neutral hadrons.
     In the \babar\ detector, about 45\% of relatively high momentum $\KL$s interacted only in the IFR system. A \KL identification capability is therefore required.

\wpsubsec{Performance Optimization}

\wpsubsubsec{Identification Technique}

  Muons are identified by measuring their penetration depth in the iron of the return yoke of the solenoid magnet.
Hadrons shower in the iron, which has a hadronic interaction length   $\lambda_I = 16.5\cm$~\cite{pdg2008ifr} so that the survival probability to a depth $d$ varies as $e^{-d/\lambda_I}$. Fluctuations in shower development and decay in flight of hadrons to final states with muons are the main sources of hadron misidentification as muons.   The penetration technique has a reduced efficiency for muons with momentum  below  1\gevc,  due to ranging out of the charged track in the absorber.  Moreover, only muons with a sufficiently high transverse momentum can penetrate the IFR to sufficient depth to be efficiently identified.

  Neutral hadrons interact in the electromagnetic calorimeter as well as in the flux return.
A \KL tends to interact in the inner section of the absorber, therefore \KL identification capability is mainly dependent on energy deposited in the inner part of the absorber, thus a fine segmentation at the beginning
of the iron stack is needed. Best performance can be obtained by combining the initial part of a shower in the electromagnetic calorimeter with the rear part in the inner portion of the IFR. An active layer between the two subsystems, external to the solenoid, is therefore desirable.

\wpsubsubsec{Baseline Design Requirements}

\begin{figure}[htb]
\begin{center}
\includegraphics[width=0.48\textwidth]{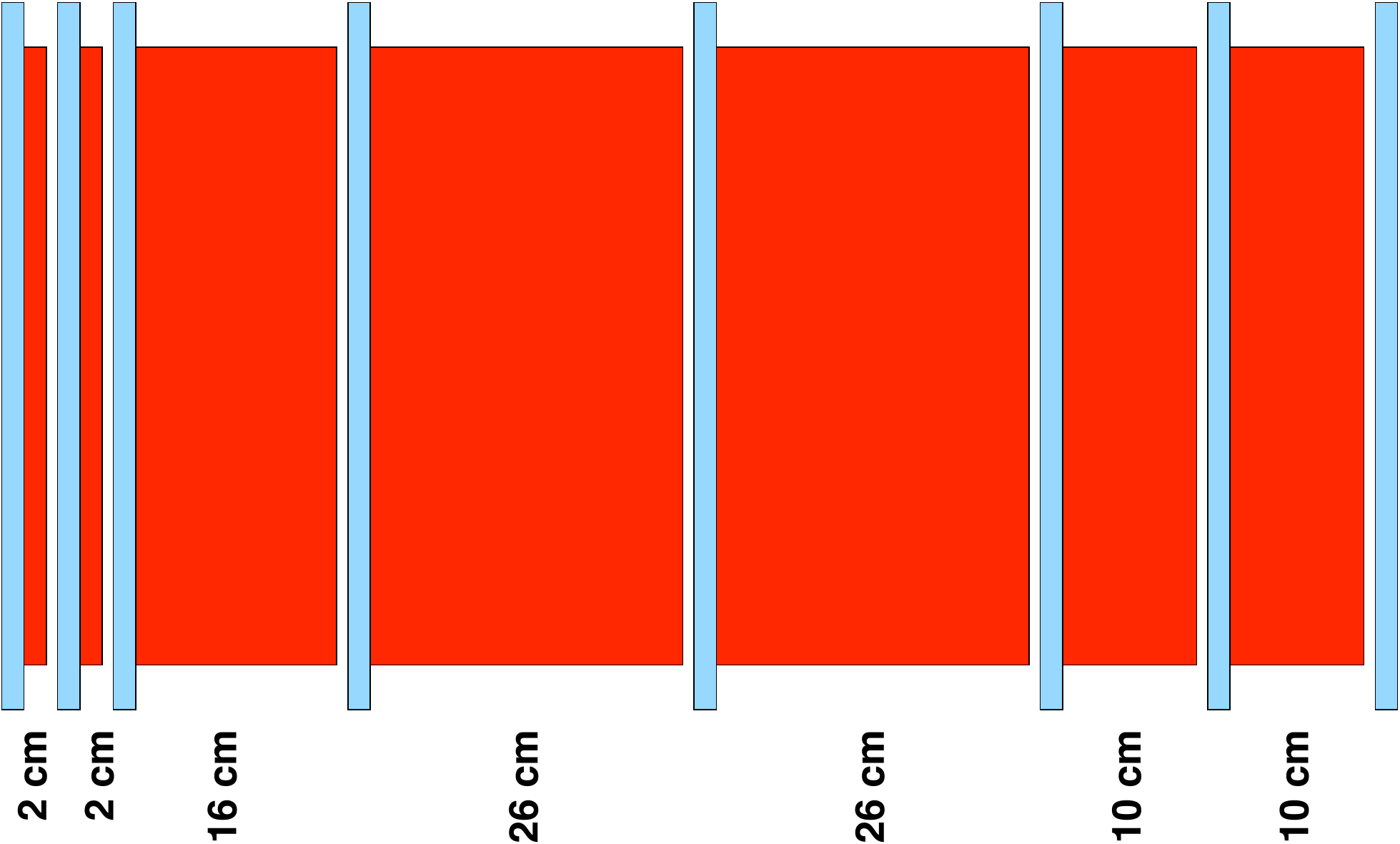}
\caption{Sketch of the longitudinal segmentation of the iron absorber (red) in the baseline configuration. Active detector positions are shown in light blue from the innermost (left) to the outermost (right) layers.}
\label{ifr:drawing}
\end{center}
\end{figure}

  The total amount of material in the \babar\ detector flux return (about 5 interaction lengths at normal incidence in the barrel region including the inner detectors) is not optimal for muon identification~\cite{Aubert:2001tu}.  Adding iron with respect to the \babar\ flux return for the upgrade to the \superb\ detector can produce an increase in the pion rejection rate at a given muon identification efficiency, and one of the goals of the simulation studies is to understand whether the \babar\ iron structure can be upgraded to match the \superb\ muon detector requirements.
A possible longitudinal segmentation of the iron is shown in Fig.~\ref{ifr:drawing}. The three inner detectors are most useful for \KL identification; the coarser segmentation in the following layers preserves the efficiency for low momentum muons.

  The layout presented in Fig.~\ref{ifr:drawing} has a total of 92\cm of iron and allows the
reuse of the \babar\ flux return with some mechanical modifications. It is our baseline configuration, although several different possible designs are under study.
The final steel segmentation will be chosen on the basis of Monte Carlo studies of muon
penetration, and charged and neutral hadron interactions. Preliminary results of these studies are shown in the next section.

\wpsubsubsec{Design Optimization and Performance Studies}

\begin{figure}[htb]
\begin{center}
\includegraphics[width=0.48\textwidth]{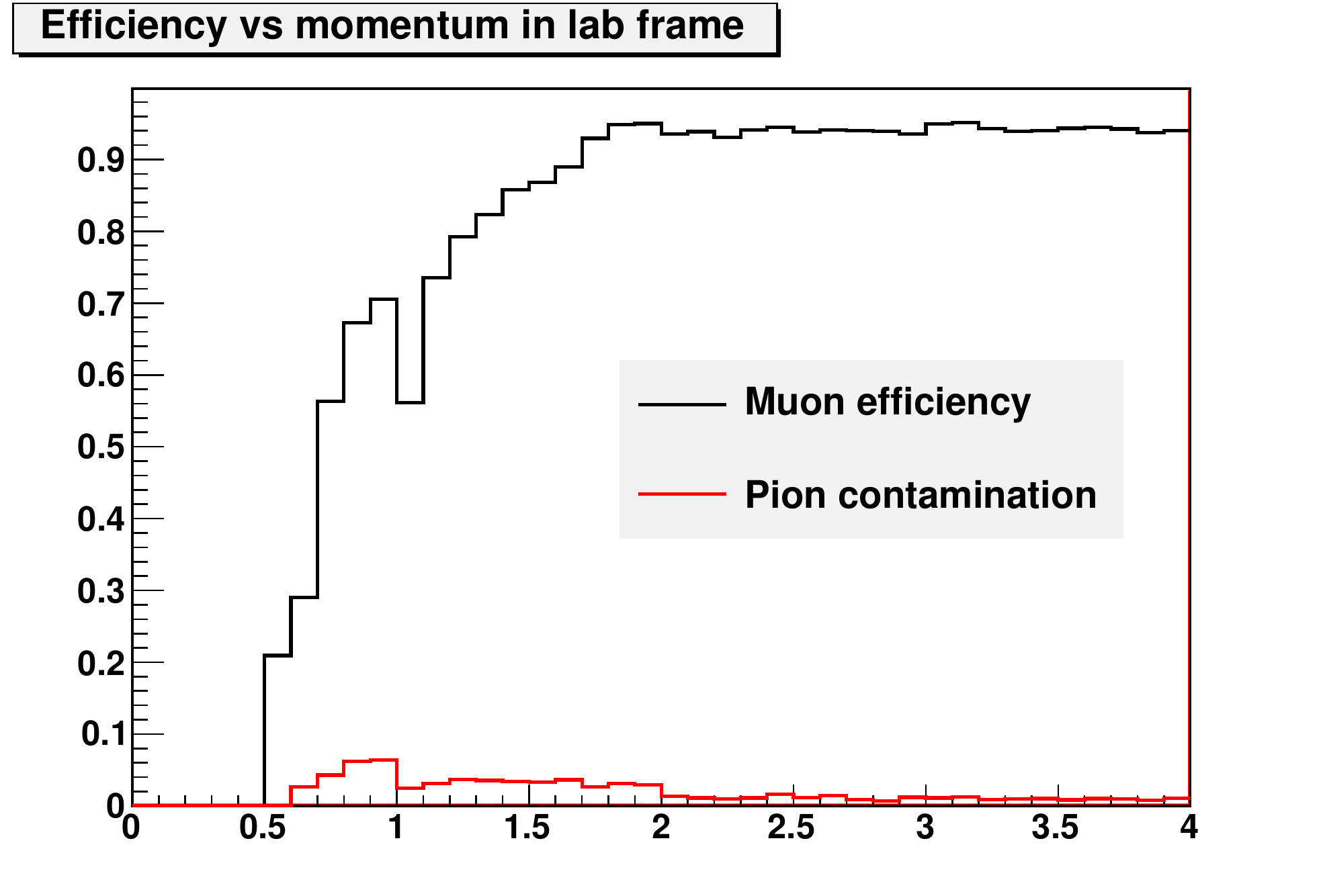}
\caption{Efficiency and misidentification probability for muons and charged pions as function of the particle momentum. Study performed with baseline detector configuration.}
\label{ifr:muonid}
\end{center}
\end{figure}

  We are performing the detector optimization by means of a Geant4 based simulation in order to have
a reliable description of hadronic showers. The simulation also includes realistic features 
derived from detector R\&D studies such as spatial resolution, detection efficiency, and electronic
noise. Single muons and pions with momentum ranging between 0.5\gevc and 4\gevc 
enter the detector and their tracks are reconstructed and analyzed to extract relevant quantities for a
cut based muon selector. Preliminary results obtained using the baseline detector configuration give 
an average muon efficiency of $\sim$ 87\% with a pion contamination of 2.1\% over the entire momentum
range. The efficiency and misidentification probability for muons and charged pions as function of the particle momentum are shown in Fig.~\ref{ifr:muonid}.

  In spite of the good results obtained with the baseline configuration, more extensive
studies are needed before making a final decision on the detector design: a careful comparison with other iron configurations will be done using a neural network algorithm for the particle identification.
Further simulation studies will also include the effects of machine background on detector performance, and a detailed investigation of neutral hadrons.

\wpsubsec{R\&D Work}

\wpparagraph{Scintillators.}

  Main requirements for the scintillator are a good light yield and a fast response. Both these requirements depend on the scintillator material characteristics and on the geometry adopted for the bar layout.
Since more than 20 metric tons of scintillator will be used in the final detector, minimizing cost is a major concern. We found the extruded scintillator produced by the FNAL-NICADD facility (also used in the MINOS experiment~\cite{ifr:minos}) suitable for our detector.

  Since the gaps between two iron absorbers are roughly 25\mm,
the bar thickness should not exceed 20\mm. The bar width is 4\cm and the fibers are placed in three holes extruded with the scintillator. We have two possible layouts for the bar:
\begin{itemize}
\item 1\cm thick, filling the gap  with two separate thin detection layers;
\item 2\cm thick, filling the gap  with only one thick  active layer.
\end{itemize}
These two scintillator layouts have been used to study different readout options: a Time readout and; a Binary readout. In the Time readout, one coordinate is determined by the scintillator position and the other by the arrival time of the signal digitized by a TDC. In this case both  coordinates are measured by the same 2\cm-thick scintillator bar, and there is therefore no ambiguity in case of multiple tracks, but the resolution of one coordinate is limited by the time resolution of the system (about 1\ns).
In the Binary readout option, the track is detected by two orthogonal 1\cm-thick scintillator bars. The spatial resolution is driven by the width of the bars (that is 4\cm as for the Time readout), but in case of multiple tracks a combinatorial association of the hits in the two views must be made.

\wpparagraph{WLS Fibers.}

The fibers are required to have a good light yield, to ensure a high detection efficiency and a time response consistent with a $\simeq1\ns$ time resolution.
WLS fibers from Saint-Gobain (BCF92) and from Kuraray (Y11-300) have been tested~\cite{ifr:fibre}.
Both companies produce multiclad fibers with a good attenuation length
($\lambda \simeq \, 3.5 \m$) and trapping efficiency ($\varepsilon \simeq 5\%$), but
Kuraray fibers have a higher light yield,  while Saint-Gobain fibers have a faster response
(with a decay time $\tau = 2.7\ns$, to be compared with Kuraray's  $\tau \simeq 9.0\ns$).

\wpparagraph{Photodetectors.}

Recently developed devices, called Geiger Mode APDs, suit rather well the needs of converting the
light signal in a tight space and high magnetic field environment.
These devices have high gain ($\simeq\,10^{5}$),
good Detection Efficiency ($\simeq 30\% $), fast response (risetime $\approx 200\ps$),
and are very small (few \mm) and insensitive to magnetic field.
On the other hand they have a rather
high dark count rate ($ \approx 1\MHz/\mma$ at 1.5 p.e.) and are sensitive to radiation.
Both  $1\times 1\mma$ SiPM,  produced by IRST-FBK, and  MPPC, produced by Hamamatsu, have been tested \cite{ifr:sipm}. The comparison between SiPMs and MPPCs showed the former to have a lower  detection efficiency, but also a faster response and  less critical  dependence on temperature and bias voltage. In order to couple the photodetector with up to four  1.0-\mm-thick fibers,  $2 \times 2\mma$ FBK and $3 \times 3\mma$ Hamamatsu devices have been tested; the latter was significantly noisier, and the SiPM is therefore currently considered to be the baseline detector.

\wpsubsubsec{R\&D Tests and Results}

R\&D Studies were performed using mainly cosmic rays, with the setup placed  inside a custom built
4\m long ``dark box"  to keep scintillators, fibers and photodetectors in a light-tight volume.

\begin{figure}[htb]
\begin{center}
\includegraphics[width= 0.42\textwidth]{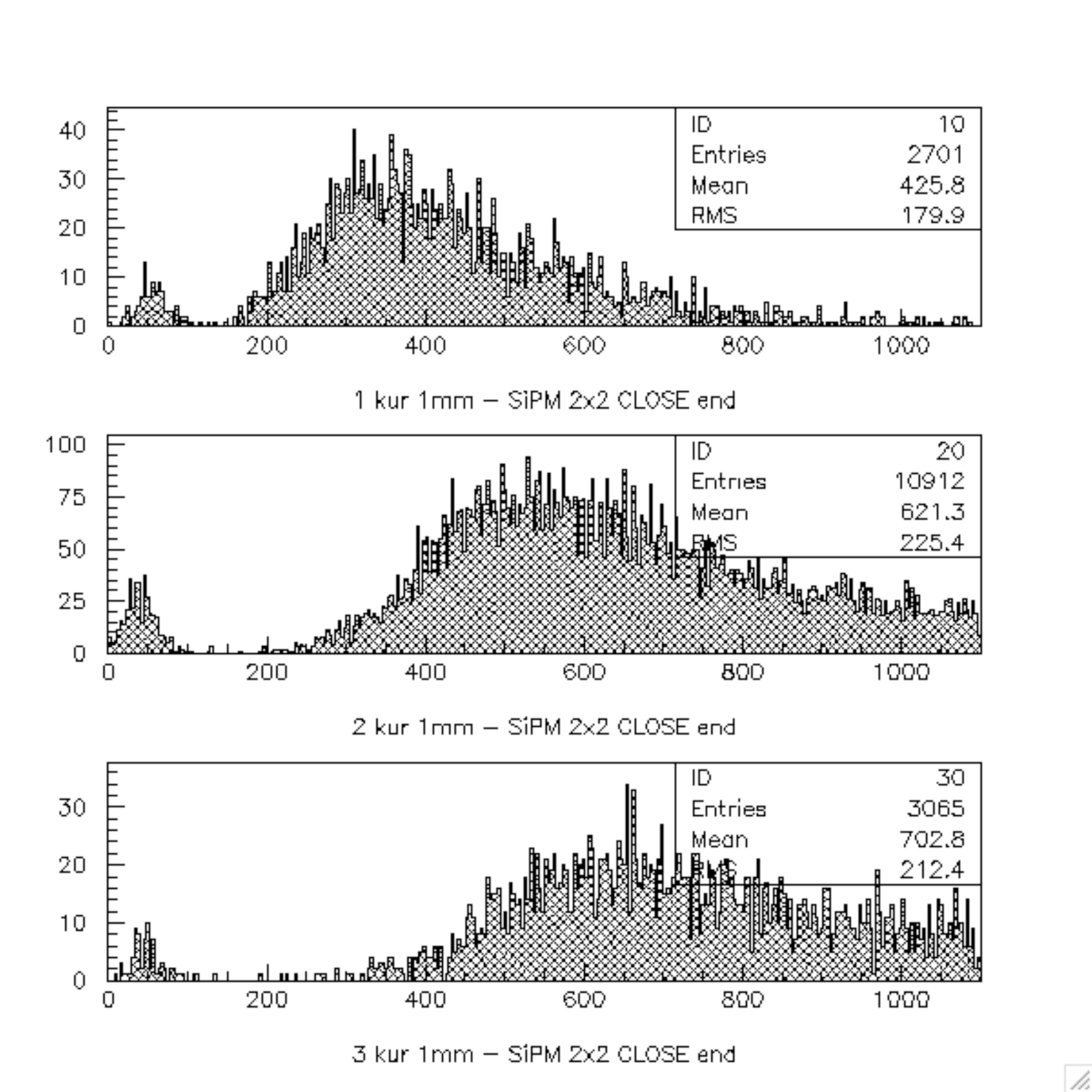}
\caption{Light collected by 1, 2 and 3 fibers coupled to a SiPM $2 \times 2\,\mma$. }
\label{ifr:123fibers}
\end{center}
\end{figure}

Given the sensitivity to radiation, the possibility of placing the SiPMs in a low radiation area outside the detector, bringing 
the light signal to the photodetectors through about 10\m of clear fibers has been studied. The light loss, expected 
to be about a factor 3 (confirmed by measurements) due to the attenuation length of the clear fiber ($\lambda \approx 10\m$), can be partially recovered by using more than 
one fiber per scintillator bar.
Figure~\ref{ifr:123fibers} shows the comparison of the collected charge in a   $2 \times 2\mma$  SiPM through 1, 2, or 3 WLS fibers.  With three fibers in the scintillator we would recover a factor 1.65, while putting a fourth fiber would add only another $10\%$ of light, insufficient to fully regain the light lost in the clear fibers, which is needed to meet the efficiency and the time resolution values discussed below.
Since the light loss is too high to bring the photodetector out of the iron,  the SiPMs must be coupled to the WLS fibers inside the detector, at the end of the scintillator bars. Appropriate neutron shields are essential to guarantee a reasonable SiPM lifetime.

\begin{figure}[htb]
\begin{center}
\includegraphics[width= 0.38\textwidth, height= 0.38\textwidth]{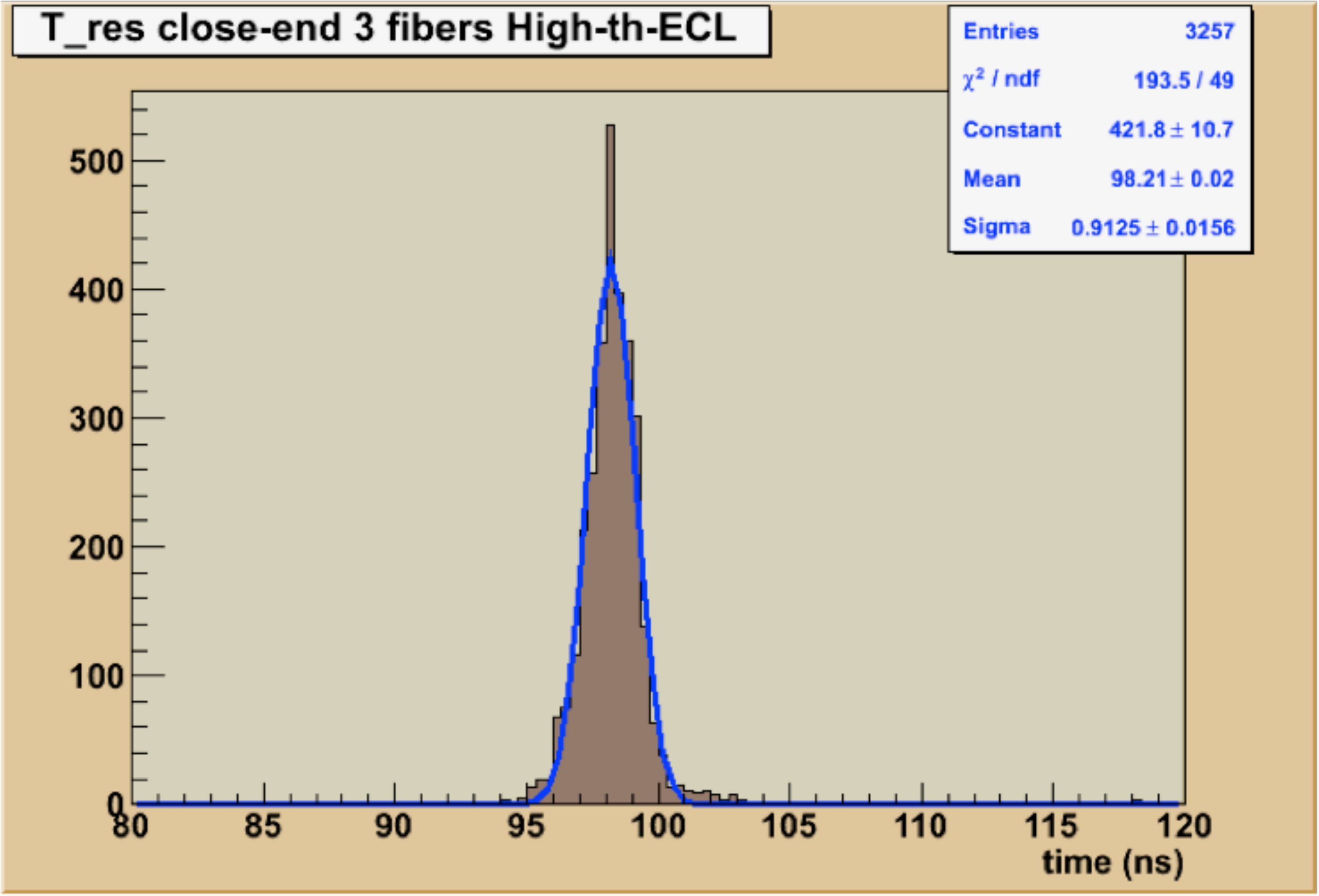}
\caption{Fit to the time distribution of the SiPM signal.}
\label{ifr:time}
\end{center}
\end{figure}

A systematic study has been performed  with the photodetectors  directly coupled to the WLS fibers.
The detection efficiency ($\varepsilon$) and the time resolution ($\sigma _T$)  have been measured in the most critical points. Figure~\ref{ifr:time} shows  a typical time distribution while all the results are reported in Table \ref{ifr:summary-results}. The goal is to have a detection efficiency  $\varepsilon > 95\%$ and, for the Time readout only, a time resolution $\sigma _T \simeq 1\,\ns$ (that would translate to a longitudinal coordinate resolution $\sigma _z \simeq 20\,\cm$). From Table \ref{ifr:summary-results}  we see that, in order to have some safety margin, the minimum of fibers to be placed inside the scintillator is three.

\begin{table*}[htb]
\begin{center}
\begin{tabular}{|c c|c|c|c|c|c|c|}
\hline
\multicolumn{8}{|c|}{ \sc{Time Readout}}  \\ \hline
\multicolumn{2}{|c|}{ }  & \multicolumn{3}{|c|}{ Time Resolution (\ns)}  & \multicolumn{3}{|c|}{ Detection Efficiency (\%) }  \\
\hline
\multicolumn{2}{|c|}{  \bf{2 fibers} }  & 1.5 p.e. &2.5 p.e.  & 3.5 p.e. &1.5p.e.  & 2.5p.e. &  3.5 p.e. \\ \hline
 & 0.3 m  &        0.91       &      0.95          &      --         &      95.4        &  98.6             &     --           \\
  &  2.2 m &         1.38      &    1.44            &      --         &    95.9          &    96.5           &      --          \\   \hline
 \multicolumn{2}{|c|}{ \bf{3 fibers} }  & 1.5 p.e. &2.5 p.e.  & 3.5 p.e. &1.5p.e.  & 2.5p.e. &  3.5 p.e. \\ \hline
 & 0.3 m  &      0.89         &      0.91          &     0.97          &   94.2           &  98.9             &   99.4             \\
  &  2.2 m &      1.16         &     1.17           &     1.26          &   95.9           &    99.1           &   99.1             \\   \hline
\hline
\hline
\multicolumn{8}{|c|}{ \sc{Binary Readout}}  \\ \hline
\multicolumn{2}{|c|}{ }  & \multicolumn{3}{|c|}{ Time Resolution (\ns)}  & \multicolumn{3}{|c|}{ Detection Efficiency (\%) }  \\
\hline
\multicolumn{2}{|c|}{ \bf{ 2 fibers} }  & 1.5 p.e. &2.5 p.e.  & 3.5 p.e. &1.5p.e.  & 2.5p.e. &  3.5 p.e. \\ \hline
  &  2.4 m &       1.87        &    2.16            &   2.14            &   98.8           &    97.4           &  91.6              \\   \hline
 \multicolumn{2}{|c|}{ \bf{ 3 fibers} }  & 1.5 p.e. &2.5 p.e.  & 3.5 p.e. &1.5p.e.  & 2.5p.e. &  3.5 p.e. \\ \hline
  &  2.4 m &        1.60       &    1.65            &   1.76            &    98.7          &   99.2            &   98.5             \\   \hline
\end{tabular}
\caption{Summary of measurements for the Time and Binary readout. The few $\%$ lowering of the detection efficiency at 1.5 p.e. threshold  is a dead time  effect due to the high rate}
\label{ifr:summary-results}
\end{center}
\end{table*}

A radiation test has also been carried out at the Frascati Neutron Generator facility (ENEA laboratory). First results (\cite{villaolmo}) show that radiation effects start from an integrated dose of $\simeq 10^{8}$ n/\cma  and remain rather stable up to a dose of  $\simeq 7 \times 10^{10}$ n/\cma; in this range, the irradiated SiPMs continue to work, although with lower efficiency and higher dark rate.

\wpsubsubsec{Prototype}
R\&D achievements will be tested on a full scale prototype that is currently in preparation and that will be used to validate the simulation results. The prototype is composed of a full stack of iron with a
segmentation which allows the study of different detector configurations. The active area is $60 \times 60 \cma$ for each gap. Scintillator slabs, full length WLS fibers and photodetectors will be located in light-tight boxes (one for each active layer) placed within the gaps. The prototype will be equipped with eight active layers: four having Binary readout and four with Time readout. A beam test will be done at Fermilab using a muon/pion beam with momentum ranging from 1\gevc to 5\gevc. Beside the muon identification capability with different iron configurations, which is the main purpose of the beam test, detection efficiency and spatial resolution of the detector will also be measured.

\wpsubsec{Baseline Detector Design}

Although the final detector design will be decided after the prototype test, a preliminary baseline layout can be
 defined from the R\&D studies, the simulation results and the experience with the \babar\ muon detector.
Binary and Time readout have pros and cons from the performance point of view, but they both match the requirements for \superb.
Mechanically, the installation of the Binary readout, with orthogonal layers of scintillator, would be rather complicated in the barrel due to the limited access to the gaps. On the other hand, the region of the endcaps at low radii is subjected to high radiation and is not a suitable location for the photodetectors. Therefore we currently plan to instrument the barrel region with Time readout, with the photodetectors on both ends of the bars, and to instrument the endcaps with Binary readout, reading the bars only on one end.
The number of fibers is three per scintillator bar for each readout mode and the photodetectors are placed inside the gaps just at the end of the bars. The signal is brought to the electronics card, placed outside the iron, by means of about 10\m of coaxial cable. A detailed description of the frontend electronics will be given in the Electronics section.

\wpsubsubsec{Flux Return}
The baseline configuration foresees reuse of the \babar\ flux return
with some mechanical modifications.
The design thickness of the absorbing material in \babar\ was 650\mm
in the barrel and 600\mm in the endcaps; in order to improve the muon
identification the thickness was then increased up to 780\mm in the
barrel and up to 840\mm in the forward endcap by replacing some active layers with
brass plates and adding a steel plate in the forward part of the endcap.
In the \superb\ baseline design, the total thickness of the absorbing
material is 920\mm, corresponding to 5.5 interaction lengths. This can be
achieved either by filling more gaps with metal plates (brass or low
permeability stainless steel), or by reusing a 100\mm steel thickness in the barrel
which was not used in \babar.
The last point requires considerable modification of the support structures
surrounding the barrel flux return and, 
due to the increased weight, a general reinforcement of the support
elements is needed.

\aftsec

\begsec
\graphicspath{{ETD/}{ETD/}}

\wpsec{Electronics, Trigger, DAQ and Online}

\newcommand{\ktilde}{$~_{\widetilde{}}\,$}

\wpsubsec{Overview of the Architecture}  

\begin{figure*}[t]
\begin{center}
\includegraphics[width=0.95\textwidth]{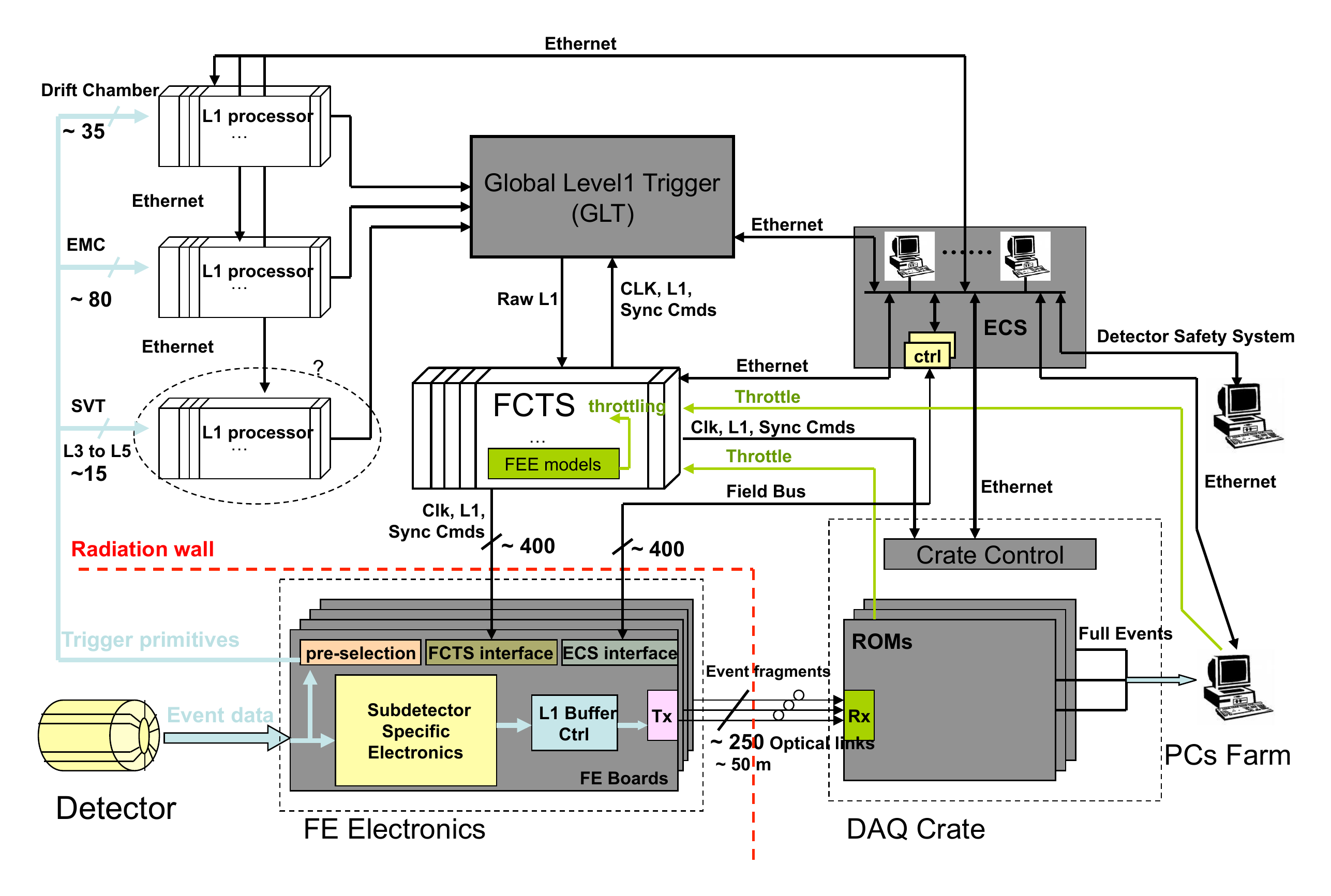}
\caption{Overview of the ETD and Online global architecture}
\label{fig:etd-overview}
\end{center}
\end{figure*}

The architecture proposed for the \superb\ ~\cite{bib:etd-SuperBCDR}
Electronics, Trigger, Data acquisition and Online systems (ETD) has
evolved from the \babar\ architecture, informed by the experience
gained from running \babar\ ~\cite{bib:etd-BaBarNIM} and building the
LHC
experiments~\cite{bib:etd-AtlasTDR},~\cite{bib:etd-CMSTDR},~\cite{bib:etd-LHCBTDR}.
The detector side of the system is synchronous and all sub-detector
readouts are now triggered, leading to improved reliability and
uniformity. In \superb, standard links like Ethernet are the default;
custom hardware links are only used where necessary. The potential for
high radiation levels makes it mandatory to design radiation-safe
on-detector electronics.

The first-level hardware trigger uses dedicated data streams of
reduced primitives from the sub-detectors and provides decisions to
the fast control and timing system (FCTS) which is the centralized
bandmaster of the system. The FCTS distributes the clock and the fast
commands to all elements of the architecture and controls the readout
of the events.

A high level trigger (HLT) processes complete events and reduces the
data stream to an acceptable rate for logging.

\wpsubsubsec{Trigger Strategy} 

The \babar\ and Belle~\cite{bib:etd-BELLENIM} experiments both chose
to use ``open triggers'' that preserved nearly 100\% of \BB\ events of
all topologies, and a very large fraction of \tautau\ and \ccbar\
events.  This choice enabled very broad physics programs at both
experiments, albeit at the cost of a large number of events that
needed to be logged and reconstructed, since it was so difficult to
reliably separate the desired signals from the \qqbar\ ($q=u,d,s$)
continuum and from higher-mass two-photon physics at trigger level .
The physics program envisioned for \superb\ requires very high 
efficiencies for a wide variety of \BB\ , \tautau, and \ccbar events,
and depends on continuing the same strategy, since few classes of the
relevant decays provide the kinds of clear signatures that allow the
construction of specific triggers.

All levels of the trigger system should be designed to permit the
acquisition of prescaled samples of events that can be used to measure
the trigger performance.

The trigger system consists of the following components \footnote{
While at this time we do not foresee a ``Level~2'' trigger that acts
on partial event information in the data path, the data acquisition
system architecture would allow the addition of such a trigger stage
at a later time, hence the nomenclature.}:

\wpparagraph{Level~1 (L1) Trigger:}

A synchronous, fully pipelined L1 trigger receives continuous data
streams from the detector independently of the event readout and
delivers readout decisions with a fixed latency.  While we have yet
to conduct detailed trigger studies, we expect the L1 trigger to be
similar to the \babar\ L1 trigger, operating on reduced-data streams
from the drift chamber and the calorimeter.  We will study the
possibilities of improving the L1 trigger performance by including SVT
information, taking advantage of larger FPGAs, faster drift chamber
sampling, the faster forward calorimeter, and improvements to the
trigger readout granularity of the EMC.

\wpparagraph{High Level Triggers (HLT)---Level~3 (L3) and Level~4 (L4):}

The L3 trigger is a software filter that runs on a commodity computer
farm and bases its decisions on specialized fast reconstruction of
complete events. An additional ``Level~4'' filter may be implemented
to reduce the volume of permanently recorded data if needed. Decisions
by L4 would be based on a more complete event reconstruction and analysis.
Depending on the worst-case performance guarantees of the
reconstruction algorithms, it might become necessary to decouple this
filter from the near-realtime requirements of L3---hence, its
designation as a separate stage.

\wpsubsubsec{Trigger Rates and Event Size Estimation} 

The present L1-accept rate design standard is 150\kHz. It has been
increased from the \superb\ CDR~\cite{bib:etd-SuperBCDR} design of
100\kHz to allow more flexibility and add headroom both to accommodate
the possibility of higher backgrounds than design (e.g.  during
machine commissioning), and the possibility that the machine might
exceed its initial design luminosity of $10^{36}\ {\rm cm}^{-2} {\rm
  sec}^{-1}$.

The event size estimates still have large uncertainties. Raw event
sizes (between front-end electronics and ROMs) are understood well
enough to determine the number of fibres required. However, neither
the algorithms that will be employed in the ROMs for data size
reduction (such as zero suppression or feature extraction) nor their
specific performance for event size reduction are yet known.  Thus,
while the 75\kbytes event size extrapolated from \babar\ for the CDR
remains our best estimate, the event size could be significantly
larger due to new detector components such as Layer~0 of the SVT
and/or the forward calorimeter. \begin{bf} \boldmath In this document
  we will use 150\kHz L1-accept rate and 75\kbytes per event as the
  baseline.
\end{bf}

With the prospect of future luminosity upgrades up to 4 times the
initial design luminosity, and the associated increases in event size
and rate, we also must define the system upgrade path, including which
elements need to be designed upfront to facilitate such an upgrade,
which can be deferred until a later time, and, ultimately, what the
associated costs would be.

\wpsubsubsec{Dead Time and Buffer Queue Depth Considerations} 

\begin{figure*}[t]
\begin{center}
\includegraphics[width=0.95\textwidth]{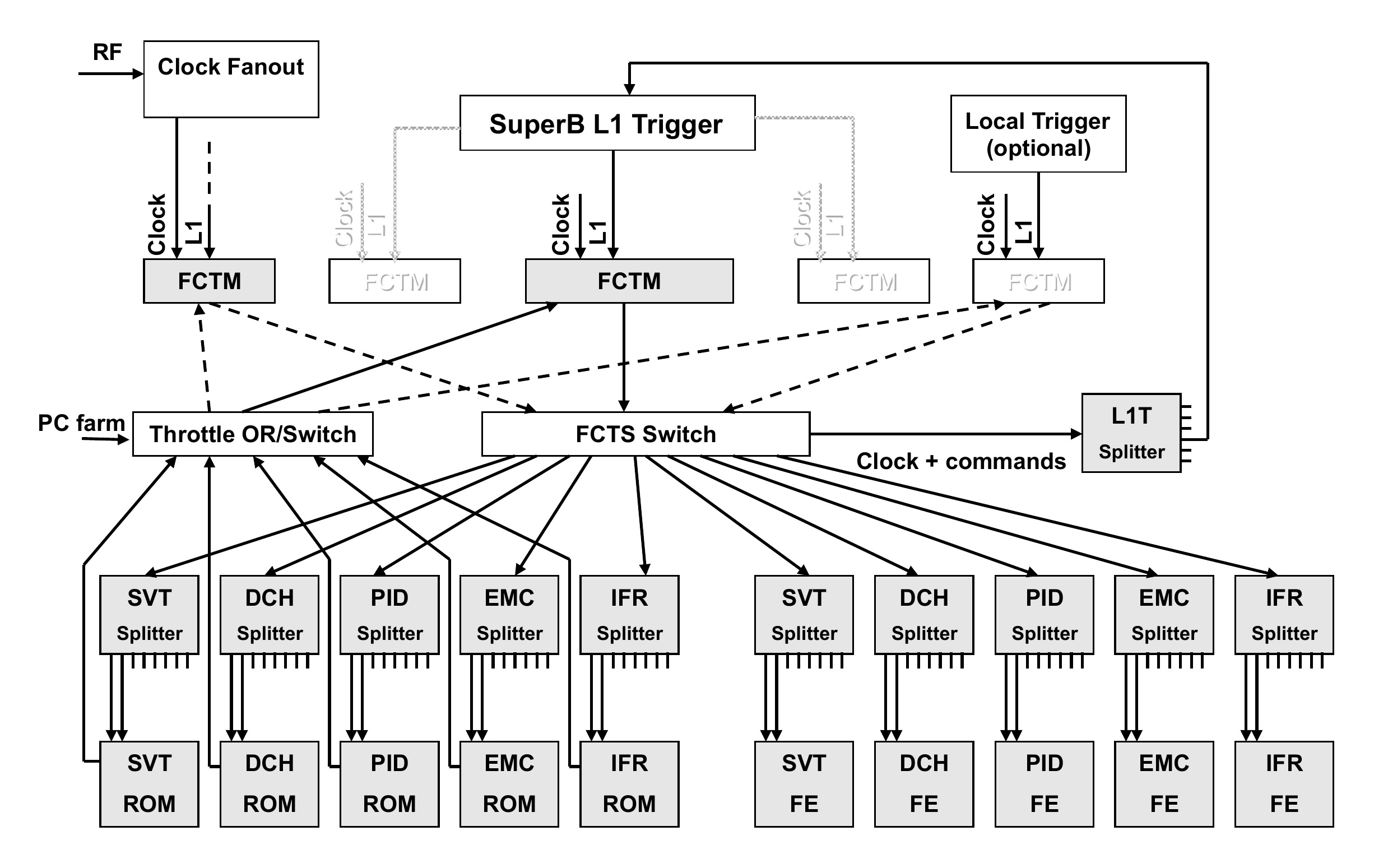}
\caption{Fast Control and Timing System}
\label{fig:etd-fcts}
\end{center}
\end{figure*}

The readout system is designed to handle an average rate of 150\kHz
and to absorb the expected instantaneous rates, both without incurring
dead time\footnote{Dead time is generated and managed centrally by the
  FCTS which will drop valid L1 trigger requests that would not fit
  into the readout system's envelope for handling of average or
  instantaneous L1 trigger rates.} of more than 1\% under normal
operating conditions at design luminosity. The average rate
requirement determines the overall system bandwidth: the instantaneous
trigger rate requirement affects the FCTS (Fast Control and Timing
System), the data extraction capabilities of the
front-end-electronics, and the depth of the de-randomization buffers.
The minimum time interval between bunch crossings at design luminosity
is about 2.1\ns---so short in comparison to detector data collection
times that we assume ``continuous beams'' for the purposes of trigger
and FCTS design. Therefore, the burst handling capabilities (minimum
time between triggers and maximum burst length) to achieve the dead
time goal are dominated by the capability of the L1 trigger to
separate events in time and by the ability of the trigger and readout
systems to handle events that are partially overlapping in space or
time (pile-up, accidentals, etc.).  Detailed detector and trigger
studies are needed to determine these requirements.

\wpsubsec{Electronics, Trigger and DAQ} 

The Electronics, Trigger and DAQ (ETD) system includes all the
hardware elements in the architecture, including FCTS,
sub-detector-specific and common parts (CFEE) of the front-end
electronics (FEE) for data readout and control, the Level~1 hardware
trigger, the Readout Module boards (ROMs), the Experiment Control
System (ECS), and the various links that interconnect these
components.

The general design approach is to standardize components across the
system as much as possible, to use mezzanine boards to isolate
sub-system-specific functions differing from the standard design, and
to use commercially available common off-the-shelf (COTS) components
where viable.

We will now describe the main components of the ETD in more detail.

\wpsubsubsec{Fast Control and Timing System} 

\begin{figure}[!b]
\includegraphics[width=0.45\textwidth]{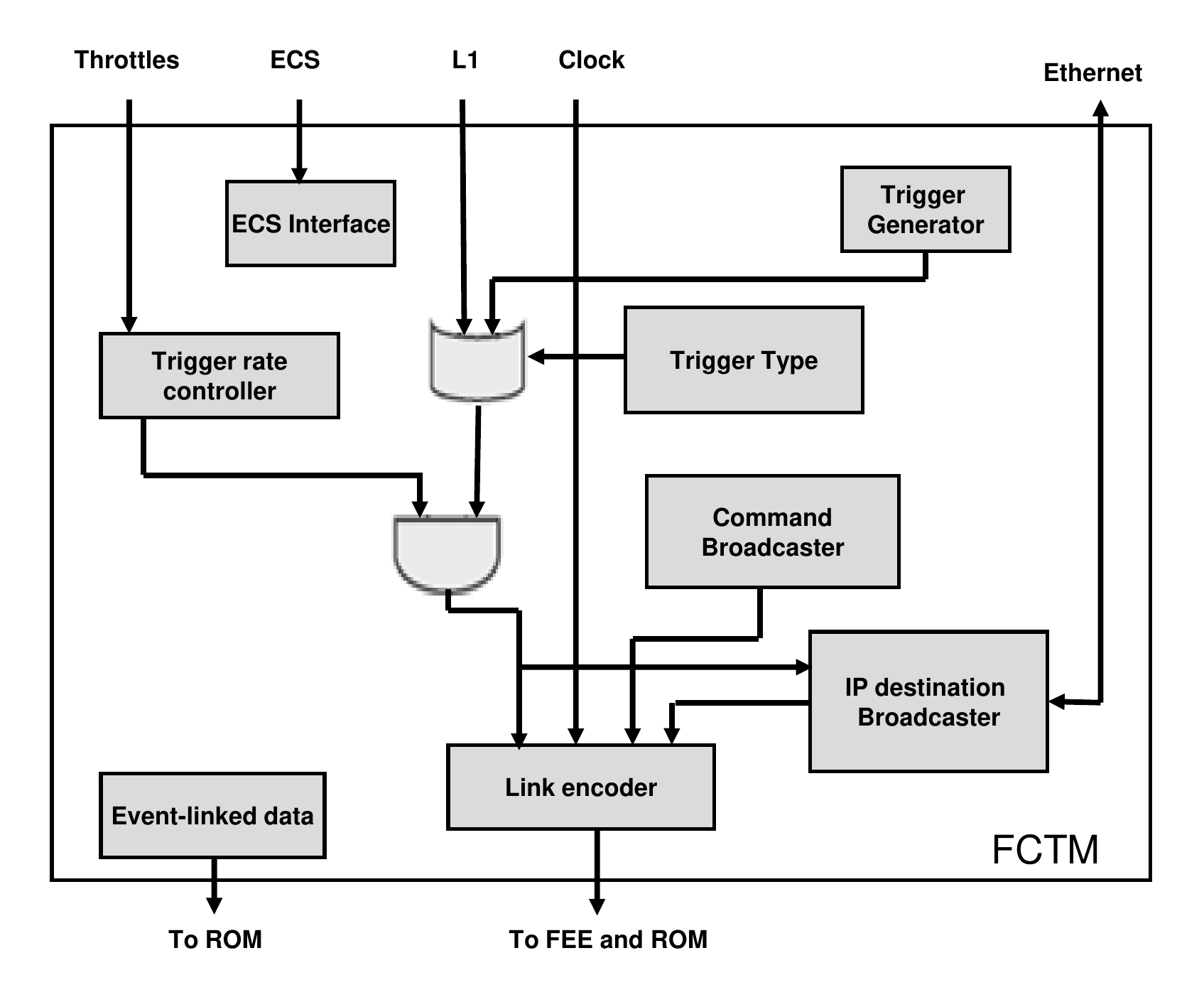}
\caption{Fast Control and Timing Module}
\label{fig:etd-fctm}
\end{figure}

The Fast Control and Timing System (FCTS, Fig.~\ref{fig:etd-fcts})
manages all elements linked to clock, trigger, and event readout, and
is responsible for partitioning the detector into independent
sub-systems for testing and commissioning.

The FCTS will be implemented in a crate where the backplane can be
used to distribute all the necessary signals in point-to-point mode.
This permits the delivery of very clean synchronous signals to all
boards---avoiding the use of external cables. The Fast Control and
Timing Module (FCTM, shown in Fig.~\ref{fig:etd-fctm}) provides the
main functions of the FCTS:

\wpparagraph{Clock and Synchronization:} The FCTS synchronizes the
experiment with the machine and its bunch pattern, distributes the
clock throughout the experiment, buffers the clock, and generates
synchronous reset commands.

\wpparagraph{Trigger Handling:} The FCTS receives the raw L1 trigger
decisions, throttles them as necessary,  and broadcasts them to the
sub-detectors.

\wpparagraph{Calibration and Commissioning:} The FCTS can trigger the
generation of calibration pulses and flexibly programmable local
triggers for calibration and commissioning.

\wpparagraph{Event Handling:} The FCTS generates event identifiers,
manages the event routing, and distributes event routing information
to the ROMs. It also keeps a trace of all of its activity, including
an accounting of triggers lost due to dead time or other sources of
throttling and event-linked data that needs to be included with the
readout data.

\wpparagraph{}

The FCTS crate includes as many FCTM boards as required to cover all
partitions. One FCTM will be dedicated to the unused sub-systems in
order to provide them with the clock and the minimum necessary
commands.

Two dedicated switches are required in order to be able to partition
the system into independent sub-systems or groups of sub-systems. One
switch distributes the clock and commands to the front-end boards, the
other collects throttling requests from the readout electronics or the
ECS. These switches can be implemented on dedicated boards, connected
with the FCTMs, and need to receive the clock. To reduce the number of
connections between ROM crates and the global throttle switch board,
throttle commands could be combined at the ROM crate level before
sending them to the global switch.

Instantaneous throttling of the data acquisition by directly
inhibiting the raw L1 trigger from the front-end electronics is not
possible because the induced latency is too long. Instead, models of
the front-ends and the L1 event buffer queues will be emulated in the
FCTM to instantaneously reduce the trigger rate if data volume exceeds
the front-end capacity.

The FCTM also manages the distribution of events to the HLT farm for
event building, deciding the destination farm node for every event.
There are many possible implementations of the event building network
protocol and the routing of events based on availability of HLT farm
machines, so at this point we can provide only a high-level
description.  We strongly prefer to use the FCTS to distribute event
routing information to the ROMs because it is simple and provides
natural synchronization. Management of event destinations and
functions such as bandwith management for the event building network
or protocols to manage the event distribution based on the
availability of farm servers can then be implemented in FCTM firmware
and/or software.

``Continuation events'' to deal with pile-up could either be merged in
the ROMs or in the high-level trigger farm, but we strongly prefer to
merge them in the ROMs. Merging them in the trigger farm would
complicate the event builder and require the FCTS to maintain event
state and adjust the event routing to send all parts of a continuation
event to the same HLT farm node.

\wpsubsubsec{Clock, Control and Data Links} 

Designing and validating the different serial links required for
\superb\ (for data transmission, timing, and control commands
distribution and read-out) will require substantial effort during the
TDR phase. Because of fixed latency and low jitter constraints, simple
solutions relying on off-the-shelf electronics components must be
thoroughly tested to validate them for use in clock and control links.
Moreover, because radiation levels on the detector side are expected
to be high, R\&D will be necessary to qualify the selected chip-sets
for radiation robustness.  Since requirements for the various link
types differ, technical solutions for different link types may also
differ.

The links are used to distribute the frequency-divided machine clock
(running at 56\MHz) and fast control signals such as trigger pulses,
bunch crossing, and event IDs or other qualifiers to all components of
the ETD system. Copper wires are used for short haul data transmission
($< 1$m), while optical fibres are used for medium and long haul. To
preserve timing information, suitable commercial components will be
chosen so that the latency of transmitted data and the phase of the
clock recovered from the serial stream do not change with power
cycles, resets, and loss-of-locks. Encoding and/or scrambling
techniques will be used to minimize the jitter on the recovered clock.
The same link architecture is also suitable for transmitting regular
data instead of fast controls, or a combination of both.

Link types can be divided into two classes:

\wpparagraph{A-Type:} The A-type links are homogeneous links with both
ends off-detector.  Given the absence of radiation, they might be
implemented with Serializer-De-serializers (SerDes) embedded in FPGAs
(Field Programmable Gate Arrays).  Logic in the FPGA fabric will be
used to implement fixed latency links and to encode/decode fast
control signals. A-Type links are used to connect the FCTS system to
the DAQ crate control and to the Global Level 1 Trigger. A-Type links
run at approximately 2.2\gbitsps.

\wpparagraph{B-Type:} The B-type hybrid links have one end on-detector
and the other end off-detector.  The on-detector side might be
implemented with off-the-shelf radiation-tolerant components---the
off-detector end might still be implemented with FPGA-embedded SerDes.
B-Type links connect the FCTS crate to the FEE and the FEE to ROMs.
The B-Type link speed might be limited by the off-the-shelf SerDes
performance, but is expected to be at least 1\gbitps for the FCTS to
FEE link and about 2\gbitsps for the FEE to ROM link.

\wpparagraph{}

All links can be implemented as plug-in boards or mezzanines, (1)
decoupling the development of the user logic from the high-speed link
design, (2) simplifying the user board layout, and (3) allowing an
easy link upgrade without affecting the host boards. Mezzanine
specifications and form-factors will likely be different for A-Type
and B-Type links, but they will be designed to share a common
interface to the host board to the maximum possible extent.

\wpsubsubsec{Common Front-End Electronics} 

\begin{figure}
\includegraphics[width=0.5\textwidth]{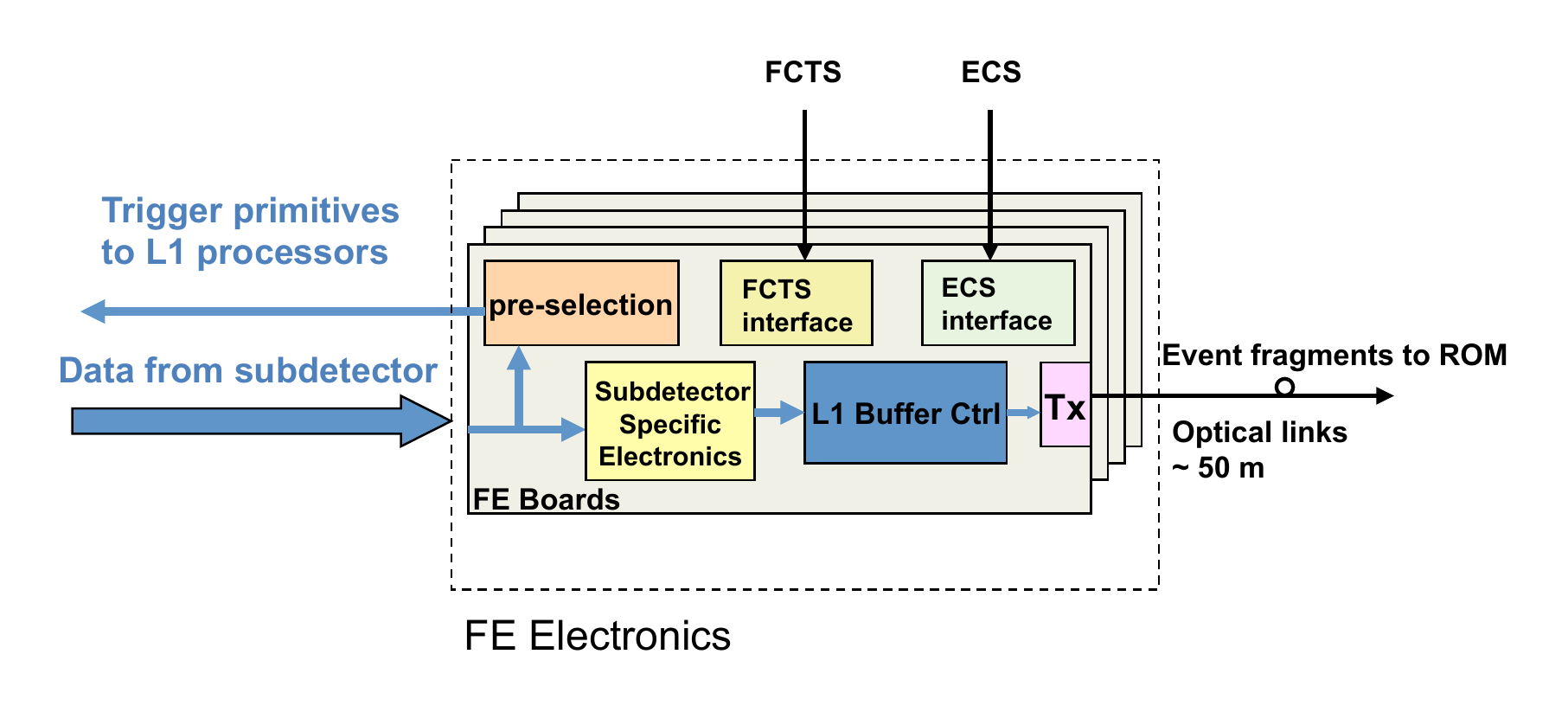}
\caption{Common Front-End Electronics}
\label{fig:etd-fex}
\end{figure}

Common Front-End Electronics (CFEE) designs and components allow us to
exploit the commonalities between the sub-detector electronics and
avoid separate design and implementation of common functions for each
sub-detector.

In our opinion, the separate functions required to drive the FEE
should be implemented in dedicated independent elements.  These
elements will be mezzanines or circuits directly mounted on the
front-end modules (which act as carrier boards) and will be
standardized across the sub-systems as much as possible. For instance,
as shown in Fig.~\ref{fig:etd-fex}, one mezzanine can be used for FCTS
signal and command decoding, and one for ECS management. To reduce the
number of links, it may be possible to decode the FCTS and ECS signals
on one mezzanine and then distribute them to the neighbouring boards.

A common dedicated control circuitry inside a radiation-tolerant FPGA
may also drive the L1 buffers. It would handle the L1 accept commands
and provide the signals necessary to manage the data transfers between
latency buffers, derandomizer buffers and the fast multiplexers
feeding the optical link serializers. If required by the system
design, it would also provide logic for special treatment of pile-up
events and/or extending the readout window back in time after a Bhabha
event has been rejected.

The latency buffers can be implemented either in the same FPGA or
directly on the carrier boards. One such circuit can drive numerous
data links in parallel, thus reducing the amount of electronics on the
front-end.

One intriguing, possible advantage of this approach is that {\em
  analog} L1 buffers might be implemented in an ASIC, though the
analog output of the ASIC then must be able to drive an internal or
external ADC that samples the signal.

Serializers and optical link drivers will also reside on carrier
boards, mainly for mechanical and thermal reasons.
Fig.~\ref{fig:etd-fex} shows a possible implementation of the L1
buffers, their control electronics (in a dedicated FPGA), and their
outputs to the optical readout links. 

All (rad-tolerant) FPGAs in the FEE have to be reprogrammable without
dismounting a board. This could be done through dedicated front panel
connectors, which might be linked to numerous FPGAs, but it would be
preferable if the reprogramming could be done through the ECS without
any manual intervention on the detector side.

Sampling of the analog signals in the FEEs is done with the global
clock or a clock signal derived from the global clock (typically by
dividing it down). To maintain the timing required by the fixed
latency design, the latency buffers in the FEEs must be read with the
same sampling frequency as they are written. In addition, when
initializing the FEE boards, care must be taken that all dividers are
reset synchronously with those of the first level trigger (by a global signal)
in order to maintain a constant phase between them.

\wpsubsubsec{Readout Module} 

\begin{figure}
\center{\includegraphics[width=0.45\textwidth]{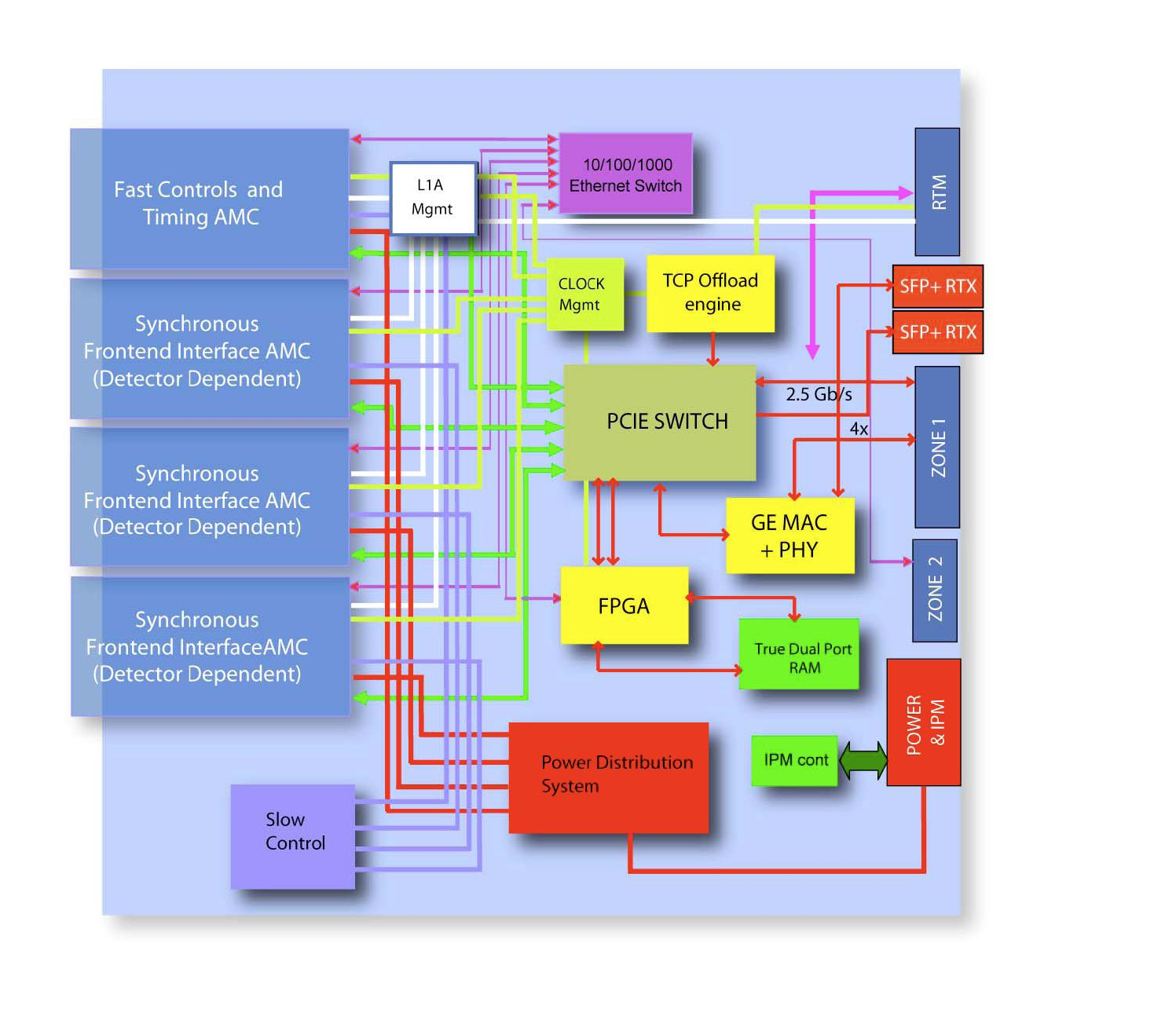}}
\caption{Readout Module}
\label{fig:etd-rom}
\end{figure}

The Readout Modules (ROM, Fig.~\ref{fig:etd-rom}) receive event
fragments from the sub-detectors' front-end electronics, tag them with
front-end identifiers and absolute time-stamps, buffer them in
de-randomizing memories, perform processing (still to be defined) on
the fragment data, and eventually inject the formatted fragment
buffers into the event builder and HLT farm. Connected to the
front-end electronics via optical fibres, they will be located in an
easily accessible, low radiation area.

A modular approach will maximize standardization across the system to
simplify development and keep costs low---different sub-detector
requirements can then be accommodated by using sub-detector-specific
``personality modules''.

On the ROM boards, signals from optical receivers mounted on mezzanine
cards will be routed to the de-serializers (in commercial FPGAs) where
data processing can take place. Special requirements from the
sub-detector systems will be accommodated by custom-built mezzanines
mounted on common carriers.  One of the mezzanine sites on the carrier
will host an interface to the FCTS to receive global timing and
trigger information.  The carrier itself will host memory buffers and
1\gbitps or 10\gbitsps links to the event building network.

A baseline of 8 optical fibres per card currently seems like a good
compromise between keeping the number of ROM boards low and adding to
their complexity. This density is sufficient so that there needs to be
only one ROM crate per sub-detector, and corresponds nicely to the
envisaged FCTS partitioning.

\wpsubsubsec{Experiment Control System} 
\label{sec:etd-ecs}

The complete \superb\ experiment (power supplies, front-end, DAQ,
etc.)  must be controlled by an Experiment Control System (ECS). As
shown in Fig.~\ref{fig:etd-overview}, the ECS is responsible both for
controlling the experiment and for monitoring its functioning.

\wpparagraph{Configuring the Front-ends:} Many front-end parameters
must be initialized before the system can work correctly. The number
of parameters per channel can range from a only a few to large
per-channel lookup tables. The ECS may also need to read back
parameters from registers in the front-end hardware to check the
status or verify that the contents have not changed. For a fast
detector configuration and recovery turnaround in factory mode, it is
critical to not have bottlenecks either in the ECS itself, or in the
ECS' access to the front-end hardware. If technically feasible and
affordable, front-end electronics on or near the detector should be
shielded or engineered to avoid frequent parameter reloads due to
radiation-induced single event upsets---reconfiguring through the ECS
should only be considered as a last resort.

\wpparagraph{Calibration:} Calibration runs require extended
functionality of the ECS. In a typical calibration run, after loading
calibration parameters, event data collected with these parameters
must be sent through the DAQ system and analyzed. Then the ECS must
load the parameters for the next calibration cycle into the front-ends
and repeat.

\wpparagraph{Testing the FEE:} The ECS may also be used to remotely
test all FEE electronics modules using dedicated software. This
obviates the need for independent self-test capability for all
modules.

\wpparagraph{Monitoring the Experiment:} The ECS continuously monitors
the entire experiment to insure that it functions properly. Some
examples include (1) independent spying on event data to verify data
quality, (2) monitoring the power supplies (voltage, current limits,
etc.), and (3) monitoring the temperature of power supplies, crates,
and modules. Support for monitoring the FEE modules themselves must be
built into the FEE hardware so that the ECS can be informed about FEE
failures. The ECS also acts as a first line of defense in protecting
the experiment from a variety of hazards. In addition, an independent,
hardware-based detector safety system (part of the Detector Control
System, see ~\ref{sec:detector_control_system}) must protect the
experiment against equipment damage in case the software-based ECS is
not operating correctly.

\wpparagraph{}The specific requirements that each of the sub-systems
makes on ECS bandwidth and functionality must be determined (or at
least estimated) as early as possible so that the ECS can be designed
to incorporate them.  Development of calibration, test, and monitoring
routines must be considered an integral part of sub-system
development, as it requires detailed knowledge about sub-system
internals.

\wpparagraph{Possible ECS Implementation:}

The field bus used for the ECS has to be radiation tolerant on the
detector side and provide very high reliability. Such a bus has been
designed for the LHCb experiment: it is called SPECS (Serial Protocol
for Experiment Control System) \cite{bib:etd-SPECS}. It is a
bidirectional 10\mbitsps bus that runs over standard Ethernet Cat5+
cable and provides all possible facilities for ECS (like JTAG (Joint
Test Action Group) and I2C (Inter IC)) on a small mezzanine. It could
be easily adapted to the \superb\ requirements.  Though SPECS was
initially based on PCI boards, it is currently being translated to an
Ethernet-based system, as part of an LHCb upgrade, also integrating
all the functionalities for the out-of-detector elements.

For the electronics located far from the detector, Ethernet will be
used for ECS communication.

\wpsubsubsec{Level~1 Hardware Trigger} 

\begin{figure}
\includegraphics[width=0.5\textwidth]{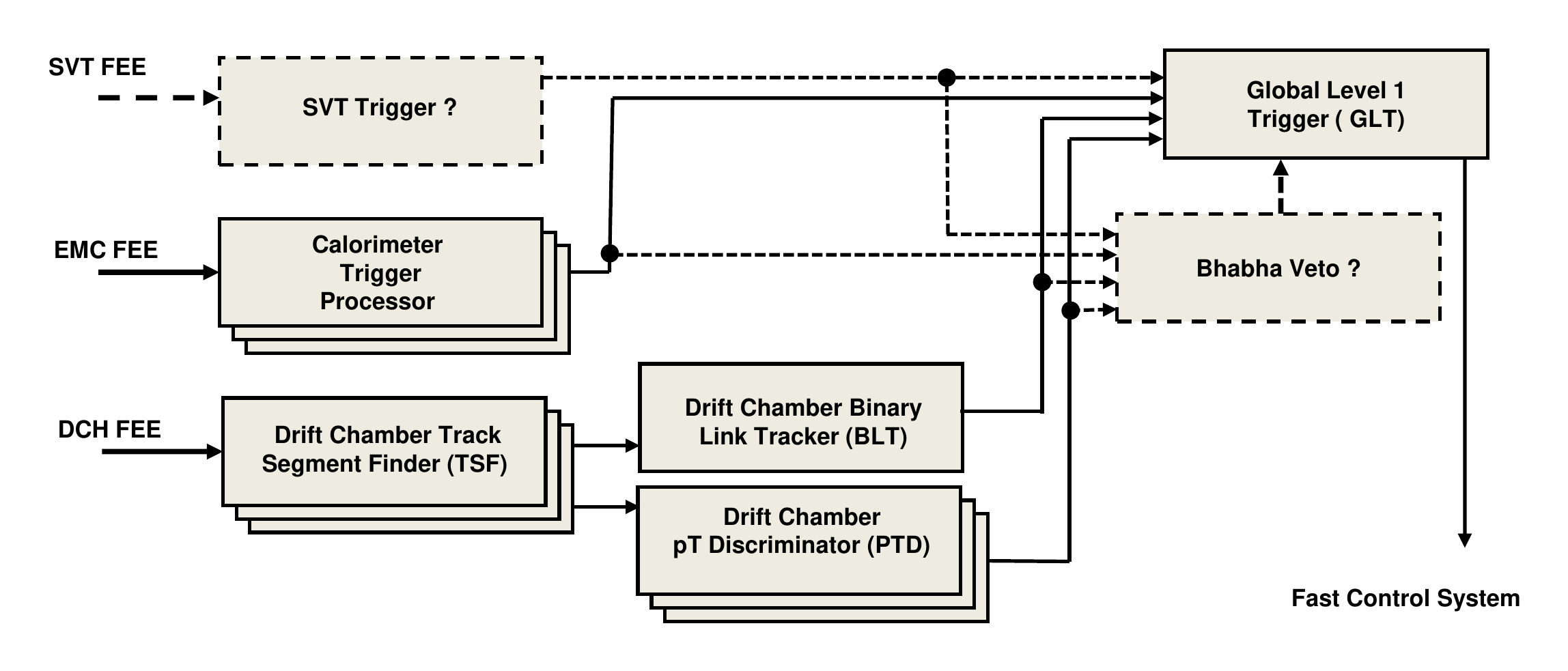}
\caption{Level~1 Trigger Overview}
\label{fig:etd-trigger-l1}
\end{figure}

The current baseline for the L1 trigger is to re-implement the \babar\
L1 trigger with state-of-the-art technology. It would be a synchronous
machine running at 56\MHz (or multiples of 56\MHz) that processes
primitives produced by dedicated electronics located on the front-end
boards or other dedicated boards of the respective sub-detector.  The
raw L1 decisions are sent to the FCTM boards which applies a throttle
if necessary and then broadcasts them to the whole experiment. The
standard chosen for the crates would most likely be either ATCA
(Advanced Telecommunications Computing Architecture) for the crates
and the backplanes, or a fully custom architecture.

The main elements of the L1 trigger are shown in
Fig.~\ref{fig:etd-trigger-l1} (see \cite{bib:etd-BaBarTRG} for
detailed descriptions of the \babar\ trigger components):

\wpparagraph{Drift chamber trigger (DCT):} The DCT consists of a track
segment finder (TSF) , a binary link tracker (BLT) and a $p_{t}$
discriminator (PTD).

\wpparagraph{Electromagnetic Calorimeter Trigger (EMT):} The EMT
processes the trigger output from the calorimeter to find clusters.

\wpparagraph{Global Trigger (GLT):} The GLT processor combines the
information from DCT and EMT (and possibly other inputs such as an SVT
trigger or a Bhabha veto) and forms a final trigger decision that is
sent to the FCTS.

\wpparagraph{}

We will study the applicability of this baseline design at \superb\
luminosities and backgrounds, and will investigate improvements, such
as adding a Bhabha veto or using SVT information in the L1 trigger.
We will also study faster sampling of the DCH and the new fast
forward calorimeter. In particular for the barrel EMC we will need to
study how the L1 trigger time resolution can be improved and the
trigger jitter can be reduced compared to BaBar. In general, improving
the trigger event time precision should allow a reduction in readout
window and raw event size.  The L1 trigger may also be improved using
larger FPGAs (e.g. by implementing tracking or clustering algorithm
improvements, or by exploiting better readout granularity in the EMC).

\wpparagraph{L1 Trigger Latency:} The \babar\ L1 trigger had 12\mus
latency. However, since the size, and cost, of the L1 data buffers in
the sub-detectors scale directly with trigger latency, it should be
substantially reduced, if possible. L1 trigger latencies of the much
larger, more complex, ATLAS, CMS and LHCb experiments range between 2
and 4\mus, however these experiments only use fast detectors for
triggering. Taking into consideration that the DCH adds an intrinsic
dead time of about 1\mus and adding some latency reserve for future
upgrades, we are currently estimating a total trigger latency of 6\mus
(or less). More detailed engineering studies will be required to
validate this estimate.

\wpparagraph{Monitoring the Trigger:}To debug and monitor the trigger,
and to provide cluster and track seed information to the higher
trigger levels, trigger information supporting the trigger decisions
is read out on a per-event basis through the regular readout system.
In this respect, the low-level trigger acts like just another
sub-detector.

\wpsubsec{Online System} 

\begin{figure}[h]
\includegraphics[width=0.45\textwidth]{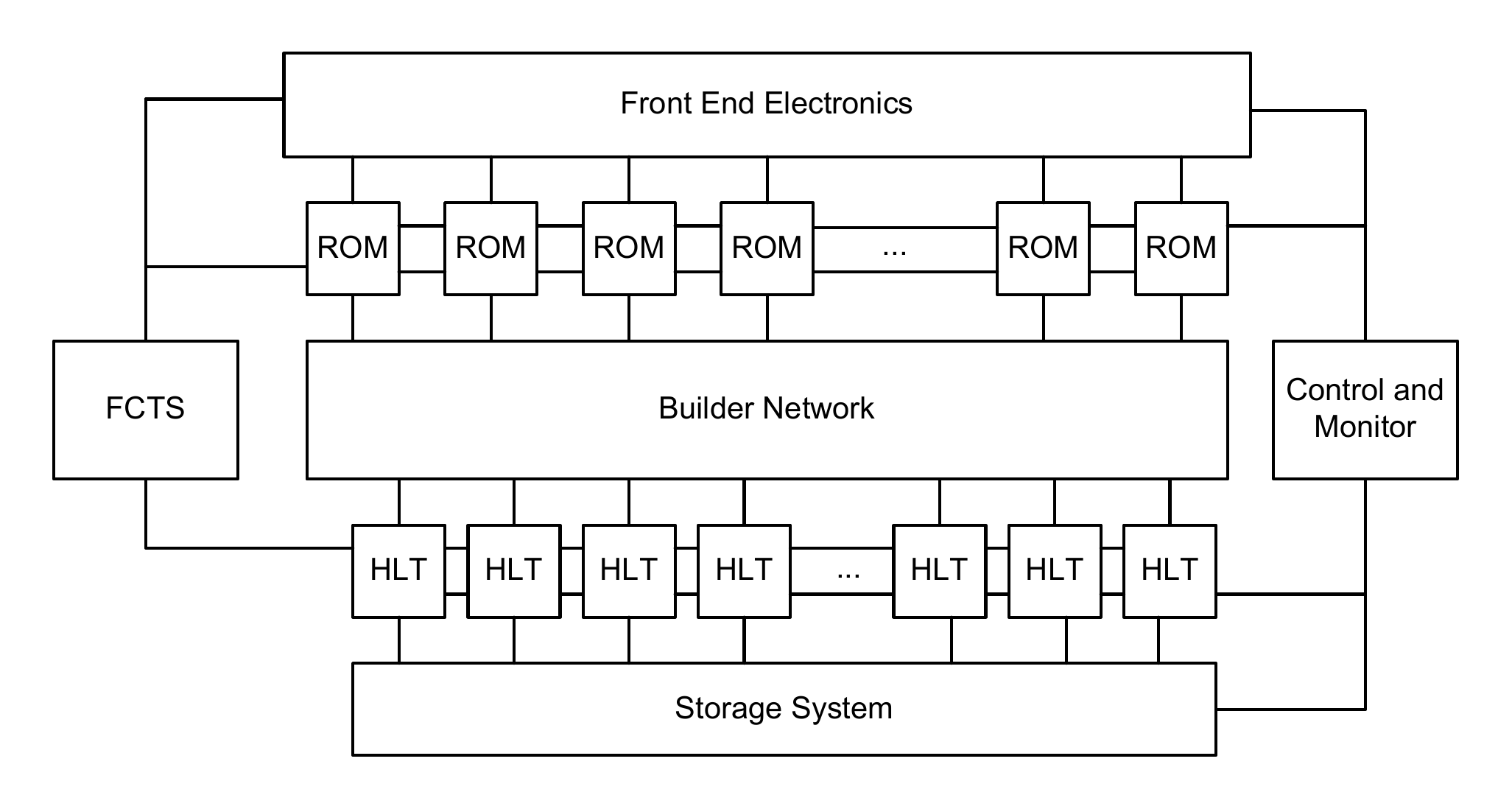}
\caption{High-level logical view of the Online System}
\label{fig:etd-online}
\end{figure}

The Online system is responsible for reading out the ROMs, building
complete events, filtering events according to their content (High
Level Triggers), and archiving the accepted events for further physics
analysis (Data Logging). It is also responsible for continuous
monitoring of the acquired data to understand detector performance and
detect detector problems (Data Quality Monitoring).  The Detector
Control System (DCS) monitors and controls the detector and its
environment.

Assuming a L1 trigger rate of 150\kHz and an event size of 75\kbytes,
the input bandwidth of the Online system must be about 12\gbytesps,
corresponding to about 120\gbitsps with overhead.  It seems prudent to
retain an additional safety factor of \ktilde2, given the event size
uncertainty and the immaturity of the overall system design.  Thus, we
will take 250\gbitsps as the baseline for the Online system input
bandwidth.

Assuming that the HLT accepts a cross-section of about 25\nb leads to
an expected event rate of 25\kHz at a luminosity of $10^{36}\ {\rm
  cm}^{-2} {\rm sec}^{-1}$, or a logging data rate of
\ktilde1.9\gbytesps.

The main elements of the Online system (Fig.~\ref{fig:etd-online})
are described in the following sections.

\wpsubsubsec{ROM Readout and Event Building} 

The ROMs read out event fragments in parallel from sub-detector
front-end electronics---buffering the fragments in deep de-randomizing
memories. The event-related information is then transferred into the
ROM memories, and sent over a network to an event buffer in one of the
machines of the HLT farm. This collection task, called event building,
can be performed in parallel for multiple events, thanks to the depth
of the ROM memories and bandwidth of the event building network switch
(preferably non-blocking). Because of this inherent parallelism, the
building rate can be scaled up as needed (up to the bandwidth limit of
the event building network).  We expect to use Ethernet as the basic
technology of the event builder network, using 1\gbitsps and
10\gbitsps links.

\wpsubsubsec{High Level Trigger Farm} 

The HLT farm needs to provide sufficient aggregate network bandwidth
and CPU resources to handle the full Level~1 trigger rate on its input
side. The Level~3 trigger algorithms should operate and log data
entirely free of event time ordering constraints and be able to take
full advantage of modern multi-core CPUs. Extrapolating from \babar,
we expect 10\ms core time per event to be more than adequate to
implement a software L3 filter, using specialized fast reconstruction
algorithms. With such a filter, an output cross-section of 25\nb
should be achievable.

To further reduce the amount of permanently stored data, an additional
filter stage (L4) could be added that acts only on events accepted by
the L3 filter. This L4 stage could be an equivalent (or extension) of
the \babar\ offline physics filter---rejecting events based either on
partial or full event reconstruction. If the worst-case behavior of
the L4 reconstruction code can be well controlled, it could be run in
near real-time as part of, or directly after, the L3 stage.
Otherwise, it may be necessary to use deep buffering to decouple the
L4 filter from the near real-time performance requirements imposed at
the L3 stage.  The discussion in the \superb\ CDR
~\cite{bib:etd-SuperBCDR} about risks and benefits of a L4 filter
still applies.

\wpsubsubsec{Data Logging} 

The output of the HLT is logged to disk storage. We assume at least a
few \tbytes of usable space per farm node, implemented either as
directly attached low-cost disks in a redundant (RAID) configuration,
or as a storage system connected through a network or SAN. We do not
expect to aggregate data from multiple farm nodes into larger files.
Instead, the individual files from the farm nodes will be maintained
in the downstream system and the bookkeeping system and data handling
procedures will have to deal with missing run contribution files.  A
switched Gigabit Ethernet network separate from the event builder
network is used to transfer data asynchronously to archival storage
and/or near-online farms for further processing.  It is not yet
decided where such facilities will be located, but network
connectivity with adequate bandwidth and reliability will need to be
provided. Enough local storage must be available to the HLT farm to
allow data buffering for the expected periods of link down-time.

While the format for the raw data has yet to be determined, many of
the basic requirements are clear, such as efficient sequential
writing, compact representation of the data, portability, long-term
accessibility, and the freedom to tune file sizes to optimize storage
system performance.

\wpsubsubsec{Event Data Quality Monitoring and Display} 

Event data quality monitoring is based on quantities calculated by the
L3 (and possibly L4) trigger, as well as quantities calculated by a
more detailed analysis on a subset of the data. A distributed
histogramming system collects the monitoring output histograms from
all sources and makes them available to automatic monitoring processes
and operator GUIs.

\wpsubsubsec{Run Control System} 

The control and monitor of the experiment is performed by the Run
Control System (RCS), providing a single point of entry to operate and
monitor the entire experiment.  It is a collection of software and
hardware modules that handle the two main functions of this component:
controlling, configuring, and monitoring the whole Online system, and
providing its user interface. The RCS interacts both with the
Experiment Control System (ECS) and with the Detector Control System
(DCS). We expect the RCS to utilize modern web technologies.

\wpsubsubsec{Detector Control System} 
\label{sec:detector_control_system}

The Detector Control System (DCS) is responsible for ensuring detector
safety, controlling the detector and detector support system, and
monitoring and recording detector and environmental conditions.

Efficient detector operations in factory mode require high levels of
automation and automatic recovery from problems. The DCS plays a key
role in maintaining high operational efficiency, and tight integration
with the Run Control System is highly desirable.

Low-level components and interlocks responsible for detector safety
(Detector Safety System, DSS) will typically be implemented as simple
circuits or with programmable logic controllers (PLCs).

The software component will be built on top of a toolkit that provides
the interface to whatever industrial buses, sensors, and actuators may
be used. It must provide a graphical user interface for the operator,
have facilities to generate alerts automatically, and have an
archiving system to record the relevant detector information.  It must
also provide software interfaces for programmatic control of the
detector.

We expect to be able to use existing commercial products and controls
frameworks developed by the CERN LHC experiments.

\wpsubsubsec{Other Components} 

\wpparagraph{Electronic Logbook:} A web-based logbook, integrated with
all major Online components, allows operators to keep an ongoing log
of the experiment's status, activities and changes.

\wpparagraph{Databases:} Online databases such as configuration,
conditions, and ambient databases are needed to track, respectively,
the intended detector configuration, calibrations, and actual state
and time-series information from the DCS.

\wpparagraph{Configuration Management:} The configuration management
system defines all hardware and software configuration parameters, and
records them in a configuration database.

\wpparagraph{Performance Monitoring:} The performance monitoring
system monitors all components of the Online.

\wpparagraph{Software Release Management:} Strict software release
management is required, as is a tracking system that records the
software version (including any patches) that was running at a given
time in any part of the ETD/Online system. Release management must
cover FPGAs and other firmware as well as software.

\wpparagraph{Computing Infrastructure Reliability:} The Online
computing infrastructure (including the specialized and
general-purpose networks, file, database and application servers,
operator consoles, and other workstations) must be designed to provide
high availability, while being self-contained (sufficiently isolated
and provided with firewalls) to minimize external dependencies and
downtime.

\wpsubsubsec{Software Infrastructure} 

The Online system is basically a distributed system built with
commodity hardware components. Substantial manpower will be needed to
design the software components---taking a homogeneous approach in
both the design and implementation phases.  An Online software
infrastructure framework will help organize this major undertaking.
It should provide basic memory management, communication services, and
the executive processes to execute the Online applications.  Specific
Online applications will make use of these general services to
simplify the performance of their functions.  Middleware designed
specifically for data acquisition exists, and may provide a simple,
consistent, and integrated distributed programming environment.

\wpsubsec{Front-End Electronics} 

\wpsubsubsec{SVT Electronics} 

\begin{figure}[b]
\includegraphics[width=0.45\textwidth]{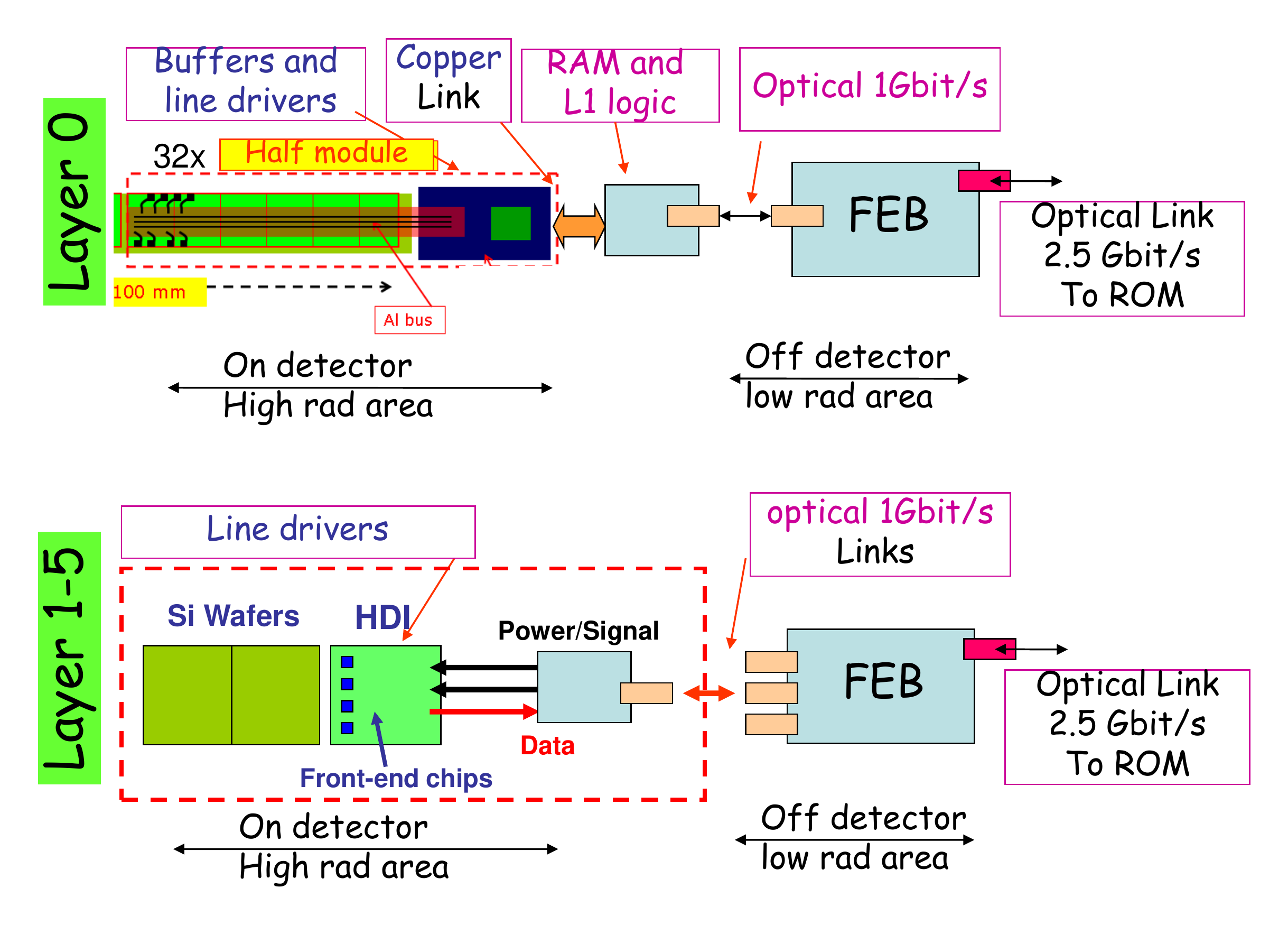}
\caption{SVT Electronics}
\label{fig:etd-svt}
\end{figure}

The SVT electronics shown in Fig.~\ref{fig:etd-svt} is designed to
take advantage, where possible, of the data-push characteristics of
the front-end chips. The time resolution of the detector is dominated
by the minimal time resolution of the FSSR2 chip, which is 132\ns.
Events are built from packets of three minimal time slices (396\ns
event time window). The readout chain in layer~0 starts from a
half-module holding two sets of pixel chips (2 readout sections, ROS).
Data are transferred on copper wires to boards located a few meters
away from the interaction region where local buffers will store the
read hits. As discussed in the SVT chapter, for layer~0, the data rate
is dominated by the background. The bandwidth needed is about
16\gbitsps/ROS.  This large bandwidth is the main reason to store hits
close to the detector and transfer only hits from triggered events.

For events accepted by the L1 trigger, the bandwidth requirement is
only 0.85\gbitsps and data from each ROS can be transferred on optical
links (1\gbitps) to the front-end boards (FEB) and then to ROMs
through the standard 2\gbitsps optical readout links. Layers~1-5 are
read out continuously with the hits being sent to the front-end boards
on 1\gbitps optical links. On the FEBs, hits are sorted in time and
formatted to reduce event size (timestamp stripping).  Hits of
triggered events are then selected and forwarded to the ROMs on
2\gbitsps standard links.

Occupancies and rates on layers 3-5 should be low enough to make them
suitable for fast track searching so that SVT information could be
used in the L1 trigger. The SVT could provide the number of tracks
found, the number of tracks not originating from the interaction
region, and the presence of back-to-back events in the $\phi$
coordinate. A possible option for SVT participation to the L1 trigger
would require two L1 trigger processing boards each one linked to the
FEBs of layers 3-5 with synchronous optical links.

In total, the SVT electronics requires 58 FEBs and 58 ROMs, 58 optical
links at 2\gbitsps, 308 links at 1\gbitps (radiation hard) and,
optionally, two L1 trigger processing boards and about 40 links at
1.25\gbitsps for L1 trigger processing.

\wpsubsubsec{DCH Electronics:} 
\label{sec:ETD_DCH}

\begin{figure*}[t]
\begin{center}
  \subfigure[Data Readout
  Path]{\label{fig:etd-dch-main}\includegraphics[width=0.45\textwidth]{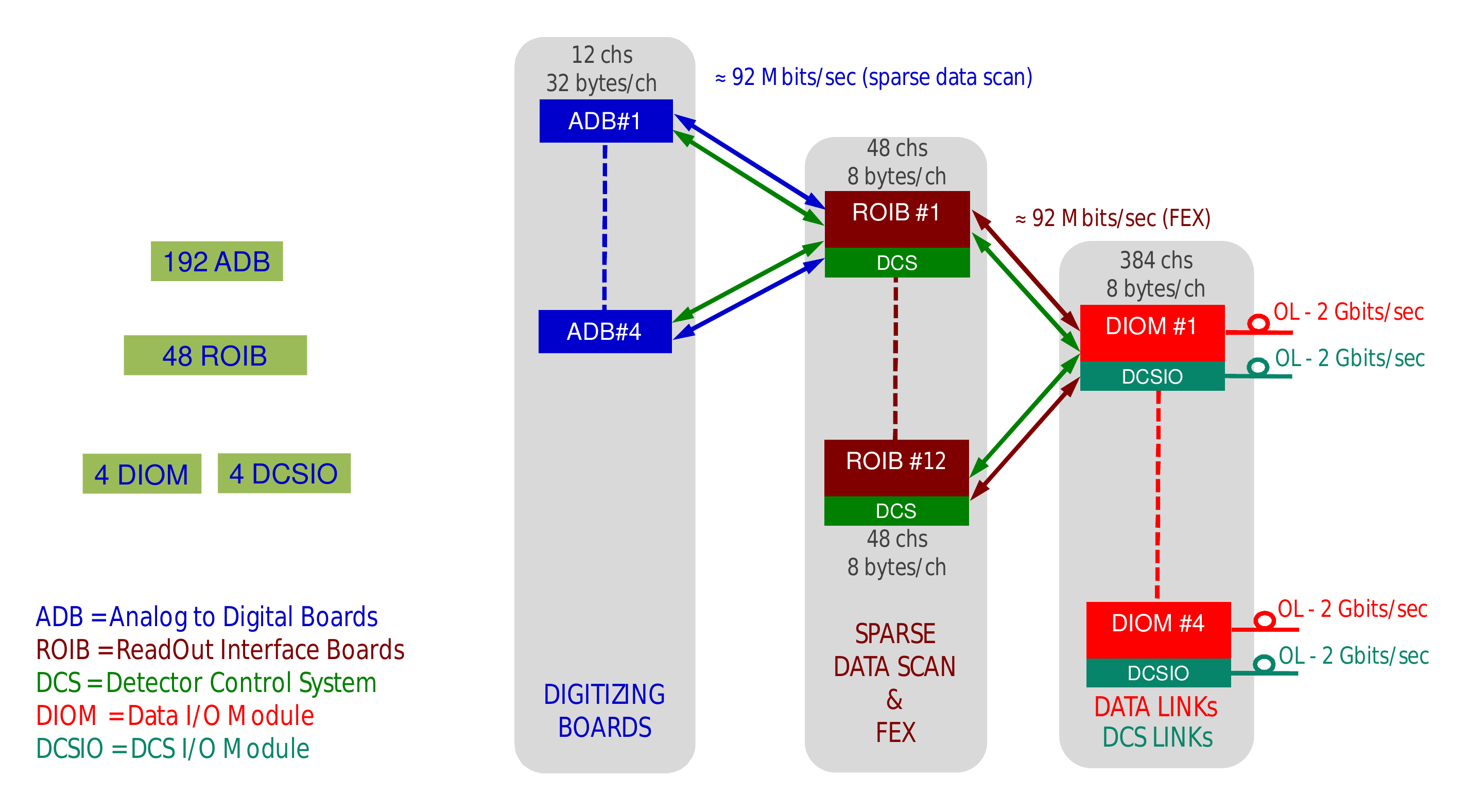}}
  \subfigure[Trigger Readout
  Path]{\label{fig:etd-dch-trigger}\includegraphics[width=0.45\textwidth]{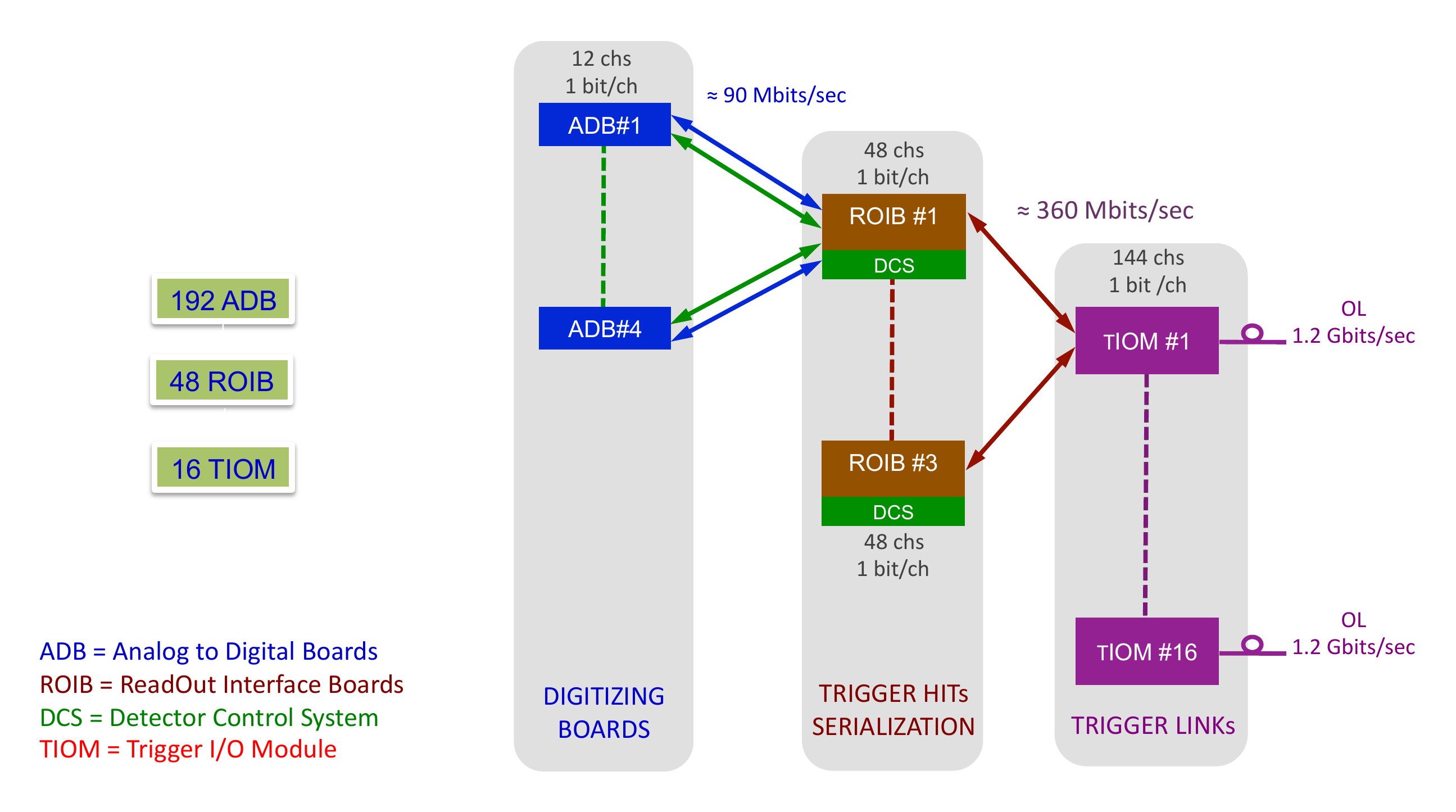}}
\end{center}
\caption{DCH Electronics}
\end{figure*}

The design is still in a very early stage, so we only provide a
baseline description of the drift chamber front-end electronics. It
does not include additional front-end features currently under study
(such as a cluster counting capability).

The DCH provides charged particle tracking, \dedx, and trigger
information. The front-end electronics measures the drift time of the
first electron and the total charge collected on the sense wires, and
generates the information to be sent to the L1 trigger.

The DCH front-end chain can be divided into three different blocks:

\wpparagraph{Very Front End Boards (VFEB):} The VFEBs contain HV
distribution and blocking capacitors, protection networks and
preamplifiers. They could also host discriminators. The VFEBs are
located on the (backward) chamber end-plate to maximize the S/N ratio.

\wpparagraph{Data Conversion and Trigger Pattern Extraction:} Data
conversion incorporates both TDCs (1\ns resolution, 10~bits dynamic
range) and continuous sampling ADCs (6~bits dynamic range). Trigger
data contain the status of the discriminated channels, sampled at
about 7\MHz (compared to 3.7\MHz in \babar). This section of the chain
can be located either on the end-plate (where power dissipation,
radiation environment, and material budget are issues) or in external
crates (where either micro-coax or twisted cables must be used to
carry out the preamplifier signals).

\wpparagraph{Readout Modules:} The ROMs collect the data from the DCH
FEE and send zero-suppressed data to DAQ and trigger.

\wpparagraph{} The number of links required for data transfer to the
DAQ system can be estimated based on the following assumptions:
150\kHz L1 trigger rate, 10k channels, 15\% chamber occupancy in a
1\mus time window, and 32 bytes per channel. At a data transfer speed
of 2\gbitsps per link, about 40 links are needed. 56 synchronous
1.25\gbitsps links are required to transmit the trigger data sampled
at 7\MHz. The topology of the electronics suggests that the number of
ECS and FCTS links should be the same as the number of readout links.

\wpsubsubsec{PID Electronics:} 

\begin{figure}[b]
\includegraphics[width=0.45\textwidth]{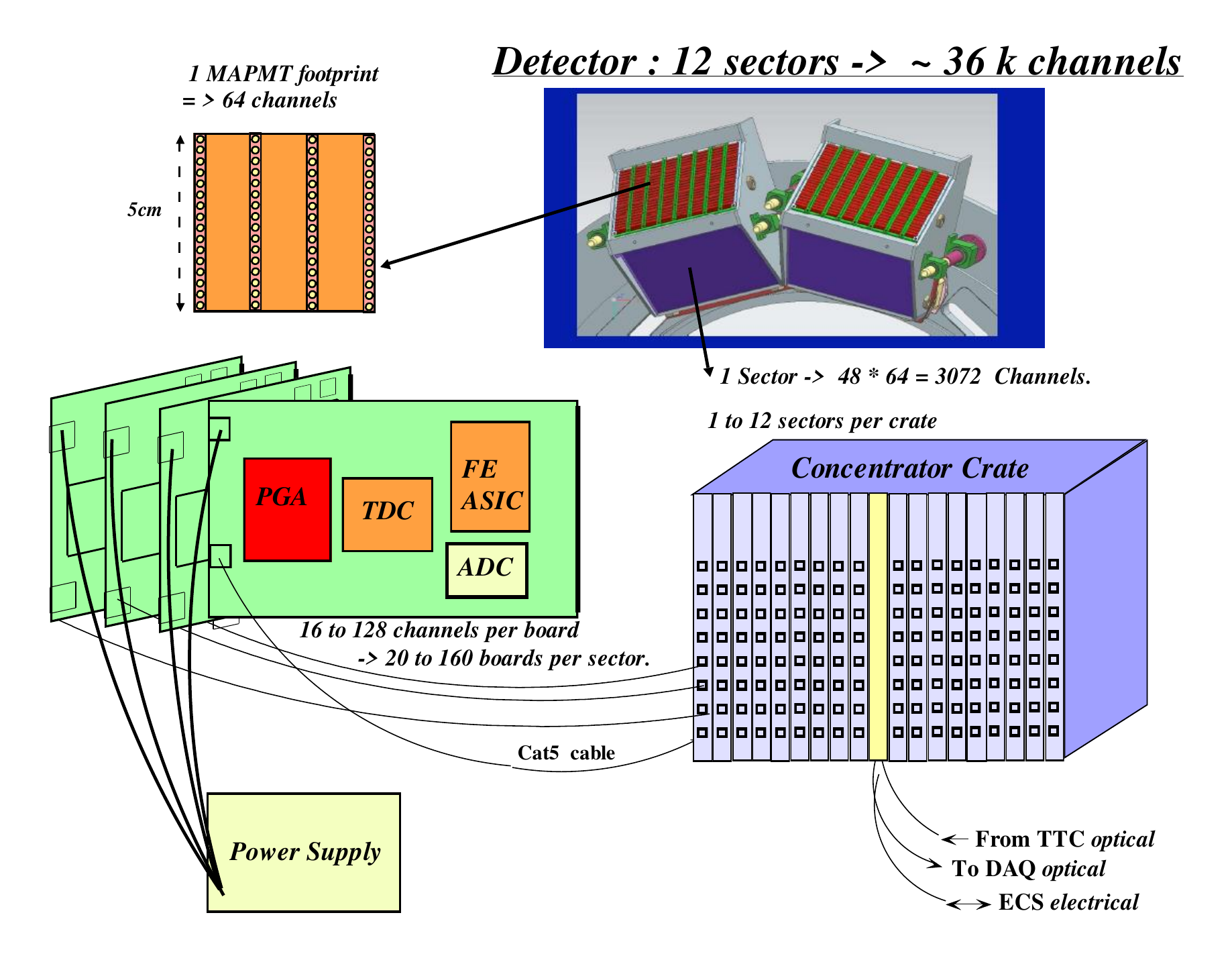}
\caption{PID Electronics}
\label{fig:etd-pid}
\end{figure}

\wpparagraph{Forward PID Option:}

There are currently two detector options be considered for the forward
PID.

The first option is to measure the time of flight (TOF) of particles
from the interaction point to the PID detector.  Two implementations
are under consideration---a pixel detector which would lead to a large
number of read-out channels (~7200), or a DIRC-like detector with
fused silica bars (plates) which would require a much smaller (~192)
channel count. Both implementations make use of fast Micro Channel
Plate PMTs (MCPPMT) and have to provide a measurement of the hit time
with a precision of \ktilde10\ps. The readout would probably use fast
analog memories which, as of today, are the most plausible solution
for a picosecond time measurement in this environment. To achieve this
time resolution, the clock distribution will have to be very carefully
designed and will likely require direct use of the machine clock at
the beam crossing frequency.

A second option is a Focusing Aerogel Cherenkov detector. Though the
timing requirements are less severe, its \ktilde 115,000 channels
would also have to come from MCPPMTs, since standard multi-anode PMTs
cannot be used in the high magnetic field where it resides. Since the
time precision needed is similar to that of the barrel, the same type
of electronics could be used. At least 50 links would be the minimum
necessary for the data readout, while the ECS and FCTS would require a
maximum of about 50 additional links.

\wpparagraph{Barrel PID:}

The barrel PID electronics must provide the measurement of the arrival
time of the photons produced in the fused silica bars with a precision
of about 100\ps rms. The \superb\ detector baseline is a focusing
DIRC, using multi-anodes photo multipliers. This optical design
(smaller camera volume, and materials) reduces the background
sensitivity by at least one order of magnitude compared to \babar\,
thus reducing the rate requirements for the front-end electronics.

The baseline design is implemented with 16-channel TDC
ASICs---offering the required precision of 100\ps rms.  A 12-bit ADC
can provide an amplitude measurement, at least for calibration,
monitoring and survey, which is transmitted with the hit time.  A
16-channel front-end analog ASIC must be designed to sample and
discriminate the analog signal.  Both ASICs would be connected to a
radiation-tolerant FPGA which would handle the hit readout sequence
and push data into the L1 trigger latency buffers.

This front-end electronics must all sit on the MAPMT base, where space
is very limited and cooling is difficult. However, crates
concentrating front-end data and driving the fast optical links can be
located outside the detector in a more convenient place where space is
available. They would be connected to the front-end through standard
commercial cables (like Cat 5 Ethernet cables). The readout mezzanines
would be implemented there, as well as the FCTS and ECS mezzanines
from where signals would be forwarded to the front-end electronics
through the same cables.

The system would be naturally divided into 12 sectors.  Using the
baseline camera with 36,864 channels, 150\kHz trigger rate,
100kHz/channel hit rate, 32 data bits/hit, and 2\gbitsps link rate,
the readout link occupancy should be only \ktilde15\%, thus offering a
pleasant safety margin. A camera using another model of PMTs with
one-half the number of channels is also being studied.

An alternative readout option would be to use analog memories instead
of TDCs to perform both time and amplitude measurements. This option
retains more information on the hit signals but would likely be more
expensive.  Its advantages and disadvantages are still under study.

\wpsubsubsec{EMC Electronics:} 

\begin{figure}
\includegraphics[width=0.45\textwidth]{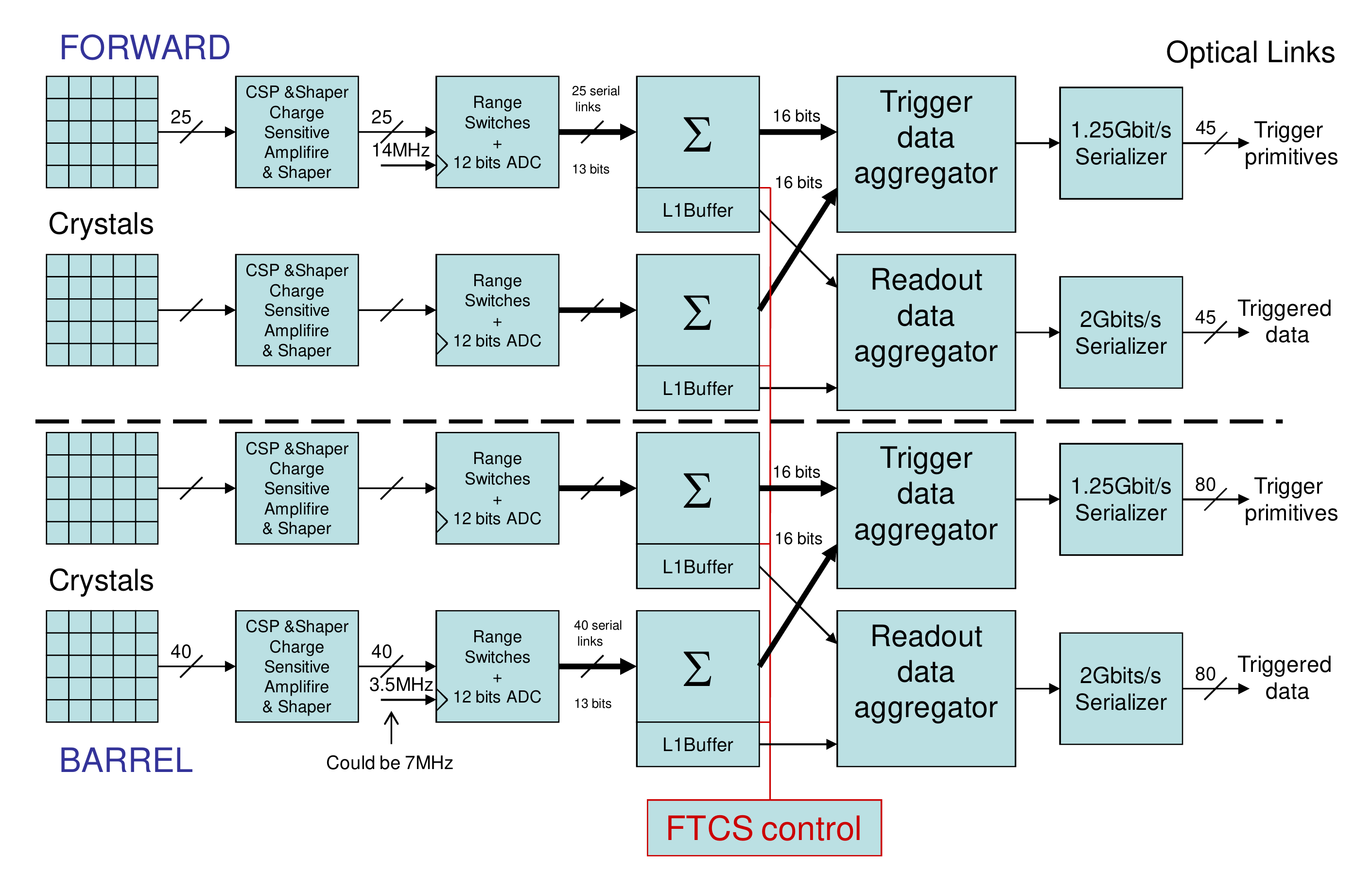}
\caption{EMC Electronics}
\label{fig:etd-emc}
\end{figure}

Two options have been considered for the EMC system design---a
\babar-like push architecture where all calorimeter data are sent over
synchronous optical 1\gbitps links to L1 latency buffers residing in
the trigger system, or a ``triggered'' pull architecture where the
trigger system receives only sums of crystals (via synchronous
1\gbitps links), and only events accepted by the trigger are sent to
the ROMs through standard 2\gbitsps optical links.

The triggered option, shown in Fig.~\ref{fig:etd-emc}, requires a much
smaller number of links and has been chosen as the baseline
implementation. The reasons for this choice and the implications are
discussed in more detail below.

To support the activated liquid-source calibration, where no central
trigger can be provided, both the barrel and the end-cap readout
systems need to support a free running ``self-triggered'' mode where
only samples with an actual pulse are sent to the ROM. Pulse detection
may require digital signal processing to suppress noisy channels.

\wpparagraph{Forward Calorimeter}

The 4500 crystals are read out with PIN or APD photodiodes. A charge
preamplifier translates the charge into voltage and the shaper uses a
100\ns shaping time to provide a pulse with a FWHM of 240\ns.

The shaped signal is amplified with two gains ($\times 1$ and $\times 64$).  
At the
end of the analog chain, an auto-range circuit decides which gain will
be digitized by a 12 bit pipeline ADC running at 14\MHz.  The 12 bits
of the ADC plus one bit for the range thus cover the full scale from
10\mev to 10\gev with a resolution better than 1\%.  A gain is set
during calibration using a programmable gain amplifier in order to
optimize the scale used during calibration with a neutron-activated
liquid-source system providing gamma photons around 6\mev.

Following the \babar\ detector design, a push architecture with a full
granularity readout scheme was first explored. In this approach, the
information from 4 channels is grouped, using copper serial links,
reaching an aggregate rate of 0.832\gbitsps per link to use up most of
the synchronous optical link's 1\gbitps bandwidth. A total of 1125
links are required. The main advantage of this architecture is the
flexibility of the trigger algorithm that can be implemented
off-detector using state of the art FPGAs without constraining their
radiation resistance. The main drawback is the large cost due to the
huge number of links.

The number of links can be reduced by summing channels together on the
detector side, and only sending the sums to the trigger. The natural
granularity of the forward detector is a module which is composed of
25 crystals. In this case, data coming from 25 crystals is summed
together, forming a word of 16 bits. Then the sums coming from 4
modules are aggregated together to produce a payload of 0.896\gbitsps.
In this case, the number of synchronous links toward the trigger is
only 45.  The same number of links would be sufficient to send the
full detector data with a 500\ns trigger window. This architecture
limits the trigger granularity, and implies more complex electronics
on the detector side, but reduces the number of links by a large
factor (from 1125 down to 90). However, it cannot be excluded that a
faster chipset will appear on the market which could significantly
reduce this implied benefit.

\wpparagraph{Barrel Calorimeter}

The EMC barrel reuses the 5760 crystals and PIN diodes from \babar,
with, however, the shaping time reduced from 1\mus to 500\ns and the
sampling rate doubled from 3.5\MHz to 7MHz. The same considerations
about serial links discussed above for the forward EMC apply to the
barrel EMC. If full granularity data were pushed synchronously to the
trigger, about 520 optical links would be necessary.

The number of synchronous trigger links can be drastically reduced by
performing sums of $4 \times 3$ cells on the detector side, so that 6 such
energy sums could be continuously transmitted through a single optical
serial link. This permits a reduction in the number of trigger links
so as to match the topology of the calorimeter electronics boxes,
which are split into 40 $\phi$ sectors on both sides of the detector.
Therefore, the total number of links would be 80 both for the trigger
and the data readout toward the ROMs, including a substantial safety
margin ($> 1.5$).

\wpsubsubsec{IFR Electronics:} 

\begin{figure}
\includegraphics[width=0.5\textwidth]{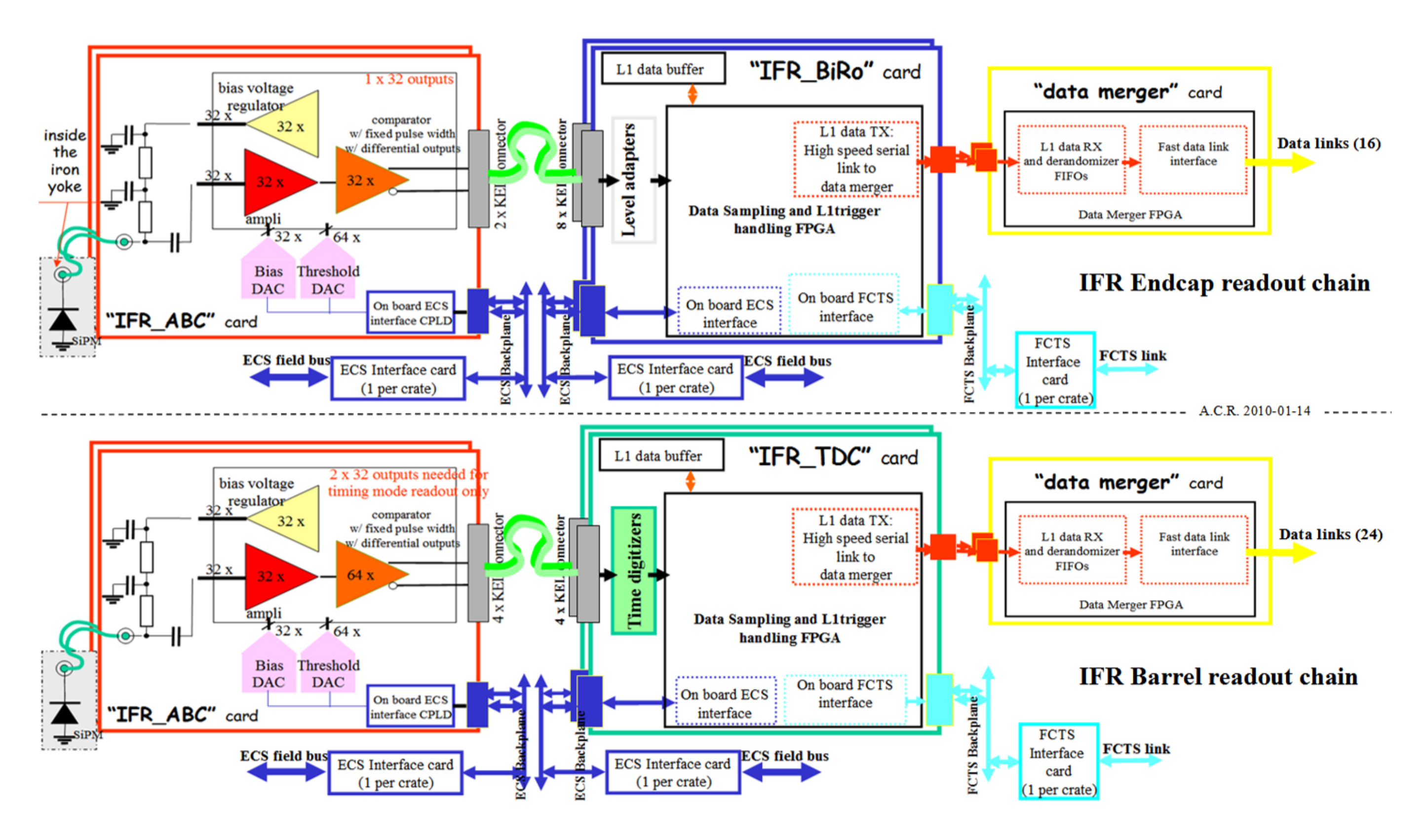}
\caption{IFR Electronics}
\label{fig:etd-ifr}
\end{figure}

The IFR is equipped with plastic scintillators coupled to wavelength
shifting fibres. Although different options have been explored, it is
currently assumed that single photon counting devices (SiPM) will be
located ``inside'' the iron, as close as possible to the scintillating
assemblies. Each SiPM will be biased and read out through a single
coaxial cable.

A schematic diagram of the IFR readout electronics is shown in
Fig.~\ref{fig:etd-ifr}.  The first stage of the readout chain is based
on the IFR\_ABC boards which provide (for 32 channels each):

\begin{itemize}
\item Amplification, presently based upon off-the-shelf components (COTS).
\item Individually programmable bias voltages for the SiPMs.
\item Comparators with individually programmable thresholds, presently
  based on COTS.
\end{itemize}

To minimize the length of the coaxial cables from the SiPMs to the
IFR\_ABC boards, these boards need to be placed as close to the iron
yoke as possible. The digital outputs of the IFR\_ABC boards will then
be processed in different ways for the IFR barrel and end-caps.

\wpparagraph{IFR Barrel} The barrel scintillation elements are mounted
parallel to the beam axis.  The time of arrival of pulses from both
ends of the scintillating elements must be recorded so that the
z-position of particle hits can be determined during reconstruction.
The signals are read out with IFR\_TDC 64-channel timing digitizer
boards.

The total TDC channel count estimate for the barrel is 14,400, which
comes from the 3600 scintillating assemblies in the barrel that are
read out at both ends with 2 comparators (with different thresholds)
per end to improve timing (and position) resolution.

\wpparagraph{IFR End-caps:} The signals from the scintillators in the
IFR end-caps (which are positioned vertically and horizontally) are
read out with IFR\_BiRO 128 channel ``Binary Readout'' boards, which
sample the status of the input lines and update a circular memory
buffer from which data are extracted upon trigger request.

The total channel count estimate for the end-caps is 9,600 BiRO
channels coming from the two end caps, each with 2,400 scintillating
assemblies in X, and 2,400 scintillating assemblies in Y read out into
a single comparator per channel.

\wpparagraph{}

The IFR\_TDC and IFR\_BiRO digitizers should be located as closely as
possible to the IFR\_ABC boards to minimize the cost of the
interconnecting cables, preferably in an area of low radiation flux.
In this case, commercial TDC ASICs could be used in the design.
Alternatively, radiation-tolerant TDCs could be used closer to the
detector. The FPGAs used in the digitizers should be protected against
radiation effects by architecture and by firmware design.

The output streams from the IFR\_TDC and IFR\_BiRO boards go through
custom ``data concentrators'' to merge the data coming from a number
of digitizers, and send the resulting output data to the ROMs via the
standard optical readout links.

In total, 225 IFR\_TDC boards (12 crates) and 75 IFR\_BiRO boards ( 4
crates) are needed. The total number of links to the ROMs is presently
estimated to be 24 for the barrel (2 links per digitizer crate), and
16 for the end-caps (4 links per digitizer crate).

To optimize the electronics topology, the number of ECS and FCTS links
should match the number of readout links.

\wpsubsec{R\&D}

For the overall EDT/Online system, substantial R\&D is needed to
better understand the global system requirements, develop solutions,
and probe the possible upgrade paths to handle luminosities of up to
$4\times10^{36}\ {\rm cm}^{-2} {\rm sec}^{-1}$ during the lifetime of
the experiment.

\wpparagraph{Data Links:}

The data links for \superb\ require R\&D in the following areas: (1)
studying jitter related issues and filtering by means of jitter
cleaners; (2) coding patterns for effective error detection and
correction: (3) radiation qualification of link components; and (4)
performance studies of the serializers/de-serializers embedded in the
new generation of FPGAs (Virtex6, Xilinx, etc.)

\wpparagraph{Readout Module:}

Readout Module R\&D includes investigation of 10\gbitsps Ethernet
technology, and detailed studies of the I/O sub-system on the ROM
boards.  The possibility of implementing the ROM functions in COTS
computers by developing suitable PCIe boards (such as optical link
boards for FCTS and FEE links, or personality cards to implement
sub-detector-specific functions) should also be investigated. 

\wpparagraph{Trigger:}

For the L1 trigger, the achievable minimum latency and physics
performance will need to be studied. The studies will need to address
many factors including (1) improved time resolution and trigger-level
granularity of the EMC and a faster DCH than \babar; (2) potential
inclusion of SVT information at L1; (3) the possibility of a L1 Bhabha
veto; (4) possibilities for handling pile-up and overlapping
(spatially and temporally) events at L1; and (5) opportunities created
by modern FPGAs to improve the trigger algorithms.

For the HLT, studies of achievable physics performance and rejection
rates need to be conducted, including the risks and benefits of a
possible L4 option.

\wpparagraph{ETD Performance and Dead Time:}

The design parameters for the ETD system are driven by trigger rates
and dead time constraints, and will need to be studied in detail to
determine the requirements for (1) trigger distribution through the
FCTS, (2) the FEE/CFEE buffer sizes, and (3) for handling pile-up and
overlapping events. Input from the L1 trigger R\&D and from background
simulation studies will be required.

\wpparagraph{Event Builder and  HLT Farm:}

The main R\&D topics for the Event Builder and HLT Farm are (1) the
applicability of existing tools and frameworks for constructing the
event builder; (2) the HLT farm framework; and (3), event building
protocols and how they map onto network hardware.

\wpparagraph{Software Infrastructure:}

To provide the most efficient use of resources, it is important to
investigate how much of the software infrastructure, frameworks and
code implementation can be shared with Offline computing. This
requires us to determine the level of reliability-engineering required
in such a shared approach. We also must develop frameworks to take
advantage of multi-core CPUs.

\wpsubsec{Conclusions} 

The architecture of the ETD system for \superb\ is optimized for
simplicity and reliability at the lowest possible cost. It builds on
substantial in-depth experience with the \babar\ experiment, as well
as more recent developments derived from building and commissioning
the LHC experiments.  The proposed system is simple and safe. Trigger
and data readout are fully synchronous---allowing them to be easily
understood and commissioned.  Safety margins are specifically included
in all designs to deal with uncertainties in backgrounds and radiation
levels. Event readout and event building are centrally supervised by a
FCTS system which continuously collects all the information necessary
to optimize the trigger rate.  The hardware trigger design philosophy
is similar to that of \babar\ but with better efficiency and smaller
latency.  The event size remains modest.

The Online design philosophy is similar---leveraging existing
experience, technology, and toolkits developed by \babar, the LHC
experiments, and commercial off-the-shelf computing and networking
components---leading to a simple and operationally efficient system to
serve the needs of \superb\ factory-mode data taking.

\aftsec

\begsec
\graphicspath{{COMP/}{COMP/}}
\wpsec{Software and Computing}\label{sec:computing}

The computing models of the \babar\ and Belle experiments have proven to be quite successful for
a flavor factory in the ${\cal L}= 10^{34}\,\rm{cm}^{-2}\rm{s}^{-1}$
luminosity regime.
A similar computing model can also work for a super flavor factory at
a luminosity of ${\cal L}= 10^{36}\,\rm{cm}^{-2}\rm{s}^{-1}$. Data volumes will be much larger, comparable in fact to those expected for the first running periods of the ATLAS and CMS experiments at the LHC, but
predictable progress in the computing industry will provide much of the performance increase needed to cope with them.   In addition, effective exploitation of computing resources on the Grid, that has become well established in the LHC era, will enable \superb\ to access a much larger set of resources than were available to \babar.

To illustrate the scale of the computing problem and how the \superb\ computing group envisage to attack it, the first part of this section contains an overview of the current \babar-inspired computing baseline model with an estimate of the extrapolated \superb\ computing requirements, followed by a description of the current development and implementation timeline.
In the current view, the design phase of the \superb\ computing model is planned to start with a dedicated R\&D program in the first year of the project and to finish with the completion of the Computing TDR by the end of the second year.

So far, the main effort of the computing group has been devoted to the development and the support of the simulation software tools and the computing production infrastructure needed for carrying out the detector design and performance evaluation studies for the Detector TDR.  Quite sophisticated and extended detector and physics studies can now be performed thanks to:
\begin{itemize}
\item
the development of a detailed Geant4-based Monte Carlo simulation (Bruno) and of a much faster parametric fast simulation (\fastsim) which can directly leverage the existing \babar\ analysis code base;
\item
the implementation of a production system for managing very large productions that can parasitically exploit the computing resources available on the European and US Grids.
\end{itemize}

A description of the tools made available and their capabilities is reported in the second part of the section.

\wpsubsec{The \superb\  baseline model}

The data processing strategy for \superb\  is envisaged to be similar to the one employed in \babar\ and can be summarized as follows.

The ``raw data'' coming from the
detector are permanently stored, and reconstructed in a two step process:
\begin{itemize}
\item a ``prompt calibration'' pass performed on a subset of the events
to determine various calibration constants.
\item  a full ``event reconstruction'' pass on all the events that uses the
constants derived in the previous step.
\end{itemize}
Reconstructed data are also permanently stored and data quality is monitored at
each step of the process.

A comparable amount of Monte Carlo simulated data is also produced in parallel and processed in the same way.

In addition to the physics triggers, the data acquisition also records
random triggers that are used to create ``background frames''.
Monte Carlo simulated data, incorporating the calibration constants and the
background frames on a run-by-run basis, are prepared.

Reconstructed data, both from the detector and from the simulation,
are stored in two different formats:
\begin{itemize}
 \item the Mini, that contains reconstructed tracks and energy clusters in the
calorimeters as well as detector information. It is a relatively
compact format, through noise suppression and efficient packing of data.
\item the Micro, that contains only information essential for physics analysis.
\end{itemize}

Detector and simulated data are made available for physics analysis in a
convenient form through the process of ``skimming''. This involves the
production of selected subsets of
the data, the ``skims'', designed for different areas of analysis.
Skims are very convenient for physics analysis, but they increase the storage
requirement because the same events can be present in more than one skim.

From time to time, as improvements in constants, reconstruction code, or
simulation are implemented, the data may be ``reprocessed'' or new
simulated data generated. If a set of new skims become available, an
additional skim cycle can be run on all the reconstructed events.

\wpsubsubsec{The requirements}
The \superb\  computing requirements can be estimated using as a basis the present experience with \babar\ and applying a scaling of about two orders of magnitude.
Fortunately, much of this scaling exercise is quite straightforward.

As a baseline, all rates are simply scaled linearly with
luminosity.  Only a few parameters have been modified to keep into account  improved  efficiency of utilization of the computing resources that are likely to be obtained with \superb, \ie:
\begin{itemize}
\item the skimmed data storage requirements have been reduced (by $\sim 40\%$), assuming a more aggressive use of event indexing techniques;
\item the CPU requirements for physics analysis are reduced by a factor of two as a result of more stringent optimization goals that can be achieved in \superb;
\item the duration of the reprocessing and simulation re-generation cycle, expected to take place once significant improvements of the reconstruction code physics performance have been obtained, has been set to two years instead of one, as it was in \babar, in view of the larger expected cost-to-benefit ratio;
\end{itemize}

The resulting CPU and storage requirements are shown in Table~\ref{tab:compreq}
for a typical year of data taking at nominal luminosity, assuming an integrated luminosity of 50\invab reached at the end of the same year.

\begin{table}[htb]
\caption{\label{tab:compreq} Summary of computing resources needed in a typical year of \superb\  data taking at nominal luminosity, assuming an integrated luminosity of 50\invab has been collected.}
\begin{center}
\begin{tabular}{lc}
\hline
\hline
Parameter & typical Year \\
\hline
Luminosity (ab$^{-1}$) &  15  \\
Storage (PB) &\\
\quad Tape & 113  \\
\quad Disk & 52 \\
CPU (KHep-Spec06) &\\
\quad Event data reconstruction &  210 \\
\quad Skimming            &  250  \\
\quad Monte Carlo         & 670 \\
\quad Physics analysis    &  570 \\
\quad Total               &   1700 \\
\hline
\end{tabular}
\end{center}
\end{table}

The total computing resources needed for one year of data taking at nominal luminosity are of
the same order as the corresponding figures estimated, in the spring of 2010, by the Atlas and CMS experiments for the
2011 running period, which amounted to
580\,KHep-Spec06 for total CPU, 60\,PB for disk space, and 47\,PB for tape space.   However \superb\  will profit from the technological advances that will take
place over a period of approximately 10 years,  and make extensive
use of distributed computing resources accessible via the Grid infrastructures. This will give an important degree of 
flexibility in providing the required level of computing resources.

\wpsubsubsec{\superb\  offline computing development}
The bulk of the \superb\ software development effort is foreseen to take place after the Computing TDR is released.
All major design choices should at that time be made, based to a large extent on the results of the R\&D activities previously carried out.

In an estimated two years a preliminary version of a fully-functional offline system can be built and
validated via dedicated data challenges, so
that the collaboration can start using it for detector and physics simulation studies in the fourth project year.

Through further extensive test and development cycles the system will be brought to its full scale size
in the following couple of years,  before the \superb\  collider is turned on.
Acquisition and deployment of dedicated computing resources will also be carried out during that period as well as
consolidation and validation of the distributed computing infrastructure that \superb\  will have to count on.

This timeline is comparable with the time needed to develop and deploy
the \babar\ offline system.

\wpsubsec{Computing tools and services for the Detector and Physics TDR studies}

\wpsubsubsec{Fast simulation}

Because the \superb\  detector and its machine environment will differ substantially from those at \babar, simple extrapolations of \babar\ measurements are not adequate to estimate the physics reach of the experiment.  Additionally, to make optimal choices in the \superb\  detector design, an understanding of the effect of design options on the final result of critical physics analyses is needed.  However, a detailed simulation of the full \superb\  detector, with its various options, carried out to the level of statistical precision needed for a relevant physics result, is beyond the capability of the current \superb\  computing infrastructure.

To address these needs, a fast simulation (\fastsim) program has been developed.  \fastsim relies on simplified models of the detector geometry, materials, response, and reconstruction to achieve an event generation rate several orders of magnitude faster than is possible with a Geant4-based detailed simulation, 
but with sufficient detail to allow realistic physics analyses. 
In order to produce more reliable results, \fastsim\ incorporates the effects of expected machine and detector backgrounds.
\fastsim\ is easily configurable, allowing different detector options to be selected at runtime, and is compatible with the \babar\ analysis framework, allowing sophisticated analyses to be performed with minimal software development.

\wpparagraph{Event generation}

Since \fastsim is compatible with the \babar\ analysis framework, we can exploit the same event generation tools used by \babar.  On-peak events ($e^+e^-\to\Upsilon(4S)\to B\bar{B}$), with the subsequent decays of the $B$ and $\bar B$ mesons, are generated through the EvtGen package~\cite{ref:evtgen}.  EvtGen also has an interface to JETSET for the generation of continuum $e^+e^-\to q\bar q$ events ($q=u,d,s,c$), and for the generic hadronic decays that are not explicitly defined in EvtGen. The \superb\ machine design includes the ability to operate with a $60-70\%$ longitudinally polarized electron beam, which is especially relevant for tau physics studies.
We generate $e^+e^-\to\tau^+\tau^-$ events with polarized beams using the KK generator and Tauola~\cite{ref:kkgen}.  Other important physics processes can be generated, such as Bhabha and radiative Bhabha scattering, and $e^+e^-\to e^+e^-e^+e^-$ or $e^+e^-\to\gamma\gamma$.  More details can be found at the end of this section where the simulation of machine backgrounds is described.

\wpparagraph{Detector description}

\fastsim\ models \superb\  as a collection of {\em detector elements} that represent medium-scale pieces of the detector.  The overall detector geometry is assumed to be cylindrical about the solenoid $\vec{B}$ axis, which simplifies the particle navigation.  Individual detector elements are described as sections of two-dimensional surfaces such as cylinders, cones, and planes, where the effect of physical thickness is modeled parametrically.  Thus a barrel layer of Si sensors is modeled as a single cylindrical element.  Intrinsically thick elements (such as the calorimeter crystals) are modeled by layering several elements and summing their effects.  Gaps and overlaps between the real detector pieces within an element (such as staves of a barrel Si detector) are modeled statistically.

The density, radiation length, interaction length, and other physical properties of common materials are described in a simple database.  Composite materials are modeled as admixtures of simpler materials. A detector element may be assigned to be composed of any material, or none.

Sensitive components are modeled by optionally adding a {\em measurement type} to an element.  Measurement types describing Si strip and pixel sensors, drift wire planes, absorption and sampling calorimeters, Cherenkov light radiators, scintillators, and TOF are available.  Specific instances of measurement types with different properties (resolutions) can co-exist.  Any measurement type instance can be assigned to any detector element, or set of elements.  Measurement types also define the time-sensitive window, which is used in the background modeling described below.

The geometry and properties of the detector elements and their associated measurement types are defined through a set of XML files using the EDML (Experimental Data Markup Language) schema, invented for \superb.

\wpparagraph{Interaction of particles with matter}

The \superb\  \fastsim\ models particle interactions using parametric functions.  Coulomb scattering and ionization energy loss are modeled using the standard parametrization in terms of radiation length and particle momentum and velocity.  Moli\`ere and Landau tails are modeled.  Bremsstrahlung and pair production are modeled using simplified cross-sections.  Discrete hadronic interactions are modeled using simplified cross-sections extracted from a study of Geant4 output.  Electromagnetic showering is modeled using an exponentially-damped power law longitudinal profile and a Gaussian transverse profile, which includes the logarithmic energy dependence and electron-photon differences of shower-max.  Hadronic showering is modeled with a simple exponentially-damped longitudinal profile tuned using Geant4 output.

Unstable particles are allowed to decay during their traversal of the detector.  Decay rates and modes are simulated using the \babar\ EvtGen code and parameters.

\wpparagraph{Detector response}

All measurement types for the detector technologies relevant to \superb\  are implemented.  Tracking measurements are described in terms of the single-hit and two-hit resolution, and the efficiency. Si strip and pixel detectors are modeled as two independent orthogonal projections, with the efficiency being uncorrelated (correlated) for strips (pixels) respectively.  Wire chamber planes are defined as a single projection with the measurement direction oriented at an angle, allowing stereo and axial layers.  Ionization measurements (\dedx) used in particle identification are modeled using a Bethe-Bloch parametrization.

The Calorimeter response is modeled in terms of the intrinsic energy resolution of clusters as a function of the incident particle energy.  Energy deposits are distributed across a grid representing the crystal or pad segmentation.

Cherenkov rings are simulated using a lookup table to define the number of photons generated based on the properties of the charged particle when it hits the radiator.  Timing detectors are modeled based on their intrinsic resolution.

\wpparagraph{Reconstruction}

A full reconstruction based on pattern-recognition is beyond the scope of \fastsim.  However, a simple smearing of particle properties is insensitive to important effects like backgrounds.  As a compromise, \fastsim\ reconstructs high-level detector objects (tracks and clusters) from simulated low-level detector objects (hits and energy deposits), using the simulation truth to associate detector objects.  Pattern recognition errors are introduced by perturbing the truth-based association, using models based on observed \babar\ pattern recognition algorithm performance.

In tracking, hits from different particles within the two-hit resolution of a device are merged, the resolution degraded, and the resultant merged hit is assigned randomly to one particle.  Hits overlapping within a region of `potential pattern recognition confusion', defined by the particle momentum, are statistically mis-assigned, based on their proximity.  The final set of hits associated to a given charged particle are then passed to the \babar\ Kalman filter track fitting algorithm to obtain reconstructed track parameters at the origin and the outer detector.  Outlier hits are pruned during the fitting, based on their contribution to the fit $\chi^2$, as in \babar.

Ionization measurements from the charged particle hits associated to a track are combined using a truncated-mean algorithm, separately for the SVT and DCH hits.  The truncated mean and its estimated error are later used in particle identification (PID) algorithms.  The measured Cherenkov angle from the DIRC is smeared according to the Kalman filter track fit covariance at the radiator.

In the calorimeter, overlapping signals from different particles are summed across the grid.  A simple cluster-finding based on a local maxima search is run on the grid of calorimeter response.  The energies deposited in the cluster cells are used to define the reconstructed cluster parameters (cluster energy and position).   A simple track-cluster matching based on proximity of the cluster position to a reconstructed track is used to distinguish charged and neutral clusters.

\wpparagraph{Machine backgrounds}

Machine backgrounds at \superb\  are assumed to be dominated by luminosity-based sources, as the \superb\  beam currents will not be much higher than at \babar, which was mostly affected by luminosity-based background.  The two dominant processes are radiative Bhabhas and QED two-photon processes ( $e^+e^-\to e^+e^-e^+e^-$).  Since the bunch spacing (a few \ns) is short relative to the time-sensitive window of most of the \superb\  detectors, interactions from a wide range of bunch crossings must be considered as potential background sources.

Understanding the effect of backgrounds on physics analyses is crucial when making detector design choices, such as the tradeoffs between spatial versus timing resolution, and for understanding the physics algorithms required to operate at ${\cal L} = 10^{36}\  \rm{cm}^{-2} \rm{s}^{-1}$.  Background effects on electronics (hit pileup) and sensors (saturation or radiation damage) are also crucial for \superb, but are best studied using the full simulation and other tools.

Background events are generated in dedicated \fastsim\ or Bruno runs.  Bruno is needed to model the effect of backgrounds coming from small-angle radiative Bhabha showers in the machine elements, as a detailed description of these elements and the processes involved are beyond the scope of \fastsim.  Large-angle radiative Bhabhas, and two-photon events, where the primary particles directly generate the background signals, are generated using \fastsim.  The same \babar\ generators are used in \fastsim\ and Bruno.

Background events are stored as lists of the generated particles, which are then
filtered to save only those which enter the sensitive detector volume.  For both the low-angle radiative Bhabha events and the two-photon events, the generated events are combined to correspond to the luminosity of a single bunch crossing at nominal machine parameters, with the actual number of combined events obtained by sampling the appropriate Poisson distribution.

Background events from all sources are overlaid on top of each generated physics event during \fastsim simulation.  The time origin of each background event is assigned randomly across a global window of 0.5\mus (the physics event time origin is defined to be zero).  Background events are sampled according to a Poisson distribution whose mean is the effective cross-section of the background process times the global time window.

Particles from background events are simulated exactly as those from the physics event, except that the response they generate in a sensitive element is modulated by their different time origin.  In general, background particle interactions outside the time-sensitive window of a measurement type do not generate any signal, while those inside the time-sensitive window generate nominal signals.  Background particle calorimeter response is modeled based on waveform analysis, resulting in exponentially-decaying signals before the time-sensitive window, and nominal signals inside.  The hit-merging, pattern recognition confusion and cluster merging described earlier are also applied to background particle signals, so that fake rates and resolution degradation can be estimated from the \fastsim output.  A mapping between reconstructed objects and particles is kept, allowing analysts to distinguish background effects from other effects.

\wpparagraph{Analysis tools}

Because \fastsim is compatible with the \babar\ analysis framework, existing \babar\ analyses can be run in \fastsim with minimal modification.  For instance, the vertexing tools and combinatorics engines used in \babar\ work also in \fastsim.

The primary difference is that only a subset of the lists of identified particles (PID lists) available in \babar\ are available in \fastsim.  The majority of the available PID lists are based on tables of purities and fake rates extracted from \babar, extended to the additional coverage of \superb.  A few PID lists based on the actual behavior of the simulated \superb detector systems (like \dedx) are available, but these are of limited utility given the lack of precise calibration or the use of sophisticated statistical techniques like neural nets used in \babar\ PID lists.

The lack of PID lists also means that the `tagging' ($B$ vs.~$\bar{B}$ meson identification) used in \babar~does not function in \fastsim\ at present.  New tagging algorithms based on  the \superb\  detector capabilities, such as the improved transverse impact parameter resolution, have not yet been developed.

The standard tool used in \babar\ to store analysis information into a ROOT tuple has been adapted to work in \fastsim,
allowing large analyses to be run in \fastsim approximately as in \babar, and to allow the use of \babar\ analysis macros.

A full mapping of analysis objects back to the particles which generated them (including background particles) is provided in \fastsim, along with the full particle genealogy.

\wpsubsubsec{Bruno: the \superb full simulation tool}

The availability of reliable tools for full simulation is crucial in the present phase of the design of both the accelerator and the detector.
For example, the background rate at the sub-detectors needs to be carefully assessed for each modification in the accelerator design and, for a given background scenario, the sub-detectors' design must be optimized.
The  full simulation tool can be used to improve the results of the \fastsim in some particular cases, as discussed in the following. The choice was made to re-write from scratch the core simulation software, aiming at having more freedom to better profit from both the \babar\ legacy and the experience gained in the development of the full simulation for the LHC experiments.
Geant4 and the  C++ programming language were therefore the natural choices as underlying technology.
While the implementation is still at a very early stage, the  software is already usable. Basic functionality is in place, and more is being added following user requests. There follows a short overview of the main characteristics, emphasizing areas where future development is planned.

\wpparagraph{Geometry description}
The need to re-use as much as possible the existing geometrical description of the \babar\ full simulation called
for some application-independent interchange format to store the information concerning the geometry and materials of the sub-detectors. Among the formats currently used in High Energy Physics applications, the Geometry Description Markup Language (GDML) was chosen because of the availability of native interfaces in Geant4 and ROOT, and the easiness of human inspection and editing provided by the XML-based structure.

\wpparagraph{Simulation input: Event generators}
The event generator code can be run either inside the Bruno executable or as a separate process.
In the latter case the results are saved in a intermediate file, which is then read by the full simulation job.
Bruno currently supports two interchange formats: a plain text file and one in ROOT format.

\wpparagraph{Simulation output}
Hits from the different sub-detectors, which represent the simulated event as seen from the detector, are saved in the output (ROOT) file for further processing. Also the Monte Carlo Truth (MCTruth), intended as a summary of the event as seen by the simulation engine itself, is saved
and can be exploited in Bruno in several useful ways, for instance to estimate
particle fluxes at sub-detector boundaries by means of full snapshots taken at different scoring volumes.

In staged simulations, snapshots of particles taken at a specific sub-detector boundaries can be saved,
read back, and used to start a new
simulation process without the need of retracking particles through sub-detectors that sit at inner positions.
This allows sub-detectors to assess the effects of layout and geometry changes without the need to
run large, computationally-heavy production jobs involving the entire detector.

\wpparagraph{Interplay with \fastsim}
The event snapshot at a specific sub-detector boundary can also be read by \fastsim, allowing a very powerful hybrid
simulation approach.
For instance the design of the interaction region, which strongly influences the background rates
in the detector, cannot be described with the required level of detail in  \fastsim,
while full simulation is not fast enough to generate the high statistics needed for physics studies.
By using Bruno to simulate background events up to and including the interaction region, and
saving a snapshot of the event without running the entire simulation, one obtains
a set of \textit{background frames}, which can be read back in  \fastsim, that then propagates  particles through the simplified detector geometry and adds the resulting hits to the ones coming from signal events.

Another aspect where the interplay between fast and full simulation is needed is the evaluation of the neutron background. The concept is to have Bruno, in addition to handling all particle interactions within the interaction
region, as explained above, also track neutrons  in the whole detector until they interact or decay, saving the products
as part of the background frame used by \fastsim.
All these functionalities are currently implemented, and have been used in recent production runs.

\wpsubsubsec{The distributed production environment}

To design the detector and to extract statistically significant results from the data analysis, a huge number of Monte Carlo simulated events is needed.
Such a production is way beyond the capacity of a single computing farm so it was decided to design, even to support the detector TDR studies, 
a distributed  model capable of fully exploiting the existing HEP world wide Grid computing infrastructure~\cite{ref:egee,ref:osg,ref:nordugrid,ref:westgrid,ref:lcg-tdr}.

The LHC Computing Grid (LCG) architecture~\cite{ref:lcg} was adopted to provide the minimum set of services and applications
upon which the \superb\  distributed model could be built, and the INFN Tier1 site located at CNAF (Bologna) was
chosen as the central site where job submission management, the Bookkeeping Data Base, and the data repository would reside.

Jobs submitted to remote sites transfer their output back to a central repository and update the Bookkeeping Data Base
containing all metadata related to the production input and output files.
The system uses standard Grid services such as WMS, VOMS, LFC, StoRM, GANGA~\cite{ref:wms,ref:voms,ref:lfc,ref:srm,ref:storm,ref:ganga}.

The distributed computing infrastructure, as of January 2010, includes several sites in Europe and North America as reported in Table~\ref{tab:gridsites}. Each site implements a
Grid flavor depending on its own affiliation and geographical position.
The EGEE Workload Manager System (WMS) allows a job's progress through the different Grid middleware flavors to be managed transparently.

\begin{table}[!t]
\caption{List of sites and Grid technologies involved in \superb\ distributed computing model as of January 2010}
\label{tab:gridsites}
\label{table_siti}
\centering
\small
\begin{tabular}{lcc}
\hline
\hline
\textbf{Site name} & \textbf{Grid flavor}\\
\hline
CNAF Tier1, Bologna, Italy & EGEE/gLite\\
Caltech,  California, USA & OSG/Condor\\
SLAC,  California, USA & OSG/Condor\\
Queen Mary, London, UK & EGEE/gLite\\
RALPP, Manchester, UK & EGEE/gLite\\
GRIF, Paris/Orsay, France & EGEE/gLite\\
IN2P3, Lyon, France & EGEE/gLite\\
INFN-LNL, Legnaro, Italy & EGEE/gLite\\
INFN-Pisa, Pisa, Italy & EGEE/gLite\\
INFN-Bari, Bari, Italy & EGEE/gLite\\
INFN-Napoli, Napoli, Italy & EGEE/gLite\\
\hline
\end{tabular}
\end{table}

\wpparagraph{Simulation production work-flow \label{sectsim}}

The production of simulated events is performed in three main phases:
\begin{enumerate}
\item Distribution of input data files to remote site Storage Elements (SE). Jobs running on each site are able to access input files from local SE avoiding a file transfer on the Wide Area Network.
\item Job submission, via the \superb\ GANGA interface at CNAF, to all available enabled remote sites.
\item Job stage out of  files to the CNAF repository.
\end{enumerate}

\begin{figure}[!t]
\centering
\includegraphics[width=0.48\textwidth]{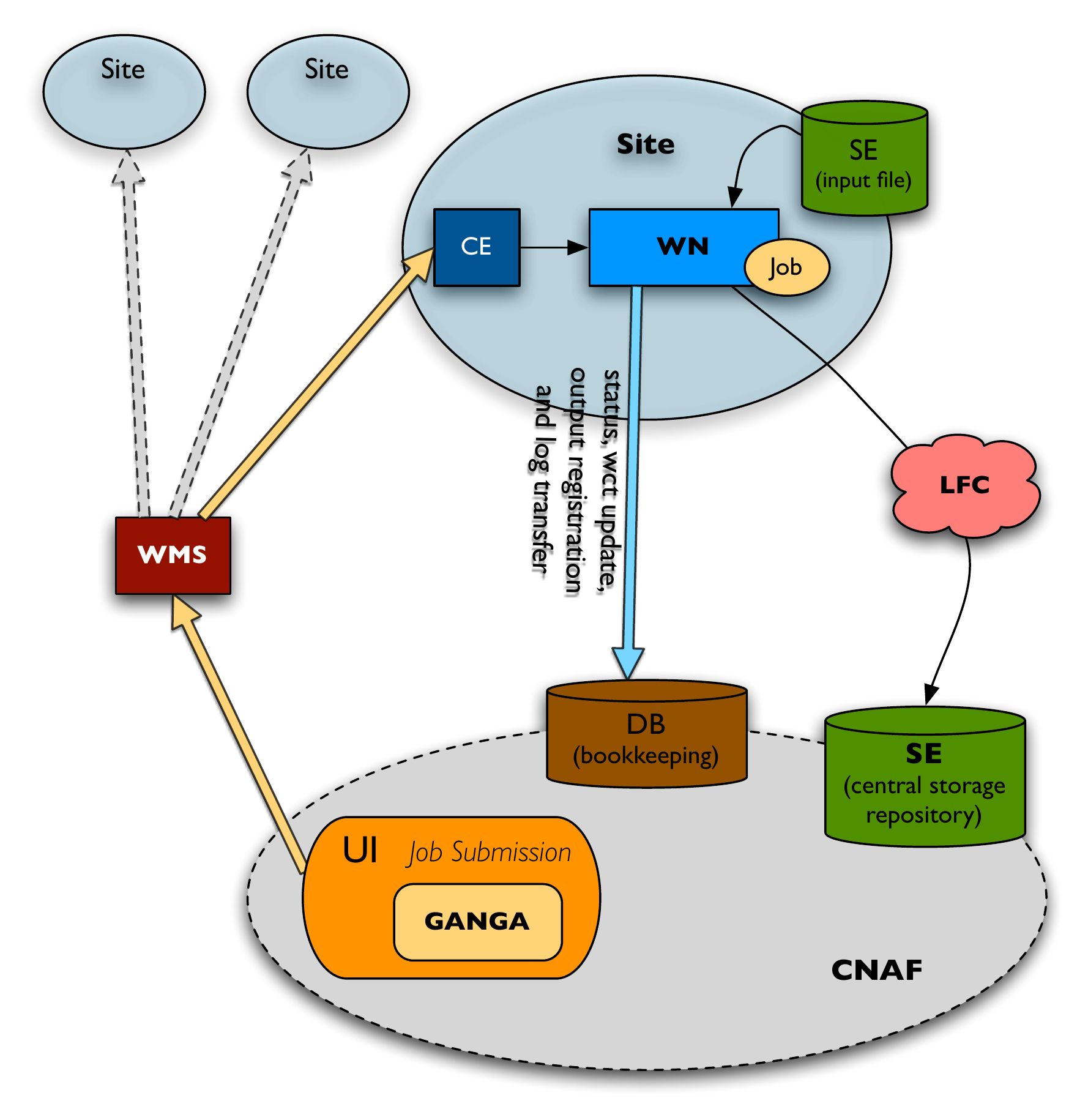}
\caption{Simulation production work-flow}
\label{fig:dist_wf}
\end{figure}

The job work-flow, shown in Fig.~\ref{fig:dist_wf}, includes also procedures for correctness check, monitoring, data handling and bookkeeping metadata communication.
Reliability and fail-over conditions have been implemented in order to maximize the efficiency for copying the output files to the CNAF repository.
A replication mechanism permits to store the job output to the local site SE.
The job input data management includes an pre-production step: the test release and background files are transferred to all involved sites
for access by the jobs at run time.
The job submission procedure includes a per site customization to adapt the job actions to site peculiarities:
\eg, file transfer to and from three
different data handling systems: StoRM~\cite{ref:storm}, dCache~\cite{ref:dcache} and DPM~\cite{ref:dpm}.

\wpparagraph{Production Tools}

Both the job submission system and the individual physicist require a way to identify interesting data files and to locate the storage holding them.
To make this possible, the experiment needs a data bookkeeping system to maintain the semantic information associated to data files and keep
track of the relation between executed jobs and their parameters and outputs.

Moreover, a semi-automatic submission procedure is needed in order to keep data consistent, to speed up production completion and to provide an
easy-to-use interface for non-expert users. To accomplish this task, a web-based user interface has been developed which takes care of the
database interactions and the job preparation; it also provides basic monitoring functionalities.

The bookkeeping database was modeled according to the requirements specified by the collaboration, and implemented adhering to the relational model with MySQL rDBMS. It was extensively tested against the most common use cases and provides a central repository of the production metadata.

A Web-based User Interface (WebUI) has been developed 
that is bound to the bookkeeping database which provides inputs for the job preparation and monitoring.

It presents two different sections for Full Simulation and  Fast Simulation,
each of which is divided into submission and monitor subsections. A basic authentication and authorization layer, based on the collaborative tools
permits the differentiation of users
and grants access to the corresponding sections of the site.

A typical production workflow consists of an initialization phase, during which the data of a bunch (or several bunches) of jobs is inserted into the database, and of the subsequent phase of submission either to a batch system or to a distributed environment. The simulation jobs interact extensively  with the database during their lifetime to update data and insert outputs and logs.

A production software layer and a database manager layer have thus been developed to interface the database with the jobs. The prototype service uses a RESTful~\cite{ref:rest} interface in order to allow the communications between centralized or distributed jobs and a centralized database. 

The WebUI provides basic monitoring features by querying the bookkeeping database.
The user can retrieve the list of jobs as a function of run number range, generator, geometry, physics list, site, status, etc. The monitor provides, for each job, the list of output files -- if any -- and direct access to the log files.
Reports on output file size, execution time, site loading, job spread over channels, and the list of the most recently completed jobs (successfully or with failures) are also provided.

\wpparagraph{Remote Grid sites}
Sites involved in the \superb\ distributed computing infrastructure need to enable
the superbvo.org Virtual Organization (VO) and install the \superb\ FastSim software
as specified in the VO Identity Card.
Currently no permanent storage is required at the sites, and memory
requirements are modest (2\gbyte RAM per core and 2\gbyte
virtual memory).

\wpparagraph{First 2010 production runs}

The production system has been used during the first large scale Bruno and \fastsim production
in January, February and March 2010.
During this production phase, over 1.7 billion simulation events, equivalent to $\sim 0.2 \invab$,
were produced using the distributed computing environment.

The Bruno production was divided in two categories: simulation of machine background frames for \fastsim
(bgframes) and full simulation for machine background studies (bgstudies).
The entire Bruno production ran at CNAF in about 12 days. Table~\ref{tab:fullfull} shows the production summary.

\begin{table}[htd]
\caption{Full Simulation Production Summary}
\begin{center}
\small
\textbf{2010\_01\_bgframes}
\vskip 0.1cm
\begin{tabular}{llrr}
\hline
\hline
Geometry &  Status & Jobs & Events \\
\hline
\superb & done & 4000 & $10^6$ \\
\superb &  failed & 1 & $250$ \\
\superb &  sys-failed & 1 & $250$ \\
\hline
\end{tabular}

\vskip 0.3cm
\textbf{2010\_02\_bgstudies}
\vskip 0.1cm
\begin{tabular}{llrr}
\hline
\hline
Geometry & Status & Jobs & Events \\
\hline
shielded &  done & 4840 & $604000$ \\
shielded &  sys-failed & 1 & $100$ \\
unshielded &  done & 785 & $196250$ \\
unshielded &  failed & 2 & $500$ \\
unshielded &  sys-failed & 13 & $3250$ \\
\hline
\end{tabular}

\end{center}
\label{tab:fullfull}
\end{table}%

The \fastsim production was divided into two categories: simulation of generic events (\BzBzb, \BpBm, \ccbar and $uds$); 
and simulation of specific decay channels (signal mode).

For generics, four generators
and three geometry configurations
were used. Events with
 and without background mixing were produced.
In the signal mode production, four physics channels events were simulated
with background mixing.
Results are summarized in Table~\ref{tab:fast}.

\begin{table}[htd]
\caption{Fast Simulation Production Summary}
\label{tab:fast}
\begin{center}
\small
\textbf{2010\_February\_Generics}
\vskip 0.1cm
\begin{tabular}{crr}
\hline
\hline
 Bkg & Jobs & Events \\
\hline

 Y & 4508 & $104.340 \times10^6$ \\
 N & 14672 & $1472.100 \times 10^6$\\
\hline
 Total  & 19180 & $\approx1.58 \times 10^9$\\
\hline
\end{tabular}

\vskip 0.3cm
\textbf{2010\_February\_Signal}
\vskip 0.1cm
\begin{tabular}{lrr}
\hline
\hline
Signal &  Jobs & Events \\
\hline
BtoTauNu & 30 & $3 \times 10^6$\\
BtoKNuNu & 60 & $6 \times 10^6$\\
BtoKstarNuNu & 60 & $6 \times 10^6$\\
BtoKplusNuNu & 688 & $68.8 \times 10^6$\\
\hline
Total & 838 & $83.8 \times 10^6$\\
\hline
\end{tabular}

\end{center}
\label{tab:fast}
\end{table}%

\begin{figure}[!tbh]
\centering
\includegraphics[width=\linewidth]{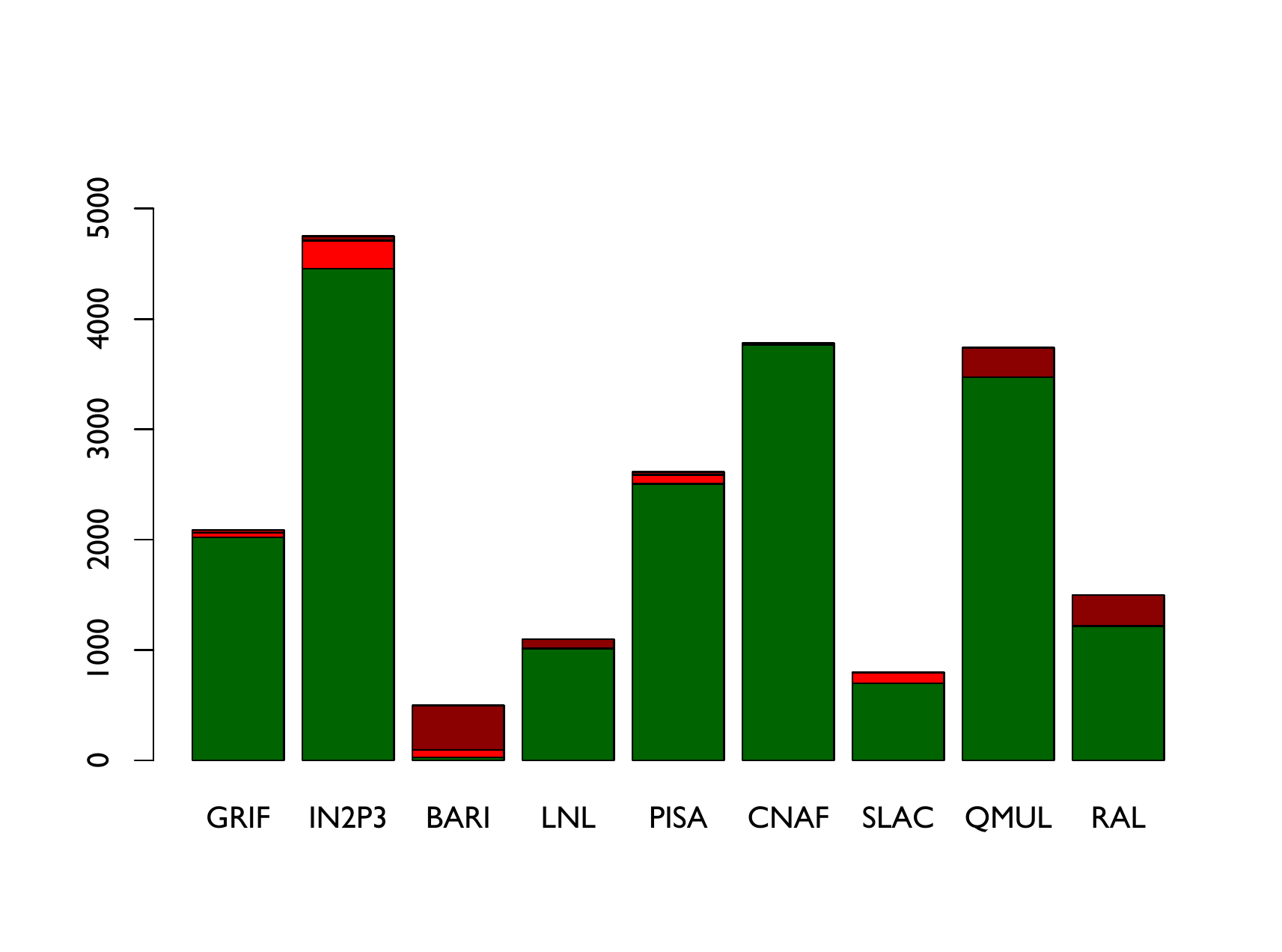}
\caption{Jobs submitted to remote sites for the Generics Production.}
\label{fig_site}
\end{figure}

\fastsim production involved nine remote sites; about $82\%$ of submitted jobs used the Grid infrastructure,
exploiting remote resources, as illustrated in Fig.~\ref{fig_site} for the generics production.

Over a period of 2 weeks, approximately 20000 jobs were completed with
an average $\sim8\%$ failure rate, mainly due to site misconfigurations
(2.6\%), proxy expiration (2.0\%), and temporary overloading of the
machine used to receive the data transfers from the remote sites at
CNAF (3\%). The peak rate reached 3200 simultaneous jobs with an
average of 500.

\wpsubsubsec{The software development and collaborative tools}

As a collaboration facing the task of preparing design documents and developing
software code, the \superb\  group needs to be supported by a set of suitable
computing tools in carrying out its day-by-day coordinated activities.
A description of currently available tools is presented here.

\wpparagraph{Directory Service}
To support the access to the collaborative tools through a unique authentication
and authorization interface, a directory service based on LDAP application protocol
was made available and it has been now actively used for more than one year.
A web interface to provide an easy way to manage the directory tree
has also been set up by modifying open source software and is ready to enter the testing
phase. In choosing all the other tools we put a great emphasis on the ability
to integrate with the directory service.

\wpparagraph{Web site}
To make the inclusion of information content easier for the collaborators
it was decided to create a web
site managed by a content management system (CMS).
After some testing phase and initial experience with Drupal, the Joomla
open source CMS~\cite{ref:joomla} was selected due to its  user-friendly interface and widespread use
in INFN. Two web sites were created, the first addressing the needs of the \superb\ group
and the second dedicated to the general public.

\wpparagraph{Wiki}
In addition to the web sites, the collaboration has realized from the
beginning the need for a Wiki site that permits the easy creation of web pages to be used as internal documentation, 
to create meaningful topic associations between different pages, and 
simple editing of existing pages, keeping track of recent changes.
The Wiki server~\cite{ref:wiki} is currently up and running,
integrated with the directory service and is used by several
sub-detector groups in the collaboration.

\wpparagraph{Software repository}
A source code management system is needed to manage the production of software code.
The Subversion~\cite{ref:subversion} open source tool was selected as the
most suitable for \superb\  at this stage of the project.
Due to the multi-developer way of producing
code within a collaboration like \superb\   the repository was the first tool
deployed and integrated with the directory service.

\wpsubsubsec{Code packaging and distribution}

As almost all the computing power available in the collaborating institutes is
made of Linux boxes with Red Hat or Scientific Linux distributions, it was
decided to build releases only for these operating systems, in particular the
builds are available for $RH/SL~4.6$ and for $RH/SL~5.1$ for 32 bit
architectures. Work is ongoing to extend the support also for 64 bit architectures.
Support for MacOSX is also under study, although it might require some time and dedicated effort,
given that this system is quite different from Linux.

\wpparagraph{Tools}

To ease distribution and installation, \superb\ software is packaged with
$RPM$~\cite{ref:rpm} and distributed with $yum$~\cite{ref:yum}, along with
external software used by it, such as $ROOT$ \cite{ref:root}, $Geant4$
\cite{ref:geant4}, $CLHEP$ \cite{ref:clhep}, $CERNLIB$ \cite{ref:cernlib} and
$Xerces\!-\!c$ \cite{ref:xercesc}. The reason to have a specific experiment
version/packaging for these tools is to avoid conflicts with other experiments
using a different release of the same software, or the same release build
with different options.\\
To improve security packages are signed in order to guarantee their origin.
To distribute the software yum repositories have been set up, one per
architecture.

\aftsec

\begsec
\graphicspath{{Integration/}{Integration/}}
\wpsec{Mechanical Integration}

\wpsubsec{Introduction}

The \babar\ detector was built to a design optimized for operations at a high luminosity asymmetric $B$ meson factory.  The detector performed well during the almost nine years of colliding beams operation, at luminosities three times the \pepii accelerator design. There are substantial cost and schedule benefits that  result from reuse of detector components from the \babar\ detector in the \superb\ detector in instances where the component performance has not been significantly compromised during the last decade of use. These benefits arise from two sources: one from having a completed detector component which, though it may require limited performance enhancements, will function well at the \superb\ Factory; and the other from requiring reduced interface engineering, installation planning and tooling manufacture (most of the assembly/disassembly tooling can be reused). The latter reduces risk to the overall success of the mechanical integration of the project. Though reuse is very attractive, risks are also introduced. Can the detector be disassembled, transported, and reassembled without compromise? Will components arrive in time to meet the project needs? 

The reuse of elements of the PID, EMC and IFR system, and associated support structures have been described in previous chapters. In this section, issues related to the magnet coil and cryostat, and the IFR steel and support structure are discussed as well as the integration and assembly of the \superb\ detector, which begins with the disassembly of \babar, and includes shipping components to Italy for reassembly there.

\wpsubsubsec{Magnet and Instrumented Flux Return}

The \babar\ superconducting coil, its cryostat and cryo-interface box, and the helium compressor and liquefier plant will be reused in whole or in part. The magnet coil, cryostat and cryo-interface box will be used in all scenarios. Use of on-detector pumps and similar components may not be cost effective due to electrical incompatibility. An initial review of local refrigeration facilities at the proposed Frascati \superb\ site suggests that there may not be sufficient capacity in that system to cool both the detector magnet and the final focus superconducting magnets. The existing \babar\ helium liquefier plant, which is halfway through its forty year service life, has sufficient capacity. The final decision about reuse of  these external service components will take into account electrical compatibility, schedule and cost. 

The initial \babar\ design contained too little steel for quality $\mu$ identification at high momentum. Additional brass absorber was added during the lifetime of the experiment to compensate  The flux return steel is organized into five structures: the barrel portion, and two sets of two end doors. Each of these is, in turn, composed of multiple structures. The substructures were sized to match the 50 ton load limit of the crane in the \babar\ hall.

Each of the end doors is composed of eighteen steel plates organized into two modules joined together on a thick steel platform. This platform rests on four columns with jacks and Hilman rollers. A counterweight is also located on the platform. There are nine steel layers of 20\mm thickness, four of 30\mm thickness, four of 50\mm thickness, and one of 100\mm thickness. During 2002, five layers of brass absorber were installed in the forward end door slots in order to increase the number of interaction lengths seen by $\mu$ candidate tracks. In the baseline, these doors will be retained, including the five 25\mm layers of brass installed in 2002, as well as the outer steel modules which can house two additional layers of detectors. Additional layers of brass or steel will be added, following the specification of the baseline design in the instrumented flux return section. A cost-benefit analysis will be performed to choose between brass and steel. The aperture of the forward plug must be opened to accommodate the accelerator beam pipe. Compensating modifications to the backward plug are likely to be necessary. These modifications to the steel may affect the central field uniformity and  the centering forces on the solenoid coil, and so must be carefully re-engineered. 

The barrel structure consists of six cradles. These are each composed of eighteen layers of steel. The inner sixteen layers have the same thicknesses as the corresponding end door plates. The two outermost layers are each 100\mm thick. The eighteen layers are organized into two parts; the inner sixteen layers are welded into a single unit along with the two side plates, and the outer two layers are welded together and then bolted into the cradle. The six cradles are in turn suspended from the double I-beam belt that supports the detector. During the 2004/2006 barrel LST upgrade, layers of 22\mm brass were installed, replacing six layers of detectors in the cradles. In the \superb\ baseline, these brass layers will be retained, as well as all the additional flux return steel attached to the barrel in the gap between the end doors and barrel. As in the end doors, additional layers of absorber will be placed in gaps occupied in \babar\ by LSTs. In order to provide more uniform coverage at the largest radius for the $\mu$ identification system, modifications to the sextant steel support mechanism are likely to be needed. Finite element analyses are in progress to confirm that deflections of the steel structure due to this alternative support mechanism do not reduce inter-plate gaps needed for the tracking detectors.

\wpsubsec{Component Extraction}

Extraction of components for reuse requires the disassembly of the \babar\ detector. This process began after the completion of \babar\ operations in April 2008. The first stage of the project was to establish a minimal maintenance state, including stand alone environmental monitoring, that preserves the assets that have reuse value. The transition was complete in August. In order to disassemble the detector into its component systems, it must be moved off the accelerator beam line where it is pinned between the massive supports of accelerator beamline elements into the more open space in the IR2 Hall. This required removal of the concrete radiation shield wall, severing of the cable and services connections to the electronics house which contained the off detector electronics, and roll-back of the electronics house 14\m. Electronics, cables and infrastructure that were located at the periphery of the detector were removed. Beamline elements close to the detector were removed to allow access to the central core of the detector by July 2009, allowing removal of the support tube, which contained the \pepii accelerator final beamline elements as well as the silicon vertex detector, the following month. 

The core disassembly sequence was optimized after the Conceptual Design Report. Completion of disassembly of the detector now requires fewer steps, less time, and poses fewer risks, with the end doors being disassembled while the detector is on beamline. As of mid-May 2010, three of the four end doors have been broken down into component parts, and the EMC forward endcap and the drift chamber have been removed, with the final end door breakdown to be completed within a month. 

This work has been accomplished as a low priority project with a small crew of engineers and technicians. Though the disassembly project is behind schedule, due to laboratory priorities, the earned value compared to actual costs indicates that the level of effort estimated to perform the disassembly is very accurate. The same methodology used to determine the level of effort needed for the \babar\ disassembly has been applied in estimating the needs for \superb\ assembly. 

The components from \babar\ that are expected to be reused in \superb\ are expected to be available for transport to Italy in mid-2011. This should present no challenge to the \superb\ project schedule. The remaining steps in the \babar\ disassembly, which represents more than half of the effort, are outlined below. Tooling that was used in the initial installation of the detector at IR2 is in the process of being refurbished and additional tooling is being fabricated for the disassembly effort. All of this tooling is available for use on \superb. 

After rolling off the beamline, the next phase of detector disassembly consists of extraction of the DIRC and its support system from the core of the detector. The SOB camera is first removed and will not be reused. In order to minimize the possibility of damage to the DIRC bar boxes and their fused silica  bar content, disassembly proceeds with removal of the bar boxes one by one from the bar box support structure inside the DIRC. The twelve bar boxes will be stored in an environmentally controlled container to await shipment to Italy. The DIRC structure is then removed from the barrel. 

The barrel EMC is then removed from the barrel steel. The solenoid is removed from the barrel steel. A temporary structure is assembled inside the barrel hexagon to support the upper cradles during disassembly. The upper half of the support belt is removed. Because of the load limitations of the IR2 crane, the six cradles must be disassembled in situ. The outer sections of the top cradles are removed, followed by the inner part of each of the three cradles. The temporary support structure is removed. The inner part of the lower cradles is removed, followed by the outer portions. The balance of the structural belt is disassembled. 

\wpsubsec{Component Transport}

The magnet steel components will be crated for transport to limit damage to mating surfaces and edges. Most, if not all, of these components can be shipped by sea. 

The \babar\ solenoid was shipped via special air transport from Italy to SLAC. It is expected that this component can be returned to Italy in the same fashion. The original transport frame needs some refurbishment. Drawings exist for parts fabrication so that only a small engineering effort is needed here.

The DIRC and barrel calorimeter present transportation challenges. In both cases transport without full disassembly is preferred. In the case of the DIRC, the central support tube will be separated from the strong support tube and transported as an assembly, assuming that engineering studies indicate that this is possible. The transport cradle for the central support tube has had no design effort yet. The bar boxes have a storage container which provides a dry environment. Whether this can be used for air transport, or a newly designed and fabricated container is required, remains to be determined. Disassembly of the bar boxes exposes the precious bars to damage; it is not considered a viable option. 

In the case of the EMC, there are two environmental constraints on shipment of the device or its components. The glue joint that attached the photodiode readout package to the back face of the crystal has been tested, in mock-up, to be stable against temperature swings of $\pm 5^{\circ} \rm{C}$. During the assembly of the endcap calorimeter, due to a failure of an air conditioning unit, the joints on one module were exposed to double this temperature swing. Several glue joints parted. The introduction of a clean air gap causes a light yield drop of about 25\%. In order to avoid this reduction in performance, temperature swings during transport must be kept small. Since the crystals are mildly hygroscopic, it is best that they be transported in a dry environment to avoid changes in the surface reflectivity, and consequent modification in the longitudinal response of the crystal. Individual \babar\ endcap modules constructed in the UK were successfully shipped to the USA in specially constructed containers that kept the temperature swings and humidity acceptably small. 

Disassembly of the barrel calorimeter for shipment presents a substantial challenge. Both the disassembly and assembly sites need to be temperature and humidity controlled. The disassembly process requires removal of the outer and inner cylindrical covers, removal of cables that connect the crystals with the electronics crates at the ends of the cylinder, splitting of the cylinder into its two component parts and removal of the 280 modules for shipment. Though much of the tooling is still in hand, the environmentally conditioned buildings used in calorimeter construction at SLAC no longer exist, though alternative facilities could be outfitted. The cooling and drying units used in the module storage/calorimeter assembly building continue to be available. 

The clear preference is to ship the barrel calorimeter as a single unit by air. With tooling support stand and environmental conditioning equipment, the load is likely to exceed 30 tons. It is anticipated that such a load could be transported in the same way as the superconducting coil and its cryostat, but verification is needed. Detailed engineering studies, which model accelerations and vibrations involved with flight that might cause the crystal containing carbon fiber modules to strike against one another, are needed to determine if the calorimeter can be safely transported. It may be that a module restraint mechanism will need to be engineered and fabricated. A transport frame must be designed and its performance modeled. Engineering studies have begun.

\wpsubsec{Detector Assembly}

Assembly of the \superb\ detector is the inverse of the disassembly of the \babar\ detector. Ease of assembly will be influenced by the facilities which are available. In the case of \babar, the use of the IR2 hall, which was  ''too small'', led to engineering compromises in the design of the detector. Assembly was made more complicated by the weight restrictions posed by the 50 ton crane. Upgrades were made more difficult because of limitations in movement imposed by the size of the hall.  

The preferred dimensions for the area around the \superb\ detector when it is located in the accelerator housing are 16.2\m transverse to the beamline, 20.0\m along the beamline, 11.0\m from the floor to the bottom of the crane hook, 15.0\m from floor to ceiling, and 3.7\m from the floor to the beamline. The increased floor to beamline height relative to \babar-\pepii, will require redesign effort for the underpinnings of the detector cradle. However, this will permit improved access for installation of cable and piping services for the detector, as well as make possible additional IFR absorption material for improved $\mu$ detector performance below the beamline.  In order to facilitate detector assembly, the preferred capacity for the two hook bridge crane is 100/25 tons.

\aftsec

\begsec
\graphicspath{{Budget/}{Budget/}}
\wpsec{Budget and Schedule}
\label{sec:Budget_and_Schedule}

The \superb\ detector cost and schedule estimates, presented in this
chapter, rely heavily on experience with the \babar\ detector at
\pepii. The reuse and refurbishing of existing components has been
assumed whenever technically possible and financially advantageous.
Though these \superb\ estimates are based on a bottoms-up evaluation
using a detailed work breakdown schedule, it should be emphasized that
the detector design is still incomplete, with numerous technical
decisions remaining to be made, and limited detailed engineering to
date, so that that cost and schedule can not yet be evaluated at the
detailed level expected in a technical design report.

The costing model used here is similar to that already used for the
\superb\ CDR.  The components are estimated in two different general
categories; (1) ``LABOR" and, (2) M\&S (Materials and Services).
M\&S cost estimates are given in 2010 Euros and include 20\% of Value Added Tax (VAT). The
``LABOR" estimates comprise two sub categories which are kept and
costed separately as they have differing cost profiles;(1) EDIA
(Engineering, Design, Inspection, and Administration) and (2) Labor
(general labor and technicians). Estimates in both categories are
presented in manpower work units (Man-Months) and not monetized, as
a monetary conversion can only be attempted after institutional
responsibilities have been identified and the project timescale is known.
The total project cost estimate can be calculated, once the responsibilities
are identified, by summing the monetary value of these three
categories.

Given the long term nature of this multi-national project, there are
several challenging general issues in arriving at appropriate costs including (1)
fluctuating currency exchange rates, and (2) escalation.  M\&S costs
and factory quotes that have been directly obtained in Euros can be
directly quoted. M\&S estimates in US Dollars are translated from
Dollars to Euros using the exchange rate as of Jun 1, 2010 (0.8198
Euros/US\$).  For costs in Euros that were obtained in earlier years,
the yearly escalation is rather small. For simplicity, we use a cost
escalation rate of 2\% per year which is consistent with the long term
HICP (Harmonized Index of Consumer Prices) from the European Central
Bank. Costs given in 2007 Euros are escalated by the net escalation
rate (1.061) for three years to arrive at the 2010 cost estimates
given here~\cite{bib:INFLATEEURO}.

For all items whose cost basis is \babar, we accept the procedure
outlined in the \superb\ CDR which arrived at the costs given there in 2007 Euros. This
procedure escalated the corresponding cost (including manpower) from
the \pepii and \babar\ projects from 1995 to 2007 using the NASA
technical inflation index~\cite{bib:NASA} and then converted from US
Dollars to Euros using the average conversion rate over the 1999---2006
period~\cite{bib:USDEURO}. The overall escalation factor in the CDR
from 1995 Dollars to 2007 Euros is thus $1.21 = 1.295 \times 0.9354$.

Similarly, the replacement values (``Rep.Val.") of the reused
components, \ie, how much money would be required to build them from
scratch, as presented in separate columns of the cost tables, have
been obtained by escalating the corresponding \babar\ project cost (including
manpower) from 1995 to 2007.  Though it is
tempting to sum the two numbers to obtain an estimate of the cost of
the project if it were to be built from scratch, this procedure yields
somewhat misleading results because of the different treatment of the
manpower (rolled up in the replacement value; separated for the new
cost estimate) and because of the double counting when the
refurbishing costs are added to the initial values.

Contingency is not included in the tables. Given the level of detail
of the engineering and the cost estimates, a contingency of about 35\% would be
appropriate.

\wpsubsec{Detector Costs}

The costs, detailed in Table\,\ref{tab:budget}, 
are presented for the detector subsystem at WBS level 3/4.

\onecolumn
{ \setlength{\tabcolsep}{4pt}  
\begin{center}
{\footnotesize
\begin{longtable}{lp{6cm}rrrr}
\caption{\superb\ detector budget. }
\label{tab:budget} \\

\hline\hline
\textbf{ } & \textbf{ }  & \textbf{EDIA} & \textbf{Labor} & \textbf{M\&S} & \textbf{Rep.Val.} \\
\textbf{WBS} & \textbf{Item}  & \textbf{mm} & \textbf{mm} & \textbf{kEuro} & \textbf{kEuro} \\
\hline
\endfirsthead

\multicolumn{6}{c}%
{\tablename\ \thetable{} -- continued from previous page} \\[3mm]
\hline\hline
\textbf{ } & \textbf{ }  & \textbf{EDIA} & \textbf{Labor} & \textbf{M\&S} & \textbf{Rep.Val.} \\
\textbf{WBS} & \textbf{Item}  & \textbf{mm} & \textbf{mm} & \textbf{kEuro} & \textbf{kEuro} \\
\hline
\endhead

 \\ \hline
  \multicolumn{6}{c}{Continued on next page} \\
\endfoot

\endlastfoot

{\bf 1} & {\bf \superb\ detector} & {\bf 4037} & {\bf 2422} & {\bf 52953} & {\bf 48922} \\
\hline
{\bf 1.0} & {\bf Interaction region} & {\bf 21} & {\bf 12} & {\bf 860} & {\bf 0} \\
{\bf 1.0.1} & {\bf Be Beampipe} & {\bf 10} & {\bf 4} & {\bf 260} & {\bf 0} \\
1.0.1.1 & Vertex chamber design & 4     & 0     & 0     & {\bf 0} \\
1.0.1.2 & Finalize Physics Req.mnts & 1     & 0     & 0     & {\bf 0} \\
1.0.1.3 & Fab method & 1     & 0     & 0     & {\bf 0} \\
1.0.1.4 & Design Review & 1     & 0     & 0     & 0 \\
1.0.1.5 & Chamber detailing & 2     & 0     & 0     & 0 \\
1.0.1.6 & Support procurement & 2     & 0     & 5     & 0 \\
1.0.1.7 & Procure Beampipe Assembly & 0     & 0     & 243   & 0 \\
1.0.1.8 & Procure Vx chamber Misc parts & 0     & 0     & 12    & 0 \\
1.0.1.9 & Assemble Vx chamber, test, clean & 0     & 2     & 0     & 0 \\
{\bf 1.0.2} & {\bf Tungsten Shield} & {\bf 9} & {\bf 6} & {\bf 540} & {\bf 0} \\
1.0.2.1 & Shield optimization & 3     & 0     & 0     & 0 \\
1.0.2.2 & Shield detailing and integration & 3     & 0     & 0     & 0 \\
1.0.2.3 & Shield procurement & 1     & 0     & 540   & 0 \\
1.0.2.4 & Shield assembly and installation & 2     & 6     & 0     & 0 \\
{\bf 1.0.3} & {\bf Radiation monitors} & {\bf 2} & {\bf 2} & {\bf 60} & {\bf 0} \\
\hline
{\bf 1.1} & {\bf Tracker (SVT + Strip + MAPS)} & {\bf 408} & {\bf 442} & {\bf 6444} & {\bf 0} \\
{\bf 1.1.1} & {\bf SVT} & {\bf 222} & {\bf 309} & {\bf 4326} & {\bf 0} \\
1.1.1.1 & Mechanical & 48    & 129   & 399   & 0 \\
1.1.1.2 & Cooling & 8     & 10    & 155   & 0 \\
1.1.1.3 & Silicon Wafers and Fanout & 24    & 120   & 2642  & 0 \\
1.1.1.4 & On-detector electronics & 72    & 42    & 1013  & 0 \\
1.1.1.5 & Detector monitoring & 4     & 4     & 92    & 0 \\
1.1.1.6 & Detector assembly & 6     & 4     & 0     & 0 \\
1.1.1.7 & System Engineering & 60    & 0     & 24    & 0 \\
{\bf 1.1.2} & {\bf L0 Striplet option} & {\bf 36} & {\bf 55} & {\bf 542} & {\bf 0} \\
1.1.2.1 & Mechanical & 12    & 30    & 60    & 0 \\
1.1.2.2 & Cooling  & 3     & 3     & 48    & 0 \\
1.1.2.3 & Silicon Wafers and Fanout & 16    & 18    & 327   & 0 \\
1.1.2.4 & On-detector electronics & 5     & 4     & 108   & 0 \\
{\bf 1.1.3} & {\bf L0 MAPS option} & {\bf 150} & {\bf 78} & {\bf 1576} & {\bf 0} \\
1.1.3.1 & Mechanical & 18    & 48    & 90    & 0 \\
1.1.3.2 & Cooling & 6     & 6     & 96    & 0 \\
1.1.3.3 & MAPS Modules Components & 126   & 24    & 1390  & 0 \\
{\it {\bf 1.1.4}} & {\it {\bf L0 Hybrid Pixel option}} & {\it {\bf 156}} & {\it {\bf 84}} & {\it {\bf 1684}} & {\it {\bf 0}} \\
{\it 1.1.4.1} & {\it Mechanical} & {\it 18} & {\it 48} & {\it 90} & {\it 0} \\
{\it 1.1.4.2} & {\it Cooling} & {\it 6} & {\it 6} & {\it 96} & {\it 0} \\
{\it 1.1.4.3} & {\it Hybrid Pixel Modules Components} & {\it 132} & {\it 30} & {\it 1498} & {\it 0} \\
\hline
{\bf 1.2} & {\bf DCH} & {\bf 165} & {\bf 139} & {\bf 3421} & {\bf 0} \\
1.2.1 & System engineering & 24    & 0     & 60    & 0 \\
1.2.2 & Endplates & 16    & 6     & 660   & 0 \\
1.2.3 & Inner cylinder & 8     & 2     & 200   & 0 \\
1.2.4 & Outer cylinder & 6     & 2     & 120   & 0 \\
1.2.5 & Wire  & 4     & 6     & 308   & 0 \\
1.2.6 & Feedthroughs & 9     & 10    & 439   & 0 \\
1.2.7 & Endplate systems & 8     & 0     & 385   & 0 \\
1.2.8 & Assembly \& Stringing & 74    & 96    & 960   & 0 \\
1.2.9 & Gas System & 10    & 8     & 240   & 0 \\
1.2.A & Test  & 6     & 9     & 48    & 0 \\
\hline
\newpage
{\bf 1.3} & {\bf PID} & {\bf 116} & {\bf 236} & {\bf 5820} & {\bf 7138} \\
{\bf 1.3.1} & {\bf DIRC Barrel (Focusing DIRC)} & {\bf 116} & {\bf 236} & {\bf 5820} & {\bf 7138} \\
1.3.1.1 & Radiator Support Structure & 4     & 4     & 10    & 2516 \\
1.3.1.2 & Radiator box/FBLOCK assembly & 14    & 40    & 2819  & 4515 \\
1.3.1.3 & New Camera mechanical boxes & 14    & 28    & 305   & 0 \\
1.3.1.4 & Photodetector assembly & 18    & 32    & 2607  & 0 \\
1.3.1.5 & Calibration System & 2     & 4     & 59    & 0 \\
1.3.1.6 & Mechanical Utilities & 4     & 8     & 20    & 107 \\
1.3.1.7 & System Integration & 60    & 120   & 0     & 0 \\
\hline
{\bf 1.4} & {\bf EMC} & {\bf 219} & {\bf 360} & {\bf 12147} & {\bf 31574} \\
{\bf 1.4.1} & {\bf Barrel EMC} & {\bf 20} & {\bf 5} & {\bf 205} & {\bf 31574} \\
1.4.1.1 & Crystal Procurement & 0     & 0     & 0     & 21742 \\
1.4.1.2 & Light Sensors \& Readout & 0     & 0     & 0     & 2654 \\
1.4.1.3 & Crystal Support Modules & 0     & 0     & 0     & 2875 \\
1.4.1.4 & Barrel Structure & 0     & 0     & 0     & 3419 \\
1.4.1.5 & Calibration Systems & 0     & 0     & 0     & 650 \\
1.4.1.6 & Project Management & 0     & 0     & 0     & 233 \\
1.4.1.7 & Barrel Transport & 20    & 5     & 205   & 0 \\
{\bf 1.4.2} & {\bf Forward EMC} & 171   & 312   & 11565 & 0 \\
1.4.2.1 & Crystal Procurement & 25    & 102   & 9403  & 0 \\
1.4.2.2 & Light Sensors \textbackslash \& Readout & 47    & 70    & 992   & 0 \\
1.4.2.3 & Crystal Support Modules & 26    & 64    & 450   & 0 \\
1.4.2.4 & Endcap Structure & 26    & 52    & 444   & 0 \\
1.4.2.5 & Calibration Systems & 24    & 24    & 156   & 0 \\
1.4.2.6 & Project Management & 24    & 0     & 120   & 0 \\
{\bf 1.4.3} & {\bf Backward EMC} & 28    & 43    & 377   & 0 \\
1.4.3.1 & Scintillator & 2     & 10    & 121   & 0 \\
1.4.3.2 & Radiator & 1     & 4     & 22    & 0 \\
1.4.3.3 & Fibers & 4     & 8     & 18    & 0 \\
1.4.3.4 & Photodetectors & 2     & 5     & 46    & 0 \\
1.4.3.5 & Mechanical support & 17    & 15    & 146   & 0 \\
1.4.3.6 & Project Management & 2     & 2     & 24    & 0 \\
\hline
{\bf 1.5} & {\bf IFR} & {\bf 37} & {\bf 184} & {\bf 1374} & {\bf 0} \\
1.5.1 & Scintillators & 0     & 0     & 266   & 0 \\
1.5.2 & WLS fibers & 0     & 0     & 362   & 0 \\
1.5.3 & Photodetectors and PCBs & 1     & 2     & 685   & 0 \\
1.5.4 & Mechanics (Production and QC) & 16    & 62    & 60    & 0 \\
1.5.5 & Module Installation & 20    & 120   & 0     & 0 \\
\hline
{\bf 1.6} & {\bf Magnet} & {\bf 93} & {\bf 59} & {\bf 3767} & {\bf 10210} \\
1.6.0 & System Management & 36    & 0     & 0     & 612 \\
1.6.1 & Superconducting solenoid & 0     & 0     & 0     & 2421 \\
1.6.2 & Mag. Power/Protection & 0     & 0     & 0     & 181 \\
1.6.3 & Cryogenics & 34    & 36    & 1753  & 0 \\
1.6.4 & Cryo monitor/Control & 17    & 11    & 214   & 0 \\
1.6.5 & Flux return & 6     & 12    & 1800  & 6481 \\
1.6.6 & Installation/test equipment & 0     & 0     & 0     & 515 \\
\hline
{\bf 1.7} & {\bf Electronics} & {\bf 994} & {\bf 342} & {\bf 9234} & {\bf 0} \\
1.7.1 & SVT   & 11    & 21    & 561   & 0 \\
1.7.2 & DCH   & 74    & 76    & 1668  & 0 \\
1.7.3 & PID Barrel (32k channels) & 136   & 18    & 612   & 0 \\
1.7.4 & EMC   & 110   & 164   & 2726  & 0 \\
1.7.5 & IFR   & 38    & 51    & 1487  & 0 \\
1.7.6 & Infrastructure & 4     & 12    & 314   & 0 \\
1.7.7 & Systems Engineering & 12    & 0     & 0     & 0 \\
1.7.8 & Hardware Trigger & 97    & 0     & 678   & 0 \\
1.7.9 & ETD (without Trigger) & 512   & 0     & 1188  & 0 \\
\hline
{\bf 1.8} & {\bf Online System} & {\bf 912} & {\bf 24} & {\bf 2074} & {\bf 0} \\
1.8.1 & Event Flow & 282   & 0     & 1676  & 0 \\
1.8.2 & Run Control / Slow Controls / ECS & 270   & 0     & 53    & 0 \\
1.8.3 & Infrastructure & 48    & 12    & 246   & 0 \\
1.8.4 & Software Triggers & 216   & 0     & 0     & 0 \\
1.8.5 & Coordination and Commissioning & 72    & 12    & 0     & 0 \\
1.8.6 & Online System R\&D & 24    & 0     & 98    & 0 \\
\hline
{\bf 1.9} & {\bf Installation and integration} & {\bf 353} & {\bf 624} & {\bf 7596} & {\bf 0} \\
1.9.1 & Disassembly & 95    & 161   & 612   & 0 \\
1.9.2 & Assembly & 222   & 463   & 3984  & 0 \\
1.9.3 & Structural analysis & 36    & 0     & 0     & 0 \\
1.9.4 & Transportation & 0     & 0     & 3000  & 0 \\
\hline
{\bf 1.A} & {\bf Project Management} & {\bf 720} & {\bf 0} & {\bf 216} & {\bf 0} \\
1.A.1 & Project engineering & 300   & 0     & 120   & 0 \\
1.A.2 & Budget, Schedule and Procurement & 300   & 0     & 48    & 0 \\
1.A.3 & ES \& H & 120   & 0     & 48    & 0 \\

\hline

\hline
\end{longtable}
}
\end{center}
}

\twocolumn

The
\superb\ detector is not completely defined: some components, such as
the forward PID, have overall integration and performance implications
that need to be carefully studied before deciding to install them; for
some other components, such as the SVT Layer0, promising new
technologies require additional R\&D before they can be definitively
used in a full scale detector. The cost estimates list the different
technologies separately, but the rolled-up value includes the baseline
detector choice that is considered most likely to be used.
Technologies that are not included in the rolled-up value are shown in
italics.

\wpsubsec{Basis of Estimate}

\wpparagraph{Vertex Detector and Tracker:} System cost is estimated
based on experience with the \babar\ detector and vendor quotes. A detailed estimate
is provided for the cost of the main detector (layers 1 to 5). The costs associated with
Layer0 are analyzed separately. The costing model assumes that a striplet detector will be installed initially -- followed by a second generation upgrade to a pixel detector, which could either be either a  MAPs or hybrid pixel device.  Substantial R\&D on these new technologies is needed in either case before such a detector can be built.
The total SVT cost is obtained by summing the baseline detector cost (with striplets for Layer0)  to
the MAPS Layer0 cost.

\wpparagraph{Drift Chamber:} The DCH costing model is based on a
straightforward extrapolation of the actual costs of  the existing \babar\ chamber to 2010,
since, as discussed in Sec.~\ref{sec:DCHmain} the main design elements
are comparable, and  many related components, such as the
length of wire, number of feedthroughs, duration of wire stringing,
etc., can be reliably estimated. Although the cell layout is still being
finalized, the total cell count will likely be about 25\% larger The endplates will be fabricated from
carbon fiber composites instead of aluminum. Though this  will require a somewhat
longer period of R\&D and engineering design, it is unlikely to
result in significantly larger production costs for the final
endplates. The DCH electronics on Table\,\ref{tab:budget}
assumes standard readout as discussed in Sec.\,\ref{sec:ETD_DCH}. A  \textit{cluster counting} readout option is under R\&D, but is not yet
sufficiently advanced that costs can be provided.

\wpparagraph{Particle Identification:} Barrel PID costs and replacement
values are derived from \babar\ costs as extrapolated to 2010, with updated quotes from vendors. The
main new component of the barrel FDIRC  is its new camera. For each module, the optical portion
consists of the focusing block (FBLOCK), an addition to the wedge (the  New Wedge) and possibly a Micro-Wedge. We have
contacted about twelve optics companies and received four preliminary bids.
We use the average bid in the present budget. We hope to continue to refine the values
through further R\&D. The photon detector cost estimate is based on the Hamamatsu
bid for 600 H-8500 MaPMTs. No
budget estimate is included at this time for a forward endcap PID. Though several options are
being studied, their performance and cost are yet to be well understood, and the overall  performance gains and losses of including a forward PID in the detector are as yet unsettled. However, as a general principle, given the limited solid angle covered by such a device, the cost of a forward PID detector must be a modest fraction of the  barrel.

\wpparagraph{Electromagnetic Calorimeter:} There are four components to
the calorimeter cost: (1) the barrel calorimeter from \babar;
(2) the forward calorimeter; (3)  the replacement of
the front-end preamps in the barrel; and (4)  the backward
calorimeter. As described in the calorimeter section, there are a
number of uncertainties remaining in the design. The present 
cost estimate is for our baseline design.

The reuse value of the barrel calorimeter is based on the actual cost
of the barrel escalated for inflation from the time of construction to
the current year. Manpower estimates for the barrel construction were
obtained by using the costs for EDIA and Labor, knowledge of the mix
of engineers and technicians who contributed to the design and
fabrication of individual components, and knowledge of their salaries.
Manpower and costs for engineering and tooling required for the
removal and transport of the barrel EMC from SLAC are engineering
estimates.

The main cost driver for the forward endcap is the cost of LYSO
crystals. This is estimated based on guideline quotes from vendors.
The next largest element is the APD photodetectors, with a cost based
on a quote from the vendor. The estimate for the crystal support
modules is based on costs for the beam test prototype.  Estimates for
the remaining smaller items are based on estimator experience and
judgment.

The cost estimate for replacing the preamplifiers in the barrel
calorimeter is based on the endcap preamplifier cost as well as the
cost of dealing with the mechanical issues.

For the backward endcap, the scintillator, lead, wavelength-shifting
fiber, and readout MPPC costs, as well
as some other minor materials, are all based on vendor quotes. Other items are based on
experience and estimator judgment.

\wpparagraph{Instrumented Flux Return:} The IFR cost is based on
quotations received for the prototype construction appropriately
scaled to the real detector dimensions. While the active part of the
detector is quite inexpensive the total cost is driven by the
electronics and the photodetectors. The current baseline design allows
the reuse of the \babar\ iron structure with some modification that
needs to be taken into account.  Manpower and cost for engineering and
module installation is based on the \babar\ experience.

\wpparagraph{Electronics, Trigger, DAQ and Online:} The cost for the
Electronics and Trigger subsystems is estimated with a combination
of scaling from the \babar\ experience and from direct estimates.  For
items expected to be similar to those used in \babar\ (such as
infrastructure, high and low voltage or the L1 trigger) costs are scaled from \babar. The same methodology is
used to estimate EDIA and Labor costs for the Online system.
However, some modifications based on ``lessons learned'' are applied.
In particular, we are including costs for development work that, in our opinion,
should have been centralized across sub-detectors in \babar\ (but
wasn't) and work that should have been done upfront but was only done
or completed as part of \babar\ Online system upgrades.

The readout systems for which the higher data rates require redesigned
electronics are estimated from the number of different components and
printed circuit boards, and their associated chip and board counts.
This methodology is also used for the possible new detectors (forward EMC, backward EMC,
forward PID) and for the elements of the overall system architecture
that are very different from \babar.

The hardware cost estimates for the Online computing system (including
the HLT farm) are, very conservatively,  based on the current prices
of hardware necessary to build the system, with the assumption that
Moore's Law will result in future systems with the same unit costs but
higher performance. This is justified by our observation that for COTS
components, constraints from system design, topology and networking
are more likely to set minimum requirements for the \emph{number} of
devices than for the per-device performance.

\wpparagraph{Transportation, installation, and commissioning: }
Installation and commissioning estimates, including disassembling and
reassembling \babar, are based on the \babar\ experience, and engineering estimates use a
detailed schedule of activities and corresponding manpower requirements. The transportation costs have been
estimated from costs associated with disassembling  and transporting \babar\ components for dispersal, if
they were not to be reused.

\wpsubsec{Schedule}

\begin{figure*}[htb]
\vspace*{-7mm}
\centering
\includegraphics[height=1.0\textwidth, angle=90]{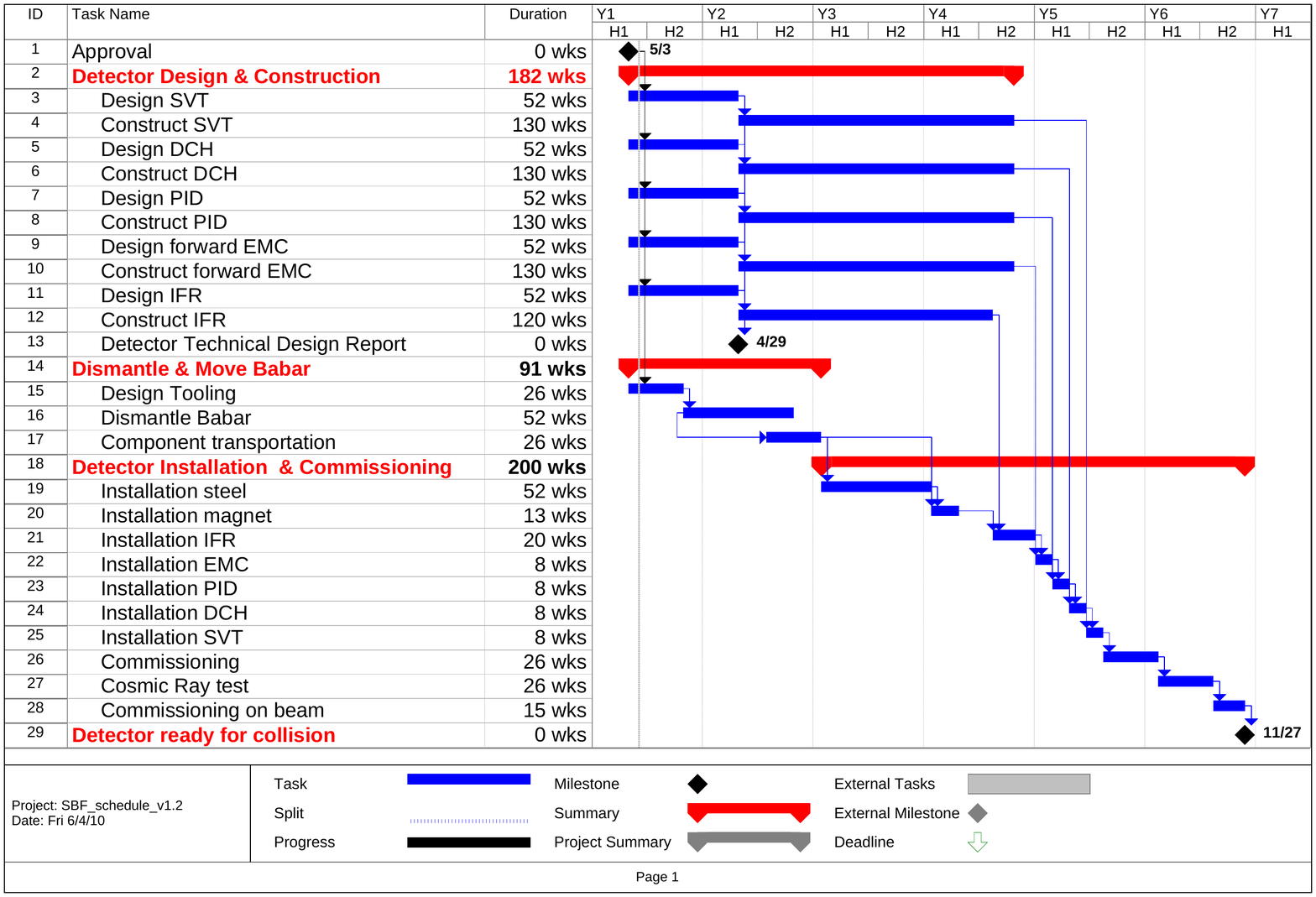}
\caption{Schedule for the construction of the \superb\ detector.}
\label{fig:schedule}
\end{figure*}

The detector construction schedule is shown in Fig.~\ref{fig:schedule}.
The construction starts with design finalization and a technical design report,
after which the fabrication of the detector subsystems can proceed in parallel.
At the same time  the \babar\ detector is disassembled, transported to the new site, and
reassembled. The detector subsystems will be installed in sequence. An extended detector
commissioning period, including a cosmic ray run, will follow to ensure proper operation
and calibration of the detector. The total construction and commissioning time is
estimated to be a little over five years.

\aftsec


\begin{thebibliography}{99}
\bibitem{bib:Fluorinert} Fluorinert is the trademark name for
polychlorotrifluoro-ethylene), manufactured by 3M Corporation,
St. Paul, MN, USA.
\end{thebibliography}
\end{document}